\documentclass[final,1p,times]{elsarticle}

\usepackage{amssymb}
\usepackage{amsmath}
\usepackage{hyperref}
\usepackage{braket, bm}
\usepackage[dvipsnames]{xcolor}
\usepackage{siunitx}
\usepackage{tabularx}
\usepackage{dutchcal}
\usepackage{multirow}
\usepackage{booktabs}
\usepackage[version=4]{mhchem}
\DeclareSIUnit\angstrom{\text {Å}}
\usepackage{threeparttable}
\usepackage{upgreek}
\usepackage{rotating}

\journal{Physics Reports}

\biboptions{sort&compress}
\begin{document}

\begin{frontmatter}

\title{Novel qubits in hybrid semiconductor-superconductor nanostructures}

\author[label1]{Marta Pita-Vidal\fnref{label6}}
\affiliation[label1]{organization={QuTech and Kavli Institute of Nanoscience, Delft University of Technology},
            addressline={2600 GA}, 
            state={Delft},
            country={The Netherlands}}
\fntext[label6]{{\it Current Address}: IBM Research Europe, 8803 Rüschlikon, Zurich, Switzerland}

\author[label2,label3]{Rubén Seoane Souto}
\affiliation[label2]{organization={Quantum Advanced Research Center (QuARC), Consejo Superior de Investigaciones Científicas (CSIC)},
            addressline={Sor Juana Inés de la Cruz 3}, 
            postcode={28049}, 
            state={Madrid},
            country={Spain}}
\affiliation[label3]{organization={Instituto de Ciencia de Materiales de Madrid (ICMM), Consejo Superior de Investigaciones Científicas (CSIC)},
            addressline={Sor Juana Inés de la Cruz 3}, 
            postcode={28049}, 
            state={Madrid},
            country={Spain}}

\author[label1]{Srijit Goswami}

\author[label1]{Christian Kraglund Andersen}
 
\author[label4]{Georgios Katsaros}
\affiliation[label4]{organization={Institute of Science and Technology Austria},
            addressline={Am Campus 1, 3400}, 
            city={Klosterneuburg},
            country={Austria}}
            
\author[label5]{Javad Shabani}
\affiliation[label5]{organization={Center for Quantum Information Physics, Department of Physics,
New York University},
            addressline={New York 10003}, 
            country={USA}}       
            
\author[label2,label3]{Ramón Aguado\corref{cor1}}
\ead{ramon.aguado@csic.es}

\begin{abstract}
In recent years, hybrid semiconductor-superconductor qubits have emerged as a compelling alternative to more traditional qubit platforms, owing to their unique combination of material properties and device-level tunability. A key feature of these systems is their gate-tunable Josephson coupling, which enables the realization of superconducting qubit architectures with full electric field control and which offers a promising route toward scalable and low-crosstalk quantum processors. One of the central motivations behind this approach is the possibility of unifying the advantages of superconducting and spin-based qubits. This approach to creating hybrid qubits can be achieved, for example, by encoding quantum information in the spin degree of freedom of a quasiparticle occupying an Andreev bound state, thereby combining long coherence times with fast and flexible control. Recent progress has been further accelerated by the bottom-up engineering of Andreev bound states into coupled quantum dot arrays, leading to the realization of novel device architectures such as minimal Kitaev chains hosting (poor man's) Majorana states, precursors of topologically protected Majorana zero modes, and, very recently, to the first coherent control of a Majorana parity qubit. In parallel, alternative strategies have focused on the design of Hamiltonian-protected hybrid qubits, including Cooper pair parity qubits and fluxon parity qubits, which aim to enhance robustness against local noise and decoherence. These approaches leverage the interplay between superconducting phase dynamics and the discrete charge or flux degrees of freedom in hybrid junctions. In this article, we review recent theoretical and experimental advances in the development of semiconductor-superconductor hybrid qubits. We provide a comprehensive overview of the underlying physical mechanisms, device implementations, and emerging architectures, with particular emphasis on their potential to enable (topologically) protected quantum information processing. While many of these qubit designs remain at the proof-of-concept stage, the rapid pace of progress suggests that practical demonstrations may be within reach in the near future.
\end{abstract}


\end{frontmatter}

\tableofcontents

\section{Introduction}
Over the past three decades, the field of quantum information science has experienced remarkable advancements, particularly in the development and control of qubits across a variety of physical platforms. These include trapped ions \cite{decross2024}, Rydberg atoms \cite{Bluvstein2024}, semiconducting devices \cite{Burkard2023}, and superconducting circuits \cite{Krantz2019}, with the latter two emerging as the leading candidates within the solid-state domain. Each platform offers unique advantages and faces distinct challenges, shaping the landscape of quantum computing research.

Within the realm of semiconducting qubits, two primary approaches have gained prominence: encoding quantum information in the spin degree of freedom of donor nuclear spins or in the spin of charge carriers confined in quantum dots \cite{Loss1998}. These systems have garnered significant interest due to their inherently long coherence times, large energy level separations that enable rapid gate operations, compact physical footprint, and compatibility with well-established semiconductor fabrication technologies~\cite{Zwerver2022}. However, the very properties that make spin qubits appealing, namely their small size and magnetic nature, also pose significant limitations. In particular, the weak coupling of spin states to electric fields complicates the implementation of long-range interactions and high-fidelity rapid readout schemes, both of which are essential for fault-tolerant quantum computation.

In contrast, superconducting qubits, such as the widely studied transmon qubit \cite{Koch2007}, encode quantum information in macroscopic voltage modes of superconducting circuits. This electric character facilitates strong qubit-qubit coupling and enables efficient readout via circuit quantum electrodynamics (cQED) techniques \cite{Blais2007,Blais2021}. These advantages have positioned superconducting qubits at the forefront of experimental quantum computing. Nevertheless, transmons are not without drawbacks. Their relatively small anharmonicity imposes constraints on gate speeds and introduces the risk of leakage into non-computational states, which can degrade quantum gate fidelity. We note that this limitation is specific to transmon-type superconducting qubits, and not to superconducting qubits in general: other designs, such as the fluxonium qubit~\cite{Manucharyan2009,Nguyen2019} discussed in Sec.~\ref{s:superconducting}, exhibit large anharmonicities, albeit with different trade-offs, such as lower qubit frequencies and more complex circuits.
\begin{figure}
    \includegraphics[width=\linewidth]{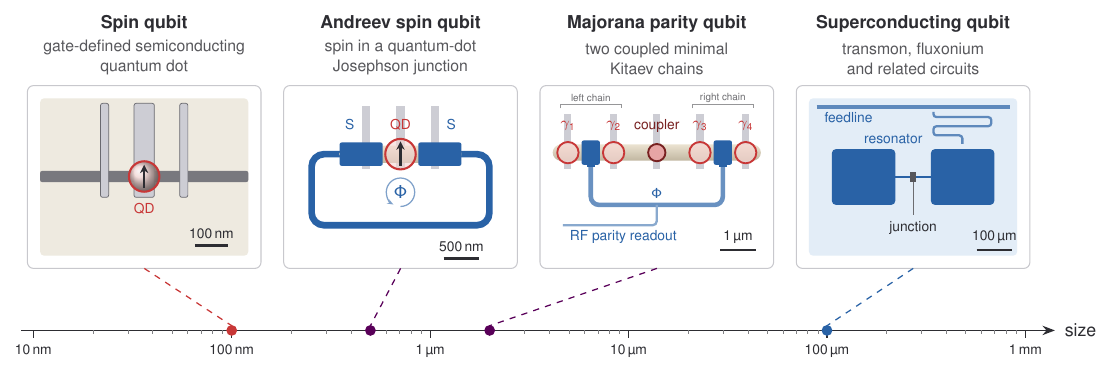}
    \caption{Schematics of solid-state qubit platforms and their typical physical sizes, indicated on a logarithmic scale. From left to right: a semiconductor spin qubit, where the spin of a single charge carrier is confined in a gate-defined quantum dot (QD) with a typical size of $\sim100$~nm; an Andreev spin qubit, where the spin is encoded in a QD Josephson junction embedded in a superconducting loop threaded by a flux $\Phi$ (junction size $\sim500$~nm); a Majorana parity qubit, where the quantum information is encoded in the fermion parity of four Majorana states, $\gamma_1$, $\gamma_2$, $\gamma_3$ and $\gamma_4$, hosted by two minimal Kitaev chains coupled through a QD (device size of a few $\mu$m); and a superconducting qubit, here a transmon coupled to a readout resonator, with sizes exceeding $100~\mu$m. Semiconductor elements (QDs) are shown in red and superconducting elements in blue. The hybrid semiconductor-superconductor qubits discussed in this review combine advantages of the two platforms, such as the small footprint of semiconductor devices and the cQED readout techniques of superconducting qubits.}
    \label{fig:qubit-size}
\end{figure}

Given the complementary strengths, different footprint (see Fig \ref{fig:qubit-size}) and weaknesses of semiconducting and superconducting qubits, a promising direction is to explore hybrid material systems that integrate both technologies within a single device. Such hybrid architectures aim to harness the best of both worlds, potentially overcoming the limitations inherent to each platform when used in isolation. As we will discuss in this review, the interplay between superconductivity and semiconducting properties gives rise to novel physical phenomena, including the emergence of topological superconductivity under appropriate conditions. The presence of a large topological gap could lead to quantum states that are inherently robust against local sources of noise and, thus, open the door to the realization of protected qubits.

From a practical standpoint, hybrid superconducting-semiconducting devices also offer new avenues for engineering qubits with enhanced functionality. For example, incorporating semiconducting Josephson junctions into superconducting qubit circuits introduces the possibility of tuning qubit frequencies via gate voltages, offering a more stable and less interference-prone alternative to magnetic flux control. Furthermore, the use of highly transparent semiconducting channels in Josephson junctions, especially when combined with strong spin-orbit coupling, enables the design of junctions with unconventional energy-phase relations. These hybrid junctions can be exploited to realize Hamiltonian-protected qubits or to encode quantum information in the spin and charge degrees of freedom of the junction itself. Two examples of such hybrid qubits are sketched in Fig.~\ref{fig:qubit-size}: the Andreev spin qubit, which encodes quantum information in the spin of a quasiparticle trapped in a QD Josephson junction, and the Majorana parity qubit, which encodes it in the fermion parity of Majorana states engineered in coupled minimal Kitaev chains. 

While numerous comprehensive reviews on superconducting qubits are available in the literature, see, for instance, Refs. \cite{Makhlin2001,Girvin2016,Krantz2019,Kjaergaard2020}, we begin this work by introducing the foundational concepts and essential tools associated with Josephson-based superconducting circuits (Sec.~\ref{s:superconducting}). This initial overview is intended to provide the necessary background and establish a common framework for readers, particularly those less familiar with the technical underpinnings of superconducting qubit architectures.
To contextualize the developments discussed in this Review, we also provide in Sec.~\ref{s:nanostructures} a brief historical overview and a discussion of the materials science advances that have enabled the rapid progress in hybrid superconducting-semiconducting technologies.

Following this, we dedicate a section to the physics of subgap states in superconducting junctions (Sec.~\ref{s:subgap}). These states, which arise within the superconducting energy gap due to the presence of weak links or impurities, play a central role in the operation and design of hybrid superconducting-semiconducting devices. Understanding their origin, behavior, and tunability is crucial for appreciating the novel phenomena and functionalities that such hybrid systems can exhibit.

Sec.~\ref{s:qubits} explores the emerging class of hybrid qubits, systems which synergistically combine the mature and highly controllable architecture of superconducting qubits with the rich and tunable physics of hybrid nanostructures. This section focuses on representative implementations of hybrid qubits, such as gatemons and Andreev qubits, emphasizing how their design strategies enable enhanced performance characteristics and unlock novel operational regimes. These architectures introduce new degrees of freedom and control modalities, offering fertile ground for both fundamental investigations and the development of practical quantum computing platforms. Furthermore, in Sec.~\ref{ss:germanium}, we discuss recent theoretical and experimental advances towards realizing hybrid qubits in hole-based systems.

Sec.~\ref{s:bottom-up} is a forward-looking section devoted to the bottom-up scaling of topological superconductivity, with a particular emphasis on minimal Kitaev chains as a foundational platform. These engineered nanostructures provide a highly controllable setting for realizing and manipulating Majorana states which, in the limit of long chains, evolve into topologically protected Majorana zero modes (MZMs)—non-Abelian quasiparticles that hold great promise for fault-tolerant quantum computation. While topologically protected Majorana qubits have yet to be demonstrated, the first coherent control of a Majorana parity qubit, based on two coupled minimal Kitaev chains, has been recently reported~\cite{Zatelli_arXiv2026}. This section surveys the state-of-the-art in both theory and experiment, highlighting the potential of these systems to implement non-Abelian braiding operations and serve as building blocks for topological quantum computing. As such, they represent a compelling direction for future research at the intersection of condensed matter physics, materials science, and quantum information. With this section, which focuses primarily on Kitaev minimal chains based on a bottom-up approach, we aim to fill a gap and present an overview of the current state of the art, as opposed to the top-down approaches that have received much more attention in numerous specialized review articles, see~\cite{Alicea_RPP2012,Leijnse_Review2012,beenakker2013search,Aguado_Nuovo2017,Lutchyn_NatRev2018,Prada_review,MarraReview_JAP}.

{The Review is written from a device-physics perspective, with the aim of explaining the physical mechanisms that define, protect, and limit each type of hybrid qubit. Secs.~\ref{s:qubits} and \ref{ss:germanium} cover the broad family of hybrid qubits, from gate-tunable transmons and fluxoniums to Hamiltonian-protected qubits and Andreev level and Andreev spin qubits. Sec.~\ref{s:bottom-up} is devoted to Majorana states and Majorana-based qubits, with particular emphasis on the bottom-up engineering of Majorana states in quantum-dot-based minimal Kitaev chains, which brings together much of the physics discussed in the previous sections: Andreev bound states, crossed Andreev reflection, and cQED and radio-frequency readout. Finally, the Outlook (Sec.~\ref{s:outlook}) provides a comparative overview of the different hybrid qubit platforms, summarizes the main open challenges in the field, and discusses system-level aspects, such as the integration of hybrid qubits into larger quantum processors.}

Taken together, the sections of this review aim to provide a comprehensive and structured overview of the rapidly evolving landscape of hybrid superconducting-semiconducting qubits. By tracing the development from foundational superconducting circuits to the frontier of topological quantum devices, we highlight the rich interplay between materials science, quantum engineering, and condensed matter physics that underpins this field. As quantum technologies continue to mature, hybrid platforms are poised to play a central role—not only by enhancing existing qubit designs, but also by enabling fundamentally new paradigms for quantum information processing. Through this review, we hope to offer both a clear snapshot of the current state of the art and a roadmap for future exploration in the pursuit of scalable, robust, and topologically protected quantum computing.

\section{Superconducting circuits and qubits}\label{s:superconducting}
\subsection{Finite-size superconducting islands}
\label{Bogoliubov QP}
Let us consider a mesoscopic superconducting island with $N$ discrete levels in a volume ${\cal V}$ described by the Hamiltonian
\begin{equation}
\label{H-mesoscopicSC}
{\cal H}=\sum_{j=1}^N\sum_{\sigma=\uparrow,\downarrow}\xi_{j}c^\dagger_{j\sigma}c_{j\sigma}+\sum_{ijkl}{\cal H}_{ijkl}c^\dagger_{i\uparrow}c^\dagger_{j\downarrow}c_{k\downarrow}c_{l\uparrow},
\end{equation}
where the operators $c_{j\sigma}^\dagger$ and $c_{j\sigma}$ create and annihilate electrons with spin $\sigma$ and energies $\xi_{j}$ (measured from the Fermi level). The second term describes electron-electron interactions through the matrix elements ${\cal H}_{ijkl}$ of the Coulomb interaction written in terms of the single-particle eigenfunctions. 

Since solving the whole problem described by Eq. (\ref{H-mesoscopicSC}) seems an impossible task, we have to make some approximations and assumptions based on physical intuition to make progress. The first step is to assume that electron-electron interactions are governed by a contact term of the form $V({\mathbf r})=(V_0/ \rho_0)\delta({\mathbf r})$, with $V_0$ being a dimensionless interaction constant, $\rho_0$ 
the density of states at the Fermi level and $\delta({\mathbf r})$ the Dirac $\delta$-function. Moreover, if $V({\mathbf r})$ contains a short-range attractive component, the three leading components of ${\cal H}_{ijkl}$ are, respectively, a Hartree term, a Fock term, and a Bardeen-Cooper-Schrieffer (BCS) term. If we also neglect the Fock term (which is equivalent to neglecting the exchange interactions, which is a good approximation in the case of non-magnetic materials), the electron-electron interaction part of the Hamiltonian takes a universal form \cite{Glazman-Catelani2021}, see also \cite{vonDelft96} and \cite{Larkin97}:
\begin{equation}
\label{H-int}
    {\cal H}_{\rm int}=E_C (\hat{N}^e-N_g)^2+\left(\frac{V_0}{{\cal V}\rho_0}\right) \hat{O}^\dagger\hat{O}\,.
\end{equation}
The first term, which originates essentially from the Hartree contribution of the Coulomb electron-electron interaction, describes the electrostatic cost of adding an additional electron to the island with charging energy $E_C$. In this term, ${\hat N}^e=\sum_{j,\sigma}c^\dagger_{j\sigma}c_{j\sigma}$ is the electron number operator in the island, while $N_g$ is a polarization charge (that can be tuned by an external gate). In the second term, the operators ${\hat O}=\sum_{|\xi_j|<\hbar\omega_D}c_{j\downarrow}c_{j\uparrow}$ already assume the short-range attractive interaction leading to superconductivity which only acts between electrons with energies within some range $|\xi_j|<\hbar\omega_D$  (with $\omega_D$ simply being the Debye frequency for the standard case of phonon-mediated superconductivity). Note that the strength of the BCS term depends on the typical spacing between adjacent energy levels in the mesoscopic island $\delta\xi=1/{\cal V}\rho_0$. 

Let us now consider an isolated superconducting island with fixed even number of electrons and therefore fixed charging energy represented by the first
term in Eq. (\ref{H-int}). The second term can be treated by performing a standard mean-field approximation with $ \Delta\equiv(V_0\delta\xi) \langle\hat{O}\rangle=(V_0\delta\xi) \sum_{|\xi_j|<\hbar\omega_D}\langle c_{j\downarrow}c_{j\uparrow}\rangle$, which results in the BCS Hamiltonian
\begin{equation}
 {\cal H}_{\rm BCS}=\sum_{j,\sigma}\xi_{j} c^\dagger_{j\sigma}c_{j\sigma}+\Delta^*\sum_{j} c_{j\downarrow}c_{j\uparrow}
 +\Delta\sum_{j} c^\dagger_{j\uparrow}c^\dagger_{j\downarrow}-\frac{1}{(V_0\delta\xi)}|\Delta|^2.
  \label{H_BCS}
\end{equation}
Since the mean-field BCS Hamiltonian is now bilinear, it can be diagonalized. The way to do the diagonalization is by the so-called Bogoliubov transformation, which is nothing but a unitary transformation of the form
\begin{equation}
    \begin{pmatrix}
    c_{j\uparrow}\\c^\dagger_{j\downarrow}
    \end{pmatrix}=\begin{pmatrix}
    u^*_{j}&v_{j}\\-v^*_{j}&u_{j}
    \end{pmatrix}\begin{pmatrix}
    \gamma_{j\uparrow}\\\gamma^\dagger_{j \downarrow}
    \end{pmatrix}.
    \label{BdG-transf}
\end{equation}
Here, $\gamma^\dagger_{n\sigma}$, $\gamma_{n\sigma}$ are creation and annihilation operators for quasiparticle excitations with spin $\sigma=\uparrow,\downarrow$.
The Bogoliubov amplitudes are complex numbers which satisfy the constraint
\begin{equation}
\label{BdG-amplitudes}
    |v_{j}|^2 = 1-|u_{j}|^2 = \frac12 \left(1-\frac{\xi_{j}}{E_{j}}\right)
\end{equation}
where $E_{j}= \sqrt{\xi_{j}^2+|\Delta|^2}$ is the quasiparticle excitation energy. 

Using the above transformation, the BCS Hamiltonian in Eq. (\ref{H_BCS}) can be diagonalized as
\begin{equation}
\label{BdGqp}
    {\cal H}_\mathrm{BdG}^\mathrm{qp} = \sum_{j,\sigma} E_{j} \gamma^{\dagger}_{j\sigma} \gamma_{j\sigma}.
\end{equation}
Eq. (\ref{BdGqp}) describes so-called Bogoliubov-de-Gennes (BdG) quasiparticles which are separated from the ground state energy (the Fermi energy) by at least an energy distance given by the superconducting gap (namely $E_{j}\rightarrow |\Delta|$ as $\xi_{j}\rightarrow 0$ \footnote{While in this discussion we are focusing on BdG quasiparticles in a superconductor with a homogeneous order parameter, the emergence of BdG quasiparticles at energies below the superconducting gap $E_{j}< |\Delta|$, generically known as subgap states, is ubiquitous in systems with spatial-dependent order parameters. Examples include subgap states trapped in magnetic flux vortices (so-called Caroli-Matricon-De
Gennes states \cite{Caroli:PL64}); Yu-Shiba-Rusinov states near magnetic impurities, \cite{Yu:APS65,Shiba:PTP68,Rusinov:SPJ69}; or at normal metal-superconductor (NS) junctions, \cite{De-Gennes:PL63,Andreev:SPJ64}, just to name a few. Typically, such subgap states are localized near the region where the order parameter has strong spatial variations, for a recent review see \cite{Sauls:PTRSA18}. We will discuss in detail the physics of such subgap states in Section \ref{s:subgap}.}). 

Importantly, the presence of BdG quasiparticles affects the energy of the island, which now depends on whether the number of electrons $N^e$ is even or odd. Indeed, if the number of electrons on the island is even, they all reside in the condensate forming Cooper pairs.
On the contrary, an odd electron in the island does not have a partner to form a Cooper pair and occupies the lowest-energy quasiparticle state (which for the simplest case discussed here would be $|\Delta|$). This so-called even-odd effect in small superconducting islands is already captured at the simplest level by writing a modified charging Hamiltonian of the form 
\begin{equation}
\label{H-classical}
    {\cal H}_{\rm C}=E_C (N^e-N_g)^2+\begin{cases}
    0 & \text{even}\\
    |\Delta| & \text{odd}.
  \end{cases}.
\end{equation}
In this language, the ground state $|GS\rangle$ of the BCS Hamiltonian is defined by the condition $\gamma_{j\sigma}|GS\rangle = 0$, which is clearly satisfied by the BCS wave function
\begin{equation}
\label{BCS-wave}
    |\phi \rangle = \prod_n \left( |u_n| + |v_n| e^{-i\phi} c^\dagger_{n\uparrow} c^\dagger_{n\downarrow}\right) |0\rangle
\end{equation}
where $|0\rangle$ is the vacuum for electronic excitations, $c_{n\sigma}|0\rangle = 0$, and the notation stresses the fact that we have chosen an arbitrary superconducting phase $\phi$ corresponding to the creation of each Cooper pair, described by the pair creation operator $P^\dagger_n\equiv c^\dagger_{n\uparrow}c^\dagger_{n\downarrow}$.

So far, the derivation we have presented for the BCS mean field solution in Eq. (\ref{H_BCS}) is essentially the same as the one for a macroscopic SC (the only difference being that we have used creation and annihilation operators for electrons in discrete levels, $c^\dagger_{j\sigma}$ and $c_{j\sigma}$, instead of operators with well-defined momentum in a translationally-invariant system, $c^\dagger_{\bold k\sigma}$ and $c_{\bold k\sigma}$). The BCS wave function in Eq. (\ref{BCS-wave}) describes a coherent superposition of states with different numbers of electron pairs, i.e., it is not an eigenfunction of the number of electrons. This conclusion seems problematic, since physical intuition tells us that the number of electrons should be well defined in an isolated island, before and after the transition to the superconducting state. In fact, the solution to this apparent conundrum is not difficult since 
one can project out a solution corresponding to a definite number of Cooper pairs $N$ from the BCS expansion in Eq. (\ref{BCS-wave}) by just multiplying by $e^{iN\phi}$ and integrating over phase.\footnote{In terms of $P^\dagger_n$, Eq.~(\ref{BCS-wave}) reads $|\phi\rangle=\prod_n\big(|u_n|+|v_n|e^{-i\phi}P^\dagger_n\big)|0\rangle$. Expanding the product, each term contains a given number $N'$ of pair creation operators, each of them accompanied by a factor $e^{-i\phi}$, so that the BCS state becomes a sum over Cooper-pair numbers,
\[
|\phi\rangle=a_0|0\rangle+e^{-i\phi}\sum_m a_m P^\dagger_m|0\rangle+e^{-2i\phi}\sum_{m<m'}a_{mm'}P^\dagger_mP^\dagger_{m'}|0\rangle+\dots\equiv\sum_{N'}e^{-iN'\phi}|N'\rangle,
\]
where the prefactors $a_0$, $a_m$, $a_{mm'}$, $\dots$ are products of Bogoliubov amplitudes, and $|N'\rangle$ contains all terms with $N'$ Cooper pairs (up to normalization). Using $\int_0^{2\pi}\frac{d\phi}{2\pi}\,e^{i(N-N')\phi}=\delta_{NN'}$, the integral in Eq.~(\ref{NP-wave}) gives $\sum_{N'}\delta_{NN'}|N'\rangle=|N\rangle$, i.e., it acts as a projector onto the subspace with exactly $N$ Cooper pairs.}  This projection was first proposed by Philip Anderson \cite{ANDERSON1967}, see also \cite{tinkham_book}, and explicitly reads
\begin{equation}
    |N\rangle = \int_0^{2\pi} \frac{d\phi}{2\pi}\, e^{iN\phi} |\phi \rangle,
    \label{NP-wave}
\end{equation}
corresponding to a definite number of Cooper pairs $N$ in the island. Note that by integrating over all values of $\phi$ in Eq. (\ref{NP-wave}), namely by projecting the wave function into a well-defined $N$, we are making the superconducting phase completely uncertain. This nicely reflects the fact that the condensate wavefunctions $|\phi \rangle$ and $|N\rangle$ form two dual bases in the Hilbert space of many-body paired electron states, similar to position and momentum in single-particle quantum mechanics. 

{This analogy can be made precise. Since Eq.~(\ref{NP-wave}) holds for any integer $N$, the states $|N\rangle$ form the number basis of the island, and the Cooper-pair number operator
\begin{equation}
  \hat N =\sum_{N} N|N\rangle\langle N|
  \label{eq:N_operator}
\end{equation}
has an integer spectrum, so that the phase states $|\phi\rangle$ and $|\phi+2\pi\rangle$ are equivalent and $\phi\in[0,2\pi)$. Equation~(\ref{NP-wave}) is thus a Fourier transform from the phase to the number basis, whose inverse gives the phase states
\begin{equation}
\label{phase-basis}
|\phi \rangle = \sum_{N}e^{-iN\phi}|N\rangle,
\end{equation}
such that
\begin{equation}
\langle \phi|N\rangle=e^{iN\phi},\qquad \int_0^{2\pi}\frac{d\phi}{2\pi}\,|\phi\rangle\langle\phi|=\sum_{N}|N\rangle\langle N|=1.
\label{eq:phase_completeness}
\end{equation}
Since the phase is compact, it enters only through $2\pi$-periodic functions, such as
\begin{equation}
e^{ip\hat\phi}=\int_0^{2\pi} \frac{d\phi}{2\pi}\, e^{ip\phi}|\phi\rangle\langle \phi|=\sum_{N}|N+p\rangle\langle N|,
\label{eq:phase_operator}
\end{equation}
with $p$ integer, which act as translation operators,
\begin{equation}
e^{ip\hat\phi}|N\rangle=|N+p\rangle,\qquad e^{-i\phi_0\hat N}|\phi\rangle=|\phi+\phi_0\rangle.
\label{eq:translations}
\end{equation}
Finally, from $\langle\phi|N\rangle=e^{iN\phi}$, the number operator acts on phase wave functions as a derivative, which gives the canonical commutation relation
\begin{equation}
\hat N=-i\frac{\partial}{\partial\phi},\qquad \left[{\hat\phi,\hat N} \right] =i ,
\label{eq:commutator}
\end{equation}
to be understood, for the compact variable $\hat\phi$, as $[e^{ip\hat\phi},\hat N]=-p\,e^{ip\hat\phi}$.}

\subsection{Josephson junctions}
\begin{figure}
    \centering
    \includegraphics[width=0.7\linewidth]{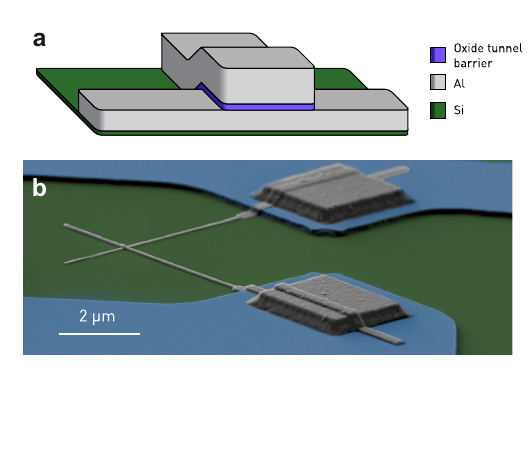}
    \caption{(a) Illustration of a Josephson junction. (b) False-colored scanning electron micrograph of a Josephson junction. Here, green indicates the silicon substrate, blue the niobium base-layer, and gray the aluminum that defined the junction where the two electrodes cross. The big squares of aluminum ensures good electric contact between the thin aluminum electrodes and the niobium base-layer. Adapted and reprinted from Ref.~\cite{colao2025mitigating}. }
    \label{fig:josephson_junction}
\end{figure}

The phase operators of Eq.~(\ref{eq:phase_operator}) change the number of Cooper pairs by one unit, Eq.~(\ref{eq:translations}), which immediately suggests that these operators can be interpreted as hopping operators in the Hilbert space of the Cooper pair number:
\begin{eqnarray}
{\hat T}=\sum_N |N+1\rangle\langle N|=e^{i\hat\phi}\nonumber\\
{\hat T}^\dagger=\sum_N |N-1\rangle\langle N|=e^{-i\hat\phi}.
\label{hatT-2}
\end{eqnarray}
If we now link two islands, the Hamiltonian describing the flow of Cooper pairs takes a general form:
\begin{equation}
{\cal H}_T=\sum_n \left(\tau_n{\hat T}_R^{\dagger n}{\hat T}_L^n+\tau_n^*{\hat T}_L^{\dagger n}{\hat T}_R^n\right),
\label{HJ1}
\end{equation}
where each term in the sum describes the transfer of $n$ Cooper pairs and where the $L$/$R$ notation means that each left/right tunneling operator changes the number of Cooper pairs in the left/right island by one.  In systems with time-reversal symmetry, the tunneling amplitudes fulfill $\tau_n=\tau_n^*$. To lowest order (tunneling limit), we can only keep the $n=1$ terms in the sum and write
\begin{equation}
{\cal H}_J=\tau_1\left({\hat T}_R^{\dagger }{\hat T}_L+{\hat T}_L^{\dagger }{\hat T}_R\right).
\end{equation}
Using the phase representation, the above tunneling term is simply
\begin{equation}
{\cal H}_J=\tau_1\left( e^{i(\hat\phi_R-\hat\phi_L)}+e^{-i(\hat\phi_R-\hat\phi_L)}\right)=2\tau_1 \cos(\hat\phi_R-\hat\phi_L),
\label{HJ2}
\end{equation}
which is nothing but the celebrated Josephson tunneling Hamiltonian describing supercurrent flow in a superconducting junction with a Josephson energy $E_J=-2\tau_1$. Experimentally, we often make Josephson junctions using Al/AlOx/Al junctions, where the AlOx serves as a tunnel-barrier between two thin films of Al; see also Fig.~\ref{fig:josephson_junction}. Along the review, we will discuss examples where the Josephson coupling deviates strongly from the tunneling limit of Eq. (\ref{HJ2}).

\subsection{Charge qubits: from the Cooper pair box to the Transmon qubit}
Let us now consider the case of a single superconducting island shunted to ground by a Josephson junction with Josephson energy $E_J$ \footnote{Using the Josephson relations \cite{tinkham_book} for the supercurrent and voltage, respectively, $I(\Phi)=I_c\sin(2\pi\frac{\Phi}{\Phi_0})$ and $V=\dot \Phi$, where $\Phi$ is the flux threading the junction and $\Phi_0=\frac{h}{2e}$ the superconducting flux quantum, one can easily find the Josephson potential of the junction as $V_J(\Phi)=\int I(t)V(t) dt=\int I(\Phi)\dot \Phi dt=-E_J\cos(2\pi\frac{\Phi}{\Phi_0})$, where $E_J=\frac{\Phi_0}{2\pi}I_c$. Note the agreement with Eq. \eqref{HJ2} in the previous subsection where we used a seemingly unrelated derivation.}. We refer to this simple construction as a Cooper pair box (CPB). In this case, we do not need to explicitly consider the number operator for ground and the only relevant phase $\hat{\phi}$ is the phase difference between the island and ground. The Hamiltonian of this system is then the combined Hamiltonian from the charging energy of the Cooper pairs on the island and the Hamiltonian of the Josephson junction. In other words, the Hamiltonian takes the simple form \footnote{This Hamiltonian is already written in terms of reduced variables $\hat N=\hat Q/2e$ and $\hat\phi=2\pi\frac{\Phi}{\Phi_0}$, where $Q$ is the charge operator and $\Phi$ is the flux operator defined as the time-integral over the voltage across the junction.}
\begin{align}
\label{SC_Hamiltonian}
\mathcal{H}_{\text{CPB}} = 4 E_C (\hat{N} - N_g)^2 - E_J \cos(\hat{\phi}).
\end{align}
In this formula, notice that we have a factor of $4$ in front of $E_C$ as compared with $\mathcal{H}_C$ in Eq.~\eqref{H-classical} since we are now using the Cooper pair number operator rather than the electron number operator. 

Let us consider the case where the charging term dominates the Josephson tunneling energy, i.e., $E_C > E_J$. Specifically, we see that when $E_J$ vanishes, the number states $\ket{N}$ are eigenstates of the system. We also notice that for $N_g = 0.5$, the states $\ket{N=0}$ and $\ket{N=1}$ are degenerate. In fact, for $N_g = N_0 + 0.5$ with $N_0$ an integer, we will have that $\ket{N_0}$ and $\ket{N_0+1}$ are the degenerate ground states of the system. Therefore, let us simply consider $\mathcal{H}_{\text{CPB}}$ in the subspace of the two lowest energy states $\ket{N_0}$ and $\ket{N_0+1}$ and consider how the Josephson junction term perturbs the system. Remember from the previous section that we can represent the Josephson junction also as
\begin{align}
\mathcal{H}_J = -\frac{E_J}{2}\sum_N \big( \ket{N}\bra{N+1} + \ket{N+1}\bra{N} \big).
\end{align}
From this expression, we immediately notice that the states
\begin{align}
\ket{\psi_\pm} = \frac{1}{\sqrt{2}}(\ket{N_0} \pm \ket{N_0+1} )
\end{align}
are eigenstates of in our subspace with eigenenergies $E_\pm = \pm E_J/2$. Other states are separated from these two states by at least $2E_C$. In other words, these two states can be operated as a qubit with energy difference of $E_J$ provided that we can fine-tune $N_g$. 

\begin{figure}
    \centering
    \includegraphics[width=\textwidth]{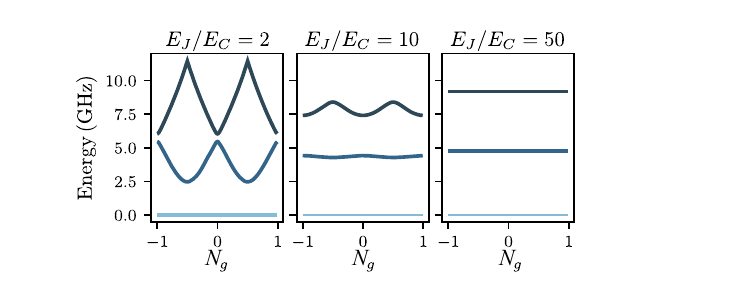}
    \caption{Frequency ($=$ Energy divided by $\hbar$) difference of the first three energy levels of the CPB Hamiltonian. For small $E_J$, we note the locally charge insensitive points at $N_g = \pm 1$, where we can operate the CPB as a qubit. For large values of $E_J$ compared to $E_C$, the energy levels become fully insensitive to $N_g$ and, thus, insensitive to charge noise. Adapted and reprinted from Ref.~\cite{Blais2021}.}
    \label{fig:transmon_charge_dispersion}
\end{figure}

Alternatively, we can consider the case where $E_J \gg E_C$ which causes the phase to be localized in the potential minimum of the Josephson potential. Since $|\langle \hat{\phi} \rangle | \ll 1$, we can Taylor expand the CPB Hamiltonian in $\hat{\phi}$ which, to second order, becomes
\begin{align}
\mathcal{H} \approx 4E_C (\hat{N} - N_g)^2 - E_J\Big( 1 - \frac{1}{2}\hat{\phi}^2\Big). \label{eq:HJ_harmonic}
\end{align}
At this point, we recall that the phase and the Cooper-pair number operators obey the canonical commutation relation of Eq.~\eqref{eq:commutator}, meaning we can analogously consider the phase and the number operators as (unitless) position and momentum operators, respectively. With this analogy in mind, we notice that Eq.~\eqref{eq:HJ_harmonic} corresponds to a quantum harmonic oscillator Hamiltonian albeit with an offset in the generalized momentum. Nonetheless, equivalently to the harmonic oscillator, we can formally offset the momentum through a simple gauge transformation, thus, the eigenenergies are, in this case, independent of the charge offset $N_g$. We are therefore left with the simpler expression (when also neglecting the constant offset):
\begin{align}
\mathcal{H} \approx 4E_c \hat{N}^2 + \frac{E_J}{2}\hat{\phi}^2. \label{eq:HJ_harmonic_simple}
\end{align}
We can solve this Hamiltonian by introducing the bosonic creation and annihilation operators $b^\dagger$ and $b$ with the commutator relation $[b, b^\dagger] = 1$ such that
\begin{align}
\hat{\phi} = \Bigg(\frac{2E_C}{E_J}\Bigg)^{1/4}(b^\dagger + b), && \hat{N} = \frac{i}{2} \Bigg(\frac{E_J}{2E_C}\Bigg)^{1/4}(b^\dagger - b),
\end{align}
which are standard harmonic oscillator operators. This can be checked by inserting these operators into Eq.~\eqref{eq:HJ_harmonic_simple}, to obtain
\begin{align}
\mathcal{H} \,&\approx -E_C \Bigg(\frac{E_J}{2E_C}\Bigg)^{1/2} (b^\dagger b^\dagger - b^\dagger b - b b^\dagger + bb) + \frac{E_J}{2} \Bigg(\frac{2E_C}{E_J}\Bigg)^{1/2} (b^\dagger b^\dagger + b^\dagger b + b b^\dagger + bb)\\
&= \sqrt{8E_J E_C} \Big( b^\dagger b + \frac{1}{2} \Big).
\end{align}
In other words, the circuit effectively implements a harmonic oscillator with an angular frequency of $\omega = \sqrt{8E_J E_C}/\hbar$. Next, we can reintroduce the nonlinearity of the Josephson junction by keeping terms up to fourth order in the Taylor expansion of the cosine of the phase in Eq. \eqref{SC_Hamiltonian} such that our Hamiltonian reads
\begin{align}
\label{expansion-fourth-order}
\mathcal{H} \approx \sqrt{8E_J E_C} \Big( b^\dagger b + \frac{1}{2} \Big) - \frac{E_J}{24} \Bigg(\frac{2E_C}{E_J}\Bigg) \Big(b^\dagger + b\Big)^4.
\end{align}
We notice that the prefactor to the nonlinear term reduces to $E_C/12$ and we recall that we are operating in the regime of $E_C \ll E_J$ which also means that $E_C \ll \sqrt{8E_J E_C}$, thus, we can treat the nonlinearity as a small perturbation that does not change the eigenstates of the system. This assumption allow us to keep only number preserving terms in the term $(b^\dagger + b)^4$. Note that this assumption is equivalent to applying the rotating wave approximation. The Hamiltonian now becomes
\begin{align}
\mathcal{H} \approx \sqrt{8E_J E_C} b^\dagger b - E_C b^\dagger b - \frac{E_C}{2} b^\dagger b^\dagger b b, \label{eq:H_transmon}
\end{align}
where we have dismissed constant terms. When the CPB is operated in a regime of $E_J \gg E_C$, we refer to the circuit as a \textit{transmon qubit}~\cite{Koch2007}. We see from Eq.~\eqref{eq:H_transmon} that the energy difference between the ground and first excited state, i.e., the qubit energy is given as
\begin{align}
E_q = E_1 - E_0 = \sqrt{8E_J E_C} - E_C,
\end{align}
which we should compare to the energy difference between the first and second excited states $E_{21} = E_2 - E_1 = \sqrt{8E_J E_C} - 2E_C$. Thus, we find that the \textit{anharmonicity} of the transmon is given as
\begin{align}
\label{anharmonicity}
\alpha = E_{21} - E_q = -E_C,
\end{align}
which is notably different from zero.  

\begin{figure}[t]
    \centering
    \includegraphics[width=0.8\linewidth]{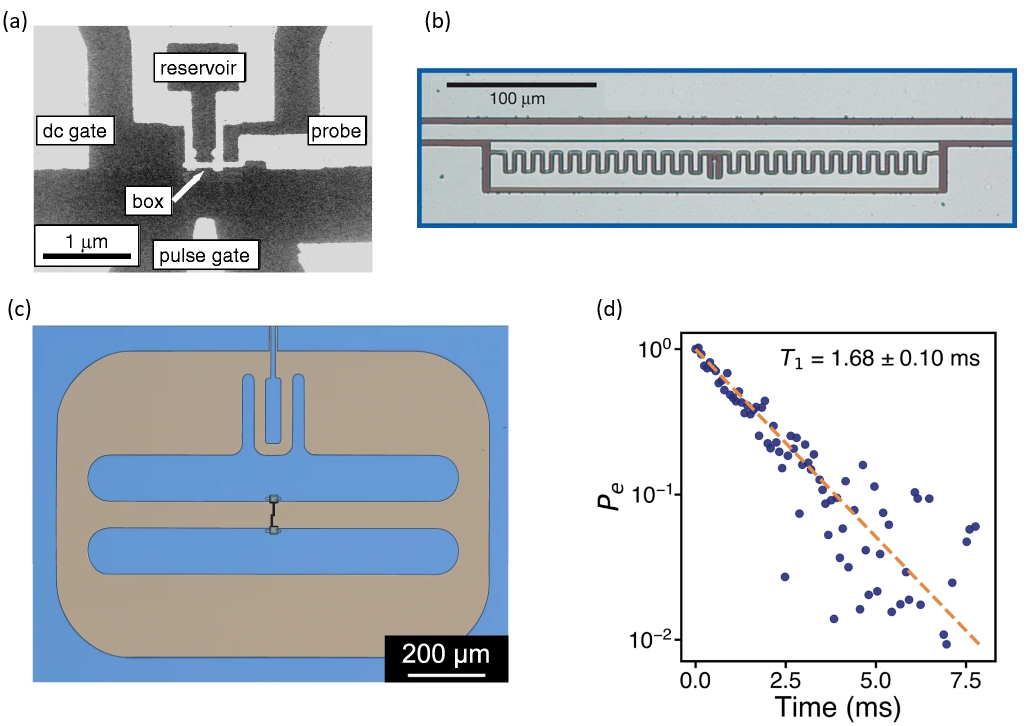}
    \caption{Progress of superconducting charge qubits over the past two decades.
(a) The experimental realization of a superconducting qubit by Nakamura et al.~\cite{Nakamura1999}, demonstrated coherent Rabi oscillations in a superconducting qubits for the first time.
(b) First-generation transmon qubit device by Houck et al.~\cite{Houck2007} operating in the regime $E_J \gg E_C$. Note that the capacitor is large compared to the Josephson junction and, therefore, the Josephson junction in the center of the transmon is barely visible in the image.
(c) Tantalum device from Place et al.~\cite{place2021new}, where coherence times were extended to several hundreds of microseconds by mitigating dielectric loss through geometric and material engineering.
(d) State-of-the-art device from Bland et al.~\cite{bland20252d}, demonstrating qubit lifetimes exceeding 1 ms. Panels (a), (b), (c) and (d) are adapted and reprinted from Refs.~\cite{Nakamura1999}, \cite{Houck2007}, \cite{place2021new} and \cite{bland20252d}, respectively.}
    \label{fig:transmon_experiments}
\end{figure}

The first experimental implementation of a charge-based superconducting qubit was demonstrated in 1999 by Nakamura et al.~\cite{Nakamura1999}, see Fig.~\ref{fig:transmon_experiments}(a). The hallmark observation in this work was the appearance of coherent Rabi oscillations. Unfortunately, due to charge fluctuations, the coherence time was limited to around 2~ns. A significant improvement in coherence times was achieved using the transmon design with $E_J\gg E_C$~\cite{Koch2007}. An example from 2007 of the first generation of devices by Houck et al.~\cite{Houck2007} is shown in Fig.~\ref{fig:transmon_experiments}(b) which demonstrated coherence times approaching microseconds (based on measurements of the linewidth). Following Wang et al.~\cite{Wang2015}, dielectric loss was identified as the main culprit in limiting the lifetime of transmon qubits. Using this insight, Place et al.~\cite{place2021new} demonstrated transmon lifetimes of several 100 $\mu$s by making the transmon qubits physically bigger to dilute the electric fields, see Fig.~\ref{fig:transmon_experiments}(c), as well as using materials with known low dielectric loss. More recently, the same group demonstrated qubit lifetimes larger than 1~ms~\cite{bland20252d}, see Fig.~\ref{fig:transmon_experiments}(d), with similar lifetimes presented in \cite{wang2022towards,kono2024mechanically,dane2025performance,tuokkola2024methods,read2023precision,bal2024systematic}. 

\begin{figure}
    \centering    \includegraphics[width=0.8\linewidth]{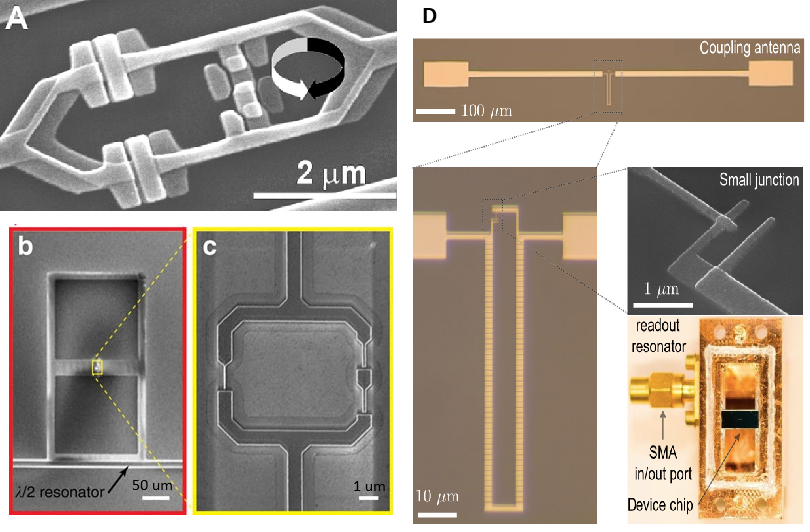}
    \caption{(a) Flux qubit from Ref.~\cite{Chiorescu2003}, which was the first flux qubit. The qubit loop is at the top of the device, indicated with an arrow, with three Josephson junctions in series. (b) and (c) Improved flux qubit from Ref.~\cite{yan2016flux} which shunted a flux loop with a large capacitor. (d) Fluxonium qubit from Ref.~\cite{Nguyen2019}. The flux loop consist of a small junction and a array of large junctions acting as a linear inductor.}
    \label{fig:flux_fluxonium}
\end{figure}

\subsection{Flux qubits and the Fluxonium qubit}\label{ss:fluxonium}
An alternative implementation of qubits using superconducting circuits relies on the phase degree of freedom to encode the qubit state as opposed to charge for the CPB. Arguably, the simplest system to obtain a qubit based on flux degrees of freedom is the rf-SQUID ~\cite{PhysRevLett.75.1614}: a superconducting loop interrupted by a single Josephson junction in series with an inductor, and threaded by an external magnetic flux $\Phi_{ext}$. At half a flux quantum ($\Phi_{ext}=\Phi_0 /2$), the potential assumes a symmetric double-well shape with two degenerate minima. In this regime, the two lowest energy eigenstates become degenerate, and their wavefunctions are localized in opposite wells, corresponding to a superposition of macroscopic circulating currents of equal magnitude flowing in opposite directions. Although conceptually elegant, the rf-SQUID faces significant practical challenges as a qubit platform. Its reliance on a single Josephson junction makes device performance highly susceptible to fabrication variations; even small deviations in junction parameters can substantially alter $E_J$
and thus modify the qubit spectrum. Moreover, achieving the required large loop inductance demands physically large qubit loops, which in turn enhances sensitivity to external flux noise, undermining coherence. 

A more advanced and versatile flux qubit design is the three-Josephson-junction circuit, which offers a compact footprint and greater parameter tunability compared to the rf-SQUID. The first version of this flux qubit was
introduced Mooij et al. in 1999~\cite{Mooij1999}, see also Ref.~\cite{PhysRevB.60.15398},  under the name of persistent-current qubit. The first experimental evidence of quantum superposition of persistent-current states was obtained via microwave spectroscopy by van der Wal et al. in 2000 \cite{Van_der_Wal2000}, while the first experimental demonstration of coherent control of a flux qubit was later reported in 2003 by Chiorescu et al.~\cite{Chiorescu2003}, see Fig.~\ref{fig:flux_fluxonium}(a). The circuit of this qubit consists
of a loop interrupted by three Josephson junctions where one of them is much smaller than the others.
An improved version of the flux qubit, using a large shunt capacitor similar to a transmon qubit was experimentally demonstrated by Yan et al.~\cite{yan2016flux}, see Fig.~\ref{fig:flux_fluxonium}(b) and (c). Other works characterizing the decoherence of flux qubits include e.g. Refs.~\cite{PhysRevLett.97.167001,Bylander2011,PhysRevLett.113.123601}. 

The flux qubit operates in the regime where the Josephson energies dominate the charging energies. This regime implies that the phase differences across the junctions, denoted as $\phi_1, \phi_2$ and $\phi_3$ are localized and thus well-defined quantum variables. The total Hamiltonian of the system can be expressed as:
\begin{align}
\mathcal{H} = \mathcal{H}_C - \sum_{i} E_{J,i} \cos(\phi_i)
\end{align}
where $\mathcal{H}_C$ describes all the charging energies and $E_{J,i}$ is the Josephson energy of the $i$th junction.

Since the three junctions form a closed superconducting loop, the flux quantization condition enforced that
\begin{align}
    \sum_i \phi_i = 2\pi n + 2\pi \frac{\Phi_{ext}}{\Phi_0}\,,
\end{align}
where $n$ is an integer. 
To simplify the analysis, we typically rewrite the Hamiltonian in terms of two independent phase variables $\phi_1$ and $\phi_2$, with the third phase determined by the quantization condition. Moreover, we will define two new variables $\phi_q = \phi_1 + \phi_2$ and $\phi_\delta = \phi_1-\phi_2$. Substituting this constraint, the potential energy of the system becomes
\begin{align}
    -2 E_J \cos (\phi_q/2) \cos (\phi_\delta/2 ) - \alpha E_J \cos(\phi_q - \phi_{ext}),
\end{align}

where we use $\alpha < 1$ to indicate the ratio between the small junction's Josephson energy relative to the two big junctions, $E_{J,3}=\alpha E_{J}$, with $E_J=E_{J,1}=E_{J,2}$, and where $\phi_{ext} = 2\pi \Phi_{ext}/\Phi_0$ refers to the reduced external phase.  When operated at $\phi_{ext} = \pi$ (and $\phi_\delta \approx 0$), the behavior of the flux qubit depends critically on the parameter $\alpha$, which shapes its energy potential. For $\alpha<0.5$, the potential has a single well, placing the system in the plasmon regime, where anharmonicity and flux dispersion remain weak. At $\alpha\simeq 0.5$, the qubit enters the so-called quarton regime, where the potential becomes approximately quartic; this allows for large anharmonicity and high coherence while preserving a practical operating frequency \cite{yan2020engineering}. Such qubits have also been employed as couplers for readout applications \cite{YufengYe2024}. Finally, for $0.5 < \alpha < 1$, the system exhibits a double-well potential in the fluxon regime, which gives rise to strong anharmonicity. In this regime, there are two (near)-degenerate minima in the $\phi_q$ variable. The values of $\phi_q$ near the minima, $\pm \phi_m$, correspond to current flowing one way or the other through the loop. We can refer to these two states as $\ket{\circlearrowright}$ and $\ket{\circlearrowleft}$. Using these two states as our basis, we can define the flux qubit Hamiltonian as
\begin{align}
    \mathcal{H}_q = -\frac{\Delta}{2} \sigma_x -\frac{\epsilon}{2} \sigma_z,
\end{align}
where $\sigma_x$ and $\sigma_z$ are the Pauli operators acting on the basis states $\ket{\circlearrowright}$ and $\ket{\circlearrowleft}$. Here, we have two energies in play, $\Delta$ corresponds to the tunneling energy between the two minima and $\epsilon$ is the effective energy difference between the basis states. While finding the precise energies analytically is daunting, we notice that following the WKB approximation, we expect $\Delta \propto e^{-\sqrt{\alpha}}$. Similarly, the energy difference scales directly with the phase offset $\epsilon \propto (\phi_{ext} - \pi)$. We find that the transition energy of the flux qubit is given by $E_{01} = \sqrt{\Delta^2 + \epsilon^2}$.

The fluxonium qubit~\cite{Manucharyan2009,Nguyen2019} is another implementation of a superconducting qubit that, like the flux qubit, uses the phase degree of freedom to encode quantum information. However, the fluxonium circuit consists of a single Josephson junction shunted by a large linear inductance, see Fig.~\ref{fig:flux_fluxonium}(c). This large inductance is typically realized using a series array of large-area Josephson junctions and is referred to as a \textit{superinductor}~\cite{Masluk2012b}. Alternative implementations of a superinductor are to use high-disorder superconducting materials such as granular aluminum (grAl)~\cite{Grunhaupt2019} or NbTiN~\cite{PitaVidal2020,Hazard2019}, which give rise to very large kinetic inductances. 

\begin{figure}
    \centering
    \includegraphics[width=\linewidth]{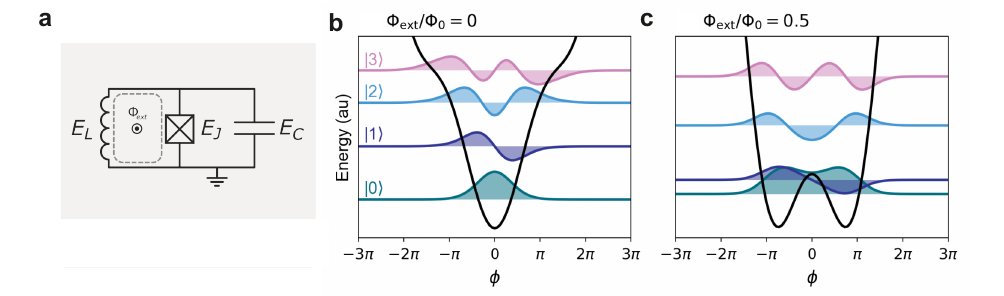}
    \caption{(a) Effective lumped element circuit diagram of a fluxonium qubit with inductive ($E_L$), Josephson ($E_J$), and capacitive energies ($E_C$) and with an external flux ($\Phi_{ext}$) threading the loop marked by the gray dashed line. (b) Wavefunctions of the first four fluxonium energy levels in the phase basis for an external flux bias of zero and (c) at half a flux quantum. Adapted and reprinted from Ref.~~\cite{StefanskiThesis}. }
    \label{fig:fluxonium}
\end{figure}

The circuit consists of a superconducting loop interrupted by a small Josephson junction with energy $E_J$, in parallel with an inductance $L$ (Fig. ~\ref{fig:fluxonium}(a)). The Hamiltonian of the fluxonium qubit is given by
\begin{align}
\mathcal{H} = 4E_C \hat{N}^2 + \frac{1}{2} E_L \hat{\phi}^2 - E_J \cos(\hat{\phi} - \phi_{\mathrm{ext}}), \label{eq:H_fluxonium}
\end{align}
where $\hat{N}$ and $\hat{\phi}$ are conjugate operators obeying $[\hat{\phi}, \hat{N}] = i$. Note that in contrast to the transmon case, the phase variable $\phi$ is now not confined to a $2\pi$ interval anymore and therefore $\hat{N}$ can take continuous values. The parameters of the fluxonium qubit are:
\begin{itemize}
    \item $E_C = \frac{e^2}{2C}$ is the charging energy associated with the junction capacitance $C$.
    \item $E_L = \left( \frac{\Phi_0}{2\pi} \right)^2 \frac{1}{L}$ is the inductive energy associated with the shunt inductance $L$.
    \item $\phi_{\mathrm{ext}} = 2\pi \frac{\Phi_{\mathrm{ext}} }{ \Phi_0 }$ is the reduced external flux threading the loop.
\end{itemize}
The fluxonium typically operates in the regime
\begin{align}
E_J \gg E_L \sim E_C,
\end{align}
which distinguishes it from both the transmon that does not include any linear inductor and the flux qubit. The potential energy of the fluxonium Hamiltonian in Eq.~\eqref{eq:H_fluxonium} is formed by the competition between the harmonic inductive term and the periodic Josephson term. This competition creates a potential landscape with multiple wells whose structure depends on the external flux $\phi_{\mathrm{ext}}$.

At zero external flux ($\phi_{\mathrm{ext}} = 0$), the potential is symmetric and has a single minimum (Fig. ~\ref{fig:fluxonium}(b)). At half a flux quantum ($\phi_{\mathrm{ext}} = \pi$), the potential becomes double-welled, with two nearly degenerate minima (Fig.~\ref{fig:fluxonium}(c)). Physically we can understand the two localized minima as currents flowing clockwise and anti-clockwise around the flux loop, similar to the flux qubit. The eigenstates of the system then correspond to symmetric and antisymmetric superpositions of localized states in each well. These two lowest energy states define the computational basis of the fluxonium qubit.

The fluxonium qubit is typically operated near $\phi_{\mathrm{ext}} = \pi$, where the energy difference between the ground and first excited state is small and tunable. This energy defines the qubit frequency:
\begin{align}
E_q = E_1 - E_0.
\end{align}
Due to the potential not being close to that of a harmonic oscillator we cannot find a simple analytical expression for the fluxonium energy. Instead, we must solve the eigenstates and energies numerically. As with other superconducting qubits, we define the anharmonicity as
\begin{align}
\alpha = (E_2 - E_1) - (E_1 - E_0).
\end{align}
In contrast to the transmon, where $\alpha \approx -E_C$, the fluxonium typically exhibits large and positive anharmonicity. This strong anharmonicity allows for selective control of transitions and for potentially faster gate operations.

One of the major advantages of the fluxonium is its reduced sensitivity to dielectric loss because of the low frequency of the computational states and because the matrix elements between the computational states remain smaller the ones in a transmon system. As a consequence, fluxonium qubits with more than 1~ms coherence times has been demonstrated~\cite{somoroff2023millisecond}. Additionally, the large anharmonicity of fluxonium qubits can be used to enable shorter single qubit gate with potentially higher fidelities~\cite{rower2024suppressing,zwanenburg2025single}. On the other hand, one important disadvantage of fluxonium qubits as compared to transmons is that they typically involve more complex circuit designs inherent to the need of large superinductance elements, such as large arrays of Josephson junctions or large kinetic inductance materials such as grAl~\cite{Grunhaupt2019,PitaVidal2020,Hazard2019}.

\subsection{Circuit QED and microwave readout}\label{ss_circuit-MW}

Circuit Quantum Electrodynamics (cQED) provides a framework for coupling superconducting qubits to quantized electromagnetic modes in microwave resonators. This architecture allows for coherent control, dispersive interactions, and quantum non-demolition (QND) measurements of qubit states using microwave photons. The cQED framework was first proposed and demonstrated in 2004 by Blais et al.~\cite{Blais2004} and Wallraff et al.~\cite{Wallraff2004}. 

A prototypical cQED system consists of a qubit, such as a transmon, fluxonium or some of the hybrid qubits that we will explore later in this work, coupled to a resonator with angular frequency $\omega_r$. Under the rotating wave approximation, the full system is described by the Jaynes-Cummings Hamiltonian:
\begin{align}
\mathcal{H} = \hbar \omega_r a^\dagger a + \frac{1}{2} \hbar \omega_q \sigma_z + \hbar g (a^\dagger \sigma_- + a \sigma_+),
\end{align}
where $a^\dagger$ and $a$ are the creation and annihilation operators for the cavity mode, $\sigma_{\pm}$ are the Pauli raising and lowering operators acting on the qubit space, $\omega_q$ is the qubit transition frequency, and $g$ is the qubit-resonator coupling strength.

We now assume that the system operates in the \textit{dispersive} regime, characterized by a large detuning
\begin{align}
\Delta = \omega_q - \omega_r, \quad |\Delta| \gg g.
\end{align}
In this limit, energy exchange between the qubit and the resonator is suppressed, but virtual transitions still lead to observable energy shifts. Our goal is to derive an effective Hamiltonian valid up to second order in $g/\Delta$ that captures these virtual processes.

To do this, we apply a unitary transformation to eliminate the interaction term to leading order. We define the transformation
\begin{align}
\mathcal{H}' = e^S \mathcal{H}_{\text{JC}} e^{-S} \approx \mathcal{H}_{\text{JC}} + [S, \mathcal{H}_{\text{JC}}] + \frac{1}{2}[S, [S, \mathcal{H}_{\text{JC}}]] + \ldots,
\end{align}
and we choose the anti-Hermitian generator $S$ such that it cancels the interaction term at first order. Let
\begin{align}
S = \frac{g}{\Delta} (a^\dagger \sigma_- - a \sigma_+).
\end{align}
This choice satisfies
\begin{align}
[S, \mathcal{H}_0] = -\hbar g (a^\dagger \sigma_- + a \sigma_+),
\end{align}
where $\mathcal{H}_0 = \hbar \omega_r a^\dagger a + \frac{1}{2} \hbar \omega_q \sigma_z$ is the unperturbed part of the Hamiltonian.

We now compute the effective Hamiltonian to second order in $g/\Delta$:
\begin{align}
\mathcal{H}' &= \mathcal{H}_0 + \frac{1}{2}[S, \mathcal{H}_{\text{int}}] \\
&= \hbar \omega_r a^\dagger a + \frac{1}{2} \hbar \omega_q \sigma_z + \frac{\hbar g^2}{\Delta} \left( \sigma_z a^\dagger a + \frac{1}{2} \sigma_z \right),
\end{align}
where we have used the identity
\begin{align}
[a^\dagger \sigma_-, a \sigma_+] = \sigma_z a^\dagger a + \frac{1}{2} \sigma_z.
\end{align}
Dropping the constant energy offset, the resulting effective Hamiltonian in the dispersive limit is
\begin{align}
\mathcal{H}_{\text{disp}} = \hbar \left( \omega_r + \chi \sigma_z \right) a^\dagger a + \frac{1}{2} \hbar \left( \omega_q + \chi \right) \sigma_z,
\label{eq:H_disp}
\end{align}
where $\chi = \frac{g^2}{\Delta}$ is the \textit{dispersive shift}. This Hamiltonian shows that the qubit imparts a frequency shift $\pm \chi$ on the cavity depending on its state, and conversely, the qubit frequency is shifted by $\chi$ per photon in the cavity.

The dispersive Hamiltonian in Eq.~\eqref{eq:H_disp} reveals an important point: although the qubit and resonator do not exchange energy in this regime, their states remain entangled via virtual interactions. The cavity frequency becomes qubit-state-dependent, enabling indirect measurement of the qubit via the cavity response. Similarly, the qubit frequency becomes photon-number-dependent, leading to phenomena such as the ac Stark shift and measurement-induced dephasing~\cite{Gambetta2006}.

The form of Eq.~\eqref{eq:H_disp} also allows us to define a number-conserving interaction term that arises from further expansions:
\begin{align}
\mathcal{H}_{\text{disp}} = \hbar \chi a^\dagger a \sigma_z + \hbar K (a^\dagger a)^2,
\end{align}
where $K$ is a small Kerr nonlinearity induced by higher-order corrections. While negligible in weakly driven systems, this term becomes relevant in strong dispersive readout or for cavity-based nonlinear optics.

While the Jaynes-Cummings model provides an accurate description of qubit-resonator interactions in the two-level approximation, the transmon qubit is inherently an anharmonic multi-level system. Its weak anharmonicity means that higher excited states, particularly the second excited state $\ket{2}$, can affect the dispersive interaction, even if they are not directly populated. For other qubits, it is always important to keep in mind the addition to the dispersive shift for higher excited states.

As a result, the dispersive shift $\chi$ experienced by the resonator due to the transmon is modified due to virtual transitions involving the $\ket{1} \leftrightarrow \ket{2}$ transition. Including the next higher level in the analysis leads to a corrected expression for the dispersive shift~\cite{Koch2007}:
\begin{align}
\chi = -\frac{E_C}{\hbar} \cdot \frac{g^2}{\Delta (\Delta - E_C/\hbar)},
\end{align}
where, $g$ is the coupling strength between the resonator and the $\ket{0} \leftrightarrow \ket{1}$ transition, $\Delta = \omega_{01} - \omega_r$ is the detuning between the transmon's fundamental transition and the resonator, $E_C$ is the charging energy of the transmon and $\omega_{01}$ and $\omega_{12} \approx \omega_{01} - E_C/\hbar$ are the transmon's transition frequencies between its lowest eigenstates.

This expression reflects the contributions of two virtual transitions:
\begin{enumerate}
    \item $\ket{0} \rightarrow \ket{1}$, detuned by $\Delta$,
    \item $\ket{1} \rightarrow \ket{2}$, detuned by $\Delta - E_C/\hbar$.
\end{enumerate}
The negative sign of $\chi$ (except for small region that we call the straddling regime) is a consequence of the transmon's weakly anharmonic ladder-type level structure. As $E_C$ decreases (i.e., in the deeply transmon regime), the denominator becomes large and the magnitude of $\chi$ becomes small, leading to reduced dispersive coupling and hence weaker measurement contrast. 

\subsubsection{Dispersive Readout}

In the dispersive regime, the measurement is performed by sending a weak coherent microwave probe near $\omega_r$ and measuring the outgoing signal—typically the transmission or reflection from the resonator. The qubit state modifies the resonator’s response: depending on whether the qubit is in $\ket{0}$ or $\ket{1}$, the resonator shifts to $\omega_r \pm \chi$, changing the amplitude and phase of the reflected or transmitted microwave field.

The measured voltage $V(t)$ from a homodyne detection scheme is directly related to the field quadrature operators of the output mode:
\begin{align}
V(t) \propto \langle \hat{X}_\theta(t) \rangle = \langle a_{\text{out}}(t) e^{-i\theta} + a_{\text{out}}^\dagger(t) e^{i\theta} \rangle,
\end{align}
where $\hat{X}_\theta(t)$ is the generalized quadrature operator at angle $\theta$, determined by the phase of the local oscillator in the homodyne detection setup, and $a_{\text{out}}(t)$ is the output field operator.

By input-output theory, the output field is related to the internal cavity mode $a(t)$ and the input drive as
\begin{align}
a_{\text{out}}(t) = \sqrt{\kappa} a(t) - a_{\text{in}}(t),
\end{align}
where $\kappa$ is the cavity decay rate, and $a_{\text{in}}(t)$ is the input field operator. Under steady-state conditions and weak drive, the intracavity field $a(t)$ can be approximated as a coherent state whose amplitude depends on the effective cavity frequency, and thus on the qubit state.

Consequently, the measured voltage $V(t)$ depends on the expectation value of $a(t)$, which in turn depends on the qubit state. This gives rise to two distinct voltage trajectories: $V_0(t)$ and $V_1(t)$, corresponding to the qubit in $\ket{0}$ and $\ket{1}$, respectively.

The signal is then integrated over a measurement time $\tau$ to yield a scalar voltage outcome:
\begin{align}
V_i = \frac{1}{\tau} \int_0^\tau V_i(t) \, dt, \quad \text{for } i \in \{0, 1\}.
\end{align}
The ability to distinguish the qubit state depends on the separation of the two voltage distributions $V_0$ and $V_1$ in the presence of noise. The signal-to-noise ratio (SNR) is defined as
\begin{align}
\text{SNR} = \frac{|\langle V_1 \rangle - \langle V_0 \rangle|}{\sqrt{\sigma_1^2 + \sigma_0^2}},
\end{align}
where $\langle V_i \rangle$ and $\sigma_i^2$ are the mean and variance of the voltage distributions conditioned on the qubit being in state $\ket{i}$. The variances arise from quantum vacuum noise, amplifier noise, and thermal fluctuations.

\begin{figure}[t]
    \centering
    \includegraphics[width=0.95\linewidth]{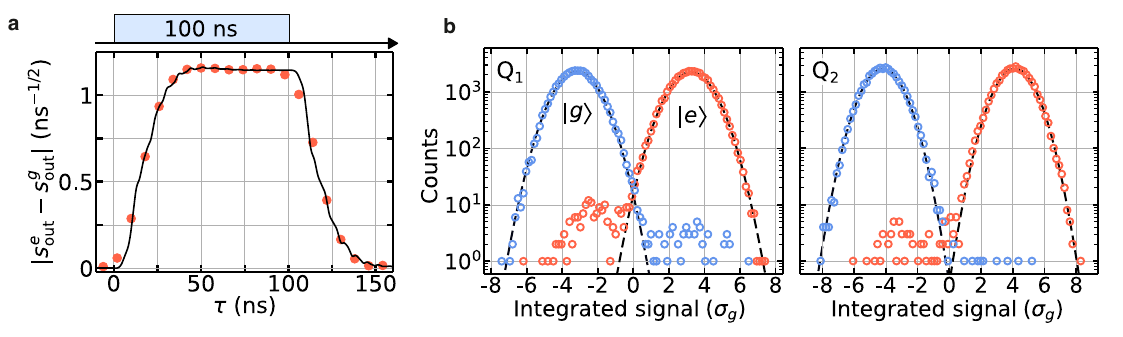}
    \caption{Readout of a transmon qubit. (a) Average of the measurement signal as a function of time, denoted as $V_1(t) - V_0(t)$ in the text, for a 100~ns readout pulse. (b) Histogram of the integrated signals $V_0$ and $V_1$ for two different qubits for a measurement time of 56~ns. The dotted lines indicate fit to a gaussian distribution. The SNR can be found as the difference between the center of each gaussian divided by the variance of the two histograms. Adapted and reprinted from Ref.~\cite{spring2025fast}.}
    \label{fig:dispersive_readout}
\end{figure}

The readout fidelity for a qubit is related to the SNR through the assignment error $P_{\text{err}}$, which for Gaussian voltage distributions is
\begin{align}
P_{\text{err}} = \frac{1}{2} \operatorname{erfc} \left( \frac{\text{SNR}}{2} \right),
\end{align}
and the single-shot readout fidelity is then $F = 1 - P_{\text{err}}$. The first dispersive readout was performed by Wallraff et al.~\cite{Wallraff2005} with recent experimental results demonstrating a readout error smaller than 0.3\% in less than 60~ns~\cite{spring2025fast}, see Fig.~\ref{fig:dispersive_readout}. 

\section{Hybrid semiconducting-superconducting nanostructures: Materials Science
advances} \label{s:nanostructures} 
\subsection{Short historical background} \label{s:history}
Experiments aimed at demonstrating electrostatic control of the Josephson effect date back several decades. While Brian Josephson’s original prediction was experimentally confirmed within a year in metal–oxide tunnel junctions \cite{PhysRevLett.10.230}, it took nearly a decade before the first convincing demonstration of supercurrents in semiconducting junctions—specifically in thinned silicon wafers—was reported \cite{doi:10.1063/1.1655388}. Among other key historical milestones (see Table \ref{TableI}) is the successful deposition of superconducting contacts on p-type InAs in the late 1970s \cite{1060276}. A bit later, Ref. \cite{doi:10.1063/1.327935} discussed the feasibility of the so-called Josephson junction field-effect transistor, arguably presenting the first explicit analysis of planar hybrid superconductor–semiconductor Josephson junctions in which the supercurrent is modulated by a gate electrode capacitively coupled to the junction via a dielectric layer or a Schottky barrier. This seminal work not only examined the physical mechanisms enabling gate-tunable Josephson coupling in hybrid devices, but also anticipated their integration into superconducting circuits, highlighting their relevance for quantum electronics, reconfigurable logic, and low-power cryogenic computing. Further progress was made in the mid-1980s with realization that the two-dimensional electron gas (2DEG) is a better candidate which led to the first demonstration of the proximity effect in the inversion layer of a p-type InAs substrate contacted by Nb electrodes \cite{PhysRevLett.54.2449}, where some degree of electrostatic modulation of the critical current $I_c$ was also observed. A similar experiment using an n-InAs/GaAs heterostructure with Nb contacts followed shortly thereafter \cite{doi:10.1063/1.97233}. A major advance occurred roughly a decade later with the first experiments on split-gate InAlAs/InGaAs semiconductor heterostructures, which provided experimental confirmation \cite{PhysRevLett.75.3533} of the theoretical prediction \cite{PhysRevLett.66.3056,PhysRevLett.67.132} of supercurrent quantization in superconducting quantum point contacts. Evidence for a proximity-induced energy gap and multiple Andreev reflections in Nb/InAs/Nb junctions was reported soon after \cite{PhysRevB.55.8457}; as well as various nonequilibrium ac Josephson effects in Nb/InAs/Nb Junctions, including half-integer Shapiro steps \cite{PhysRevLett.82.1265}. Other important attempts include detailed studies of Nb/InAs(2DEG)/Nb Josephson field-effect transistors \cite{RichterSST199}.

\begin{table*}
\begin{center}
\begin{tabular}{ |p{0.7cm}||p{12cm}|p{12cm}|  }
 \hline
 Year & Milestone  \\
 \hline
 1975  & First demonstration of Josephson effect in a semiconductor-based junction (thinned silicon wafers), Ref. \cite{doi:10.1063/1.1655388}   \\
\hline
1979 &Superconductor contacted to p-InAs,  Ref. \cite{1060276} \\
\hline
1980 &First paper discussing the feasibility of hybrid Josephson field effect transistors, Ref. \cite{doi:10.1063/1.327935} \\
\hline
1985 & 2DEG (inversion layer on a p-type InAs substrate) with Nb electrodes, some electrostatic gating of $I_c$ is observed, Ref. \cite{PhysRevLett.54.2449}  \\ \hline
1986 & SNS junction based on a n-InAs/GaAs heterostructure contacted to Nb electrodes, Ref. \cite{doi:10.1063/1.97233} \\ \hline
1995 & Supercurrent quantization in a split-gate InAlAs/InGaAs semiconductor heterostructure with Nb electrodes.  Ref. \cite{PhysRevLett.75.3533} \\ \hline
1997 & Evidence for a proximity-induced energy gap and multiple Andreev reflections in Nb/InAs/Nb junctions, Ref. \cite{PhysRevB.55.8457}  \\ \hline
2005 &First demonstration of gate tunability of $I_c$ in a Josephson junctions based on semiconductor nanowires (Al/InAs) , Ref. \cite{Doh272}  \\ \hline
2006& First hole-based hybrid device (Josephson junction based on Ge/Si core-shell nanowires), Ref.~\cite{Xiang2006}\\ \hline
2015 &Epitaxial growth of a thin aluminium shell on InAs, Refs. \cite{KrogstrupNM:15,ChangNN:15,shabani2016}. \\ \hline
2015 & First demonstration of NW-based gate tunable transmon qubits (gatemons), Refs. \cite{Larsen2015,deLange2015}. \\ \hline
2020 &Epitaxial growth of other combinations of superconductors and semiconductors is demonstrated, Refs. \cite{Carrad2020,Khan2020}. \\ \hline
2021&Shadow-wall lithography and Single-Shot Fabrication of hybrid Semiconducting–Superconducting Devices, Refs. \cite{Heedt2021,Borsoi2021}. \\ \hline
2021-2023& First demonstrations of spin qubits in superconducting circuits, so-called Andreev spin qubits, Refs. \cite{Hays2021,PitaVidal2023}.
\\ \hline
2024-2025& First demonstrations of experimental Kitaev chains based on minimal arrays of hybrid semiconductor-superconductor quantum dots, Ref.~\cite{Dvir2023}.\\ \hline
\end{tabular}

\caption{The fifty years old road towards semiconductor-based superconducting qubits.}
\end{center}
\label{TableI}
\end{table*}

However, this rapid progress encountered a serious bottleneck in improving interface transparency, as reviewed in Ref. \cite{Schapers:01}: After a relatively successful development of Josephson junction concepts, the field entered a ``valley of death'', with little advancement in the 2DEG-superconductor platform. For example, experiments with ballistic InAs 2DEGs showed very low critical currents, attributed to overall diffusive superconducting transport \cite{PhysRevB.60.13135}. These observations highlighted the difficulty in achieving high interface transparency required for robust proximity effects. Despite extensive efforts, only a handful of techniques successfully enhanced junction transparency. Notably, silicon engineering of Al/InGaAs heterostructures has been shown to eliminate the native Schottky barrier, greatly improving transparency and critical current characteristics compared to traditional approaches \cite{doi:10.1063/1.122926}.

One major impediment to progress and reproducibility was the fabrication strategy. Theory suggested that the optimal Josephson junction would require a high-mobility 2DEG, a superconductor with a large coherence length, and short superconductor contact separation, alongside highly transparent ohmic contacts. However, the early efforts mentioned before often focused on "deep" quantum wells to boost carrier mobilities, making reliable contacts technologically challenging. The need to partially etch or expose the buried 2DEG for contacting introduced variability, sensitivity to etching and cleaning protocols, and yielded a low fraction of working junctions. The standard figure of merit for JJs is the product $I_cR_N$, where $R_N = G_N^{-1}$ is the normal resistance of the junction. This product is related to the superconducting gap by $I_cR_N = \eta\Delta/e$, where the prefactor $\eta$ ranges from $\pi$ (ballistic limit) to $\pi/2$ (diffusive limit) for transparent junctions \cite{RevModPhys.76.411,RevModPhys.51.101}. Typical values in these early experiments were significantly below these upper bounds, indicating an imperfect proximity effect (i.e., poor Andreev reflection at the interface). Typical supercurrents were typically $I_{c} = $10-100 nA yielding $I_{c}R_{N}$ values about 20 times less than superconducting gap $\Delta$. Furthermore, the success of this process was tightly linked to the material stack: in lattice-matched InAs/GaSb heterostructures, etch selectivity \cite{PhysRevB.93.075302,KROEMER2004196} and undercut control critically determined device yield. More detailed discussions about such crucial steps can be found in e.g. Refs.\cite{Mayer2018,Chakraborti_2018}.

By this time, many of the exciting possibilities of proximitized 2DEG were explored, but the difficulty in reproducibility prevented further complicated devices. A few years later, the field was revitalized when attention shifted to semiconductor nanowires (NWs) coupled to superconductors. The first measurements demonstrating gate tunability of $I_c$ in JJs based on semiconductor NWs (Al/InAs) were published in 2005 \cite{Doh272}. 
Since then, various material combinations have been explored, including Al/SiGe \cite{Xiang2006}, Al/InSb \cite{NilssonNL:12}, and Nb/InAs \cite{Gharavi_2017}. 

\subsection{\label{sec:materials} Epitaxial Superconductor-Semiconductor Interface}

The design of hybrid superconductor-semiconductor nanostructures involves several critical factors. Chief among them is the need for a high-mobility two-2DEG and a nearly transparent interface with the superconductor. As previously discussed, however, achieving such transparency proved to be challenging. Semiconductors like Ge, InAs, and InSb are among the few viable candidates for these hybrid systems due to their low or negligible Schottky barriers. This is crucial because even an atomically perfect contact between a superconductor and Si typically results in poor electron transmission, owing to Schottky barrier formation caused by Fermi level pinning at the surface. In contrast, InAs facilitates ohmic contact as electrons are attracted to the metal, while InSb and Ge possess sufficiently low Schottky barriers that can be overcome by inducing carriers beneath the metal. A major breakthrough in improving interface transparency occurred around 2015 with the fabrication of heterostructures featuring a thin epitaxial aluminium shell grown on InAs, both in nanowires (NWs) \cite{KrogstrupNM:15,ChangNN:15} and in 2DEGs \cite{shabani2016}.

Crucially, the development of these high-quality hybrid structures was made possible by Molecular Beam Epitaxy (MBE). High-mobility semiconductors such as InAs and InSb are typically grown using MBE, a state-of-the-art technique that offers atomic-layer control and extremely high purity for the fabrication of epitaxial thin films of both semiconductors and metals \cite{Kroemer96}. In this process, beams of elemental atoms or molecules are directed onto a heated substrate under ultra-high vacuum conditions, ensuring minimal contamination and precise stoichiometry in the resulting material layers. The ability of MBE to grow atomically flat and highly controlled heterostructures has made it a cornerstone in semiconductor materials science \cite{MBE2002} and, in general, has played a pivotal role in advancing many modern technologies.

Among the many technological domains impacted by MBE are high-efficiency photovoltaic devices \cite{Melloch1990,Geisz2007}, semiconductor lasers \cite{CAPASSO1997231}, quantum dots \cite{10.1063/1.110199,10.1063/5.0012066}, and nanowires \cite{10.1063/1.3525610,KrogstrupNM:15}. Moreover, it has enabled groundbreaking investigations at the intersection of condensed matter and particle physics, such as the study of emergent quasiparticles that exhibit exotic behaviors \cite{PhysRevLett.59.1776,Chung21}. A notable example is the use of MBE to explore topological materials \cite{Moore2010} and, in general, the possibility to engineer quantum heterostructures hosting novel phenomena.
This exemplifies how MBE serves not only as a fabrication tool but also as a bridge between traditionally distinct domains of physics. In light of its broad and sustained impact, it is indeed difficult to identify a subfield of physics that has not been influenced by the capabilities and applications of MBE \cite{Mccray2007MBEDA}.

Using MBE, one can start from a substrate of choice and stack various materials, typically with different band gaps to create a two-dimensional electron gas confined to a quantum well. Depending on commercially available substrates or appropriate materials for barriers, there could be lattice mismatch to the active quantum well e.g. InAs or Ge. The typical misfit dislocations introduce surface corrugation that requires metamorphic or strained growth to achieve the highest quality 2DEGs. This is where layer-by-layer growth, such as MBE or CVD techniques, provides excellent control.

\begin{figure}[t]
    \centering
   \includegraphics[width=\linewidth]{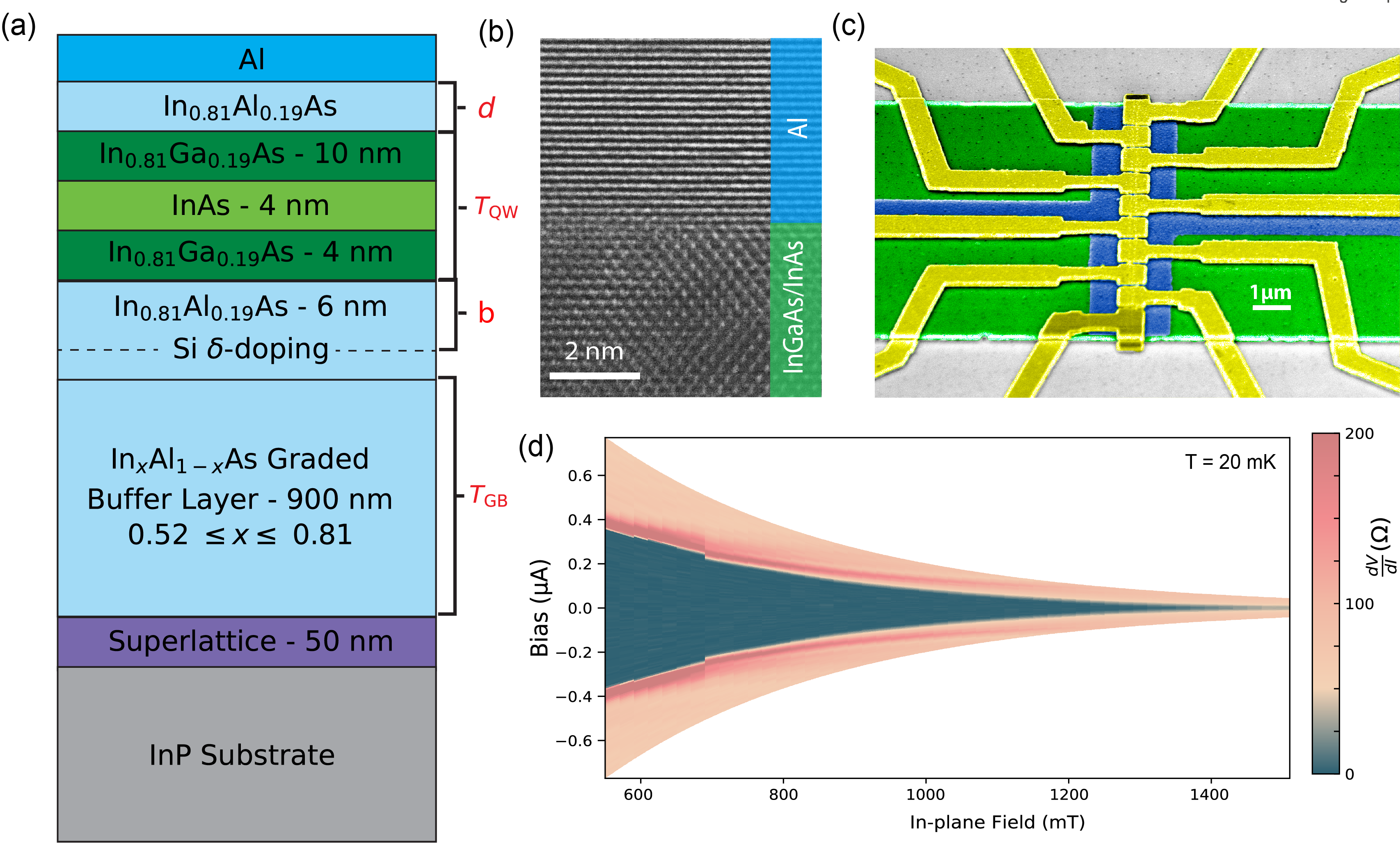}   
   \caption[clean super-semi interface]{{Engineering clean superconductor--semiconductor interfaces.} (a) An InGaAs/InAs/InGaAs quantum well is grown atop an InP substrate by lattice matching via a step-graded \ce{In_xAl_{1-x}As} buffer layer.  A thin \ce{In_{0.81}Al_{0.19}As} layer serves as a diffusion barrier between the Al and the quantum well \cite{sarney2018,sarney2020}.  (b) Transmission electron micrograph of an atomically clean superconductor--semiconductor interface enabling a hard induced gap.  (c) Gated, planar Josephson junction fabricated on the heterostructure in (a). (d) Differential resistance vs.\ in-plane field showing the large in-plane critical field of the junction due to the thin superconducting film. Adapted and reprinted from Ref.~\cite{Schiela2024}.}
    \label{fig:disorder:S-Sm}
\end{figure}

Arguably, the most developed epitaxial superconductor-semiconductor system is based on the well-established epitaxial relationship between Al and InAs quantum wells \cite{shabani2016}. Fig.~\ref{fig:disorder:S-Sm}(a) shows a typical InAs quantum well material stack on an InP (100) substrate. The stack starts with a step graded buffer layer of In$_{x}$Al$_{1-x}$As that is grown at low temperature, typically $T_\text{GB} \sim$\SI{360}{\celsius}, to minimize dislocations forming due to the lattice mismatch between the active region and the InP substrate.  The quantum well is typically grown at a higher temperature $T_\text{QW} \sim \SI{420}{\celsius}$ and consists of a \SI{4}{nm} layer of InAs grown on a \SI{4}{nm} layer of In$_{0.81}$Ga$_{0.19}$As. A top layer, typically \SI{10}{nm} of In$_{0.81}$Ga$_{0.19}$As, is grown on the strained InAs quantum well as shown. An extra layer of In$_{0.81}$Al$_{0.19}$As, $d \approx \SI{1}{nm}$ can be grown on top to reduce the coupling of the superconductor and the 2DEG if desired. The structure is usually delta doped with Si $b = \SI{6}{nm}$ below the quantum well. After the quantum well is grown, the substrate is cooled to promote the growth of epitaxial Al (111) \cite{shabani2016}.  Fig.~\ref{fig:disorder:S-Sm}(b) shows a high-resolution transmission electron microscope (TEM) image of this interface between Al and In$_{0.81}$Ga$_{0.19}$As, with atomic planes of both crystals clearly visible.  A 2-inch wafer with epitaxial Al is processed by selectively removing the Al using Transene Type-D Al etchant to define device features. Fig.~\ref{fig:disorder:S-Sm}(c) shows a false colored scanning electron micrograph of a Josephson junction formed by removing \SI{100}{nm} by \SI{8}{\micro m} of Al. Using the layer by layer growth of epitaxial Al, it is possible to stop the growth at ultra thin thicknesses (considering the Al oxidation of \SIrange{1}{2}{nm} after exposure to ambient conditions). It is routine to target an Al thickness of \SI{10}{nm} or less for enhanced in-plane critical magnetic field. Fig.~\ref{fig:disorder:S-Sm}(d) shows supercurrent as a function of applied parallel magnetic field exhibiting \SI{1.4}{T} critical field.

These proximitized quantum wells offer a number of improvements as well as challenges when compared to nanowires.  For example, these systems are more amenable to top-down design and nanofabrication using well-established wafer-scale lithographic techniques, enabling more flexible, scalable, and creative device geometries (which can be useful in e.g. the design of novel Josephson junction concepts \cite{Schiela2024}). However, the material processing involved, particularly wet etching and dielectric growth, introduces additional disorder to the system. The necessity of a substrate material for growth and the resulting misfit and dislocations in quantum wells also create unwanted interface and surface roughness. In addition, the thickness of the top barrier of the quantum well provides a natural and important experimental tuning knob which is absent in the nanowire system. The top barrier thickness controls the strength of the proximity coupling between the semiconducting 2DEG and the superconductor.

Although the majority of hybrid experiments conducted so far have focused on group III-V semiconductors—such as InAs and InSb, as previously discussed—there is a growing shift in attention toward group IV semiconductors. Among these, germanium stands out due to its unique electronic properties. In particular, the structure of its valence band enables efficient Fermi-level pinning with a wide range of metals, which is a key factor in facilitating strong coupling and proximity-induced phenomena in hole systems. Moreover, germanium offers a favorable environment for quantum coherence, as its nuclear spin-free isotopes significantly reduce hyperfine interactions, a major source of decoherence in many semiconductor platforms. This characteristic makes germanium a highly promising platform for exploring novel quantum effects, device architectures, and long-lived qubits~\cite{Scappucci2021}. Recent years have witnessed a series of important experimental breakthroughs using germanium-based hybrids, underscoring its potential for future applications in quantum technologies. In light of this growing interest and the substantial progress already achieved, we dedicate an entire chapter to this topic in Section \ref{ss:germanium}.

\subsection{\label{sec:materials-properties} Relevant parameters in hybrid Superconductor-Semiconductor devices}

Semiconductors such as InAs, InSb, and Ge are particularly suited for the hybrid qubits discussed in this review—namely, superconducting spin qubits (Section \ref{s:qubits}) and Majorana-based topological qubits (Section \ref{s:bottom-up}). This suitability stems from their favorable intrinsic properties, especially strong spin-orbit coupling and, in the case of InAs and InSb, a large g-factor. Consequently, the performance of a semiconductor-superconductor heterostructure depends on a hierarchy of parameters, including the semiconductor's chemical potential, g-factor, spin-orbit coupling strength, and mobility, as well as the superconductor's critical temperature, critical field, and gap.

However, predicting and controlling these parameters is profoundly challenging. Modeling these heterostructures must account for the complex interplay of multiple physical effects—from proximity-induced superconductivity and electrostatics to orbital, Zeeman, and disorder interactions. This theoretical effort has revealed a critical and often counterintuitive trade-off: the coupling strength at the interface, while essential for inducing superconductivity, can be detrimental if excessively strong. A key finding is that such strong coupling renormalizes the very semiconductor properties that make these materials attractive, namely the g-factor and spin-orbit interaction \cite{PhysRevB.96.014510,PhysRevB.97.165425,PhysRevX.8.031040,PhysRevX.8.031041,PhysRevB.99.245408}.

This renormalization originates from the hybrid band structure at the interface. Therefore, the degree of subband hybridization—tunable by gate voltages and the superconductor layer's thickness—governs a delicate balance. It must be sufficient to generate a robust superconducting pairing, yet not so severe that it triggers a "metallization" of the semiconductor, where profound band shifts effectively destroy its semiconducting character \cite{PhysRevB.97.165425}. 

Ultimately, the central challenge is to engineer an interface that achieves the right balance of optimized properties while preserving a good proximity effect, as we discuss in what follows.

\subsubsection{\label{sec:materials:gap}Superconducting Gap: Enhancement vs.\ Preservation} The fundamental materials challenge of engineering hybrids with the right balance of optimized properties has directed the search for superconductors as the parent material to induce
superconductivity in the semiconductor. 
To date, only a few have been successfully integrated, with epitaxial aluminum being the most prominent. Its advantages include a self-limiting native oxide and the ability to induce a hard superconducting gap in proximate semiconductors \cite{KrogstrupNM:15,ChangNN:15,shabani2016}. Al is also widely known to exhibit charge
parity stability demonstrated by the 2e charging in small
islands \cite{Albrecht2016,Shen_NatComm2018}. However, Al's limitations—a relatively small superconducting gap, critical temperature, and critical field \cite{ChangNN:15}—restrict its use in devices requiring high magnetic fields. While reducing the Al thickness can enhance its critical field, this often introduces increased surface scattering and nonuniformity, which reduce the mean free path and coherence length \cite{Mazur2022,Levajac2023}.

Driven by these limitations with Al, there is strong interest in broadening the scope of superconductor-semiconductor hybrids. Promising alternatives include Sn, Pb, and Nb compounds grown on InAs and InSb nanowires, which exhibit a large, hard induced gap that remains resilient to high magnetic fields and elevated temperatures \cite{Pendharkar2021,Kanne2021,Drachmann2017}. These new hybrids have also demonstrated the coveted two-electron charging effect in small islands \cite{Pendharkar2021,Kanne2021}, indicating a charge parity stability comparable to Al \cite{Albrecht2016,Shen_NatComm2018}. Recent progress has further expanded the materials toolkit, with demonstrations 
of induced superconductivity in hole-based semiconductors, like e.g. a hard gap induced in germanium by superconducting PtSiGe \cite{tosato2023}, as we discuss in great detail in Section \ref{ss:germanium}, and in InSbAs with epitaxial Al \cite{Sestoft2018,Moehle2021}, see subsection \ref{sec:new-hybrids}.

\subsubsection{\label{sec:materials:soc}Spin--Orbit Coupling in Proximity Devices}
The strength of spin-orbit coupling (SOC) is a pivotal parameter in hybrid semiconductor systems, as it directly sets the upper bounds for critical quantum properties. Most notably, it limits the achievable zero-field splitting in Andreev spin qubits and the size of the topological gap in engineered superconducting phases—relationships we will explore in detail in Sections \ref{s:qubits} and \ref{s:bottom-up}, respectively. In these materials, SOC primarily arises from a broken spatial inversion symmetry. This asymmetry, which can be intrinsic to the crystal lattice or engineered via a heterostructure, generates an effective electric field, $\bm{E}$.
The interaction of an electron's spin with this field yields a spin-orbit term of the general form $\bm{E}\cdot\left(\bm{\sigma}\times\bm{p}\right)$ where $\bm{\sigma}$ is the Pauli vector of spin operators and $\bm{p}$ is the electron momentum. In a quantum well with structural asymmetry along the growth direction, the dominant contribution is often the Bychkov-Rashba effect. Here, the inversion asymmetry of the heterostructure itself produces an average electric field $\bm{E} \parallel \hat{\bm{z}}$, leading to the characteristic Rashba Hamiltonian, which is indispensable for spin-momentum locking in two-dimensional electron gases \citep{PismaZhETF.39.66,fabian2007review,Winkler:684956}:
\begin{equation}
    \mathcal{H}_\text{R} = \alpha\left(\sigma_xp_y - \sigma_yp_x\right).
    \label{eq:spin--orbit-interaction}
\end{equation}
The strength of this Rashba coupling, parameterized by
$\alpha$, is determined by the electric field in the valence band of the quantum well and, crucially, can be tuned in situ via electrostatic gating, offering dynamic control over spin-orbit effects \citep{fabian2007review}. It is important to note that additional contributions, such as Dresselhaus SOC arising from bulk inversion asymmetry of the crystal itself, may also be present and can interfere with the Rashba term, its linear contribution reads \cite{Winkler:684956}
\begin{equation}
   \mathcal{H}_\text{D} = \beta\left(\sigma_xp_x - \sigma_yp_y\right).
    \label{eq:Dreselhaus-spin--orbit-interaction}
\end{equation}

The primary effect of the Rashba spin-orbit interaction, as described in Eq.~\eqref{eq:spin--orbit-interaction}, is to lift the Kramers degeneracy of the electronic bands, splitting the spin-up and spin-down states for all wavevectors except at the time-reversal invariant point $\bm{p}=0$. To lift this remaining degeneracy and create spin-polarized bands—a necessity for engineering topological superconductivity with effective p-wave pairing (see Section \ref{s:bottom-up}) —time-reversal symmetry must be broken. This is conventionally achieved by applying an external magnetic field $\bm{B}$ with the resulting Zeeman interaction
\begin{equation}
    \mathcal{H}_\text{Z} = -\frac{1}{2}g^*\mu_\text{B}\bm{B}\cdot\bm{\sigma},
    \label{eq:zeeman-interaction}
\end{equation}
where $g^*$ is the effective $g$-factor and $\mu_\text{B}$ the Bohr magneton.  The direction of the magnetic field should be chosen not parallel to the spin--orbit field and therefore depends on both the relative strength of Rashba and Dresselhaus spin--orbit interactions and the crystallographic orientation of the system \citep{Scharf2019,Pakizer2021,Pekerten2022}. 

However, engineering a system with optimal SOC involves navigating significant trade-offs. Crucially, strategies to enhance SOC can adversely affect other key parameters, particularly the electronic g-factor. For instance, narrowing a quantum well enhances the interfacial contribution to the Rashba effect and increases the linearized Dresselhaus coefficient \cite{Mayer2020,PhysRevB.86.195309}, but the associated confinement effects—such as an increased band gap and nonparabolic dispersion—can substantially suppress the g-factor \cite{Mayer2020,PhysRevB.35.7729,PhysRevB.101.205310}. Conversely, at low electron densities, many-body electron-electron interactions become more relevant and can have the opposite effect, enhancing the g-factor while suppressing SOC \cite{Maryenko2021}. A deliberate trade-off between a large g-factor and strong spin-orbit coupling must therefore be made when designing a quantum well, with SOC typically being maximized to the extent that the g-factor remains sufficient for a sizable Zeeman splitting.
Beyond well width, all available strategies—from fundamental material choice to electrostatic gate-tunability \cite{Wickramasinghe2018,Farzaneh2024} and device geometry—should be leveraged to maximize SOC.
Given the complex interplay of material and structural factors influencing the effective SOC, rigorous characterization of each semiconductor wafer is a prerequisite for understanding devices fabricated from it. Weak antilocalization (WAL) measurements in a Hall bar geometry provide a straightforward method to estimate the SOC strength. The WAL signature—a sharp peak in the magnetoconductance near zero magnetic field—arises from the Berry phase acquired by electrons traversing closed paths in the presence of a spin-orbit field. This phase leads to destructive interference that suppresses backscattering, and the magnitude and shape of the peak are direct proxies for the SOC strength. A key application of this technique is the extraction of the Rashba parameter $\alpha$ by fitting WAL peaks \cite{WAL1996}. Specifically, by measuring the WAL peak as a function of a gate voltage, one can directly probe how the structural inversion asymmetry and the resulting SOC are modulated by the electric field \cite{Wickramasinghe2018,Farzaneh2024}. 

Furthermore, WAL studies under an in-plane magnetic field reveal an anisotropy that serves as a fingerprint for the relative strength of Dresselhaus SOC. Quantifying this Rashba-Dresselhaus anisotropy is critical, as it has direct implications for e.g. determining the optimal in-plane field angle to induce topological superconductivity in planar Josephson junctions \cite{Scharf2019,Pakizer2021,Pekerten2022,Farzaneh2024}. Despite its utility, extracting quantitatively accurate SOC values that can be compared across disparate material systems from WAL alone remains challenging, often requiring complementary measurement techniques.

\subsubsection{Importance of \label{sec:g-factor}$g$-factor}

As we already mentioned, the effective Landé $g$-factor of a proximitized 2DEG is a critical material parameter for hybrid semiconductor-superconductor devices operating under a magnetic field. A large Zeeman energy is often a prerequisite for target phenomena, such as topological superconductivity. However, applying a strong external magnetic field presents a fundamental conflict: it simultaneously induces the desired Zeeman effect in the semiconductor while suppressing the parent s-wave superconducting gap through the orbital effect and the Zeeman-driven Chandrasekhar-Clogston limit. Consequently, a large semiconductor $g$-factor ($g^*$) is highly desirable, as it enables the attainment of significant Zeeman energies at lower magnetic fields, thereby preserving superconductivity.

\begin{figure}
    \centering
    \includegraphics{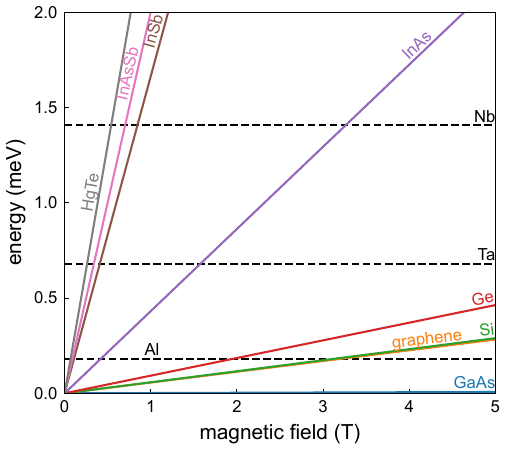}
    \caption[Zeeman]{{Material Selection for Zeeman-dominated Hybrid Devices}  This plot compares the achievable Zeeman energy $|{g^*}|\mu_\text{B} B / 2$ for various semiconductors (colored solid lines) in various semiconductors (colored solid lines) with the superconducting gaps $\Delta_S$ of several metals (black dashed lines). The large g-factor imbalance in systems like InAs/Al allows the semiconductor to reach a significant spin-splitting energy at a moderate magnetic field, where the superconducting gap remains largely unaffected. This illustrates the design principle of using a large $g^*$ semiconductor to generate large Zeeman effects without suppressing superconductivity via the Chandrasekhar-Clogston limit. Adapted and reprinted from Ref.~\cite{Schiela2024}.}
    \label{fig:material-choices}
\end{figure}

Material selection, however, is complicated by two key competing factors. First, the bulk $g^*$-factors of promising semiconductors (see Table~\ref{tab:semiconductor-material-parameters}) are often renormalized downward by electrostatic confinement and proximity-induced interactions in a quantum well \cite{Mayer2020}, reducing the effective Zeeman in the semiconducting channel. Second, and more fundamentally, one must contend with the properties of the parent superconductor itself, which typically possesses a much smaller intrinsic $g$-factor ($g_S \ll g^*$). This disparity is critical because superconductivity is destroyed when the Zeeman energy in the metal exceeds the Chandrasekhar-Clogston limit, $\frac{1}{2}g_S \mu_B B > \Delta_S / \sqrt{2}$, where $\Delta_S$ is the superconducting gap. This imposes a strict upper bound on the magnetic field, dictated by the weaker link in the heterostructure.

Consequently, the optimal design strategy involves engineering a heterostructure with a highly imbalanced $g$-factor ratio, $|g^*/g_S| \gg 1$. This asymmetry effectively localizes the Zeeman field's impact, allowing the semiconductor to reach the high spin-splitting energies required for applications like topological superconductivity, while the superconductor remains well below its destructive Chandrasekhar-Clogston limit. We illustrate this principle in Fig.~\ref{fig:material-choices}, plotting the effective Zeeman energy for various semiconductors against the magnetic field. The plot demonstrates that materials like InAs can generate Zeeman energies exceeding the gap of a conventional superconductor like Al at applied fields that remain safely below the superconductor's in-plane critical field \cite{doi:10.1126/sciadv.adf5500}, thereby fulfilling the core requirement for a viable hybrid system.

\begin{table*}[tbp]
    \centering
    \begin{threeparttable}
\begin{tabular}{lSS[table-format=+1.5]S[table-format=3.3]SSS[table-format=+2.2]S}
\toprule
{material}  & {$\alpha$ (\si{meV.\angstrom})} & {$\beta$ (\si{meV.\angstrom})} & {$E_\text{so}$ (\si{\micro eV})} & {$g^*$} & {$E_\text{Z}/B$ (\si{\micro eV/T})} \\
\midrule
Si  &  &  &  & 2.0 & 57 \\
Ge &  &  &  & -3.2\tnote{b} & -93 \\
AlAs  & 0.43 & 7.91 & 20.1 & 1.9 & 56 \\
GaAs  & 4.72 & 16.75 & 8.1 & 0.0 & -1 \\
GaSb  & 35.52 & 122.35 & 255.1 & -8.8 & -253 \\
InP  & 1.57 & -7.09 & 1.6 & 1.3 & 38 \\
InAs  & 112.49 & 33.33 & 133.9 & -14.9 & -431 \\
InSb  & 534.21 & 324.60 & 2516.6 & -57.5 & -1664 \\
InAsSb  & 325\tnote{a} &  & 305.0 & -69.2 & -2001 \\
ZnSe  & 1.057 &  & 0.0 & 1.3 & 38 \\
CdTe  & 6.930 &  & 1.1 & -1.2 & -34 \\
HgTe  &  &  &  & -27.2\tnote{c} & -786 \\
\bottomrule
\end{tabular}
\begin{tablenotes}
    \item[a] Refs.~\cite{Mayer2020,Moehle2021}.
    \item[b] At the $\Gamma$ point.  $g^*\lesssim 4$ in the L valley \cite{Roth1959,Roth-Lax1959,Watzinger2016}.
    \item[c] Up to -90 observed \cite{Jiang2023}.
\end{tablenotes}    \end{threeparttable}
    \caption{Material parameters for various zinc-blende semiconductors. Rashba parameters $\alpha$ and linearized Dresselhaus parameters $\beta$ are results of (extended) Kane model calculations from Refs.~\cite{fabian2007review,Winkler:684956}.  The spin--orbit energy $E_{so}=2m^*\left((|{\alpha}|+|{\beta}|)/\hbar\right)^2$ is estimated for a Fermi level within the $p=0$ Zeeman gap.  The effective g-factor, from which the Zeeman energy per unit field  $E_\text{Z}/B = g^*\mu_\text{B}/2$ is also calculated, is estimated from the Roth formula \cite{Mayer2020,Roth1959} $g^*=g-\frac{2}{3}(\frac{1}{E_{g}}-\frac{1}{E_{so}-E_{g}})E_{p}$, where $g\approx 2$ is the free electron g-factor, while $E_{g}$ and $E_{p}$ are the band gap and the Kane energy, respectively (we use band parameters from Refs.~\cite{Adachi2005,Adachi2009}). Adapted and reprinted from Ref.~\cite{Schiela2024}.}
    \label{tab:semiconductor-material-parameters}
\end{table*}

\subsubsection{Recent progress in hybrids with new materials combinations \label{sec:new-hybrids}}

\begin{figure}[t]
\centering
   \includegraphics[width=0.6\linewidth]{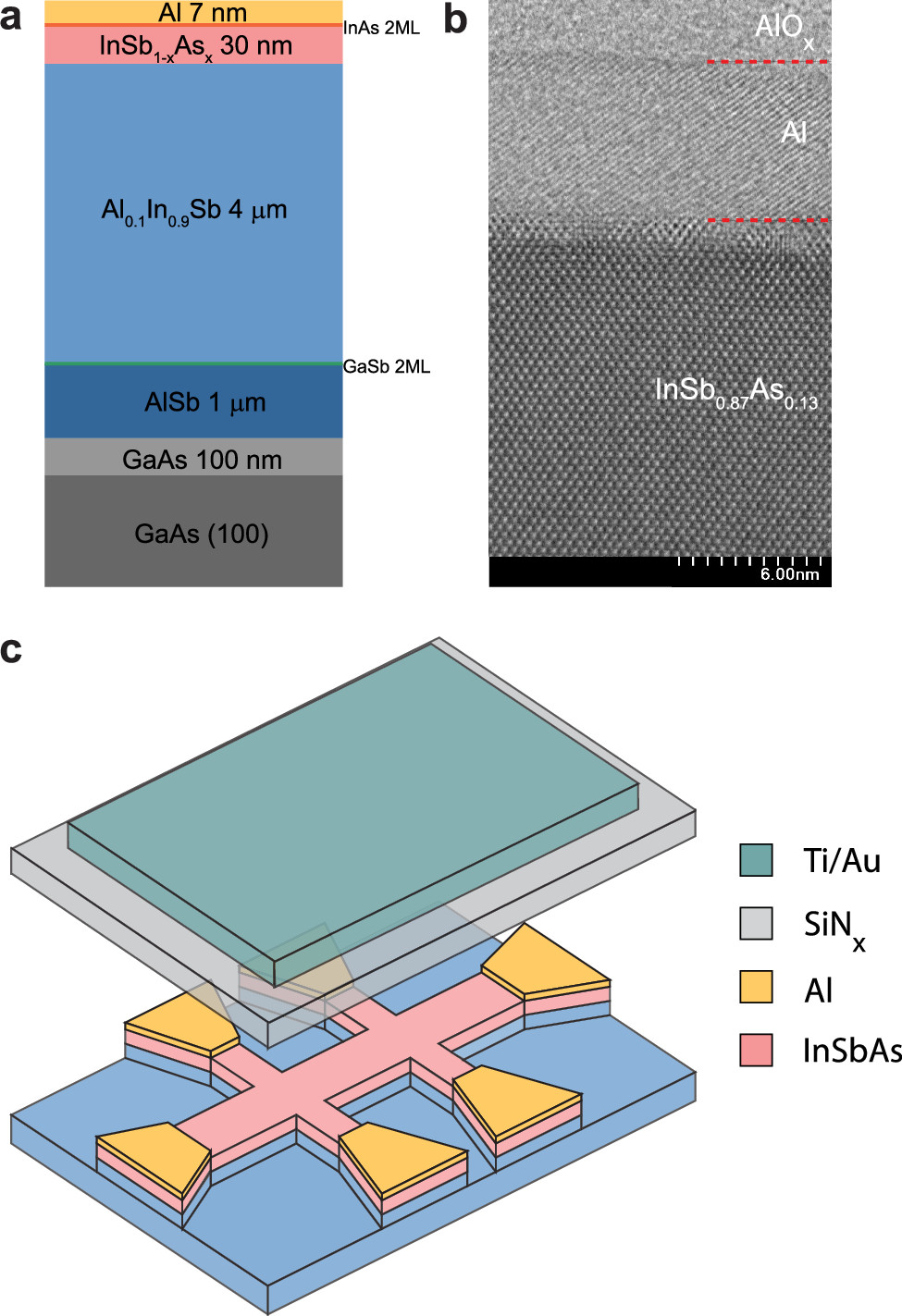}   
   \caption{{Hybrid Al/InSbAs heterostructures}. (a) Layer stack of the Al/InSb${}_{1-x}$As${}_{x}$ hybrid heterostructures. (b) Bright-field scanning transmission electron micrograph of the Al-InSb$0.870$As$0.130$ interface along the [110] zone axis of the semiconductor. Red lines indicate the boundaries of the aluminum. (c) Schematic of a Hall bar that is used to extract the 2DEG properties. Adapted and reprinted from Ref.~\cite{Moehle2021}.}
    \label{fig:InSbAs}
\end{figure}

The recent InSbAs coupled to in situ grown Al system \cite{Moehle2021} provides a compelling case study to illustrate the ideas presented in this Section. It features a high g-factor ($g^*\approx$55), spin-orbit coupling stronger than in InAs or InSb, and a hard induced superconducting gap thanks to its pristine interface. These combined properties were instrumental in realizing a minimal Kitaev chain in a 2DEG, a key achievement detailed in Section \ref{s:bottom-up}.
\begin{figure}[t]
    \centering
   \includegraphics[width=\linewidth]{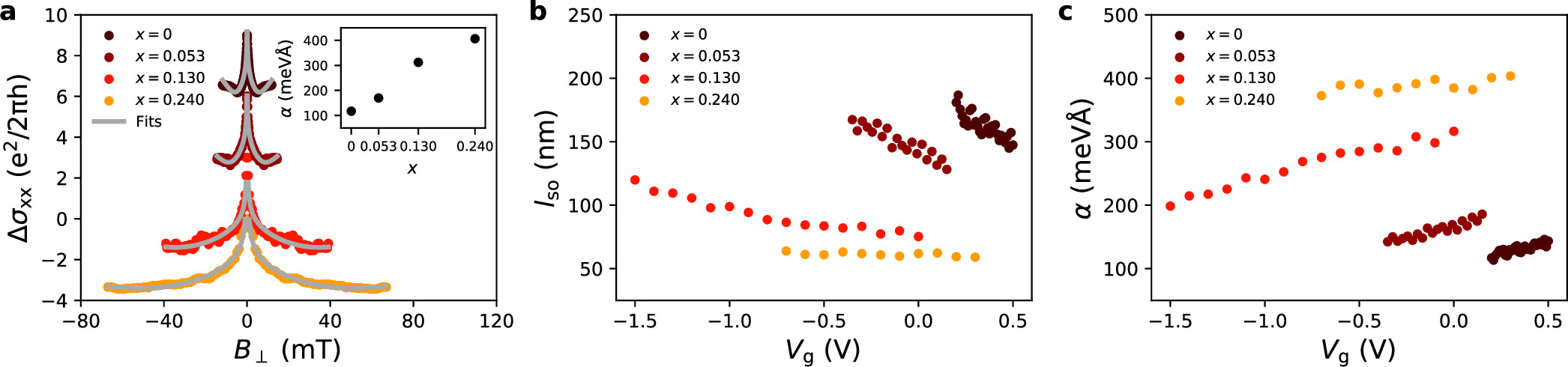}   
\caption{{Large and tunable spin orbit coupling}.(a) Magneto-conductivity correction at $V_g$ = 0 V for different InSb$_{1–x}$As$_x$ 2DEGs. The $x$ = 0 curve is measured at $V_g$= 0.2 V due to a high resistance at $V_g$ = 0 V. The gray lines are theory fits \cite{WAL1996} to the weak antilocalization data. In the inset, the extracted linear Rashba coefficient is plotted for the four As concentrations, showing a monotonic increase with increasing As concentration. For the higher As concentrations, $\alpha$ is 3-4 times larger than the value for pure InSb ($x$ = 0). (b) Spin orbit length plotted against $V_g$. (c) Spin orbit coupling $\alpha$  as a function of $V_g$. The extracted spin orbit length $l_{so}$ ($\alpha$) decreases (increases) with increasing As concentration when compared at a fixed gate voltage. Adapted and reprinted from Ref.~\cite{Moehle2021}.}
    \label{fig:SO-InSbAs}
\end{figure}
Specifically, InSb$_{1–x}$As$_x$ 2DEGs with varying As concentration are grown by MBE on undoped, semi-insulating GaAs(100) substrates (see Fig. \ref{fig:InSbAs}a for a schematic of the layer stack). The growth starts with a 100 nm GaAs buffer layer, directly followed by a 1~\si{\micro\meter} thick AlSb nucleation layer~\cite{Goldammer1999} and a 4~\si{\micro\meter} thick Al0.1In0.9Sb layer. The latter forms a closely matched pseudosubstrate for the InSb$_{1–x}$As$_x$ growth and the bottom barrier of the quantum well~\cite{Lehner_PRM2018}. The As concentration in the InSb$_{1–x}$As$_x$ is controlled by the growth temperature and the As flux. In Ref. \cite{Sestoft2018,Moehle2021}, heterostructures with x = 0, 0.053, 0.080, 0.130, 0.140, and 0.240 are grown. The semiconductor growth is terminated by the deposition of two monolayers (ML) of InAs, serving as a screening layer to prevent intermixing between the semiconductor structure and the superconducting Al layer. After the semiconductor growth, the heterostructures are transferred under ultrahigh vacuum to a second MBE chamber to deposit 7 nm of Al, using methods described in Ref.~\cite{Thomas_PRM2019}. Figure \ref{fig:InSbAs}b displays a bright-field scanning transmission electron micrograph focusing on the InSb$_{1–x}$As$_x$ interface for x = 0.130. The interface appears sharp with a slight change of atomic contrast that is attributed to the relaxed InAs screening layer. The semiconducting properties of the InSb$_{1–x}$As$_x$ 2DEGs can be characterized by removing the Al in the active device area to fabricate Hall bars. After the Al removal, the 2DEG is etched in unwanted areas, followed by the deposition of a SiNx dielectric layer. Lastly, a Ti/Au top gate is evaporated and used to control the electron density in the 2DEG (see Fig. \ref{fig:InSbAs}c for a schematic), which allows to measure peak mobilities of $\approx$20000–28000 cm2/Vs. 
Interestingly, weak antilocalization in magneto-conductivity measurements allows to demonstrate  
large and tunable spin–orbit coupling in these 2DEGs, which can be related to the concentration of As. A representative example is shown in Fig. \ref{fig:SO-InSbAs}. The systematic increase in spin orbit coupling with As concentration can arise from a combination of several effects. First, band structure calculations of InSb$_{1–x}$As$_x$ show that the Rashba parameter is strongly influenced by the As concentration \cite{Mayer2020,Winkler2016},  which has been observed in experiments on ternary nanowires \cite{Sestoft2018}. Second, electric fields across the 2DEG can also influence the spin–orbit interaction \cite{Metti2023}.
\begin{figure}[t]
    \centering
   \includegraphics[width=\linewidth]{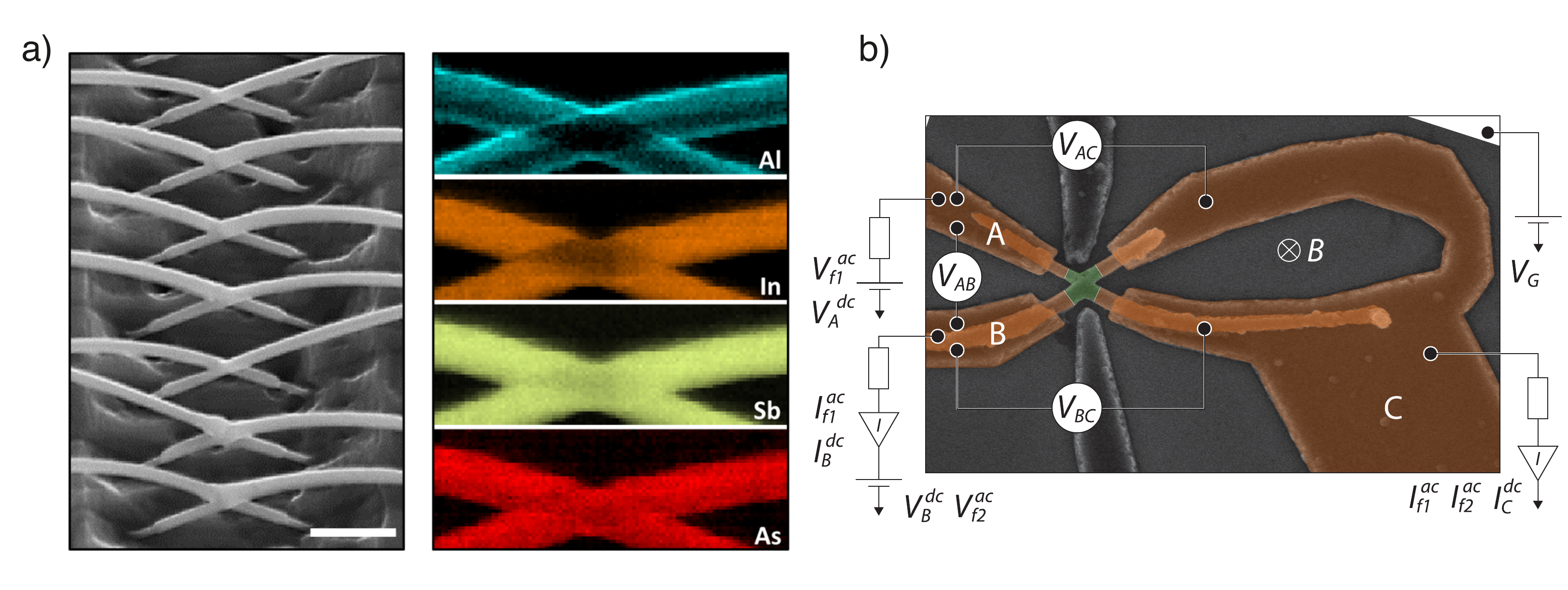}   
\caption{{Formation of the ternary network and multiterminal Josephson device}. (a) Scanning electron microscopy of InAsSb-Al nanocrosses grown from the V-grove trenches. To the right, Electron Energy Loss Spectroscopy elemental maps show the spatial distribution of In, As, Sb and Al. (b) False colored SEM of a typical multiterminal Josephson device. Superconducting contacts are in orange, and exposed semiconducting InAsSb in green. The multiterminal configuration allows for simultaneous measurements of the three differential resistances $R_{AB}\equiv dV_{AB}/dI_{AB}$, $R_{AC}\equiv dV_{AC}/dI_{AC}$ and $R_{BC}\equiv dV_{BC}/dI_{BC}$ as a function of different dc currents $I_A$, $I_B$ and $I_C$. Tuning of the superconducting phase difference $\phi$ is performed by a perpendicular magnetic field $B$ that threads a superconducting loop of area $A\sim 5\mu^2$, such that changes in the field by $\delta B = \Phi_0/A \sim 320 \mu T$ modulate $\phi$ by $2\pi$. Adapted and reprinted from Ref.~\cite{kahn2025}.}
    \label{fig:nanocrosses-InSbAs}
\end{figure}
A groundbreaking method for nanoscale integration has been established through the molecular beam epitaxy of hybrid InAsSb–Al nanocrosses \cite{kahn2025}. This synthesis leverages an As-assisted fusion process where oppositely oriented InAsSb nanowires, grown from precisely aligned Au catalysts, merge seamlessly at their junctions. Critical to this process is the initial growth of an InAs stem, which enables subsequent Sb incorporation to form the ternary InAsSb composition. Following growth, a few-nanometer-thick Al layer was deposited on three facets. Fig. \ref{fig:nanocrosses-InSbAs}a shows the resulting ordered arrays, with electron energy-loss spectroscopy maps confirming the distinct spatial distribution of In, As, Sb, and Al. Furthermore, the underlying fusion mechanism was successfully modeled with a temperature-dependent phase diagram and directly confirmed by atomic-resolution imaging at the junctions.

This controlled fabrication of complex hybrid architectures opens a path for advanced quantum device engineering. The true potential of these nanocrosses is realized when they are fabricated into multiterminal Josephson junction devices (Fig. \ref{fig:nanocrosses-InSbAs}b). Moving beyond conventional two-terminal junctions, this multi-terminal configuration could enable a new class of quantum components. Potential applications include the precise, multi-directional control of superconducting quantum states and the execution of advanced gate operations that are currently challenging with standard architectures. Looking forward, this sophisticated platform opens yet another compelling path toward the braiding of Majorana quasiparticles, a pivotal step in the pursuit of topologically protected fault-tolerant qubits using the top-down approach detailed in Section \ref{s:bottom-up}.

Other recent materials combinations include platforms based on an InAs 2DEG combined with in-situ deposited Nb and NbTi superconductors, which offer a larger operating range in temperature and magnetic field due to their larger superconducting gap. In particular, the induced gap is approximately five times larger than the values reported for Al-based hybrid materials and indicates the formation of highly-transparent interfaces that are required in high-quality hybrid material platforms \cite{Telkamp2025}.

\section{Subgap states in superconducting nanostructures  \label{s:subgap}}
Superconducting junctions exhibit new properties when the insulating layer (I) in a standard Josephson junction (SIS) is replaced by a normal metal (N), forming an SNS junction within a superconducting circuit.
When only a few subgap levels, generically known as Andreev bound states (ABSs), are present in such an SNS junction, 
the Josephson potential acquires a higher harmonic content 
\begin{equation}
\label{higher-harmonic_Josephson}
V_J(\phi)=\sum_m E_{J,m}\cos(m\phi),
\end{equation}
with $E_{J,m\neq 1}$ being higher Fourier coefficients.
Interestingly, the replacement $-E_J \cos (\phi)\rightarrow V_J(\phi)$, Eq. (\ref{SC_Hamiltonian}),
opens up a great deal of possibilities for various implementations of new qubits, including novel forms of protection against different sources of noise. 
In what follows, we discuss phase-dispersing bound states emerging at subgap energies as the key mechanism behind the formation of such Josephson potentials in SNS junctions.
\subsection{Andreev bound states and supercurrent in weak links}
A unifying microscopic description of the Josephson effect in generic weak links considers the induced superconducting pairing in the non-superconducting regions due to the proximity effect \footnote{Early theoretical treatments of the dc Josephson effect in superconductor-normal metal-superconductor (SNS) junctions date back to the 1970s, with seminal contributions from Kulik \cite{Kulik1969}, Ishii \cite{Ishii1970}, and Bardeen and Johnson \cite{PhysRevB.5.72}. A major conceptual shift occurred following the scattering-matrix formulation of Andreev conductance at NS interfaces, introduced by Blonder, Tinkham, and Klapwijk in 1982 \cite{PhysRevB.25.4515}. Building on this framework, Furusaki and Tsukada later derived a general expression for the Josephson current in terms of Andreev reflection probability amplitudes \cite{FURUSAKI91}. The early 1990s then saw a series of pivotal works that firmly established the modern interpretation of the Josephson effect in terms of ABSs, including Furusaki and Tsukada \cite{PhysRevB.43.10164}; Beenakker \cite{PhysRevLett.67.3836}, see also \cite{BeenakkervanHouten1992}; and Bagwell \cite{PhysRevB.46.12573}, among many others.}. Specifically, at highly transparent NS interfaces, processes known as Andreev reflections may take place, whereby electrons are coherently retro-reflected back as holes with inverted spin and momentum, while transferring a Cooper pair into/from the superconductor; see, {\it e.g.} , Ref. \cite{tinkham_book}. The phase difference between the incoming electron and the reflected hole during an Andreev reflection depends on its energy ($\epsilon$) as compared to the pairing amplitude $\Delta$ in the superconductor side (which acts as a potential barrier) as
\begin{equation}
\label{Andreev-phase}
\phi_A=-\arccos \left(\frac{\varepsilon}{\Delta}\right)\pm \frac{\phi}{2},
\end{equation}
where the sign $\pm$ depends on whether we consider right or left moving electrons reflecting in a superconductor with phase $\pm \frac{\phi}{2}$. 

In SNS junctions, the constructive interference between Andreev processes at both NS interfaces leads to a coherent electron-hole superposition. Such superpositions result in standing waves with quantized energy leading to subgap states at energies below the superconducting gap, known as Andreev bound states (ABSs). ABSs are a special form of BdG quasiparticles, Eq. (\ref{BdGqp}), which can be viewed as the electron-hole counterparts of particle-in-a-box states of a quantum well where the boundaries are replaced by superconducting walls \footnote{An interesting connection between ABSs and odd-frequency pairing has been proposed in Refs.\cite{Tanaka2012-hx,Tanaka2024-fk}.}. In this process, we also need to consider the phase acquired by the electron-hole pair over a round trip in the normal region 
\begin{equation}
\label{normal-phase}
\phi_N=(k_e-k_h)L_N=\frac{2\varepsilon L_N}{\hbar v_F},
\end{equation}
where we have used the so-called Andreev approximation $k_F\gg k_e,k_h$, where one can linearize the electron/hole dispersion relation such that $k_{e/h}=k_F\pm\frac{\varepsilon}{\hbar v_F}$. Using Eqs. (\ref{Andreev-phase}) and (\ref{normal-phase}), the ABS energies are just the solutions of the transcendental equation:
\begin{equation}
\pm \phi-2\arccos \left(\frac{\varepsilon}{\Delta}\right)+(k_e-k_h)L_N=2\pi n,\; n\in N,
\end{equation}
which can be rewritten as
\begin{equation}
\label{ABS-energy}
\pm \phi-2\arccos \left(\frac{\varepsilon}{\Delta}\right)+2\frac{\varepsilon}{\Delta}\frac{L_N}{\xi}=2\pi n,\; n\in N,
\end{equation}
where in the last step we have used the definition of coherence length in the ballistic limit $\xi=\hbar v_F/\Delta$.
Restricting ourselves to the phase interval $\phi\in[0,2\pi]$, Eq. (\ref{ABS-energy}) can be rewritten as
\begin{equation}
\label{ABS-energy2}
\frac{\varepsilon}{\Delta}=\pm 
cos \left(\lambda\frac{\varepsilon}{\Delta}\pm\frac{\phi}{2}\right)=\pm cos \left(\lambda\frac{\varepsilon}{\Delta}\right)cos \left(\frac{\phi}{2}\right)\mp sin \left(\lambda\frac{\varepsilon}{\Delta}\right)sin \left(\frac{\phi}{2}\right),
\end{equation}
with $\lambda\equiv\frac{L_N}{\xi}$. Physically, the parameter $\lambda$ can also be related to the dwell time that electrons/holes spend in the normal region, which is of order $t_{dwell}\approx \frac{L_N}{v_F}$, such that $\lambda=t_{dwell}\frac{\Delta}{\hbar}=\frac{t_{dwell}}{t_{A}}$, with $t_{A}=\frac{\hbar}{\Delta}$ being the typical time scale involved in an Andreev reflection. Namely, $\lambda$ is the ratio between the dwell time spent in the normal region over the typical Andreev time. Alternatively, we can interpret it  as the ratio $\lambda=\frac{\Delta}{E_T}$, with $E_T=\frac{\hbar}{t_{dwell}}=\frac{\hbar v_F}{L_N}$ being the Thouless energy of the junction.

Although Eq. (\ref{ABS-energy2}) does not admit analytical solutions in general, we may find approximate expressions either in the limit $\frac{\varepsilon}{\Delta}\ll 1$ or $\lambda\ll 1$. In this latter case, we may expand it as:
\begin{equation}
\label{ABS-energy3}
\frac{\varepsilon}{\Delta}=\pm cos \left(\frac{\phi}{2}\right)\mp \lambda\frac{\varepsilon}{\Delta} sin \left(\frac{\phi}{2}\right),
\end{equation}
from which we can extract the solution
\begin{equation}
\label{ABS-energy4}
\frac{\varepsilon}{\Delta}=\pm \frac{cos \left(\frac{\phi}{2}\right)}{1+ \lambda sin \left(\frac{\phi}{2}\right)}=\pm \frac{cos \left(\frac{\phi}{2}\right)}{1+ \frac{t_{dwell}}{t_{A}} sin \left(\frac{\phi}{2}\right)}=\pm \frac{cos \left(\frac{\phi}{2}\right)}{1+ \frac{\Delta}{E_T} sin \left(\frac{\phi}{2}\right)}.
\end{equation}

\subsubsection{Short junction limit \label{sub-sec_short-junction}}
The simplest limit is the so-called short junction limit ($L_N\ll \xi$), where one can neglect the $sin \left(\frac{\phi}{2}\right)$ in the denominator of Eq. (\ref{ABS-energy4}) by just taking the limit $\lambda\rightarrow 0$ (which, physically, means zero dwell time $t_{dwell}=0$ inside the junction or, alternatively, $\frac{\Delta}{E_T}\rightarrow 0$) and obtain \cite{PhysRevB.43.10164,PhysRevLett.67.3836}.

\begin{equation}
\label{ABS-ballistic}
\varepsilon_A(\phi)=\pm\Delta\cos(\phi/2).
\end{equation}
In a realistic SNS junction, we need to take into account deviations from perfect ballistic transport due to back-scattering by impurities or Fermi velocity mismatch between the S and N regions which likely cause the formation of tunneling barriers at the NS interfaces. All these effects can be captured by the normal-state
junction transparency $\tau<1$, which is in turn related to the NS Andreev conductance \cite{PhysRevB.25.4515,PhysRevB.46.12841}. Within a scattering approach, one can still obtain an analytic solution for the ABSs energies with $\tau\neq 1$  \cite{PhysRevB.43.10164,PhysRevLett.67.3836} \footnote{We cite here the papers where this specific form  was first obtained, but note that the existence of such ABSs is implicit in much earlier studies of Josephson currents through weak links, see e.g. \cite{Kulik77}}:
\begin{equation}
\label{ABS}
\varepsilon_A(\phi)=\pm\Delta\sqrt{1-\tau \sin^2(\phi/2)}.
\end{equation}
Alternatively, this formula can also be obtained from Eq. (\ref{ABS-energy}) by including an extra phase shift $\phi_B$ due to tunneling through a potential barrier in the normal region \cite{PhysRevB.46.12573} with
\begin{equation}
\label{Bagwell-finite-tau}
\cos(\phi_B)=\tau\cos(\phi)+(1-\tau)\cos\left[\frac{(L_N-2x_0)}{\xi}\frac{\epsilon}{\Delta}\right],
\end{equation}
with $x_0$ being the potential barrier position.

\begin{figure*}
\begin{center}
\includegraphics[width=0.75\textwidth]{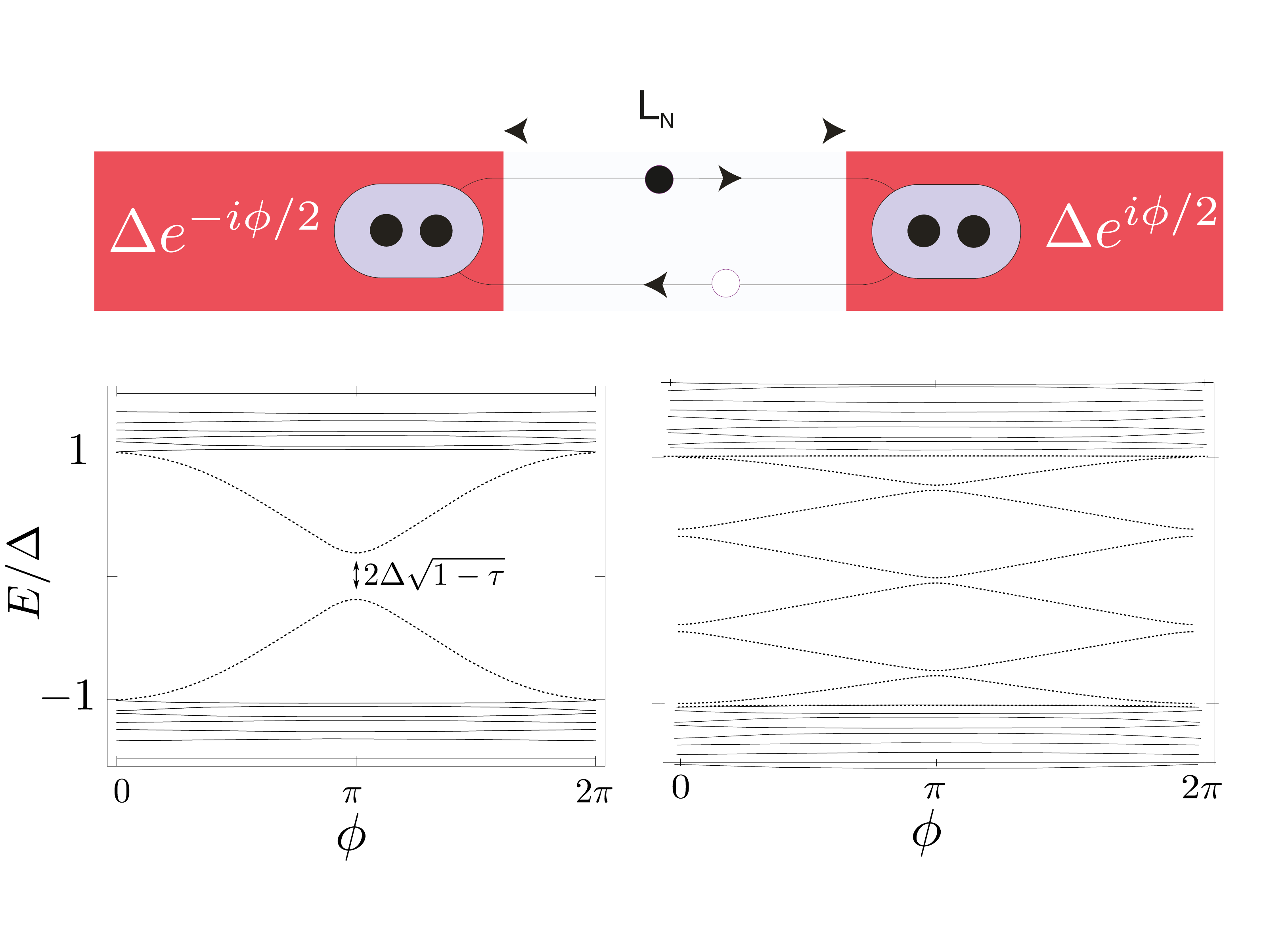}
\end{center}
\caption{Top: Schematic of a single channel SNS junction with transparency $\tau$, normal length $L_N$ and superconducting phase difference $\phi$. Bottom left: the subgap ABS spectrum in the short junction limit with $L_N\ll \xi$, c. f. Eq. (\ref{ABS}). Bottom right: ABS spectrum in the long junction limit with $L_N\gg \xi$.} 
\label{Fig:1}
\end{figure*}

The bottom left panel of Fig. \ref{Fig:1} shows a typical short junction ABS spectrum with its characteristic avoided crossing  of order $2\Delta\sqrt{1-\tau}$ near phase $\phi=\pi$ \footnote{Note that such anticrossing at phase $\phi=\pi$ is still finite even for a ballistic junction with $\tau=1$ when the so-called Andreev limit, namely the assumption that the chemical potential is much larger than the superconducting gap $\mu\gg\Delta$, is not fulfilled. Deviations from such Andreev limit are relevant for semiconducting junctions near depletion $\mu\approx 0$.}. 

In the context of this review, it is important to note that ABSs in a short junction are
an interesting realization of a two-level system with well-characterized parameters \cite{IvanovFeigelman1999,PhysRevB.64.140511,Zazunov2003}. Indeed, they were proposed as a solid
state realization of a qubit \cite{Zazunov2003} many years before the experimental demonstrations that we will discuss in detail in section \ref{s:qubits}. This can be formalized by understanding the ABS energies in Eq. \eqref{ABS} as the eigenvalues of the Hamiltonian
\begin{equation} 
\label{ABS-qubit1}
\mathcal{H}_{A} = \Delta[\cos(\phi/2)\tilde\sigma_z+\sqrt{1-\tau}\sin(\phi/2)\tilde\sigma_x],
\end{equation}
with Pauli matrices $\tilde\sigma$ written in the so-called ballistic basis of transparent channels (i. e. the one describing the ABSs in Eq. \eqref{ABS-ballistic}), while it reads 
\begin{equation} 
\label{ABS-qubit2}
\mathcal{H}_{A} = \varepsilon_A(\phi)\sigma_z,
\end{equation}
with the Pauli matrix now written in the basis of the Andreev states of Eq. \eqref{ABS}. 
\subsubsection{Long junction limit \label{sub-sec_long-junction}}
In a long junction, the ratio $\frac{L_N}{\xi}$ is no longer negligible and Eq.~(\ref{ABS-energy}) admits multiple solutions.
While in the short junction limit, the formation of ABSs requires that the solution of Eq.~(\ref{ABS-energy}) is bounded by the superconducting gap, $|\varepsilon|=\Delta$, the dwell time in a long junction is no longer zero, and the relevant energy scale is instead the Thouless energy, $E_T=\frac{\hbar}{t_{dwell}}=\frac{\hbar v_F}{L_N}<\Delta$. In such situation, the number of ABSs per conducting channel roughly equals $\lambda=\frac{t_{dwell}}{t_A}=\frac{\Delta}{E_T}=\frac{L_N}{\xi}$. Therefore, as the junction becomes longer, the Thouless energy (dwell time) becomes smaller (longer), such that many ABSs appear inside the gap, as illustrated in the bottom right panel of Fig \ref{Fig:1}.

If we keep $\lambda\neq 0$ in Eq. (\ref{ABS-energy2}), we may still find analytical solutions in the limit where we only consider low-lying ABSs $\frac{\varepsilon}{\Delta}\ll 1$. This can be easily obtained by expanding $\arccos \left(\frac{\varepsilon}{\Delta}\right)\approx \frac{\pi}{2}-\frac{\varepsilon}{\Delta}$, such that Eq. (\ref{ABS-energy2}) can be rewritten as:
\begin{equation}
\label{ABS-energy-long}
\frac{\varepsilon}{\Delta}= \frac{ (2n+1)\pi\mp\phi}{2(1+\lambda)}.
\end{equation}
Except close to the gap boundaries, where the approximation $\frac{\varepsilon}{\Delta}\ll 1$ is no longer valid, Eq. (\ref{ABS-energy-long}) describes very well the exact ABS spectrum of a long junction, in particular the almost linear phase dispersion near $\phi\approx\pi$.

In terms of the dwell times, Eq. \eqref{ABS-energy-long} can be nicely rewritten as:
\begin{equation}
\label{ABS-energy-long2}
\frac{\varepsilon}{\Delta}= \frac{\hbar}{2\Delta}\frac{ (2n+1)\pi\mp\phi}{(t_A+t_{dwell})}.
\end{equation}
Alternatively, we can write \eqref{ABS-energy-long2} in terms of the relevant energy scales, which gives
\begin{equation}
\label{ABS-energy-long3}
\frac{\varepsilon}{\Delta}= E_T\frac{ (2n+1)\pi\mp\phi}{2(E_T+\Delta)}\approx\frac{\xi}{2L_N}[(2n+1)\pi\mp\phi],
\end{equation}
where in the last step we have assumed  $\Delta\gg E_T$. Following the previous analogy, where ABSs can be understood as the electron-hole counterparts of particle-in-a-box states of a quantum
well, their energy decreases with the well length (the normal region), as expected.
\subsubsection{Current phase relations and critical currents in the short junction limit}
Andreev reflections, that lead to the formation of ABSs, are the intrinsic microscopic mechanism that mediates the supercurrent flow in the Josephson effect, as first discussed by \cite{Kulik1969} (see also Refs. \cite{Ishii1970,PhysRevB.5.72,FURUSAKI91}).
For the simple short junction case discussed here, it is a good approximation to neglect the contribution from states above the gap, which allows us to write the finite-temperature supercurrent in a very compact form as \cite{PhysRevLett.67.3836}\footnote{For further detailed discussions of novel current-phase relationships beyond the standard tunneling formula and including different pairing symmetries, see e.g. Refs.\cite{PhysRevB.56.892,Tanaka2024-fk}.}
\begin{equation}
\label{currrentABS_Temp}
I_J^{T\neq 0}(\phi)=\sum_i I_i(\phi)=\frac{e\Delta}{2\hbar}\sum_i\frac{\tau_i\sin\phi}{\varepsilon(\phi)}\tanh{\left(\frac{\varepsilon_i(\phi)}{2k_B T}\right)},
\end{equation}
where $T$ is the temperature and $\varepsilon_i$ is the energy of the subgap states, given by Eq.~\eqref{ABS}. This case considers the multichannel situation with $\tau_i$ being the transmission of each channel, that relates to the normal conductance of the junction as $G_N=2e^2/h\sum_i \tau_i$. For $T=0$, the expression simplifies as
\begin{equation}
\label{currrentABS}
I_J(\phi)=-\frac{2e}{\hbar}\frac{dV_J(\phi)}{d\phi}=-\frac{2e}{\hbar}\sum_i\frac{d\varepsilon_i(\phi)}{d\phi}=\frac{e\Delta}{2\hbar}\sum_i\frac{\tau_i\sin(\phi)}{\sqrt{1-\tau_i \sin^2(\phi/2)}},
\end{equation}
where we have defined the multichannel generalization of a short-junction Josephson potential 
\begin{equation}
\label{Josephson-short-multi}
V_J(\phi)=-\Delta\sum_i\sqrt{1-\tau_i \sin^2(\phi/2)}\,.
\end{equation}
An interesting consequence of Eq. (\ref{currrentABS}) is the discretization of the critical current in the ballistic limit, where $N_0$ perfectly-conducting spin-degenerate channels contribute with a critical current $I_c=N_0\frac{e\Delta}{\hbar}$ \cite{PhysRevLett.66.3056}, independent of the properties of the junction and, in particular, its width\footnote{Deviations from this perfect short junction limit result in critical currents containing geometry-dependent factors, see \cite{PhysRevLett.67.132}.}. This result is the superconducting analogue of the quantized conductance of a quantum point contact $G_N=N_0\frac{2e^2}{h}$ \cite{PhysRevLett.60.848}\footnote{This phenomenon of quantized conductance in superconducting junctions is by no means unique to transparent NS junctions. Other relevant systems with perfect NS conductance quantization include resonant Andreev reflection due to MZMs \cite{PhysRevLett.103.237001}, tunneling spectroscopy in d-wave superconductors \cite{PhysRevLett.74.3451}, diffusive junctions with triplet superconductors \cite{PhysRevB.70.012507,PhysRevB.71.094513}, etc.}. 

Depending on details of the SNS junction (number of channels, transparency), the Josephson current-phase relationship (CPR) in Eq. (\ref{currrentABS}) may strongly deviate from the standard Josephson effect in a SIS junction, see e.g. \cite{RevModPhys.76.411}.
This can be understood by writing the supercurrent as a Fourier expansion of the form $I_J(\phi)=\sum_n c_n(\tau_i)\sin(n\phi)$, where $n$ represents the number of the simultaneously tunneling Cooper pairs across the junction. From this expression, the sinusoidal limit for the current is recovered by assuming that only the first term contributes ($E_{J,m}=0$ for $m>1$ and $E_J\equiv E_{J,1}$),
\begin{equation}
\label{SIScurrent}
 I_J(\phi)=I_c \sin (\phi)=\frac{2e}{\hbar}E_J \sin (\phi),
\end{equation}
with $I_c$ being the maximum supercurrent of the tunnel junction and $E_J$ the Josephson energy. The tunnel SIS limit in Eq. \eqref{SIScurrent} can be recovered from the general SNS expression in Eq. \eqref{currrentABS} by assuming $N_0\gg 1$ poorly transmitting channels with $\tau_i\ll1$, in which case $I_c=\frac{\pi\Delta}{2e}G_N$ is the Ambegaokar-Baratoff formula \cite{PhysRevLett.10.486}, with $G_N=\frac{2e^2}{h}\sum_{i=1}^{N_0} \tau_i$ being the normal conductance of the junction\footnote{Apart from this tunneling limit, Eq. (\ref{currrentABS}) contains other interesting limits including the result first obtained by \cite{Kulik77} for a point contact in a disordered superconductor, which is essentially the classical limit of Eq. (\ref{currrentABS}) by considering many ballistic channels, see also \cite{Kulik78}.}. Equivalently, the tunnel limit is described by a Josephson potential
\begin{equation}
\label{SISJosephson}
V_J(\phi)=-E_J\cos(\phi)=-\frac{\Delta}{4}\sum_{i=1}\tau_i \cos(\phi),
\end{equation}
which can be obtained by expanding Eq. \eqref{Josephson-short-multi} as $V_J(\phi)\approx -\Delta\sum_{i}(1-\frac{\tau_i(1-\cos(\phi))}{4})$.

While Eq. (\ref{SISJosephson}) nicely captures all the complications of a mesoscopic multichannel junction into the single effective parameter $E_J$, recent experiments \cite{Pop-NaturePhysics2024} have shown that this standard expression usually fails to accurately describe the energy spectra of transmon qubits, fabricated with aluminum--aluminium oxide Josephson junctions, since the contributions from higher Josephson harmonics cannot be neglected. Physically, this implies that processes from the tunneling of $n>1$ Cooper pairs, that lead to contributions to the supercurrent of the form $\sin(n\phi)$, cannot be neglected, see also the discussion after Eq. (\ref{HJ1}). Later in the review, we will discuss again the role of higher harmonics in the context of parity-protected qubits (section \ref{s:protected-qubits})\footnote{ The consequences of deviating from this simple tunneling-limit formula are, of course, not confined to qubit physics; they can also be relevant in other contexts, such as multiple Andreev reflection experiments (see, e.g., Ref. \cite{lt7q-6ff3}). However, a detailed discussion of this lies beyond the scope of the present review.}.

Finally, let us finish this subsection by mentioning that, in general, the electrodynamics of a Josephson weak link in a circuit depends not only on the ABSs energy but also on their occupation. This can be included either as finite temperature or nonequilibrium population effects. 

\subsection{Subgap levels in quantum dot Josephson junctions: from Andreev to Yu-Shiba-Rusinov}
In the previous discussion about ABSs, we have assumed that the channels connecting the superconducting electrodes allow for coherent transport through ballistic segments. In contrast, it is quite common that SNS junctions defined in semiconductors near depletion form quantum dots (QDs). These QDs can also form by electrostatic confinement using gates. In this situation, charges localize in the channel and the effects of confinement and electrostatic repulsion (charging effects) become important. 

In the simplest approximation of a QD orbital being close to the superconductor's Fermi level, the standard theoretical model that describes a Josephson junction containing a QD is the single impurity Anderson model tunnel coupled to two superconducting reservoirs, so-called superconducting Anderson model. Despite its apparent simplicity, this model correctly captures the interplay between different physical mechanisms that appear owing to the competition between electron repulsion, due to Coulomb interactions, and electron pairing, coming from the coupling to the superconductor. The Hamiltonian of the superconducting Anderson model is
\begin{equation}
    \mathcal{H} = \mathcal{H}_{QD} +\mathcal{H}^L_S +\mathcal{H}^R_S + \mathcal{H}_T^S\,.
    \label{Eq:H_SAnderson}
\end{equation}
Here, $\mathcal{H}_{QD}$ refers to the decoupled QD Hamiltonian given by
\begin{equation} 
\mathcal{H}_{QD} =  \sum_\sigma \epsilon_\sigma d^\dagger_\sigma d_\sigma + U n_\uparrow n_\downarrow,
\end{equation} 
where $d_\sigma$ ($d_\sigma^\dagger$) annihilates (creates) an electron with spin $\sigma=\{ \uparrow, \downarrow \}$ and energy $\epsilon_\sigma = \epsilon_0 \mp V_Z$, where $V_Z=g\mu_BB/2$ is the Zeeman splitting due to the external magnetic field $B$. Here, $g$ and $\mu_B$ are the gyromagnetic factor and the Bohr's magneton, respectively. Electron-electron correlations in the dot occupations $n_\sigma = d^\dagger_\sigma d_\sigma $ are included by means of the charging energy $U$. 

At the mean-field level, the BCS pairing Hamiltonian at each lead is given by
\begin{equation} 
\label{Eq:H_leads_SQDS}
\mathcal{H}^{L/R}_{S} =  \sum_{ k_S \sigma}  \varepsilon^{L/R}_{k_S} c^\dagger_{k_S \sigma} c_{k_S \sigma} + \sum_{k_S} \Delta^{L/R}\left(c^\dagger_{k_S \uparrow} c^\dagger_{-k_S \downarrow} + \mbox{H.c.}\right),
\end{equation} 
where $c^\dagger_{k_S}$ ($c_{k_S}$) denotes the creation (destruction) operator in the lead with momentum $k_S$, spin $\sigma$ and energy $\varepsilon_{k_S}$, which is measured with respect to the chemical potential of the superconductor. The complex  superconducting order parameter on each lead is $\Delta^{L/R}= |\Delta^{L/R}|e^{i\phi_{L/R}}$, where $\phi_{L/R}$ is the superconducting phase of the left/right lead. The remaining two terms in the Hamiltonian are the couplings between the QD and the left/right superconducting lead
\begin{equation} 
\mathcal{H}_{T}^S =  \sum_{\alpha\in L,R}\sum_{k_S \sigma} \left(V^\alpha_{k_S} c^\dagger_{k_S \sigma} d_\sigma + \mbox{H.c}. \right),
\label{Eq:H_tunnel_SQDS}
\end{equation}  
which define the two tunneling rates: 
\begin{equation}
\Gamma_{L/R} = \pi \sum_{k_{S} }\left|V^{L/R}_{k_{S}}\right|^2 \delta(\omega - \epsilon_{k_{S}})=\pi\left|V^{L/R}_{{S}}\right|^2\rho^{L/R}_{{S}},
\label{rates_SQDS}
\end{equation} 
with $\rho^{L/R}_{{S}}$ being the normal density of states of the S leads evaluated at the Fermi energy \footnote{This so-called wideband approximation assumes that both the tunneling amplitudes and the density of states in the reservoir are frequency-independent (roughly given by their value evaluated at $k_F$),  which allows to write the tunneling couplings as constants, see e.g. Ref. \cite{PhysRevLett.68.2512}. The inclusion of Non-Markovian corrections due to the frequency dependence of the reservoir, see e.g. Ref. \cite{Marcos2011}, is beyond the scope of this discussion.}. The total coupling between the QD and the superconducting electrodes is $\Gamma=\Gamma_{L}+\Gamma_{R}$ and the superconducting phase bias is $\phi=\phi_R-\phi_L$.

The physics of the superconducting Anderson model is very rich, in particular when interactions are included.  This leads to an interesting competition between different physical regimes: A large charging energy favors single-electron doublet occupancy of the QD and thus a spin doublet, $ |D\rangle$, with spin $\frac{1}{2}$, while a strong coupling to the superconducting leads favors double occupancy in a singlet configuration $|S\rangle$ with zero spin. A quantum phase transition between the singlet and doublet ground state can occur as system parameters such as the QD energy level and the coupling strength are varied. The latter also controls the nature of the singlet ground state, which can be of either BCS or Kondo
type, depending on the ratio between the induced pair correlations and interaction strength, $\Delta/U$. Transitions between the ground state and the first excited state of the system, i.e., between a doublet and a singlet state or vice-versa, are manifested as a subgap resonance at energies $\pm\varepsilon_A$. Changes in the parity of the ground state of the system appear as points in parameter space where $\varepsilon_A$ changes sign (at the quantum phase transition point, where the two ground state parities are degenerate, this corresponds to a zero-energy excitation, namely a zero-energy crossing of the subgap states). In what follows, we discuss in detail the physics that governs such subgap states in the different regimes (for an early review on the topic, see Ref. \cite{Martin-Rodero-Levy-Yeyati2011}).
\subsubsection{Non-interacting limit (resonant level model)}
Before explaining the interacting model, it is very useful to understand in detail the non-interacting case with $U=0$, where
 the resonant state inside the QD dominates the physical properties of the junction \cite{10.1007/978-3-642-77274-0_20}. Specifically, in the single resonant level limit, one can still use Eq. \eqref{ABS} for the resulting ABSs
\begin{equation}
\label{ABS_QD}
\varepsilon_{A}(\phi)=\pm\tilde\Delta\sqrt{1-\tau \sin^2(\phi/2)},
\end{equation}
where 
\begin{equation}
\tau=\frac{\Gamma^2}{\epsilon^2_0+\Gamma^2}   
\end{equation}
 is now the transmission through a Breit-Wigner resonance of energy $\epsilon_0$ and width $\Gamma=\Gamma_L+\Gamma_R$ corresponding to the total tunneling rate between the two superconducting electrodes and the resonant
level. The energy scale $\tilde\Delta$ corresponds to the position of the subgap states at $\phi=0,2\pi$ and can be obtained from the transcendental equation:
\begin{eqnarray}
 \label{ABSphi0}
\tilde\Delta^2 \left[ 1+\tau\frac{\Delta^2-\tilde\Delta^2}{\Gamma^2}+2\tau\sqrt{\frac{\Delta^2-\tilde\Delta^2}{\Gamma^2}}\right]=\Delta^2,
\end{eqnarray}
which defines various interesting limits. Specifically, it goes from $\tilde\Delta\approx\Delta$  for $\Gamma\gg\Delta$ and $\tilde\Delta\approx\Gamma$ in the opposite large gap $\Delta\gg\Gamma$ limit. For $|\epsilon_0|\ll\Delta$, Eq. \eqref{ABS_QD} can be conveniently rewritten as \cite{Tanaka_2007,PhysRevB.104.174517,Vakhtel2023} 
\begin{equation}
\label{ABS_QD2}
\varepsilon_{A}(\phi)=\pm\frac{\Delta}{\Delta+\Gamma}\sqrt{\epsilon_0^2+\Gamma^2\cos^2(\phi/2)+(\Gamma_L-\Gamma_R)^2\sin^2(\phi/2)},
\end{equation}
which is valid for $\Delta\gg\Gamma$ at any $\phi$ or $\Gamma\gtrsim\Delta$ provided that $|\pi-\phi|\ll 1$. The prefactor $\frac{\Delta}{\Delta+\Gamma}$  characterizes the extent to which
the wave function of the ABS is localized at the quantum
dot. In the strong coupling limit $\Gamma\gg\Delta$, the support of the ABSs' wave function is mostly in the leads.
In the opposite weak coupling limit, $\Gamma\ll\Delta$, the wave function is predominantly localized at the
QD and $\frac{\Delta}{\Delta+\Gamma}\approx 1$. In this weak coupling limit, 
Eq. \eqref{ABS_QD2} can be rearranged as \cite{10.1007/978-3-642-77274-0_20,PhysRevB.104.174517,Vakhtel2023,Vakhtel2024}
\begin{equation}
\label{ABS_QD3}
\varepsilon_{A}(\phi)=\pm\tilde\Gamma\sqrt{1-\tau\sin^2(\phi/2)},
\end{equation}
with $\tilde\Gamma^2=(\epsilon_0^2+\Gamma^2)$, and $\tau=1-|r|^2$ the transparency of the junction, controlled by a reflection coefficient $r=(\epsilon_0+i\delta\Gamma)/\tilde\Gamma$  which depends on both the level position $\epsilon_0$ and the rate asymmetry $\delta\Gamma=(\Gamma_L-\Gamma_R)$. 

Note that the spectrum in Eq. \eqref{ABS_QD3} is similar to that of a SNS in Eq. \eqref{ABS}, but, importantly, the ABSs in this case
are detached from the continuum, and the wavefunctions are different, see e.g. Ref.\cite{PhysRevB.104.174517}. An extreme situation occurs for $\epsilon_0=\delta\Gamma=0$, which results in perfect transparency $\tau=1$, and hence
\begin{equation}
\label{ABS_QD4}
\varepsilon_{A}(\phi)\approx\pm \Gamma\cos(\phi/2).
\end{equation}
This ABS spectrum results in $4\pi$-periodic branches with a zero-energy level crossing at $\phi=\pi$ \footnote{Other examples of 4$\pi$-periodic subgap spectra include junctions with MZMs, see section \ref{s:bottom-up}, and in junctions with surface bound states due to unconventional superconductors, see e.g. Ref. \cite{Kashiwaya2000-we}}. Note that for $\Gamma\ll\Delta$ the subgap levels lie very deep inside the gap of the superconductor even at phases $\phi=2n\pi$.

Another important observation is that the spectrum in Eq. (\ref{ABS_QD3}) no longer depends on $\Delta$. In fact,
it directly follows from an effective Hamiltonian of the form
\begin{equation}
\label{ABS_QD3-effec}
\mathcal{H}_{\Delta\gg\Gamma} =  \sum_\sigma \epsilon_0 d^\dagger_\sigma d_\sigma-\left(\Gamma_\phi d^\dagger_\uparrow d^\dagger_\downarrow+ \mbox{H.c.}\right),
\end{equation} 
where $\Gamma_\phi=\Gamma_Le^{i\phi_L}+\Gamma_R e^{i\phi_R}$. Eq. (\ref{ABS_QD3-effec}) is purely local in terms of QD operators, which allows to rewrite it as a two-level system expressed in terms of Pauli matrices $\tilde\sigma$, acting on the Hilbert space
corresponding to the resonant level being empty or doubly
occupied:
\begin{equation} 
\label{ABS_QD3-qubit}
\mathcal{H}_{\Delta\gg\Gamma} = -\Gamma \cos(\phi)\tilde\sigma_x-\delta\Gamma\sin(\phi)\tilde\sigma_y-\epsilon_0\tilde\sigma_z,
\end{equation}
while in the eigenbasis of Eq. \eqref{ABS_QD3} it can simply be written as:
\begin{equation} 
\label{ABS-qubit}
\mathcal{H}_{\Delta\gg\Gamma} =\varepsilon_{A}(\phi)\sigma_z.
\end{equation}
The above expressions are fully equivalent to the ABS models in Eqs. \eqref{ABS-qubit1} and \eqref{ABS-qubit2} where the single channel is now replaced by the resonant level. 

\subsubsection{Large gap limit $\Delta\gg U$: BCS-like charge singlets \label{InteractingQDS}}
For finite charging energy $U\neq 0$, but still in a large gap limit with $\Delta/U\gg 1$ limit, we can write a Hamiltonian similar to the one in Eq. (\ref{ABS_QD3-effec}) with the only inclusion of the $U\neq 0$ term
\begin{equation} 
\label{SC-QD}
\mathcal{H}^S_{QD} =  \sum_\sigma \epsilon_\sigma d^\dagger_\sigma d_\sigma-\left(\Gamma_\phi d^\dagger_\uparrow d^\dagger_\downarrow+\mbox{H.c.}\right)+ U n_\uparrow n_\downarrow.
\end{equation}

In this so-called atomic limit \cite{PhysRevB.68.035105,Tanaka_2007,Bauer_2007,PhysRevB.79.224521}, the quasiparticles of the superconducting electrodes are removed from the problem \footnote{Note that the problem no longer depends on $\Delta$, but on the induced pairing $\Gamma_\phi$.} and the coupling to the superconductor only induces local superconducting correlations in the quantum dot \footnote{Technically, this is achieved by performing a low frequency limit in the tunneling self-energies resulting from the QD-SC coupling. This ultimately gives a purely local off-diagonal anomalous contribution parametrized by $\Gamma_\phi$.}. Since there are no quasiparticles involved, the Hilbert space is just four dimensional spanned by 
even fermionic parity sector $|0\rangle$, $|2\rangle$ we already used in 
Eq. (\ref{ABS_QD3-qubit}), and the odd fermionic parity sector spanned by the doublet $|\uparrow\rangle$, $|\downarrow\rangle$. 
Using this four dimensional Hilbert space, the Hamiltonian in Eq. \eqref{SC-QD} can be written in a compact form as \footnote{While, for simplicity, we are assuming $\epsilon_\uparrow=\epsilon_\downarrow=\epsilon_0$, this discussion can be trivially extended to the case where time reversal symmetry is broken, $\epsilon_\uparrow\neq\epsilon_\downarrow$ by e.g. an external Zeeman field.}:
\begin{equation}
\mathcal{H}^S_{QD} =\begin{pmatrix}
0 & \Gamma_\phi & 0 & 0\\
\Gamma_\phi & 2\epsilon_0+U & 0 & 0\\
0 & 0 & \epsilon_0 & 0\\
0 & 0 & 0 & \epsilon_0
\end{pmatrix}\,,
\label{matrix-SC-QD}
\end{equation}
which is block-diagonal in the even and odd fermion parity subspaces. Clearly, the odd-parity sector remains unaffected by superconductivity, which is a trivial consequence of the absence of quasiparticles in this large gap limit. The even sector, on the other hand can be readily diagonalized by performing a Bogoliubov transformation, see Eq. (\ref{BdG-transf})
\begin{eqnarray}
\label{BdG-QD}
\gamma^\dagger_{\uparrow}&=&ud^\dagger_\uparrow-v d_\downarrow,\nonumber\\  
\gamma^\dagger_{\downarrow}&=&ud_\uparrow+v d^\dagger_\downarrow,  
\end{eqnarray}
which results in the eigenstates
\begin{eqnarray}
\label{BCS-singlets}
&&|+\rangle=u|2\rangle+v^*|0\rangle,\nonumber\\
&&|-\rangle=-v^*|2\rangle+u|0\rangle,
\end{eqnarray}
with Bogoliubov amplitudes, c.f. Eq. (\ref{BdG-amplitudes}):
\begin{eqnarray}
u^2&=&\frac{1}{2}\left[1+\frac{\epsilon_0+U/2}{\sqrt{(\epsilon_0+U/2)^2+\Gamma^2\cos^2(\phi/2)}}\right],\nonumber\\
v^2&=&\frac{1}{2}\left[1-\frac{\epsilon_0+U/2}{\sqrt{(\epsilon_0+U/2)^2+\Gamma^2\cos^2(\phi/2)}}\right],
\end{eqnarray}
and energies 
\begin{equation}
E_\pm(\phi)=\epsilon_0+U/2\pm\sqrt{(\epsilon_0+U/2)^2+\Gamma^2\cos^2(\phi/2)}
\label{ABS-BCS1},  
\end{equation}
where, for simplicity, we have already considered symmetric couplings $\Gamma_L=\Gamma_R$, such that $\Delta_\phi=\Gamma_Le^{-i\phi/2}+\Gamma_Re^{i\phi/2}=\Gamma \cos(\phi/2)$.
Note that Eq.~\eqref{ABS-BCS1} can be rearranged like the resonant level model in Eq. \eqref{ABS_QD3},%
\begin{equation}
\label{ABS-BCS2}
E_\pm(\phi)=\epsilon_0+U/2\pm\tilde\Gamma\sqrt{1-\tau\sin^2(\phi/2)},
\end{equation}
with $\tilde\Gamma^2=(\epsilon_0+U/2)^2+\Gamma^2$ and 
\begin{equation}
\tau=\frac{\Gamma^2}{(\epsilon_0+U/2)^2+\Gamma^2}.  
\end{equation}
For $\epsilon_0=-U/2$ the system exhibits particle-hole symmetry
with $u^2=v^2$ (namely, singlet states which  are a perfect combination of the $|0\rangle$ and $|2\rangle$ charge states). 
Moreover, the system behaves as a perfect resonant channel with $\tau=1$. As expected for this resonant condition, the energy difference between the two singlet states reads $E_+(\phi)-E_-(\phi)=2\Gamma \cos(\phi/2)$.

Owing to the two possible fermionic parities in the problem, there are two possible ground states depending on whether $E_-<\epsilon_0$ (singlet ground state) or the opposite $\epsilon_0<E_-$ (doublet ground state)\footnote{Note that since $E_+>E_-$, the singlet $|+\rangle$ is always a high-energy state and does not enter this comparison.}. When $E_-=\epsilon_0$ there is a quantum phase transition between the two possible ground states of different fermionic parity. Specifically, the transition happens at
\begin{equation}
\label{boundary-singlet-doublet}
(\epsilon_0+U/2)^2+\Gamma^2\cos^2(\phi/2)=U^2/4,
\end{equation}
which defines a boundary between singlet and doublet ground states (for an example, see Fig.~\ref{Fig:ABS}c).

Physically, the quasiparticle operators in Eq. (\ref{BdG-QD}) describe transitions between the lower singlet state and the doublets $\gamma^\dagger_{\uparrow}|-\rangle=|\uparrow\rangle$ and $\gamma^\dagger_{\downarrow}|-\rangle=|\downarrow\rangle$ with energies $\epsilon_\uparrow-E_-$ and $\epsilon_\downarrow-E_-$, respectively. Such quasiparticle \emph{excitations} are the ones typically seen in experiments as subgap peaks in e.g. tunneling spectroscopy. Here, it is important to stress that since Eq. (\ref{boundary-singlet-doublet}) defines the condition for a singlet-doublet phase transition, across this boundary the energy of such excitations \emph{goes to zero}.

Finite $\Delta$ corrections to the above atomic limit are captured by the so-called generalized atomic limit where the original parameters of Eq. (\ref{SC-QD}) are rescaled by the quantity 
$1/(1+\Gamma/\Delta)$ \cite{Zonda1,Zonda2,Zonda3}.  
Remarkably, this rescaling results in a singlet-doublet boundary 
\begin{equation}
\label{boundary-singlet-doublet-finitegap}
(\epsilon_0+U/2)^2+\Gamma^2\cos^2(\phi/2)=\left[U/2(1+\Gamma/\Delta)\right]^2,
\end{equation}
which agrees very well with results obtained from Numerical Renormalization Group (NRG) calculations to a quantitative degree even for large $U>\Delta$ \cite{Zonda3}. Other extensions to include finite gap corrections include a self-consistent expansion of the tunneling self-energies around the low-frequency expansion from which Eq. (\ref{SC-QD}) is derived. This allows to extend the effective
local Hamiltonian to superconducting leads with a finite electronic bandwidth \cite{PhysRevB.79.224521}, which also results in very good agreement with NRG numerics. 
\subsubsection{Large charging limit $U\gg\Delta$: Yu-Shiba-Rusinov singlets}
In the opposite limit, $U\gg\Delta$, the QD is in the so-called Coulomb-dominated regime, where the number of electrons in the QD is well-defined, except to close to its ground state transitions~\cite{Balatsky2006}. In the limit of well-separated energy levels in the QD, the properties of the system can be understood by considering one spinful dot level coupled to a superconducting reservoir, (S-QD system). For a single dot occupation, the situation is identical to having an isolated magnetic impurity in a superconductor. For a purely classical magnetic impurity, the energy of the resulting subgap states, so-called Yu-Shiba-Rusinov (YSR) states \cite{Yu:APS65,Shiba:PTP68,Rusinov:SPJ69}, is well known since the 60s 
\begin{equation}
\label{YSR}
    E_{YSR}=\pm\frac{1-(JS\pi\rho_0/2)^2}{1+(JS\pi\rho_0/2)^2}.
\end{equation}
Here $J$ is the magnetic moment, $S$ the spin, and $\rho_0$ the DOS of the normal state. Within the classical YSR picture, the wavefunction of the bound state is localized near the impurity site with typical length scale 
$r_0\sim \xi_0/\sqrt{1-E^2_{YSR}}$ where $\xi_0$ is the superconducting coherence length \cite{Balatsky2006}.

Quantum fluctuations of QD's spin lead to Kondo-like correlations, mediated by the BdG quasiparticles outside the gap of the superconductor. These correlations leads to an exchange interaction $J\sim 2\Gamma/U$. This exchange interaction originates Kondo-like YSR singlet states, beyond the classical limit of Eq. (\ref{YSR}), which makes the problem rather involved,   see e.g. ~\cite{PhysRevB.91.045441}. Unlike the conventional Kondo effect, the superconducting electrodes host no states inside the gap. Hence, Kondo screening is incomplete. The YSR regime, $U\gg\Delta$, has two main differences, compared to the conventional BCS regime ($U\ll\Delta$): i) the YSR singlets have a nonlocal character corresponding to Kondo-like superpositions between the spin doublet and Bogoliubov quasiparticles in the superconductor, unlike the fully local BCS singlets. The screening length of the QD's spin depends on several properties, including the superconducting phase gradient~\cite{PhysRevB.109.125131}; ii) the boundary of the singlet-doublet transition in Eq. (\ref{boundary-singlet-doublet}) gets replaced by the condition $T_K\sim 0.3\Delta$, where $T_K$ is the Kondo temperature of the problem. Early work on hybrid QDs indicated the importance of Kondo-like correlations \cite{Eichler2007,Grove-Rasmussen2009}, while more recent experimental work has provided precise boundaries for the transition \cite{PhysRevB.95.180502} and many other aspects \cite{PhysRevLett.118.117001,Estrada-2020,GarciaCorral2020,PhysRevLett.130.136004}.

While a full solution of this intricate problem needs sophisticated numerics, such as e.g. the numerical renormalization group method \cite{PhysRevB.91.045441,Bulla2008}, deep in the Kondo limit, with $U\rightarrow\infty$, one can use renormalised parameters using e.g. slave boson theory and write an effective resonant level model as
\begin{eqnarray}
 \label{slave}
\varepsilon_A \approx\pm\tilde\Delta\sqrt{ 1-\tilde\tau sin^2(\phi/2)},
\end{eqnarray}
with $\tilde\tau=\frac{\tilde\Gamma^2_S}{\tilde\epsilon^2_0+\tilde\Gamma^2_S}$ and renormalised parameters defined through the constraint $\sqrt{\tilde\epsilon^2_0+\tilde\Gamma^2_S}=T_K\approx De^{-\frac{\pi |\epsilon_0|}{\Gamma_S}}$, with $D$ a high-energy bandwidth cutoff \cite{aguadoslave,alfredoslave}. In the unitary limit $\tilde\epsilon_0\rightarrow 0$ and $\tilde\Gamma_S\rightarrow T_K$, we can write the ABSs as \cite{alfredoslave,PhysRevB.75.045132}:
\begin{eqnarray}
 \label{Kondo}
\varepsilon_A \approx\pm\Delta\left[1-2\left(\frac{\Delta}{T_K}\right)^2\right]\cos(\phi/2),
\end{eqnarray}
which is essentially Eq. \eqref{ABS_QD3} for a zero-energy resonant level model written in terms of Kondo parameters.

\subsubsection{Intermediate regimes: the zero-bandwidth approximation}\label{InteractingQDS_numerics}
The interacting problem becomes analytically unsolvable in the limit where there is not a clear separation between the different energy scales. In this limit, only a numerical treatment can provide an exact solution to the Hamiltonian in Eq.~\eqref{Eq:H_SAnderson}. In equilibrium conditions, the numerical renormalization group method leads to an exact description of the low-energy properties of the system. The method is based on a logarithmic discretization of the superconductor's band that has an exponential resolution at low energies, close to the superconductor's Fermi level. Precise results on the energy of the ABSs, the phase diagram, and the supercurrent have been reported, see for example Refs.~\cite{Yoshioka_JPSJ2000,Choi_PRB2004,Oguri_JPSJ2004,Tanaka_2007,Lim_JPCM2008,Bauer_2007,Karrasch_PRB2008,Rodero_JPCM2012,Pillet_PRB2013,PhysRevB.91.045441}. Quantum Monte Carlo~\cite{Siano_PRL2004,Luitz_PRB2010,Luitz_PRL2012,Pokorny_PRR2021} and functional renormalization~\cite{Kadlecova_PRB2017,Karrasch_PRB2008} group are alternative numerical methods. The system under non-equilibrium conditions has also been studied by means of exact numerical techniques, like time-dependent density matrix renormalization group~\cite{Souto_PRB2021}.

Apart from the previously mentioned numerical methods, there exists a series of approximations that allow for a simpler treatment of the problem. These methods are based on reducing the number of relevant degrees of freedom, so the system can be diagonalized exactly, or perturbation expansions on some of the parameters. The simplest way of making the problem exactly solvable is to replace the sum over $k$ describing the electronic bath in Eq.~\eqref{Eq:H_leads_SQDS} by just one orbital.
Under this so-called zero-bandwidth approximation \cite{PhysRevB.68.035105,Bauer_2007,Affleck2000}, the resulting leads Hamiltonian is given by 
\begin{equation}
\mathcal{H}^{L/R}_{S} =  \varepsilon^{L/R} c^\dagger_{\sigma} c_{\sigma} + \Delta^{L/R}(c^\dagger_{\uparrow} c^\dagger_{\downarrow} + \mbox{H.c.}),
\end{equation} 
that describes two quasiparticle states at energies $\pm\Delta$. Consistently, the hopping between the QD and the superconductors has the same form as in Eq.~\eqref{Eq:H_tunnel_SQDS}. The model provides a good qualitative description of the system properties in a broad range of parameters, including the singlet to doublet ground state transition described before.
However, the absence of a real continuum in the superconductor imposes constraints on the validity range of the approximation. For instance, the approximation fails at describing the regime where the QD is strongly coupled to the superconductor near a phase difference $\phi\approx\pi$. This regime can be described by reformulating the superconducting Anderson model in the basis of symmetry-adapted
orbitals, using a set of states that permits to account
for the finite bandwidth. Using these ideas, even a minimal option consisting of a minimal basis set with a single orbital directly coupled to the dot, plus three additional indirectly coupled orbitals, allows to account for the quasiparticles in the vicinity
of the dot as well as those further away in the leads. This multi-orbital approximation provides insights into the spin-screening mechanisms and the nature of the doublet state for all coupling strengths and all values of phase bias, including near $\phi\approx\pi$, where a doublet ground state, even for large couplings, arises due to residual quasiparticles located close to the QD \cite{PhysRevB.109.125131}. Other extensions along similar lines include the surrogate model solver \cite{PhysRevB.108.L220506}, which is based on the exact diagonalization of a single-impurity Anderson model with discretized superconducting reservoirs including only a small number of effective levels. This surrogate model method has been shown to capture impurity-induced superconducting subgap states in quantitative agreement with the numerical renormalization group and quantum Monte Carlo simulations. 

\subsection{Spin splitting}\label{ss:spin-splitting}
One crucial condition to implement a superconducting spin qubit is to have phase-dependent spin split subgap levels in the junction. This condition is needed to have a finite spin-supercurrent coupling which can be subsequently used in cQED schemes, as we will later explain in subsection \ref{s:ASQ}. 
\subsubsection{Spin split Andreev levels in the long junction limit}
Finite length junctions allow to have spin-split levels, even in the absence of external magnetic fields, due to the finite dwell time that electrons spend in the finite junction. Physically, this can originate, for example, from a finite spin-orbit length leading to spin-dependent phase shifts obtained during propagation, due to a spin-dependent Fermi velocity in Eq.~(\ref{normal-phase}).

In order to illustrate this idea, let us consider a simple case where mixing between the two lowest transverse subbands may produce a strongly spin-dependent Fermi velocity  $v^\uparrow_F\neq v^\downarrow_F$, and hence coherence lengths $\xi_{\uparrow/\downarrow}=\frac{\hbar v^{\uparrow/\downarrow}_F}{\Delta}$, which leads to two spin-dependent quantization conditions according to Eq. (\ref{ABS-energy}). In the ballistic limit, and assuming that both $\lambda_{\uparrow/\downarrow}=L_N/\xi_{\uparrow/\downarrow}\ll 1$, we can write two spin-dependent copies of Eq. \eqref{ABS-energy4} as \cite{Park:PRB17} 
\begin{equation}
\frac{\varepsilon_{\uparrow/\downarrow}(\phi)}{\Delta } = \pm  \frac{\cos(\phi/2)}{1+\lambda_{\uparrow/\downarrow} \sin(\phi/2)}.
\label{ABS-SO2}
\end{equation}
The spin splitting between ABSs then reads
\begin{equation}
\varepsilon_\uparrow(\phi)-\varepsilon_\downarrow(\phi) = \frac{\Delta(\lambda_\uparrow-\lambda_\downarrow)\sin(\phi)}{2[1+\lambda_\uparrow \sin(\phi/2)][1+\lambda_\downarrow \sin(\phi/2)]}.
\label{ABS-SO3}
\end{equation}
This phase-dependent spin splitting is finite for $\phi\neq 0,\pi$, and comes from the difference in coherence lengths and Fermi velocities. Spin-degeneracy at $\phi=0$  and $\phi=\pi$ is protected by time-reversal symmetry. Deviations from perfect ballistic behavior with $\tau\neq 1$ can also be included by using Eq.~(\ref{Bagwell-finite-tau}) for each of the spin species, to obtain~\cite{Tosi2019} 
\begin{equation}
\label{Tosi-two-subbands}  
\frac{\tau\,\varepsilon}{\Delta}\cos(\lambda_\uparrow-\lambda_\downarrow)\mp\phi+(1-\tau)\cos\left[(\lambda_\uparrow+\lambda_\downarrow)\frac{2x_0}{L_N}\right]=\cos\left[2\arccos\left(\frac{\varepsilon}{\Delta}\right)-(\lambda_\uparrow+\lambda_\downarrow)\frac{\varepsilon}{\Delta}\right].
\end{equation}

\subsubsection{Spin split subgap levels in quantum dot Josephson junctions}
In the quantum dot limit, one can write a minimal extension of the single impurity Anderson model in Eq. \ref{Eq:H_SAnderson} that exhibits spin split subgap levels. This can be accomplished by a modified non-interacting part of the Hamiltonian of the form
\begin{eqnarray}
\label{H_spin}
\mathcal{H}_0
&=& \sum_\sigma \epsilon_\sigma d^\dag_\sigma d_\sigma 
+ \sum_{\alpha\in L,R}\sum_{k_S \sigma} \epsilon^\alpha_{k_S} c^\dag_{\alpha,{k_S}\sigma} c_{\alpha,{k_S}\sigma}  
+\sum_{\alpha\in L,R}\sum_{k_S} \Delta^\alpha \left(e^{i\phi_\alpha} c^\dag_{\alpha,{k_S}\uparrow}  c^\dag_{\alpha,{k_S}\downarrow} + \text{H.c.} \right)\nonumber\\
&+&\sum_{\alpha\in L,R}\sum_{k_S \sigma} \left( V^\alpha_{k_S} c^\dag_{\alpha,{k_S}\sigma} d_\sigma + \text{H.c.} \right)
+ \sum_{\alpha\in L,R}\sum_{k_S \sigma} \left( i W^\alpha_{k_S} c^\dag_{\alpha,{k_S}\sigma} d_{\bar{\sigma}} + \text{H.c.}\right)\nonumber\\
&+& \sum_{k_S,k_S',\sigma} \left( t_{LR} c^\dag_{{\rm L},{k_S}\sigma} c_{{\rm R},{k'_S}\sigma} + \text{H.c.} \right),
\end{eqnarray}
where in comparison with Eq. \eqref{Eq:H_tunnel_SQDS} there are two extra tunneling terms: The term proportional to the amplitudes $W^\alpha_{k_S}$ describes spin-flip tunneling between the impurity and the leads (the notation $\bar{\sigma}$ denotes spin inversion, $\bar{\uparrow} = \downarrow$, $\bar{\downarrow} = \uparrow$), which is physically motivated by the presence of a finite spin-orbit coupling in the normal region defining the quantum dot \footnote{Such spin-flip tunneling is expected in normal regions with lengths of the order of the spin-orbit length or longer.}, and defines two spin-flip (sf) tunneling rates which, within the wideband limit, read
$\Gamma_{\rm L}^{\rm sf}=\pi \rho |W_{{\rm L}}|^2$ and $\Gamma_{\rm R}^{\rm sf}=\pi \rho |W_{{\rm R}}|^2$. Finally, 
the term proportional to $t_{LR}$ is an additional direct tunneling term between the leads which can be formally obtained by integrating out all the possible cotunneling paths through high-lying levels, which provides  another conduction pathway through the dot. Formally, it reads
\begin{equation}
 t_{LR}= \sum_{k_S,k_S',j} \frac{V^{{\rm L}*}_{k_S;j} V^{{\rm R}}_{k'_S;j}}{\delta\epsilon_j},   
\end{equation}
where we sum over all cotunneling paths through levels of energy $\delta \epsilon_l$, and with  tunneling amplitudes $V^{{\rm L/R}}_{k_S;j}$. Note that this
 inter-lead hopping term makes the model resemble those for a QD embedded in a nanoscopic Aharonov-Bohm ring \cite{Karrasch2009} and importantly, as we will see in what follows, introduces a nontrivial dependence on the superconducting phase difference $\phi$. This model breaks down if the typical level spacing $\delta \epsilon_l$ becomes small as compared to the typical tunneling rates: in that case one should use explicitly a multi-orbital Anderson impurity model instead.
 
Interestingly, the generalized Anderson model resulting from Eq. (\ref{H_spin}) can be exactly solved in the doublet limit $U \to 0$, $\Delta_{\rm L, R} \to \infty$. Specifically, the effective Hamiltonian of the doublet sector in the ${\uparrow,\downarrow}$-basis is given by
\begin{equation}\label{eq:Heff}
\mathcal{H}_{\rm eff} = \begin{pmatrix} 
\epsilon_0 + E_{z}/2 & \left(E_{x} -i E_{y}\right)/2 \\
\left(E_{x} + i E_{y}\right)/2 & \epsilon_0 - E_{z}/2
\end{pmatrix} - \frac{2\pi \rho}{1+2\tilde{t}_{LR}^2\cos{\phi}+\tilde{t}_{LR}^4} \begin{pmatrix} 
a & c \\
c & a 
\end{pmatrix},
\end{equation}
where the first term explicitly considers the possibility of an arbitrary external Zeeman field $\vec{E}_{\rm Z}= (E_x, E_y,E_z)$, while the second term describes the intricate phase dependence of the spin-split Anderson model (even in this simplified version) through the coefficients 
\begin{equation}
\begin{split}
a &= (V_{\rm L} V_{\rm R}+W_{\rm L} W_{\rm R})\tilde t_{LR}(\tilde t_{LR}^2+\cos\phi), \\
b_i &= V_i^2 + W_i^2, \\
b &= e^{-i\phi/2}(b_{\rm R}+\tilde t_{LR}^2 b_{\rm L}) + e^{+i\phi/2}(b_{\rm L} + b_{\rm R} \tilde t_{LR}^2), \\
c &= (V_{\rm R} W_{\rm L} - V_{\rm L} W_{\rm R}) \tt\sin\phi, 
\end{split}
\end{equation}
where $\tilde t_{LR} = \pi \rho t_{LR}$, with $\rho$ being the normal-state
density of states. 
Assuming $V_{\rm L}=V_{\rm R} \equiv V$, $W_{\rm L} = -W_{\rm R} \equiv W$\footnote{Note that this sign for $W_i$ choice merely reflects the sign convention in the Hamiltonian and actually corresponds to the symmetric situation with the same amplitude for the left SC to QD and for the QD to right SC spin-flip tunneling.}, such that $\Gamma_{\rm L}=\Gamma_{\rm R}=\Gamma$ and $\Gamma_{\rm L}^{\rm sf}=\Gamma_{\rm R}^{\rm sf}=\Gamma^{\rm sf}$, 
the second term of Eq.~\eqref{eq:Heff} can be written as 
\begin{equation}
\frac{2\pi \rho}{1+2\tilde{t}_{LR}^2\cos{\phi}+\tilde{t}_{LR}^4} \begin{pmatrix} 
\left(\Gamma-\Gamma^{\rm sf}\right)\left(\tilde{t}_{LR}^2+\cos{\phi}\right) & 2\sqrt{\Gamma\Gamma^{\rm sf}}  \sin{\phi} \\
2\sqrt{\Gamma\Gamma^{\rm sf}}  \sin{\phi} & \left(\Gamma-\Gamma^{\rm sf}\right)\left(\tilde{t}_{LR}^2+\cos{\phi}\right)
\end{pmatrix}.
\label{eq:H_effectiveWVt}
\end{equation}
Assuming that $\tilde{t}_{LR}\ll 1$, we can perform a series expansion to obtain
\begin{equation}
\label{SO-QD-effective}
 \begin{pmatrix} 
E_{t} \cos{\phi} & E_{\rm SO} \sin{\phi} \\
E_{\rm SO} \sin{\phi} & E_{t} \cos{\phi} 
\end{pmatrix},
\end{equation}
where we have defined the effective parameters \begin{equation}
\label{eq:ESO}
E_{t} = 2\tilde{t}_{LR} \left(\Gamma-\Gamma^{\rm sf}\right),
\quad
E_{\rm SO} = 4\tilde{t}_{LR} \sqrt{\Gamma\Gamma^{\rm sf}}.
\end{equation}

It is important to stress that, within the limit considered here, we find that the effective $E_{\rm SO}$ depends on each of the three types of coupling in the model: spin-conserving, spin-flipping, and direct lead-lead tunneling. All three have to be present for the spin splitting to occur. Interestingly, this spin splitting has the same phase dependence $\propto sin (\phi)$ as the one we discussed for long junctions in Eq. (\ref{ABS-SO3}). Finally, we mention in passing that, given the potential cancelation of the $E_t$ term, it is prudent to include in the model an additional term of the form $E_D \cos(\phi)$ from processes that are higher-order in hybridization. This term will combine with $-E_t \cos(\phi)$ to produce a $+E_0 \cos(\phi)$ potential with $E_0=E_D-E_t$ in Eq.~\eqref{SO-QD-effective}.
Note that $E_0$ can take either a positive or negative sign. As we will discuss further in the review, the effective term in Eq.~\eqref{SO-QD-effective} forms the basis of the so-called Andreev spin qubit (see subsection \ref{s:ASQ}) when incorporated into a SC qubit model. Recently, Eq.~\eqref{H_spin} has also been used in Ref. \cite{ilicin2026} to explore the role of charging energy and SO coupling in the even sector, i. e. in the Andreev pair qubit (subsection \ref{s:ALQ}).

It is instructive to evaluate the eigenvalues of \eqref{eq:Heff} in the regime where ~\eqref{SO-QD-effective} holds. These are given by \footnote{Note that these levels can give rise to an anomalous Josephson effect: a shift of the
energy-phase relation minimum to a value $\phi_0$ different
from $\phi=0$ or $\pi$, or, equivalently, the presence of a nonzero
equilibrium supercurrent at $\phi=0$. For more details, see Ref. \cite{Bargerbos2023b}.} 
\begin{equation}
\varepsilon_{\uparrow/\downarrow}(\phi)  = E_{0} \cos{\phi} \pm \frac{1}{2}\sqrt{E_{y}^2+E_{z}^2 + \left(E_{x} - 2 E_{\rm SO} \sin{\phi}\right)^2}.
\end{equation}
For $E_{y} = E_{z} = 0$, this simplifies to 
\begin{equation}
\varepsilon_{\uparrow/\downarrow}(\phi)  = E_{0} \cos{\phi} \pm \left(E_{x}/2 - E_{\rm SO} \sin{\phi}\right),
\end{equation}
which results in a spin splitting 
\begin{equation}
\varepsilon_\uparrow(\phi)-\varepsilon_\downarrow(\phi) = E_{x} - 2E_{\rm SO} \sin{\phi}
\end{equation}
which is linear in the applied Zeeman field. Setting $E_{x} = E_{y} = 0$ instead, we find 
\begin{equation}
\varepsilon_{\uparrow/\downarrow}(\phi) = E_{0} \cos{\phi} \pm \frac{1}{2}\sqrt{E_{z}^2 + 4 E_{\rm SO}^2 \sin^2{\phi}},
\end{equation}
 resulting in a spin splitting 
\begin{equation}
\varepsilon_\uparrow(\phi)-\varepsilon_\downarrow(\phi) = \sqrt{E_{z}^2 + 4 E_{\rm SO}^2 \sin^2{\phi}}.
\end{equation}
Note that presence of the $\sin^2{\phi}$ term results in the doubling in periodicity against superconducting phase which, again, exemplifies the intricacies of the Josephson effect beyond the simple $\propto \cos{\phi}$ of a SIS junction.

The above ideas can be generalised to write a Josephson potential model \cite{Padurariu2010},

\begin{equation}
U_A(\phi) = E_0\cos\left(\phi\right) - E_{\rm SO}\, \vec{\sigma} \cdot \vec{n}\,\sin\left(\phi\right) +\frac{1}{2} \vec{E}_{\rm Z} \cdot\vec{\sigma}\,,
\label{eq:ESOpotential}
\end{equation}
where $\vec{n}$ is a unit vector along the spin-polarization direction set by
the spin-orbit interaction. It is important to stress that even in the absence of external magnetic field $\vec{E}_{\rm Z}=0$, the two spin branches are non-degenerate (except for $\phi=\pm n\pi$, with $n\in 0,1,...$), and given by 
\begin{equation}
\label{spin-Josephson}
\varepsilon_{\uparrow/\downarrow}(\phi) = E_{0} \cos{\phi} \pm E_{\rm SO}\sin{\phi}.
\end{equation}
Eq. \eqref{spin-Josephson} describes two sinusoids with an amplitude of $\sqrt{E^2_{0}+E^2_{\rm SO}}$ which, physically, can be interpreted as two spin-dependent Josephson potentials, with spins separated in phase space with minima at 
\begin{equation}
\phi_{\uparrow/\downarrow}^{\rm min}=\pi\pm\phi_{\rm SO}\,,\qquad \phi_{\rm SO}=\arctan\left(\frac{E_{\rm SO}}{E_{0}}\right),
\label{eq:ASQ_minima}
\end{equation}
Such displaced minima allow us to define an effective separation between the spin wells, $\phi^{\rm ASQ}=\min\{|\phi_{\uparrow}^{\rm min}-\phi_{\downarrow}^{\rm min}|,\,2\pi-|\phi_{\uparrow}^{\rm min}-\phi_{\downarrow}^{\rm min}|\}\leq\pi$. As we will discuss in subsection \ref{s:ASQ}, Eq. \eqref{spin-Josephson} forms the basis of the so-called Andreev spin qubit (ASQ). 
\subsection{Measuring subgap states}\label{s:spectroscopy}
If the reader will pardon the pun, distinguishing features originating from subgap states from subgap features related to other phenomena can be a nontrivial task. An obvious example are subgap transport features associated with multiple Andreev reflection (MAR) at an SNS junction that are not associated with transport through discrete quasiparticle excitations, such as an ABS or YSR subgap state. Arguably, the first clear tunnelling spectroscopy of individually resolved ABS
was reported in three experiments from 2010 by Deacon et al \cite{Deacon2010}, Dirks et al \cite{Dirks2011} and Pillet et al \cite{Pillet2010}, in QD-based hybrids fabricated with self-assembled InAs, graphene and nanotubes, respectively. Previous attempts, see e.g. Refs. \cite{Eichler2007,Grove-Rasmussen2009}, were less clear, with Kondo features, MAR and subgap states appearing in a somewhat complicated and complex mix. Since these pioneering experiments, several measurement techniques have been developed
to obtain information about ABSs in hybrid
junctions \footnote{Note that given the focus of this review we are purposefully omitting the vast literature on STM-based spectroscopy of subgap states, starting from pioneering works on the subgap structure of vortices to the observation of YSR states of individual magnetic atoms or chains on top of superconducting surfaces. For a recent review, see Ref. \cite{LoConte2024}.}. Here we briefly discuss the two more broadly used: tunneling
spectroscopy and microwave spectroscopy. 
\subsubsection{Tunneling spectroscopy}\label{s:tunneling-spectroscopy}
\begin{figure*}[h!]
\centering
\includegraphics[width=0.9\textwidth]{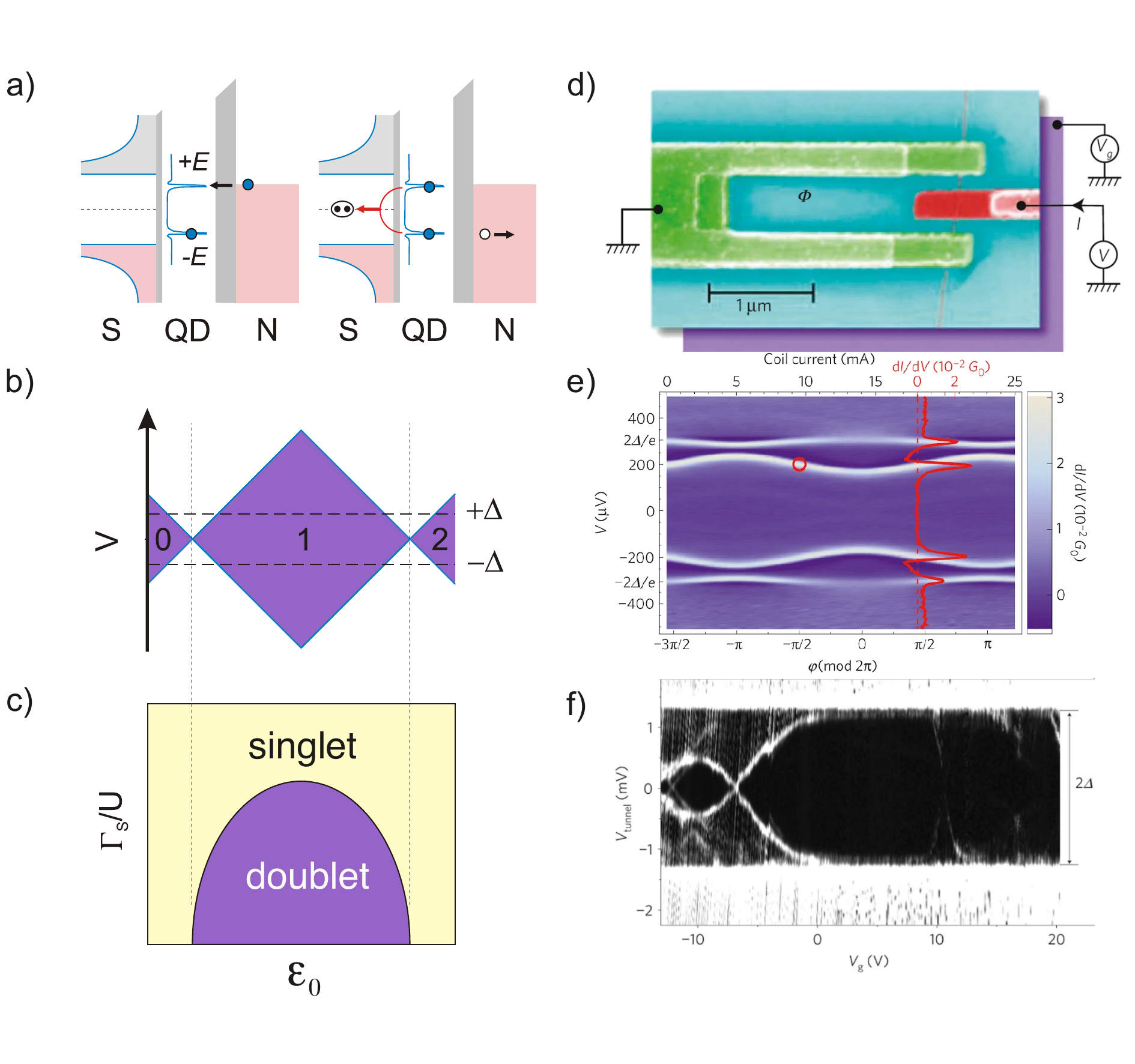}
\caption{a) Transport spectroscopy of ABSs formed within the SC gap by the Yu-Shiba-Rusinov (YSR) mechanism, where the confined spin (impurity) is screened by quasiparticles in the superconducting (S) lead. Resonant $dI/dV$ peaks in tunneling conductance are observed when the chemical potential of the N probe matches the bound state energy $\pm E$, which represents the excitation energy from the ground state of the QD-SC system to an excited state. The transport cycle first involves the tunneling of an electron (hole) to the QD-SC system, changing its parity, followed by an Andreev reflection process whereby a Cooper pair is formed (broken) in SC and a hole (electron) is reflected to the probe. b) Charge stability diagram of a normal QD as a function of the bias voltage, V , and the gate voltage, Vg . The dot occupation (0, 1 or 2) is well-defined inside the Coulomb diamonds. Bottom panel: phase diagram of a hybrid QD as a function of the QD level position $\epsilon_0$ and the QD-SC coupling, $\Gamma_S$ , normalized to the charging energy $U$. In the weak coupling limit, $\Gamma_S\ll U$ , the ground state is a spin-doublet when the dot is occupied by an odd number of electrons. Conversely, for  $\Gamma_S\gg U$, the ground state is a spin-singlet irrespective of the dot occupancy. The precise boundary between both states can be obtained by experimentally tuning the ratio $\Gamma_S/U$ in very good agreement with theoretical results obtained by a superconducting analog of the Anderson model, as we discuss in the main text. (d) Scanning electron microscopy image of a carbon nanotube contacted with a superconducting loop for resolving ABS via tunneling spectroscopy. (e) Phase dependence of the ABS spectrum.  (f) $dI/dV$  versus backgate voltage for a graphene device containing a QD reveals two regimes with doublet and singlet ground states. Adapted and reprinted from Ref.~\cite{Pillet2010} (panels d and e) and Ref.~\cite{Dirks2011} (panel f).} 
\label{Fig:ABS}
\end{figure*}

The first technique is based on direct quasiparticle tunneling which proves
the ABS spectrum. 
These experiments typically utilize a normal probe junction defined by e.g. a gate-defined depleted section near the hybrid nanostructure or an in-situ grown tunnel barrier. This measurement geometry allows for the characterization of energy
spectra in proximitized semiconductor quantum dots, see Figs. \ref{Fig:ABS} (a-c), like in the experiments previously mentioned, or more generally normal-superconducting junctions. 

\begin{figure*}[h!]
\begin{centering}
\includegraphics[width=\textwidth]{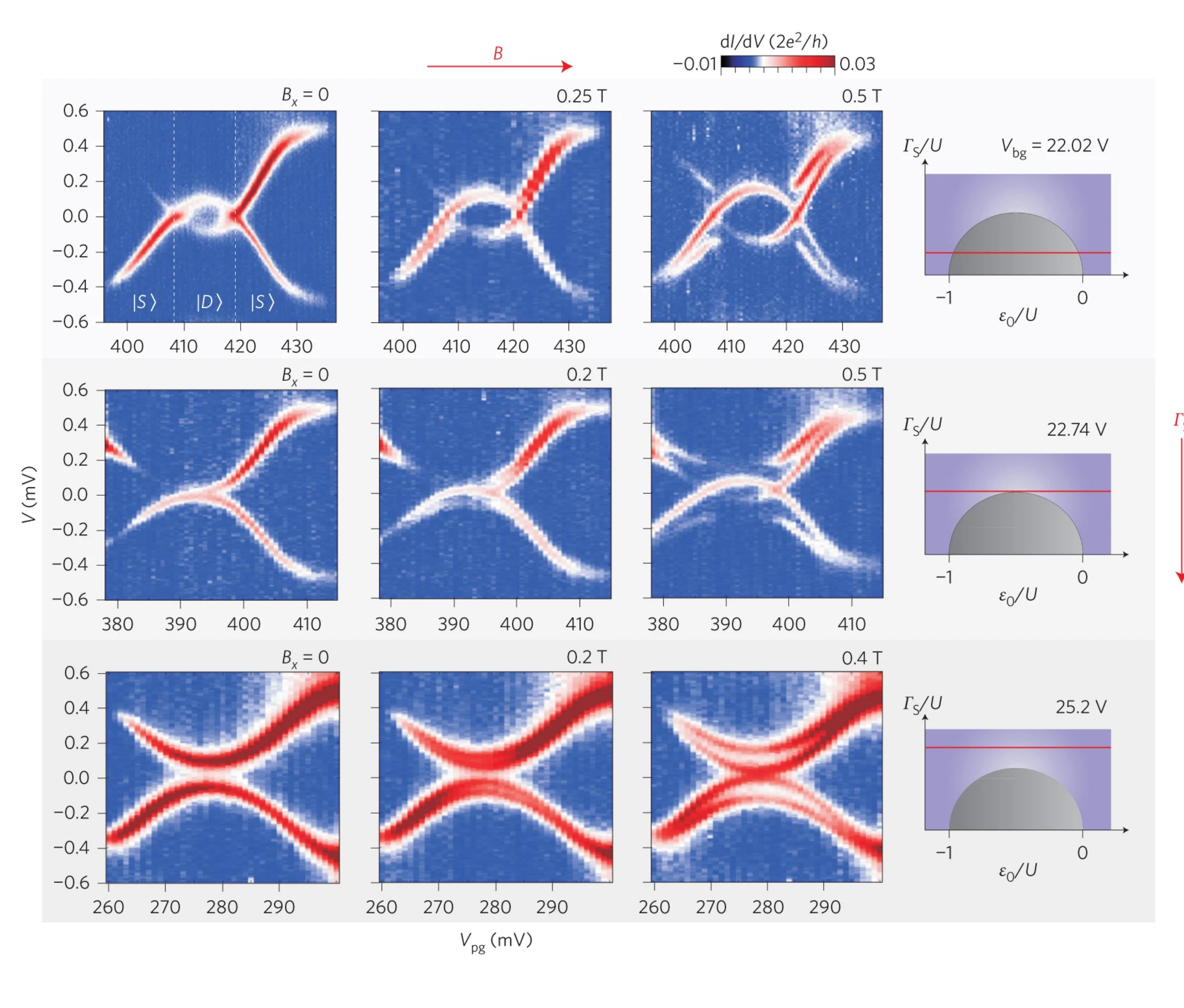}
\end{centering}
\caption{ Tunneling spectroscopy measurements for an InAs/InP nanowire quantum dot. Experimental plots of $dI/dV$ versus ($V_{pg}$, $V$) for different QD–S couplings, $\Gamma_S$ (increasing from top to bottom) and magnetic fields parallel to the NW axis ($B_x$ increasing from left to right) reveal the Zeeman spin-splitting when the system is in the singlet ground state (since the measured tunneling peaks correspond to doublet excitations). Conversely, inside the doublet ground state loop no Zeeman splitting occurs. Adapted and reprinted from Ref.~\cite{Lee2014}.} 
\label{Fig:ABS2}
\end{figure*}
The first tunneling spectroscopy measurements of individually resolved ABSs were reported in 2010. Pillet et al. \cite{Pillet2010} used a carbon nanotube device. The carbon nanotube was contacted with a Ti/Al loop and an additional Ti/Al section was used as tunnel probe, Fig. \ref{Fig:ABS}d. The highly doped silicon substrate on which the device was fabricated was used as a gate. By applying a magnetic flux through the loop a phase difference across the CNT nanotube was created. This phase difference was used to change the energy of the ABS as demonstrated in the differential conductance measurement versus phase  (see Fig.~\ref{Fig:ABS}e). Furthermore, by changing the backgate voltage the charge occupation of the CNT was changed. This allowed to tune the device between the singlet and doublet ground state, 
when a single orbital level was occupied by two or one charges, respectively. More complex features attributed to the presence of two serially connected QDs were also observed.  Deacon et al. \cite{Deacon2010} used a self-assembled InAs QD grown on a GaAs substrate to study Andreev levels. The QD was contacted with one superconducting Ti/Al and one normal metallic Ti/Au lead. By strongly coupling to the superconducting lead and weakly to the normal one, tunneling spectroscopy measurements could reveal the singlet and doublet phases. In 2011, Dirks et al. \cite{Dirks2011} observed discrete ABS-like features in a graphene quantum dot device (Fig.~\ref{Fig:ABS}f). Critical examination, however, suggests the states likely originated in an unintended quantum dot within the superconducting probe, not in the graphene channel.  The first unambiguous spectroscopy of intrinsic graphene ABSs came several years later, when Bretheau et al. \cite{Bretheau2017} used tunnelling spectroscopy on superconductor–graphene–superconductor junctions to map the phase-dependent quasicontinuum of states, a hallmark of the proximity effect in these diffusive junctions.

In 2014 Lee et al. \cite{Lee2014}, used an InAs/InP nanowire contacted with a Ti/V/Al film to study spin-split Andreev levels as with such a superconducting contact tunneling spectroscopy measurements could be performed up to about 2 Tesla. As in the previous experiments, here the QD system also exhibits a competition between superconductivity (which favors even electron numbers and singlet states) and Coulomb blockade (which promotes discrete electron occupancy, including odd numbers). Depending on the tunnel coupling, superconducting gap, QD energy level, and charging energy, the ground state of the QD–S system was either a spin-singlet (even parity) or a spin-doublet (odd parity). Different from the previous experiments, a magnetic field was used. Such a field lifts the spin degeneracy due to the Zeeman effect. For a singlet ground state, the magnetic field causes a Zeeman splitting of the Andreev levels (since they correspond to \emph{doublet excitations}). On the other hand, for a doublet ground state, the levels shift but do not split (see Fig..~\ref{Fig:ABS2}). This experiment demonstrated magnetic-field-induced quantum phase transitions between singlet and doublet GSs, visible as zero-bias peaks in differential conductance measurements. 

\subsubsection{Circuit QED and Microwave readout}\label{s:microwave-spectroscopy}
We can probe the Andreev levels of a SNS junction using microwave techniques~\cite{Tosi2019,Wesdorp2024}. While these techniques are very general, let us for simplicity consider the short junction limit discussed in subsection \ref{sub-sec_short-junction}.
For a single mode with transmission $\tau$, the ABS energies are given by Eq. \eqref{ABS}. 
Each of these states are spin-degenerate in the absence of magnetic fields and can be occupied by zero, one, or two quasiparticles, giving rise to a four-level system: The \emph{even} (ground) state: $\ket{0}$ (no quasiparticles), the \emph{odd} states: $\ket{\uparrow}$, $\ket{\downarrow}$ (single quasiparticle with spin) and the \emph{even} (excited) state: $\ket{2}$ (doubly occupied). Following Eq. \ref{currrentABS_Temp}, each occupied ABS contributes to the Josephson current:
\begin{align}
\label{ABS-supercurrent-prefactor}
I_A(\phi) = \frac{2e}{\hbar} \frac{\partial \varepsilon_{A}(\phi)}{\partial \phi} \tanh \left( \frac{\varepsilon_{A}(\phi)}{2k_B T} \right)
\end{align}
In thermal equilibrium, only the even-parity ground state is significantly populated at low temperature, resulting in a supercurrent carried predominantly by the lowest ABS \footnote{We have changed the notation from the one used in Eq. \ref{currrentABS_Temp} $I_J\rightarrow I_A$ to stress the fact that we are discussing single ABS contributions to the supercurrent.}. 

Using microwave techniques, we can probe transitions between these ABS occupation states. For example, transitions from $\ket{0} \to \ket{\uparrow}$ or $\ket{\downarrow}$ correspond to single-quasiparticle excitations at frequency:
\begin{align}
hf = \varepsilon_{A}(\phi).
\end{align}
Transitions between $\ket{0}$ and $\ket{2}$ occur via two-quasiparticle (even parity) processes and require $hf = 2\varepsilon_{A}(\phi)$.

These transitions manifest as resonance features in the microwave transmission spectroscopy, allowing one to extract both $E_{\text{ABS}}(\phi)$ and the coherence properties of the ABS. To probe the dynamics of ABSs in a hybrid junction, one typically embeds the junction in a superconducting circuit that allows for microwave spectroscopy (subsection \ref{ss_circuit-MW}). A common method is to galvanically connect the junction to a lumped-element microwave resonator. The resonator consists of an inductance $L_r$ and capacitance $C_r$ to ground, forming a harmonic oscillator with resonance frequency:
\begin{align}
\omega_r = \frac{1}{\sqrt{L_r C_r}}.
\end{align}
The Hamiltonian of the resonator is then
\begin{align}
\mathcal{H}_R = \hbar \omega_r a^\dagger a,
\end{align}
where $a$ ($a^\dagger$) are the annihilation (creation) operators of the resonator mode.

To probe the ABS, a SNS junction is placed in series with a portion of the resonator inductance typically in an rf-SQUID geometry. This geometry creates a direct galvanic coupling between the resonator and the ABS, modifying the total phase drop across the junction. If the bare ABS Hamiltonian is denoted by $\hat{H}_A(\phi)$, where $\phi$ is the superconducting phase difference across the junction, then the resonator's zero-point fluctuations perturb this phase as
\begin{align}
\phi \rightarrow \phi + \delta\hat{ \phi}, \quad \text{with} \quad \delta \hat{\phi} = p \frac{\Phi_{\mathrm{zpf}}}{\phi_0}(a + a^\dagger),
\end{align}
where, $\phi_0 = \hbar/2e$ is the reduced flux quantum, $\Phi_{\mathrm{zpf}} = \sqrt{\hbar Z_r/2}$ is the zero-point flux fluctuation of the resonator, with $Z_r = \sqrt{L_r/C_r}$ the resonator impedance, $p \approx L_s / (L_s + L_r)$ is the participation ratio (the fraction of resonator inductance dropping across the junction) and $L_s$ is the series inductance in the loop with the junction.

\begin{figure*}[ht!]
\label{fig:shadow-shift}
\begin{center}
\includegraphics[width=0.78\textwidth]{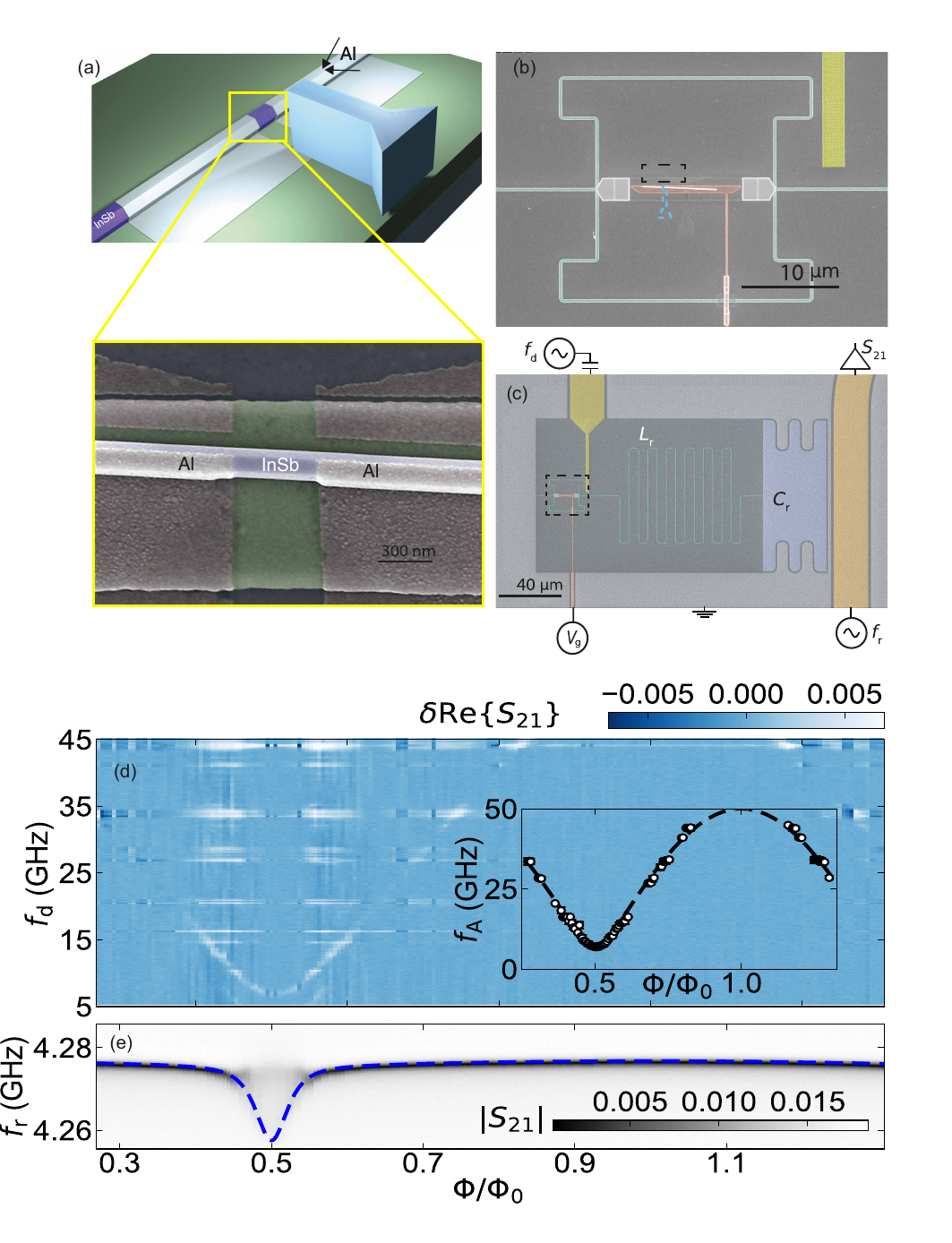}
\caption{(a) Illustration of shadow-wall lithography for fabricating pristine Josephson junctions. A metal, such as Aluminium (grey) is deposited under an angle (arrows) on a semiconducting nanowire, e. g. InSb (purple). A fabricated wall (light blue) creates a shadow  during the deposition of the metal, thus defining a junction region. The yellow blow-up shows an image of a real nanowire Josephson junction (Al-InSb) fabricated using the shadow-wall technique. (b-c)False colored scanning electron micrographs of the full circuit used for microwave readout. The zoom in (b) shows the SQUID loop used for flux control, while (c) shows the complete circuit with a resonator consisting of a capacitor $C_r$ (purple) that is connected via an inductor $L_r$ (green) via the SQUID loop to the ground plane. The resonator is capacitively coupled to the feedline (amber), while the hybrid Josephson junction is placed in the central arm of the SQUID and is tunable with an electrostatic gate (red). (d) Two tone spectroscopy versus flux. The inset shows a fit of the extracted even pair transition frequency which, using the single channel ABS formula in Eq. \eqref{ABS}, $\varepsilon_A(\phi\equiv2\pi\Phi/\Phi_0)=\pm\Delta\sqrt{1-\tau \sin^2(\pi\Phi/\Phi_0)}$ with $\Delta/h=25GHz$, reveals a highly transparent single-channel junction with $\tau=0.983$. (e) Using this single channel ABS fit, the measured flux dependent resonator shift agrees almost perfectly with Eq. \eqref{resonator-shift} without additional fit parameters (blue dashed line). Adapted and reprinted from Ref.~\cite{Wesdorp-Thesis}.
} 
\end{center}
\end{figure*}

The total Hamiltonian of the coupled system becomes:
\begin{align}
\mathcal{H} = \mathcal{H}_R + \mathcal{H}_A\left( \phi + \hat{\delta \phi} \right).
\end{align}
We now expand the ABS Hamiltonian in powers of $\hat{\delta \phi}$ \footnote{For details going beyond the discussion presented here, see Ref. \cite{PhysRevLett.125.077701}, where a full theory bridging the large detuning limit, where the discussion in terms of inductive shifts is valid, and the small detuning limit where the exchange of virtual microwave photons is relevant. }:
\begin{align}
\mathcal{H} &\approx \hbar \omega_r  a^\dagger a  + \mathcal{H}_A(\phi) + \frac{\partial \mathcal{H}_A}{\partial \phi} \, \delta \hat{ \phi} + \frac{1}{2} \frac{\partial^2 \mathcal{H}_A}{\partial \phi^2} \, \delta \hat{\phi}^2. \label{eq:H_coupled}
\end{align}
After a rotating wave approximation \footnote{This approximation neglects fast two-photon rotating terms $a^\dagger a^\dagger$ and $aa$.}, this expansion leads to two physically distinct effects:
(i) The quadratic term leads to a shift of the resonator frequency due to the ABS acting as a nonlinear inductive element. This results in an effective frequency:
\begin{align}
\omega_r^{\text{eff}} \approx \omega_r + \chi_{\mathrm{ind}}, \quad \text{with} \quad \hbar \chi_{\mathrm{ind}} = \langle \hat{L}_A^{-1} \rangle p^2 \Phi_{\mathrm{zpf}}^2,
\end{align}
where the inverse inductance operator of the ABS is defined as:
\begin{align}
\label{inverse-inductance}
\hat{L}_A^{-1} = \frac{1}{\phi_0^2} \frac{\partial^2 \mathcal{H}_A}{\partial \phi^2}.
\end{align}
(ii) The linear term in Eq.~\eqref{eq:H_coupled} represents a dipole-like interaction between the resonator mode and the ABS current operator:
\begin{align}
\label{ABS-resonator-coupling}
\hat{I}_A = \frac{1}{\phi_0} \frac{\partial \mathcal{H}_A}{\partial \phi}, \quad \Rightarrow \quad \mathcal{H}_{\text{int}} = \hat{I}_A \, p \, \Phi_{\mathrm{zpf}} (a + a^\dagger).
\end{align}
This term leads to a Jaynes-Cummings-type coupling when the ABS is truncated to a two-level system. The current operator is generally not diagonal in the eigenbasis of the ABS Hamiltonian and includes Pauli operators (e.g., $\sigma_x$), giving rise to resonant and dispersive coupling regimes depending on the qubit-resonator detuning. Specifically, by using the two-level hamiltonian in Eq. (\ref{ABS-qubit2}) it can be shown that the current and inductance operators read, respectively:
\begin{eqnarray}
 \hat{I}_A &=&I_A(\phi)[\sigma_z+\sqrt{1-\tau}\tan(\phi/2)\sigma_x\nonumber\\
 \hat{L}_A^{-1}&=&L_A^{-1}(\phi)[\sigma_z-\frac{2\sqrt{1-\tau}\sin(\phi)}{\tau(2-\tau)\cos(\phi)}\sigma_y],
\end{eqnarray}
with $I_A(\phi)$ being the supercurrent expression in Eq. \eqref{ABS-supercurrent-prefactor} and $L_A^{-1}(\phi)=\frac{1}{\phi_0}I_A(\phi)\frac{\tau(2-\tau)\cos(\phi)}{2\sin(\phi)}.$
Apart from the inductive shift $\chi_{\mathrm{ind}}$, the direct coupling between the resonator and the ABS in Eq. \ref{ABS-resonator-coupling} leads to a dispersive shift 
\begin{align}
\chi_{\mathrm{dis}}(\phi)=\frac{p^2 \Phi_{\mathrm{zpf}}^2I^2_A(\phi)(1-\tau)\tan^2(\phi/2)}{\hbar^2(\omega_A(\phi)-\omega_r)},
\end{align}
such that the resonator experiences a total (inductive+dispersive) shift due to the ABS of the form:
\begin{align}
\omega_r \rightarrow \omega_r + \delta\omega_r(\phi), \quad \text{with} \quad \delta\omega_r(\phi) = \chi_{\mathrm{ind}}(\phi)+\chi_{\mathrm{dis}}(\phi).
\label{resonator-shift}
\end{align}
While the majority of early implementations of semiconductor-superconductor hybrids in superconducting circuits have relied on InAs-Al nanowires with etched junctions, recent advances have introduced shadow lithography \cite{Heedt2021,Borsoi2021,Goswami-shadow23} as a promising alternative (see next subsection). In this technique, the junction region remains pristine, as it is never exposed to chemical etchants, resulting in highly transparent conduction channels with minimal disorder. This fabrication approach not only preserves the intrinsic properties of the hybrid interface but also enables the realization of more versatile and precisely defined junction geometries through the use of on-chip shadow walls. When combined with cQED techniques, these cleaner and more controllable junctions provide an ideal platform for testing theoretical predictions under near-optimal experimental conditions. A representative example of such advanced design setup and the good agreement with the theory is illustrated in Fig.~\ref{fig:shadow-shift}.

Finally, it is important to emphasize that the discussion presented above, beginning with Eq. \eqref{eq:H_coupled}, is by no means restricted to Andreev bound states. More broadly, this approach can be applied to any phase-dependent Hamiltonian embedded within a microwave cQED architecture, which makes it a versatile tool for exploring a wide range of quantum materials. For example,
the inductive shift, which has been recently observed in InAs-Al hybrid systems—in applications such as fast parity readout of superconducting islands \cite{PRXQuantum.5.030337}—represents the magnetic dual of quantum capacitance \cite{10.1063/5.0088229} measured in charge-based systems $C_q\sim (\frac{\partial^2 \mathcal{H}}{\partial q^2})^{-1}$ \cite{PhysRevApplied.11.044061}, a quantity typically measured in charge-sensitive platforms. While quantum capacitance reflects the system’s response to variations in electrostatic potential, the inductive shift captures its sensitivity to changes in magnetic flux or phase bias. This duality highlights the rich interplay between charge and flux degrees of freedom in mesoscopic superconducting devices, and underscores the potential of inductive measurements as a complementary probe for quantum state detection and material characterization in cQED architectures. This latter aspect, in particular, is emerging as a powerful technique for probing variations in the kinetic inductance of quantum materials. These shifts can serve as sensitive indicators of underlying physical phenomena, such as the presence of Bogoliubov Fermi surfaces in two-dimensional Al-InAs hybrid structures subjected to finite magnetic fields \cite{PhysRevLett.128.107701} or the superfluid stiffness of magic-angle twisted graphene \cite{Tanaka-Wang2025,Banerjee2025}.
\begin{figure*}[ht!]
\includegraphics[width=0.95\textwidth]{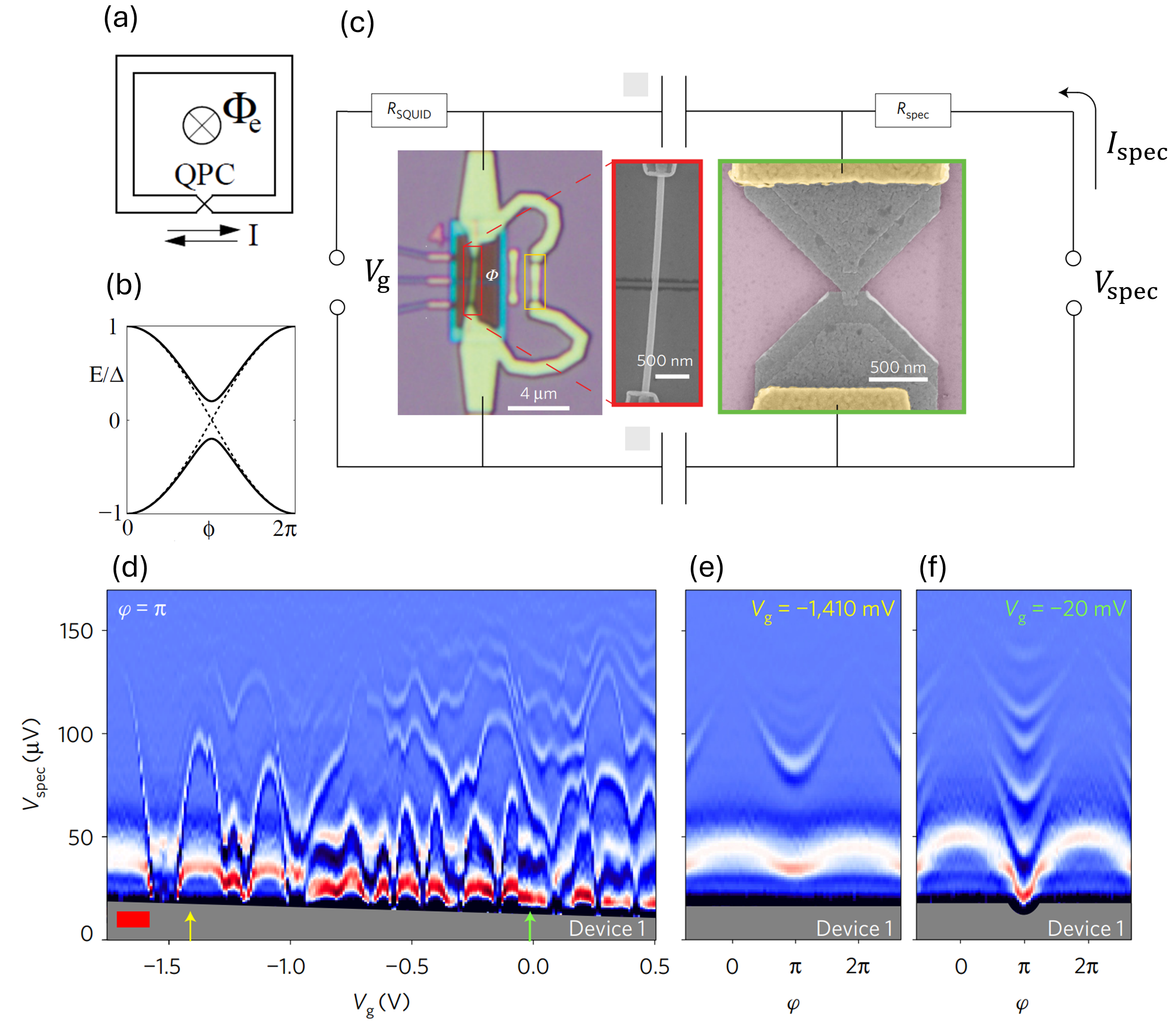}
\caption{ 
Microwave spectroscopy of Andreev levels. 
(a) Diagram of an Andreev level two level system implemented in a quantum point contact (QPC). 
The QPC is embedded in a superconducting loop through which an external magnetic flux, $\Phi_{\rm e}$, is threaded.
(b) Energies of the ground and excited even Andreev bound states of the QPC, $E$, as a function of the superconducting phase drop across the QPC, $\phi$. 
The energies are plotted in units of the superconducting gap, $\Delta$. 
The continuous and dashed lines correspond to ABS transparencies of 0.96 and 1, respectively. 
(c) Experimental implementation of a device to investigate the spectral response of the Andreev bound states in a semiconducting Josephson junction. 
The optical image shows a hybrid SQUID consisting of the parallel combination of one InAs semiconductor nanowire weak link (SEM enlargement in the red box) and an Al/AlOx/Al Josephson junction (yellow box). 
The flux through the SQUID is denoted as $\Phi$, with $\varphi=2\pi\Phi/\Phi_0$ being the reduced flux.
The SQUID is capacitively coupled to a second Al/AlOx/Al Josephson junction (SEM enlargement in the green box) that serves as a spectrometer. 
The InAs Josephson junction is tuned by a gate voltage, $V_{\rm g}$.
(d-f) Spectroscopy of the SQUId subgap states. 
The colormap displays $-d^2I_{\rm spec}/dV_{\rm spec}^2$ of measured across the spectrometer junction.
(d) Gate, $V_{\rm g}$, dependence at $\varphi=\pi$.
(e-f) Flux dependence of the subgap transitions at two gate points, indicated with color-matching arrows in (d).
Adapted and reprinted from Ref.~\cite{Zazunov2003} and \cite{vanWoerkom2017}.
} 
\label{fig:ALQ}
\end{figure*}
Apart from using direct resonator-ABS couplings in cQED, another interesting alternative, demonstrated by van Woerkom {\it et al.} \cite{vanWoerkom2017}, is to employ a spectroscopy method similar to that developed in the atomic contact experiments \cite{Bretheau2012}, is
to investigate the spectral properties of the Andreev levels in a hybrid nanowire-based junction by using a second junction as a microwave spectrometer. The basic concept in such microwave measurements is illustrated in  Fig.~\ref{fig:ALQ} a and b,
which show a cartoon of an Andreev level two level system implemented in a quantum point contact (QPC). The QPC is embedded in a superconducting loop through which an external magnetic flux, $\Phi_{\rm e}$, is threaded.  In this approach, a second tunnel junction—fabricated using conventional Al/AlOx/Al technology is used as a microwave spectrometer (green in (Fig.~\ref{fig:ALQ}(c)) \cite{vanWoerkom2017}. 
By applying a DC voltage to this spectrometer junction, it emits electromagnetic radiation with a frequency proportional to the voltage bias. 
When the emitted photons match the transition energy between occupancies of the Andreev states in the nanowire junction, absorption occurs and is detected through a change in the spectrometer’s differential conductance (Fig.~\ref{fig:ALQ}(d-f)). 

This spectroscopic method has proven to reveal key features of the Andreev level spectrum in InAs nanowire-based Josephson junctions. 
Fig.~\ref{fig:ALQ}(d) displays the gate-gate dependence of the junction spectrum. 
The transition frequency disperses non-monotonically with gate voltage, evolving from a single-channel regime at low voltages  (Fig.~\ref{fig:ALQ}(e)) to a multi-channel regime at higher voltages (Fig.~\ref{fig:ALQ}(f)), consistent with theoretical predictions for short junctions. These results confirmed that nanowire JJ devices can be operated in the short-junction limit with a reduced number of active channels and provided a stepping stone for future time-domain experiments and scalable qubit architectures. 

In Sec.~\ref{s:ALQ}, we will further discuss how microwave spectroscopy techniques in cQED setups can be used to probe the spectrum of hybrid Josephson junctions and enable qubits based on Andreev levels.
\section{Semiconductor-based hybrid superconductor qubits }\label{s:qubits} 
In recent years, hybrid qubits have emerged as a promising approach, combining the techniques of superconducting qubits (Sec.~\ref{s:superconducting}) with the distinctive properties of superconducting-semiconducting hybrid nanostructures (Secs.~\ref{s:subgap} and \ref{s:nanostructures}). 
This section reviews key examples of these devices, emphasizing how their design combines the strengths of their components to achieve enhanced performance and unique properties. 
Our focus is on hybrid devices that represent novel types of qubits.

This approach inevitably excludes other important developments at the interface of superconducting and semiconducting technologies, such as gate-tunable parametric amplifiers based on either SNS Josephson junctions~\citep{Butseraen2022,Sarkar2022,Phan2023,Hao2024} or superconducting-semiconducting hybrids~\citep{Splitthoff2024}, gate-tunable resonators that incorporate an SNS Josephson junction~\citep{Strickland2023}, hybrid two-dimensional material structures for low-loss and small footprint superconducting qubits~\citep{Antony2021,Wang2022b}, 
and high-impedance resonators for coupling spin qubits over long distances~\citep{Samkharadze2018,Mi2018,Landig2018,Landig2019,Clerk2020,Burkard2020,HarveyCollard2022,Yu2023,Ungerer2024}. 
Readers interested in these topics are encouraged to explore the cited references for further insights.

\subsection{Gate-tunable superconducting qubits}

One approach to creating hybrid superconducting-semiconducting qubits is to replace the non-tunable SIS Josephson junction or flux-tunable SQUID of a conventional superconducting qubit with a gate-tunable SNS Josephson junction~\citep{Aguado2020}. 

This modification introduces several advantages. 
From the perspective of superconducting qubits, gate-tunability allows for frequency tuning without relying on flux lines, which can induce unwanted cross-talk. From a semiconducting standpoint, hybrid circuits offer a powerful platform for probing the dynamics of semiconducting junctions with superior time and energy resolution compared to conventional low-frequency transport techniques. This enhanced capability is particularly valuable for investigating hybrid nanostructures as foundational elements for qubits operating in novel regimes—such as Hamiltonian-protected qubits (see Sec.~\ref{s:protected-qubits}); various types of Andreev qubits (see Secs.~\ref{s:ALQ} and \ref{s:ASQ}); and parity qubits based on MZMs in minimal Kitaev chains.
The latter approach presents a compelling framework for encoding nonlocal topological qubits by constructing bottom-up arrays of semiconducting quantum dots interconnected via hybrid superconducting segments, as detailed in Sec.~\ref{s:bottom-up}. Another key advantage of these hybrid systems is their compatibility with cQED readout techniques (see Sec.~\ref{ss_circuit-MW}), while still retaining the tunability inherent to semiconductors.

\begin{figure*}[h!]
\includegraphics[width=\textwidth]{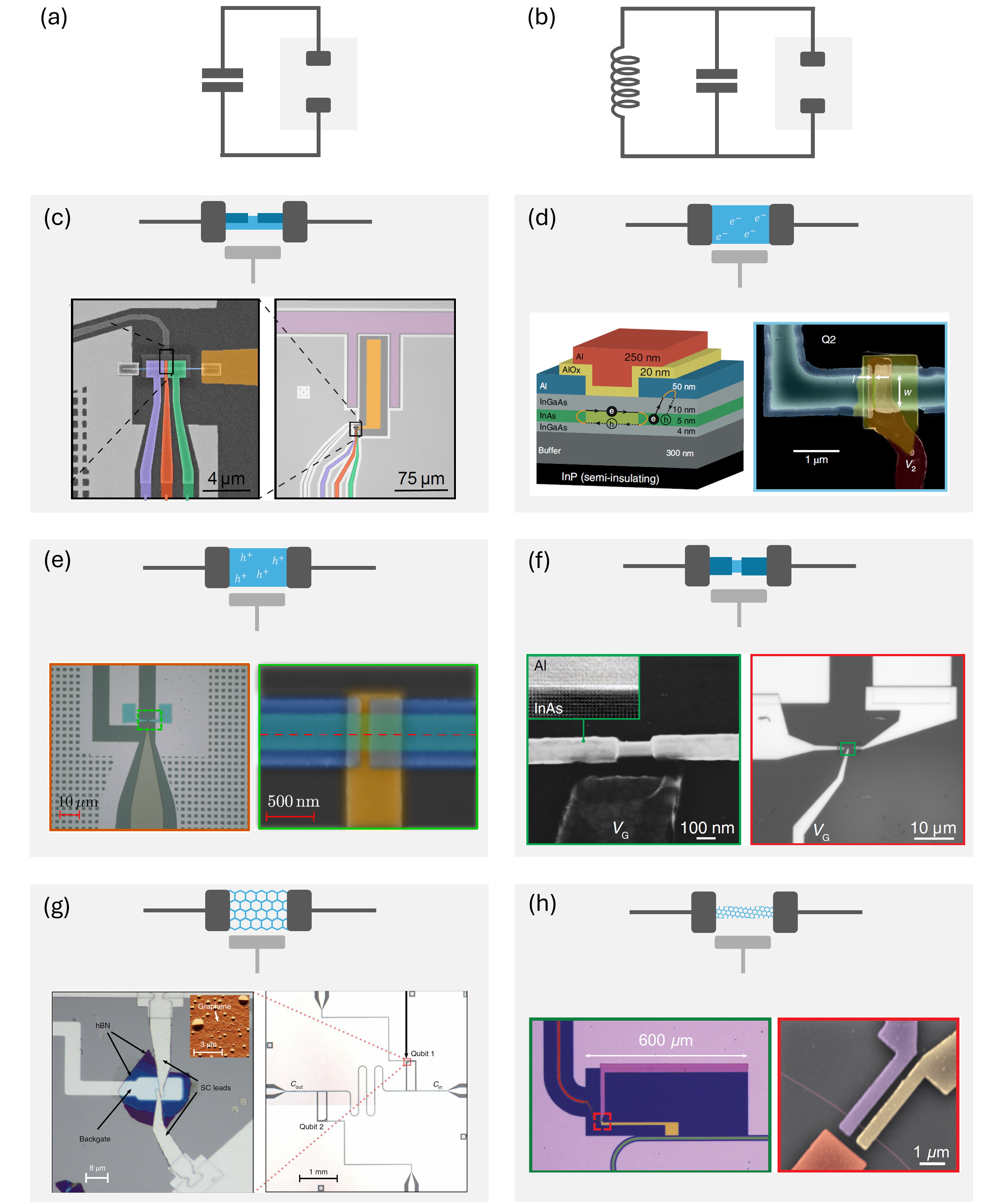}
\caption{ Circuit diagrams of (a) a gatemon (gate-tunable transmon) and (b) a gate-tunable fluxonium. 
In both cases, the Josephson junction of a conventional superconducting qubit is replaced by an electrostatically tunable semiconducting Josephson junction. 
(c)-(h) Examples of nanostructures used as the weak link of gatemons in the literature. 
Adapted and reprinted from Ref.~\citep{Uilhoorn2021}, \citep{Casparis2018}, \citep{Sagi2024}, \citep{Larsen2015}, \citep{Wang2019} and \citep{Mergenthaler2021}, respectively.
(c) Half-shell superconductor-proximitized nanowire.
(d) Two-dimensional electron gas.
(e) Two-dimensional hole gas.
(f) Full-shell superconductor-proximitized nanowire.
(g) Two-dimensional van der Waals heterostructure.
(h) Carbon nanotube.
} 
\label{fig:gatemon}
\end{figure*}

The two simplest gate-tunable superconducting qubit geometries are a gate-tunable transmon (gatemon, Fig.~\ref{fig:gatemon}(a)) and a gate-tunable fluxonium (gatemonium,  Fig.~\ref{fig:gatemon}(b)).  
Fig.~\ref{fig:gatemon}(c-h) showcase experimental implementations of such devices.
The first gatemon implementations in the literature used InAs nanowires as the gate-tunable junction \citep{Larsen2015,deLange2015}. 
Since then, a wide variety of materials and configurations have been explored, including bare, full-shell, and half-shell superconductor-proximitized InAs nanowires~\citep{Casparis2016,Luthi2018,Kringhoj2018,Kringhoj2020b,Sabonis2020,Bargerbos2020,Kringhoj2020,Kringhoj2021,Uilhoorn2021,Bargerbos2022,Danilenko2023,feldsteinbofill2024}, InAs two-dimensional electron gases (2DEGs)~\citep{Casparis2018,Strickland2024}, germanium two-dimensional hole gases (2DHG)~\citep{Sagi2024,Kiyooka2024} and Ge/Si core/shell nanowires \citep{Zhuo2023,Zheng2024}.
Beyond standard semiconductors, gatemons have been demonstrated in graphene flakes~\citep{Wang2019,Kroll2018}, MoTe$_2$ flakes~\citep{Chiu2020},(Bi, Sb)$_2$Te$_3$ selective area grown (SAG) structures~\citep{Schmitt2022} or carbon nanotubes~\citep{Mergenthaler2021,Riechert2025}. Similarly, gate-tunable fluxoniums have been realized using Josephson junctions based on InAs nanowires~\citep{PitaVidal2020} and InAs 2DEGs~\citep{Strickland2024}.

\begin{figure*}[t]
\includegraphics[width=\textwidth]{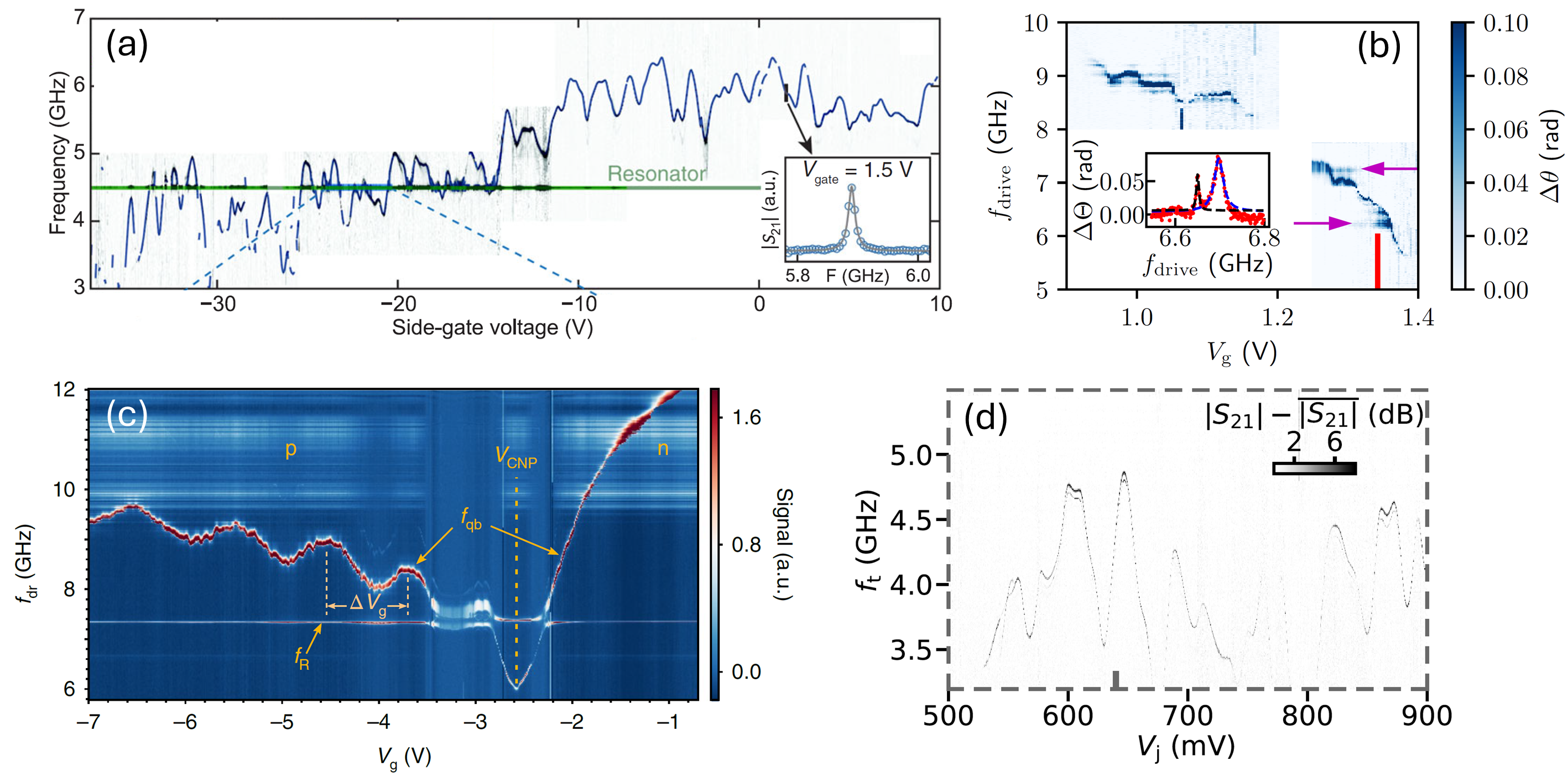}
\caption{ (a-d) Two-tone spectroscopy measurements of the gatemon frequency as a function of the applied electrostatic gate voltage for gatemons implemented with (a) InAs nanowire, (b) Ge quantum well on a SiGe heterostructure, (c) graphene and (d) aluminum-proximitized InAs nanowire.
Adapted and reprinted from Ref.~\citep{deLange2015}, \citep{Kiyooka2024}, \citep{Wang2019} and \citep{Bargerbos2022}, respectively.} 
\label{fig:gatemon-spectroscopy}
\end{figure*}

As shown in Fig.~\ref{fig:gatemon-spectroscopy}, tuning the electrostatic gate voltage enables the gatemon frequency to vary over several gigahertz. 
This gate-tunability avoids the use of flux lines and, notably, exhibits sweet spots at different frequencies. 
This permits choosing the qubit frequency while keeping it, to first order, resilient to voltage noise, making such devices promising candidates for scalable quantum computing architectures.
Gate-tunable transmon and fluxonium qubits also demonstrate compatibility with magnetic fields up to the order of \SI{1}{T}~\citep{Kroll2018,PitaVidal2020,Uilhoorn2021,Kringhoj2021,Danilenko2023}, a feature that sets them apart from traditional superconducting qubit platforms. 
This magnetic field tolerance allows investigation of qubit performance under magnetic field noise \citep{Luthi2018}, as well as the integration of superconducting and spin qubit architectures \citep{Landig2019}.
Moreover, experiments conducted under magnetic fields as high as several Tesla enable the exploration of energy-phase relations that emerge under broken time-reversal symmetry, such as the anomalous Josephson effect, as well as the investigation of topological superconducting states in the high-frequency domain. 
Hybrid circuits have been proposed to detect topological superconductivity, as they can be used to spectroscopically observe the $4\pi$ Josephson effect, a hallmark of topological superconducting states \citep{Hassler2010,Hassler2011,Pekker2013,Muller2013,Ginossar2014,Vayrynen2015,Avila2020a,Avila2020}. 
Furthermore, these circuits can mediate coupling between topological qubits and superconducting qubits, enabling the implementation of hybrid gate operations \citep{Pekker2013}. Further details about topological hybrids will be discussed in Sec. \ref{s:bottom-up}.

Hybrid circuits have also been used to probe highly transparent channels in semiconducting Josephson junctions, due to the  nonsinusoidal character of the Josephson current-phase relation resulting from the high transparency. 
The fine energy resolution of cQED techniques facilitates the observation of such channels, which, for a gatemon circuit, manifest experimentally as reduced anharmonicity and diminished charge dispersion \citep{Kringhoj2018,Bargerbos2020,Kringhoj2020,Vakhtel2023}. 
Similarly, signatures of highly transparent channels have been spectroscopically observed using a fluxonium circuit and have been suggested to enable novel types of protected qubits \citep{Vakhtel2024}.

\subsubsection{Role of junction properties on qubit anharmonicity}\label{ss:lenght-anhar}

Apart from the gate tunability of the qubit frequency, the replacement $-E_J \cos (\phi)\rightarrow V(\phi)$ in Eq. (\ref{SC_Hamiltonian}) has strong consequences on qubit anharmonicity, a quantity which, as we discussed in section \ref{s:superconducting}, is very relevant for designing selective control of transitions and gate operations. This physics can be nicely understood by expanding a generic Josephson potential as a  polynomial expansion about the minimum as \footnote{This discussion focuses on the ground state of weak links without charging energy and with time reversal
symmetry, like e.g. the resonant level model in Eq. \eqref{ABS_QD}, in so that the minimum is always at $\phi= 0$ and symmetric, resulting in $c_n
= 0$ for odd $n$.}
\begin{equation}
\label{higher-harmonic_Josephson2}
V(\phi)=\sum_{n=0}^\infty c_n \phi^n,
\end{equation}
with coefficients $c_n$ to which we can give a physical meaning (for example, and following the same reasoning that led us to the Eq. \eqref{inverse-inductance}, the coefficient $c_2$ can be identified with the inverse linear inductance of the junction.

For a tunnel junction shunted by a capacitor which provides a charging energy $E_C$, the fourth-order expansion in Eq. \eqref{expansion-fourth-order} gives $c_4=-E_J/24$ and an anharmonicity $\alpha=12E_Cc_4/c_2=-E_C$ (see Eq. \eqref{anharmonicity}). As pointed out by Kringhoj {\it et al.} in Ref. \cite{Kringhoj2018}, this anharmonicity can get reduced by varying the transparency of a short junction, a result that can be obtained by just expanding the Josephson potential in Eq. \eqref{Josephson-short-multi} to fourth order in the phase as:
\begin{equation}
\label{short-junction-Josephson-fourth}
V(\phi)\approx E_J\frac{\phi^2}{2}-E_J(1-\frac{3\sum_i \tau^2_i}{4\sum_i \tau_i})\frac{\phi^4}{24},
\end{equation}
with $E_J=\frac{\Delta}{4} \sum_i \tau_i$, resulting in a qubit transition frequency
\begin{equation}
 hf_{01}\approx\sqrt{8E_CE_J} = \sqrt{2E_C\Delta \sum_i \tau_i},   
\end{equation}
and anharmonicity
\begin{equation}
\alpha\equiv h(f_{12}-f_{01})\approx -E_C(1-\frac{3\sum_i \tau^2_i}{4\sum_i \tau_i}).
\end{equation}
In the tunnel limit, where $\tau_i\rightarrow 0$ for all $i$, we recover the standard transmon case with $\alpha=-E_C$, while for maximally transparent junctions with $\tau_i\rightarrow 1$, one gets $\alpha=-E_C/4$, giving a reduced qubit nonlinearity.

\begin{figure*}[t]
\begin{center}
\includegraphics[width=0.8\textwidth]{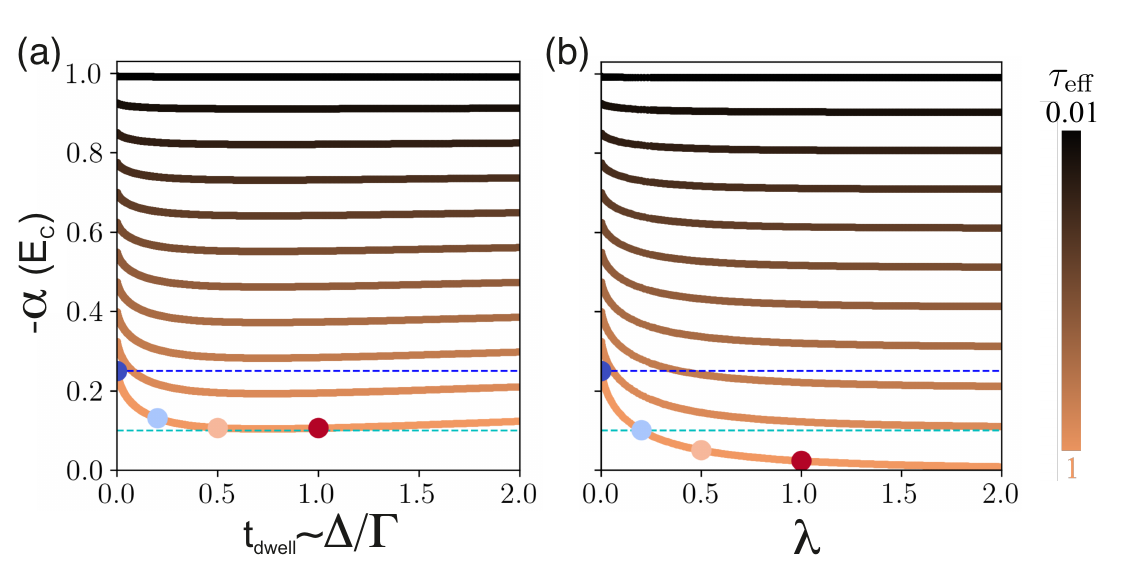}
\end{center}
\caption{(a) Relative anharmonicity $-\alpha$ for a resonant level model, for
evenly-spaced transparencies (colour bar), with the lower axis the parameter proportional to the
dwell time in the circuit. (b) Same as (a) but for a finite-length SNS junction.
Adapted and reprinted from Ref.~\citep{10.21468/SciPostPhys.18.3.091}.} 
\label{fig:anharmonicity}
\end{figure*}
Including the finite length of the weak link (see subsection \ref{sub-sec_long-junction}) leads to further suppression of the nonlinearity, a relevant aspect in the context of gate-tunable superconducting qubits that has been pointed out by Fatemi {\it et al.} in Ref. \cite{10.21468/SciPostPhys.18.3.091}. Physically, this can be understood since the junction becomes progressively more harmonic as $\lambda=\frac{t_{dwell}}{t_A}=\frac{L_N}{\xi}$ grows \footnote{Equivalently, in the long-length limit $\lambda\rightarrow\infty$, the current-phase relation of a ballistic device approaches a sawtooth function.}. 

Interestingly, the resonant level model \cite{10.1007/978-3-642-77274-0_20}, c.f. Eq. \eqref{ABS_QD}, smoothly interpolates between
an ideal short junction and a junction with a large dwell time by varying the coupling between the resonant level and the superconducting leads. This can be understood by noting that the model captures the effect of a finite dwell time $t_{dwell}\propto \Delta/(\Gamma_L+\Gamma_R)$ in the junction, but it remains “short” in the sense that it always contains a single ABS as compared to e.g the SNS long junction in  Eq. \eqref{ABS-energy-long2}. 

Another interesting option to enhance the anharmonicity is to exploit interference effects of the higher-order harmonics of $V(\phi)$ by using flux frustration in split junction geometries. This effect has been explored in split-junction gatemon devices based on InAs/Al 2D heterostructures, where large anharmonicities can be routinely achieved at the half-integer flux sweet-spot without any need for electrical gating or excessive sensitivity to the offset charge noise \cite{liu2025stronglyanharmonicgatelessgatemonqubits}. Such anharmonic qubits can be driven with Rabi frequencies reaching more than 100 MHz, enabling gate operations much faster than what is possible with traditional gatemons and transmons. 

\subsubsection{Reduced charge dispersion by increasing junction transparency}\label{ss:reduced-charge dispersion}

A key source of dephasing in superconducting qubits is charge noise, which couples to the qubit transition frequency through the residual dependence of the energy spectrum on the offset charge. 
In the transmon regime and for tunnel junctions, this charge dispersion is exponentially suppressed as the $E_J/E_C$ ratio increases. 
In gatemons, however, it can be further reduced in the presence of Josephson junctions with highly transparent conduction channels. 
In the limit of near-unity transmission, the qubit frequency becomes nearly independent of the offset charge and is found to be well-described by a resonant-level junction model
\begin{equation}
\label{Eq:Htransparency}
\mathcal{H} = 4E_C (i \delta_\phi - n_{\rm g})^2 + \Delta \begin{pmatrix}
                                                \cos(\phi/2)  & \sqrt{1-T^2}\sin(\phi/2)  \\
                                                \sqrt{1-T^2}\sin(\phi/2) & -\cos(\phi/2) 
                                                \end{pmatrix},
\end{equation}
where $E_C$ and $n_{\rm g}$ represent the charging energy and charge offset of the transmon island, respectively, and $\Delta$ and $T$ represent the effective superconductive gap and the transmission amplitude of the junction, respectively.

\begin{figure*}[t]
\includegraphics[width=\textwidth]{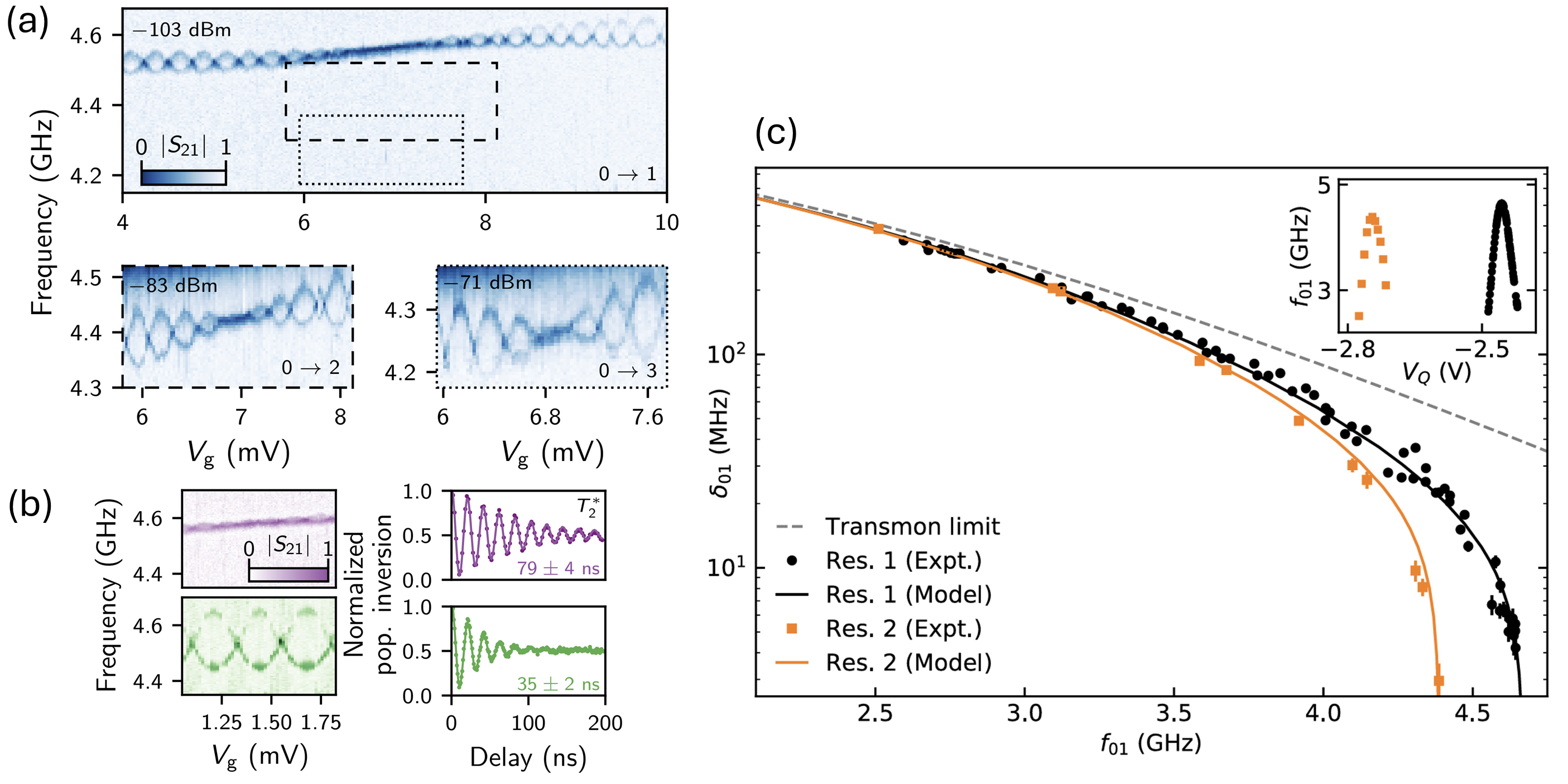}
\caption{ 
Suppression of charge dispersion under high junction transparency.
(a) Normalized two-tone spectroscopy measurement of the single-photon $0\leftrightarrow 1$, two-photon $0\leftrightarrow 2$ and three-photon $0\leftrightarrow 3$ transitions over a range of gate voltage encompassing many periods in $n_{\rm g}$. 
The charge dispersion suppresses down to the linewidth and subsequently recovers, while the qubit frequency increases over the entire gate range. 
Adapted and reprinted from  \citep{Bargerbos2020}.
(b) Extracted charge dispersion amplitude (markers) and numerical fit results to Eq.~\ref{Eq:Htransparency} (curves) as a function of the qubit frequency.
Different colors represent two different datasets at different regions in gate space.
(Inset) Measured qubit frequency as a function of gate voltage.
Adapted and reprinted from \citep{Kringhoj2020}.
(c) Two-tone spectroscopy (left) and time-resolved $T^*_2$ Ramsey experiments for setpoints with $T\rightarrow 0$ (purple) and  $T \simeq 0.8$ (green).
Adapted and reprinted from Ref.~\citep{Bargerbos2020}.} 
\label{fig:supression_dispersion}
\end{figure*}

This effect was experimentally investigated in Refs.~\citep{Bargerbos2020} and \citep{Kringhoj2020} using gatemon qubits that incorporate a gate-tunable Al-InAs-Al Josephson junction.
The authors measured the frequency response in different junction transparency regimes, observing drastic variations of the charge dispersion that are not correlated with changes in qubit frequency (see Fig.~\ref{fig:supression_dispersion}(a)).
In gate regimes when the Josephson junction hosts a channel with transmission approaching one, the charge dispersion of the gatemon is strongly suppressed and even vanishes within experimental resolution.
To measure the charge dispersion amplitude beyond the spectroscopic resolution, one approach is to measure multiphoton transitions between the ground state and excited states beyond the first one. 
For these transitions, the amplitude of the charge dispersion is larger than that of the first transmon transition and can therefore be resolved down to lower values (see bottom panels in Fig.~\ref{fig:supression_dispersion}(a)).

For certain parameter regimes, the extracted charge dispersion as a function of gate voltage is not well fit by the expected exponential dependence of a tunnel junction transmon (dashed gray line in Fig.~\ref{fig:supression_dispersion}(b)).
Instead, the model of Eq.~\ref{Eq:Htransparency} is well fit to the data, indicating the presence of highly transparent channels.
Fits to charge dispersion data were found to reach transparencies of up to 0.9996 \cite{Bargerbos2020}.

Beyond spectroscopy, time-domain experiments directly show how reduced charge dispersion effectively eliminates charge noise as a decoherence channel.
In particular, Ramsey measurements show that $T_2^*$ is significantly increased in transparency regimes where the charge dispersion is strongly suppressed (see Fig.~\ref{fig:supression_dispersion}(c)), reflecting the drastic reduction in sensitivity to charge noise.

These results highlight the central role of junction transparency in mitigating charge noise in gatemon qubits.
By pushing to the high-transmission regime, gatemons can suppress charge dispersion beyond what is achievable in conventional transmons, opening a path to enhanced dephasing times without relying solely on large shunting capacitances. 

\subsection{Hamiltonian-protected hybrid qubits}\label{s:protected-qubits}
\begin{figure*}[h!]
\includegraphics[width=\textwidth]{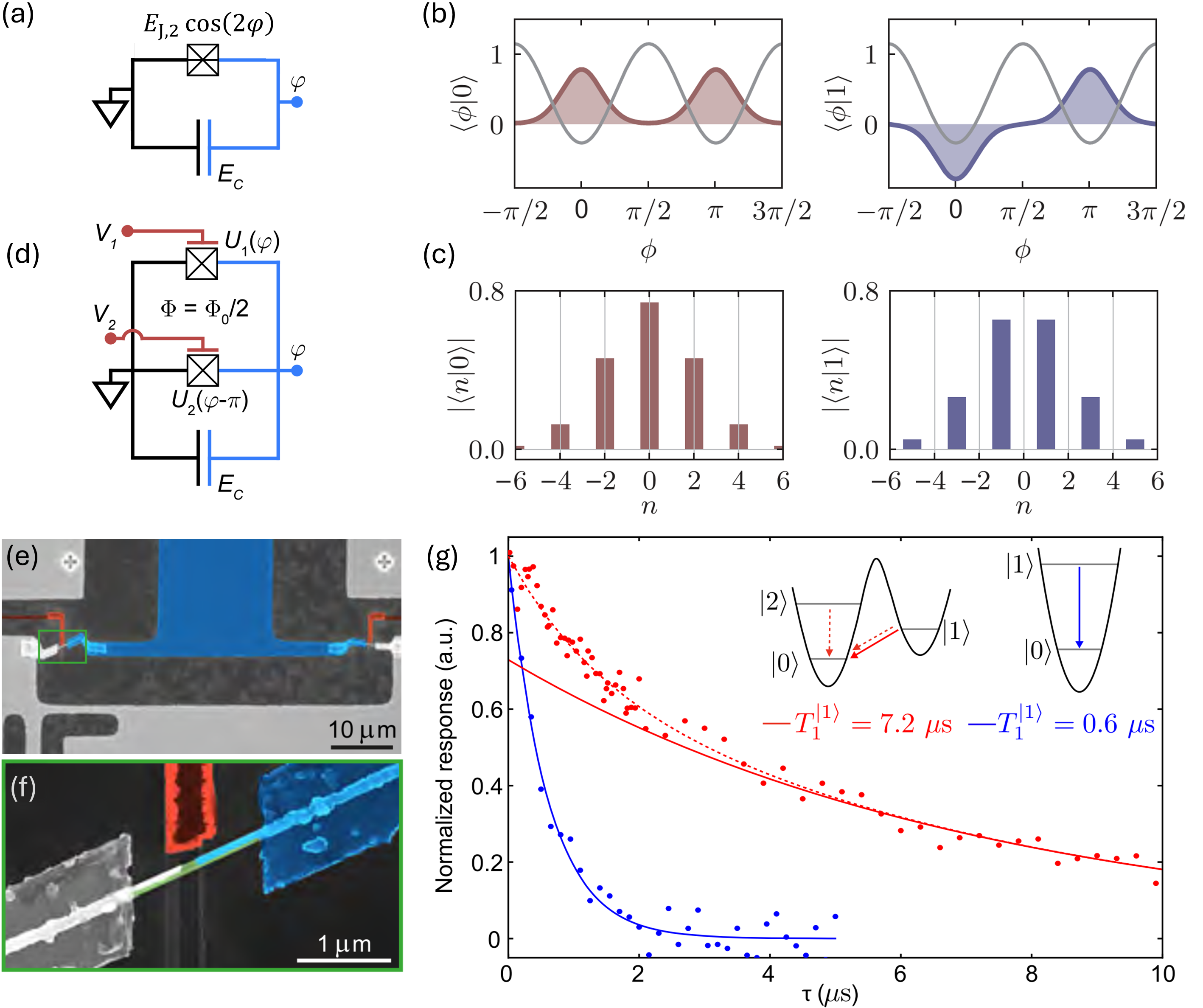}
\caption{ 
Hamiltonian-protected $\cos(2\varphi)$ qubit.
(a) Circuit model of an ideal $\cos(2\varphi)$ qubit composed of a superconducting island (blue) connected to ground (black) by the parallel combination of a $\pi$-periodic Josephson junction with effective Josephson energy $E_{\rm J,2}$ and a capacitor with charging energy $E_C$.  The superconducting phase on the island is denoted by $\varphi$.
(b) Wavefunctions of the qubit states in phase space (blue and red) plotted over the $\cos(2\varphi)$ potential (grey).
(c) Wavefunctions of the qubit states in charge space.
(d) Same as (a) but with the $\cos(2\varphi)$ element implemented by the parallel combination of two high-transparency few-channel semiconductor Josephson junctions. 
A magnetic flux $\Phi$ is thread through the SQUID loop defined by the two junctions.
(e) False-color scanning electron microscopy image of the superconducting island (blue) and the two aluminum-proximitized InAs nanowire junctions. 
(f) Enlargement of the region indicated with a green rectangle in (e). 
The Josephson junction is formed by etching away a section of the aluminum shell of the nanowire. 
An electrostatic gate is indicated in red.
(g) Lifetime measurements of the device shown in (e), (f) for $\Phi=0$ (blue) and for $\Phi=0.512\Phi_0$(red).
Adapted and reprinted from Refs.~\citep{Schrade2022} and \citep{Larsen2020}.} 
\label{fig:protected-qubits}
\end{figure*}

In the pursuit of robust quantum information processing, Hamiltonian-protected qubits have emerged as a promising avenue. A notable example is a $\cos(2\varphi)$ qubit, which can be described as a transmon qubit incorporating a Josephson junction with a $\pi$-periodic, instead of the typical $2\pi$-periodic, potential \citep{Kitaev2006,Brooks2013}, see Fig.~\ref{fig:protected-qubits}(a).

This configuration effectively creates a potential landscape with two energy minima, as shown by the grey line in Fig.~\ref{fig:protected-qubits}(b).
In phase space, the two resulting lowest energy eigenstates of the system are symmetric and antisymmetric superpositions of the wavefunctions localized in each of these two wells.
In charge space, their wavefunctions are localized in even and odd Cooper pair charge subspaces (Fig.~\ref{fig:protected-qubits}(c)).
This results from the fact that a junction with a $\cos(2\varphi)$ potential only permits the tunneling of {\it pairs} of Cooper pairs.
As a result, both the charge and phase transition matrix elements between the two qubit states vanish, providing inherent protection against energy decay.
If the $E_{\rm J, 2}\cos(2\varphi)$ element is moreover shunted with a large capacitor with charging energy $E_C$, much smaller than the effective Josephson energy $E_{\rm J, 2}$, the qubit states become delocalized in charge space, and the qubit, in turn, becomes protected against local charge noise perturbations. 

Traditional SIS Josephson junctions, however, exhibit a $2\pi$-periodic current-phase relationship, which limits their application in creating $\pi$-periodic potentials necessary for $\cos(2\varphi)$ qubits. 
To engineer a $\cos(2\varphi)$ Hamiltonian using SIS junctions, researchers have developed circuits that exploit symmetries to effectively modify the periodicity of the potential \citep{Groszkowski2018,DiPaolo2019,Gyenis2021,Smith2020}. 
However, these methods often require intricate designs and precise control over circuit parameters.

Recent experimental efforts have focused on the realization of the $\cos(2\varphi)$ qubit by exploiting higher harmonics in the current-phase relation of semiconducting Josephson junctions \citep{Larsen2020,shagalov2025}. 
This can be achieved using a superconducting quantum interference device (SQUID) formed by two high-transparency semiconductor nanowire Josephson junctions (Fig.~\ref{fig:protected-qubits}(d,e,f)).
When the junctions are gate tuned so that the $2\pi$-periodic components of their potentials are equal and the SQUID loop is frustrated by half a flux quantum, the $\cos(\varphi)$ terms of the potential cancel out.
As a result, the second harmonic $\cos(2\varphi)$ term becomes dominant, thereby approximating the desired $\pi$-periodic potential.
As shown in Fig.~\ref{fig:protected-qubits}(g), the decay time shows an improvement by a factor of ten around half a flux quantum (close to the protected regime), with respect to the same measurement at zero flux \citep{Larsen2020}.  Alternatively, two Josephson junctions in series can lead to a current-phase relation similar to a high-transparent junction~\cite{Bozkurt_SciPost,Banszerus_PRL2024}, that can also be used to realize a $\cos(2\varphi)$ qubit~\citep{Banszerus_PRX2025}. Further control can be achieved in the so-called  "harmonic parity qubit"; an element that combines two aluminum-oxide tunnel junctions in parallel with a gate-tunable InAs/Al nanowire junction forming a SQUID \cite{feldsteinbofill2026controlledparitycooperpair}.

Schrade {\it et al.} expanded on this idea and proposed a protected qubit based on a modular array of superconducting islands connected by such semiconductor Josephson interferometers. 
They showed that the protection from noise and offsets offered by the array improves with the number of elements, enhancing the robustness of the qubit against decoherence \citep{Schrade2022}. 
Additionally, proposals for two-qubit gates have been put forward to facilitate interactions between these protected qubits. 
Leroux and Blais introduced a single qubit $Z$ gate inspired by the noise-bias-preserving gate of the Kerr-cat qubit. 
This scheme relies on a $\pi$ rotation in phase space via a beam splitter-like transformation between a qubit and an ancilla qubit, implemented by adiabatically changing the potential energies of the two qubits to preserve a double-well potential at all times. 
This gate constrains the dynamics within the subspace of a $\cos(2\varphi)$ qubit, yielding a high-fidelity operation while preserving the qubit coherence \citep{leroux2023}.

Beyond the $\cos(2\varphi)$ qubit, other circuit designs have been proposed to implement intrinsically protected qubits. 
One such design involves a fluxonium qubit that incorporates a Josephson junction with a $4\pi$-periodic potential, achieved through a highly transmissive channel. 
This approach is akin to the bifluxon and blochnium \citep{Kalashnikov2020,Pechenezhskiy2020,Chirolli2023} qubits but leverages hybrid superconducting elements to enhance protection.

{This idea has been analyzed in detail by Vakhtel {\it et al.}~\cite{Vakhtel2024}, who studied a fluxonium-like circuit, namely an inductively shunted junction described by Eq.~\eqref{eq:H_fluxonium}, in which Cooper pairs tunnel across the junction through a single resonant level, the so-called Josephson resonant level. Treating the junction as a two-level system (resonant level empty or doubly occupied), the Josephson potential takes the matrix form
\begin{equation}
V(\hat\phi)=-\Gamma\cos(\hat\phi/2)\,\tilde\sigma_x-\delta\Gamma\sin(\hat\phi/2)\,\tilde\sigma_y-\epsilon_0\,\tilde\sigma_z\,,
\label{eq:Josephson_resonant_level}
\end{equation}
where $\Gamma=\Gamma_L+\Gamma_R$ and $\delta\Gamma=\Gamma_L-\Gamma_R$ are the sum and difference of the pair-tunneling rates between the level and the two leads, $\epsilon_0$ is the energy of the level, and the Pauli matrices $\tilde\sigma_i$ act on the two junction states, compare with the resonant-level model of Eq.~\eqref{Eq:Htransparency}. In the inductive loop, the persistent-current states in which the phase winds by $2\pi m$ (fluxons) are coupled by quantum phase slips at the junction. For a tunnel junction, $2\pi$ phase slips ($\Delta m=1$) dominate, while $4\pi$ phase slips ($\Delta m=2$) are a much weaker, higher-order process. At resonance, $\epsilon_0=\delta\Gamma=0$, the junction becomes fully transparent, the two branches of the potential become the $4\pi$-periodic functions $\mp\Gamma\cos(\phi/2)$, and $2\pi$ phase slips are suppressed, so that $4\pi$ phase slips become the dominant coupling between fluxons. Fluxons with even and odd $m$ then decouple, and the fluxon parity becomes an approximately conserved quantity that can be used to encode a qubit: states of opposite fluxon parity are quasi-degenerate and have disjoint support in phase space.

This is precisely the protection mechanism of the bifluxon qubit~\cite{Kalashnikov2020}, where resonant Cooper-pair tunneling is obtained with two tunnel junctions in series, separated by a small island tuned close to charge degeneracy; deep in the charging regime of the island, the potential in Eq.~\eqref{eq:Josephson_resonant_level} also describes the bifluxon~\cite{Vakhtel2024}. In hybrid devices, the resonant level can instead be a gate-tunable QD in a semiconductor junction. The suppression of $2\pi$ phase slips close to resonance has been observed in gatemons as a suppression of charge dispersion (Sec.~\ref{ss:reduced-charge dispersion})~\cite{Bargerbos2020,Kringhoj2020}, but the $4\pi$-dominated regime has not yet been reached~\cite{Vakhtel2023}. Vakhtel {\it et al.} showed that this regime should be observable in the transition spectrum of a circuit with typical fluxonium parameters: for $E_c/h=2.5$~GHz, $E_L/h=0.25$~GHz and $\Gamma/h\approx5$~GHz, the $4\pi$ phase-slip splitting is $\approx5$~MHz at resonance. The regime is, however, narrow, requiring $\epsilon_0/\Gamma$ and $\delta\Gamma/\Gamma$ of order $10^{-3}$ or smaller. In the bifluxon, the asymmetry $\delta\Gamma$ is fixed by fabrication, whereas in semiconductor junctions both $\epsilon_0$ and $\delta\Gamma$ are gate tunable, at the price of a sensitivity to charge noise. An advantage of semiconductor junctions is that the coupling of the level to the leads can be increased without compromising the two-level description of the junction.

Interestingly, when the inductive energy is much smaller than the plasma frequency of the junction, the low-energy Hamiltonian of the circuit is dual to that of a topological superconducting island, such as the Majorana transmon of Sec.~\ref{Sec:topoQubits}: the fluxon parity of the loop maps onto the fermion parity of the island, and $2\pi$ ($4\pi$) phase slips map onto single-electron (Cooper-pair) tunneling~\cite{Vakhtel2024}. This duality also clarifies the limitations of the fluxon-parity qubit. In the topological island, the parity-breaking coupling between Majoranas can be suppressed exponentially without fine tuning, while a large Josephson energy is provided by a separate junction. In the inductive loop, in contrast, the $2\pi$ phase-slip amplitude can only be suppressed by tuning the junction to resonance, and reducing it away from resonance also reduces the $4\pi$ amplitude, so that the protected regime always requires fine tuning. On the other hand, the analog of quasiparticle poisoning in the fluxon-based design, namely incoherent $2\pi$ phase slips in the loop, may be easier to control than poisoning of a Majorana island, although poisoning of the resonant level itself must still be minimized~\cite{Vakhtel2024}.}

An alternative direction explores multiterminal Josephson junctions, in which the resulting ABSs govern the global circuit behavior \cite{PRXQuantum.5.020340}. Remarkably, generalizing these circuits —by endowing leads with finite capacitance or embedding them in inductive loops— provides a broader design principle for noise protection, effectively realizing bifluxon or $0-\pi$-type qubits in the tunneling limit.

\subsection{Andreev level qubit}\label{s:ALQ}

A second approach for implementing qubits using hybrid superconducting-semiconducting devices consists of exploiting a distinct Andreev bound state (ABS) localized in a superconducting-semiconducting-superconducting Josephson junction to define the qubit. 
As opposed to the approaches that we just discussed, such Andreev pair qubits or Andreev level qubits (ALQs) are microscopic in nature. 

The concept of utilizing an ABS for quantum computation was first proposed by Zazunov {\it et al.} in 2003 \citep{Zazunov2003}. 
They introduced the idea of a superconducting weak link qubit such as the one shown in Fig.~\ref{fig:ALQ}(a). 
The quantum two-level system is formed by the occupation states of an ABS in the superconducting weak link. 
As shown in Fig.~\ref{fig:ALQ}(b), the energies of these two states disperse differently as a function of the superconducting phase drop across the weak link. 
This, in turn, results in distinct circulating supercurrents and junction inductances depending on the ABS occupation (i.e., depending on the state of the Andreev level qubit).
Such state-dependent supercurrent and inductance can then be used to address and readout the qubit.

This theoretical framework laid the groundwork for subsequent experiments investigating the spectroscopic and coherence properties of these systems. The first direct photon-absorption spectroscopy was achieved using a mechanically controllable atomic contact—a superconducting weak link formed on a flexible substrate and narrowed to just a few atoms. In such a device, Bretheau {\it et al.} directly observed the energy-phase dependence of the pair qubit transition \citep{Bretheau2012,Bretheau2013b,Gustavsson2013}. To further simplify the spectrum, the contact can be adjusted until the discrete Andreev levels within the superconducting gap collapse into a single conductive channel. In this single-channel regime, because microwave photons excite quasiparticles in pairs, two types of transitions become possible: pair transitions, which place both quasiparticles into a discrete level, and transitions to odd states, in which one quasiparticle occupies a discrete level and the other is excited into the continuum. The latter transition was demonstrated in subsequent experiments \cite{PhysRevX.3.041034}, where quasiparticle excitations were induced by a broadband on-chip microwave source and detected via changes in the supercurrent flowing through the atomic contact. Crucially, while absorption spectroscopy cannot detect these configurations, supercurrent spectroscopy provides direct access to the long-lived odd states.

The first demonstration of coherent manipulation of the transition between the two even occupancies of an Andreev state was also demonstrated in a superconducting atomic contact by the same group \citep{Janvier2015}.
By embedding the atomic contact within a superconducting loop and coupling it to a microwave resonator, single-shot qubit readout was achieved.
This cQED architecture, together with precise timing of the drive pulse, resulted on the observation of Rabi oscillations between the ground and excited even states.
The energy decay time found in this weak link implementation was of \SI{4}{\micro s} and the Ramsey and echo coherence times were measured to be \SI{38}{ns} and \SI{565}{ns}, respectively.

\begin{figure*}[h!]
\includegraphics[width=0.9\textwidth]{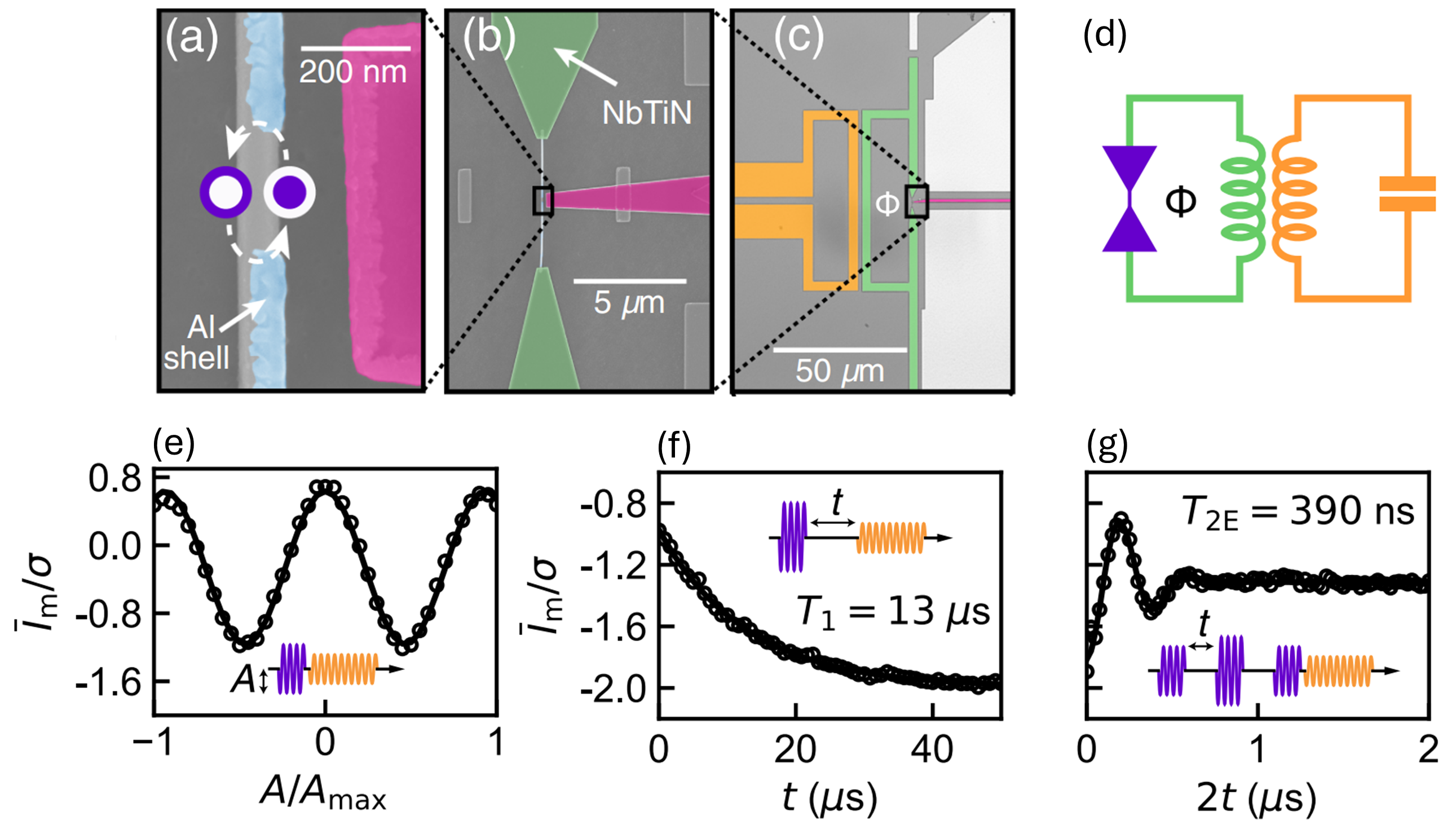}
\caption{
Coherent manipulation of an Andreev level qubit.
(a-c) Scanning electron micrographs showing different levels of enlargement of the device. 
An InAs nanowire Josephson junction (grey) proximitized by aluminium (blue) is situated next to a NbTiN electrostatic gate for ALQ transition frequency control (pink).
The JJ is connected in parallel to an inductive NbTiN loop (green), thus defining a loop through which a magnetic flux, $\Phi$, is threaded for phase control.
The loop is inductively coupled to a readout resonator (yellow).
(d) Circuit diagram of the SQUID loop (green) coupled to the readout resonator (yellow).
(e) Measurement (markers) and fit (line) of the Rabi oscillations between the two qubit states as a function of the amplitude of the applied pulse.
(f), (g)  Measurement (markers) and fit (line) of the decay time and the echo coherence time, respectively.
Adapted and reprinted from Ref.~\citep{Hays2018}.
} 
\label{fig:ALQ-coherence}
\end{figure*}

The precise control of the ABS transparency in a superconducting atomic contact, however, requires the application of a mechanical force, which brings complexity to the setup and hinders scalability.
An alternative tunability method consists of instead using a Josephson junction implemented in a ballistic semiconducting nanowire (Fig.~\ref{fig:ALQ}(c) and Fig.~\ref{fig:ALQ-coherence}(a)).
This implementation permits using a gate electrode to electrostatically control the transparency of the Andreev level and, in turn, the ALQ frequency.

Although the photon-assisted tunneling technique from Refs.~\citep{vanWoerkom2017} and \citep{Bretheau2012} allows for broadband spectroscopy (up to tens of GHz, limited by the superconducting gap of the spectrometer junction), it is not well suited for quantum control experiments due to its continuous-wave nature and lack of fast, time-resolved control.
Building on these spectroscopic insights, and following the techniques used for weak links experiments \citep{Janvier2015}, Hays {\it et al.} \citep{Hays2018} demonstrated the coherent manipulation of even-parity Andreev states in a hybrid InAs/Al nanowire Josephson junction embedded in a cQED architecture (Fig.~\ref{fig:ALQ-coherence}).
This experiment marked the first time-domain demonstration of an Andreev level qubit in a semiconducting weak link, confirming the feasibility of transferring the ALQ concept from atomic contacts to more tunable semiconducting devices.

The experimental setup integrated the semiconducting junction (Fig.~\ref{fig:ALQ-coherence}(a)) into a superconducting loop (green in Fig.~\ref{fig:ALQ-coherence}(c, d)) and capacitively coupled it to a microwave resonator (yellow in Fig.~\ref{fig:ALQ-coherence}(c, d)) for dispersive readout.
The application of resonant microwave pulses then allows for coherently driving of transitions between the ground and excited Andreev level qubit states and observing Rabi oscillations (Fig.~\ref{fig:ALQ-coherence}(e)).
The frequency of the transition can be controlled electrostatically by varying the gate voltage, which tunes the transmission properties of the junction.

This work revealed relaxation and echo times of \SI{13}{\micro s} and \SI{390}{ns}, respectively (Fig.~\ref{fig:ALQ-coherence}(f-g)), likely limited by both charge noise and quasiparticle poisoning \footnote{Relaxation limited by quasiparticle poisoning is compatible with earlier experiments in Ref. \cite{PhysRevLett.106.257003}, where it was shown that quasiparticle trapping in ABSs formed at atomic contacts can exhibit lifetimes exceeding hundreds of microseconds (see also Ref. \cite{PhysRevB.89.104504})}.
These times are comparable to those of early transmon qubits and on par with the atomic-contact implementation \citep{Janvier2015}, demonstrating the potential of nanowire-based ALQs as coherent two-level systems and opening a promising path for integrating ALQs into larger superconducting quantum circuits.
Nevertheless, further work is needed to address the sources of relaxation and dephasing in ALQs to improve their coherence times in future device generations.

\begin{figure*}[h!]
\includegraphics[width=\textwidth]{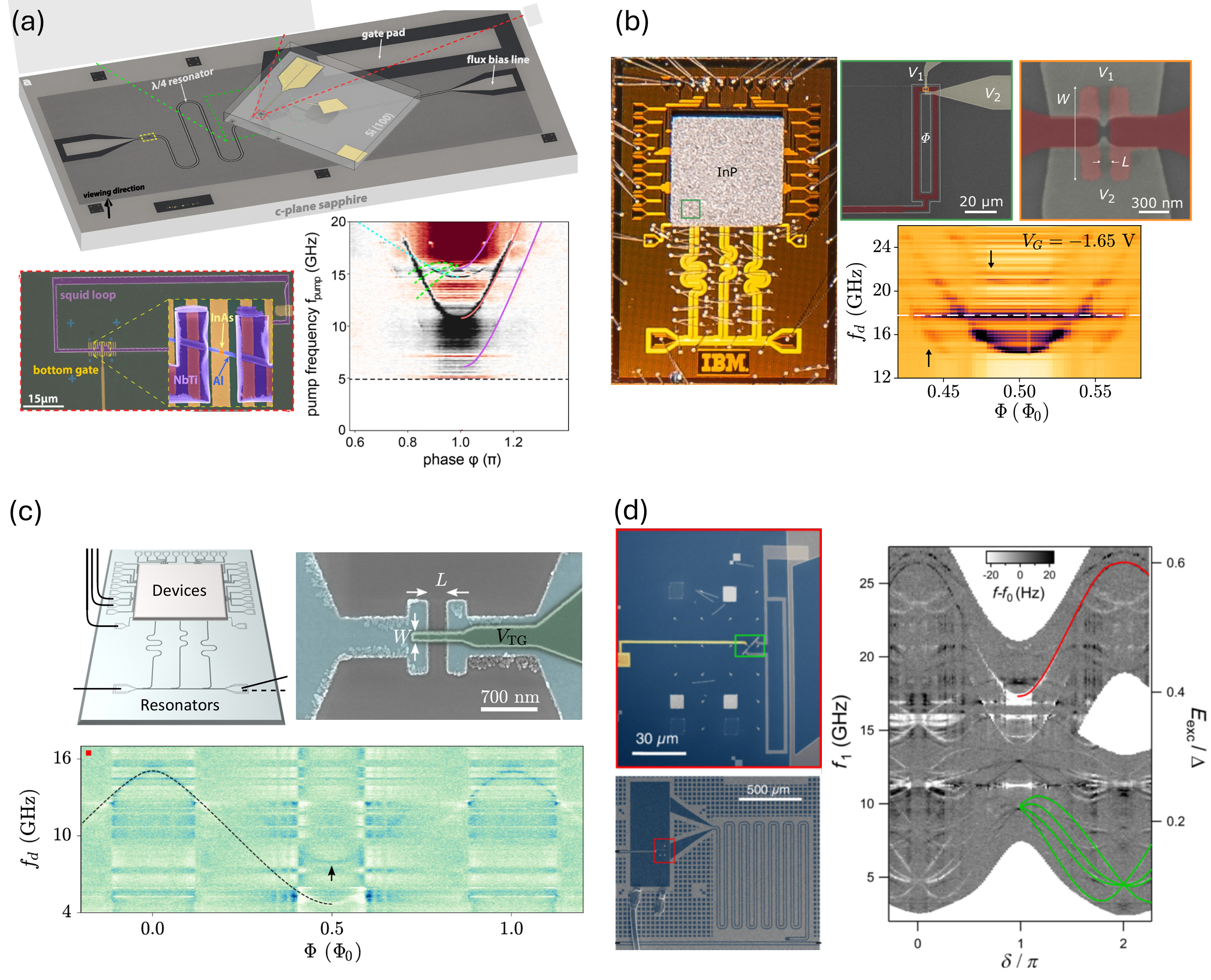}
\caption{ Spectroscopy of the Andreev pair transition.
(a-d) Top: Different devices for the measurement of the microwave response of ABSs in semiconducting Josephson junctions.
Bottom: Two-tone spectroscopic measurements of the Andreev pair transition versus phase in the single- or few-channel regime.
(a), (b) and (c) show flip-chip configurations for InAs nanowire, InAs 2DEG and Ge 2DHG Josephson junctions, respectively.
In all cases, the JJ is incorporated in an RF SQUID loop, for flux control, which is in turn inductively coupled to a readout resonator.
Adapted and reprinted from Ref.~\citep{Zellekens2022}, \citep{Hinderling2023}, \citep{Hinderling2024} and \citep{Tosi2019}.
} 
\label{fig:ABS-spectroscopy}
\end{figure*}

Beyond coherent Andreev manipulation, several recent experiments have expanded our understanding of Andreev bound states in different hybrid semiconducting-superconducting junctions.
Tosi {\it et al.} \citep{Tosi2019} performed microwave spectroscopy of Andreev transitions in gate-tunable InAs nanowire Josephson junctions, probing their dependence on phase and gate voltage. 
Apart from observing the even parity transition (red in Fig.~\ref{fig:ABS-spectroscopy}(d)), they investigated the energy splitting of spinful Andreev states in the presence of spin-orbit interaction, as well as inter-level transitions (see next section).

The future implementation of Andreev level qubits on a scaled-up architecture, however, calls for their definition on planar Josephson junctions on semiconducting heterostructures.  
Various groups have gone in this direction in the past few years.
Ref.~\citep{Chidambaram2022}, for example, found a strong variation of ABS transparency with gate in InAs planar junctions, reaching ABS transparencies above 0.99. 
However, ABS coherence times below the nanosecond range limited the experimental signatures to single-tone spectroscopy observation of the dispersive shift induced on the coupled resonator.

A challenge when combining planar materials with microwave circuits is that the substrates where these heterostructures are grown are typically lossy in the microwave regime, compared to other substrates typically used for microwave experiments, such as high resistivity silicon or sapphire. 
This challenge has been recently addressed by separating the qubit and readout parts of the device on different chips.
References~\citep{Zellekens2022}, Ref.~\citep{Hinderling2023} and \citep{Hinderling2024} employed flip-chip resonator coupling to perform gate- and flux-tunable spectroscopy of ABSs implemented in InAs nanowires, InAs 2DEGs and germanium 2DHGs, respectively (Fig.~\ref{fig:ABS-spectroscopy}(a-c)).
These flip-chip architectures open the possibility of implementing Andreev level qubits in materials that require substrates with high losses in the microwave regime, since they permit implementing the superconducting readout circuitry on a separate chip with better microwave performance. 

\subsection{Andreev spin qubit}\label{s:ASQ}

Andreev spin qubits (ASQs)—also referred to as superconducting spin qubits—represent another variant of Andreev qubits in which quantum information is encoded in the microscopic degrees of freedom of Andreev states localized within a Josephson junction.
Unlike Andreev level qubits that utilize the even-parity manifold, ASQs operate within the odd-parity sector, and encode their state on the spin degree of freedom of a single quasiparticle occupying an Andreev bound state. 
The feasibility of this approach relies on the spin-splitting of ABS energies, which arises from the interplay of spin–orbit coupling and the presence of multiple transport channels, as discussed in Sec.~\ref{ss:spin-splitting}.

The basic concept of encoding and manipulating quantum information in the spin subspace of ABSs was introduced by Chtchelkatchev and Nazarov \citep{Chtchelkatchev2003}. 
They outlined a full operational cycle for such a qubit, including initialization, coherent control, and readout. 
Since the odd states are excited states of the superconducting-semiconducting-superconducting junction, the qubit can be initialized in its computational spin-1/2 subspace by irradiating it with microwaves: a Cooper pair is broken, with one quasiparticle escaping to the continuum and the other remaining trapped in the ABS. The resulting singly occupied ABS serves as the qubit.
For readout, they proposed detecting the spin state via the associated spin-dependent supercurrent, which would generate a measurable magnetic flux difference in a superconducting loop.
Qubit manipulation was addressed through two schemes: (i) coherent spin rotations via a static magnetic field applied perpendicular to the spin quantization axis, enabling natural Rabi oscillations between the spin states, and (ii) resonant driving using an oscillating magnetic field with frequency matched to the spin splitting.

\begin{figure*}[t]
\includegraphics[width=\textwidth]{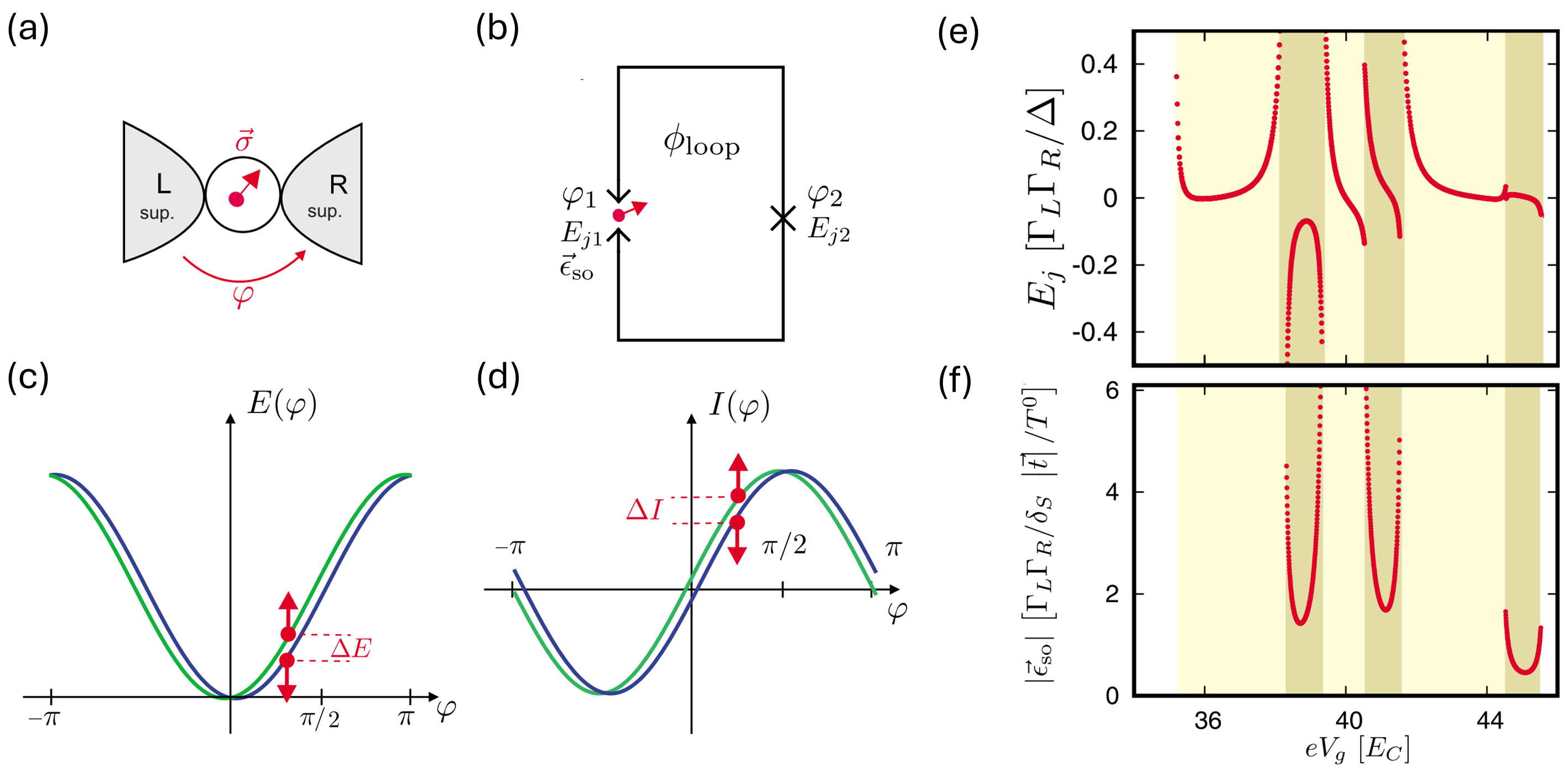}
\caption{ Model of a superconductor-quantum dot-superconductor-junction-based ASQ.
(a) A quantum dot (white circle) occupied by an odd number of electrons and thus resulting in a single spin degree of freedom (red arrow). 
The quantum dot is tunnel-coupled to two superconducting leads (grey shaded regions) with a phase offset $\varphi$ between them.
(b) The Andreev spin qubit quantum dot junction from (a), characterized by spin-independent and spin-dependent Josephson energies $E_{j1}$ and $\vec{\epsilon_{\rm so}}$, respectively, is connected in parallel to a conventional Josephson junction with Josephson energy $E_{j2}$.
A magnetic flux through the loop $\phi_{\rm loop}$ fixes the relation between the phase drops across each of the two junctions, $\varphi_1$ and $\varphi_2$, respectively, to $ \varphi_1+\varphi_2 = 2\pi\phi_{\rm loop}/\Phi_0$, where $\Phi_0$ is the superconducting flux quantum.
(c) and (d) Energy $E$ and supercurrent $I$, respectively, as a function of the phase drop on the junction. 
The two possible spin states are represented with green and blue lines. 
(e) and (f) Spin-independent, $E_j$ and spin-dependent $|\vec{\epsilon_{\rm so}}|$ Josephson energies, respectively, calculated from numerical diagonalization of the system shown in (a). 
$E_C/\Delta= 10$, $\delta/E_C= 1.5$. 
Regions for which the ground state has an even or odd number of electrons in the junction are represented by lighter or darker shaded regions. 
Adapted and reprinted from Ref.~\citep{Padurariu2010}. For details on the numerical analysis, see \citep{Padurariu2010}.
} 
\label{fig:ASQ-theory}
\end{figure*}

Building on this foundation, Ref.~\citep{Padurariu2010} further developed the idea, introducing a method for stabilizing the ASQ junction in its computational manifold. 
They suggested to implement the ASQ on a Josephson junction with a strong charging energy, $E_C$, and tuning its chemical potential to energetically favour the odd-parity manifold, so that the spinful doublet states become the two lowest energy eigenstates of the system.  
This removes the need for active parity initialization and allows the qubit to remain passively confined within the computational subspace.
This system can be modelled with a quantum dot tunnel-coupled to two superconducting lead, illustrated in Fig.~\ref{fig:ASQ-theory}(a), and, under the presence of spin-flipping tunneling terms, results on a spin split energy phase relation such as that shown in Fig.~\ref{fig:ASQ-theory}(c).
The spin-dependent and spin-independent Josephson energies can be varied by electrostatically tuning the chemical potential of the quantum dot (Fig.~\ref{fig:ASQ-theory}(e,f)).
The authors moreover introduced the notion of manipulating the ABS spin degree of freedom through the superconducting phase difference. 
While these ideas remained theoretical for several years, they laid the groundwork for the experimental implementations and coherence investigations discussed in the following paragraphs.

\begin{figure*}[t]
\includegraphics[width=\textwidth]{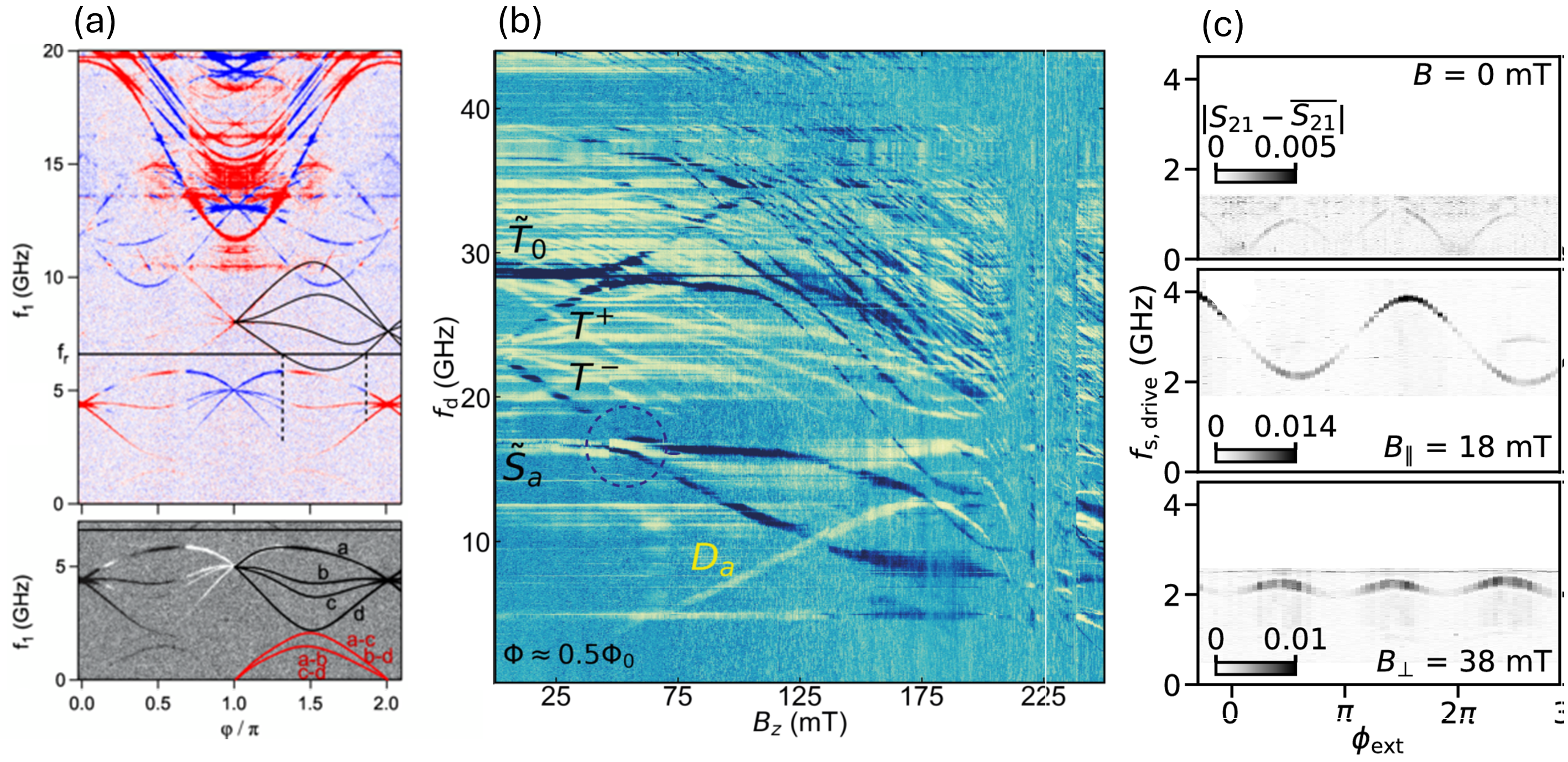}
\caption{ Spectroscopy of spinful Andreev states in InAs-nanowire Josephson junctions.
(a) Two-tone spectrum of an SNS junction versus external flux, in the absence of a magnetic field.
The transitions around 5~GHz correspond to inter-manifold transitions.
The two spin-flip transitions within the two lowest doublet manifolds are indicated with red lines in the bottom panel.
(b) Spectroscopy of singlet, doublet and triplet Andreev states on an SNS junction versus parallel magnetic field and in the vicinity of $\Phi_{\rm ext} = \Phi_0/2$.
The spin-flip transition of the lowest doublet is indicated as $D_a$.
(c) Direct spin-flip spectroscopy for an SQDS junction with a doublet ground state. 
The transition is measured as a function of external flux and for no (top), parallel (center), and perpendicular (bottom) magnetic field relative to the spin direction.
Adapted and reprinted from Ref.~\citep{Metzger2021,Wesdorp2024,Bargerbos2023b}.
} 
\label{fig:ASQ-spect}
\end{figure*}

Following the theoretical proposals for Andreev spin qubits, several experiments have provided spectroscopic evidence for spin-split Andreev bound states and transitions between them. 
As discussed in the previous section, spectroscopic experiments on nanowire-based Josephson junctions can reveal the presence of spin-split states in a very direct manner.
In Fig.~\ref{fig:ABS-spectroscopy}(d), spectroscopy performed on an InAs nanowire junction shows the phase-dispersion of multiple transitions involving spin-orbit-split Andreev levels \citep{Tosi2019}. 
The spectrum shows particle parity-conserving transitions within both even and odd parity manifolds. 
In particular, green transitions correspond to an existing quasiparticle in one ABS being excited to a different ABS, indicating the presence of multiple bound states with spin-dependent structure.

Clear evidence for control of the spin degree of freedom emerged from spectroscopy experiments that resolved spin-flip transitions within the same ABS.
These transitions, highlighted in Fig.~\ref{fig:ASQ-spect}(a), were also observed via microwave spectroscopy in InAs-based Josephson junctions \citep{Metzger2021,Canadas2022,Wesdorp2024}. 
In these works, the spin-flip transitions appeared as a distinct spectroscopic line which, in the absence of a magnetic field, has a frequency of the order of a GHz, strongly dispersing with flux. 
As expected from the Zeeman effect, the spin-flip frequency is found to increase linearly with magnetic field Fig.~\ref{fig:ASQ-spect}(b).

In devices where the junction hosts a quantum dot, charging energy can shift the odd-parity manifold to lower energies, stabilizing the spinful ABS ground states. 
Such effects were investigated using spectroscopic techniques in superconductor–quantum dot–superconductor junctions, where the interplay between Coulomb blockade and superconductivity results in tunable ground state parity \citep{Bargerbos2022,Bargerbos2023b}. 
As shown in  Fig.~\ref{fig:ASQ-spect}(c), the spin-flip frequency, $f_{\rm s}$, has a $\propto |\sin(\phi_{\rm ext})|$ dependence at zero magnetic field.
When a magnetic field is applied parallel to the spin direction, the spin-flip frequency increases by a constant Zeeman offset, preserving its $\propto \sin(\phi_{\rm ext})$ flux dependence.
If a perpendicular magnetic field is applied, however, the two spin degrees of freedom hybridize, reducing as a consequence the flux-dependence of $f_{\rm s}$ .
Similar phenomenology has also been observed in junctions without explicitly defined quantum dots, where charging effects arising from unintentional barriers or material inhomogeneity lead to the stabilization of odd-parity ABS ground states and facilitate the observation of spin-flip transitions \citep{Sahu2024,PhysRevB.104.174517,Fatemi2022,lu2025}. 

The first experimental demonstration of a coherently manipulated Andreev spin qubit was achieved by Hays {\it et al.} (2021)~\citep{Hays2021}, who demonstrated coherent manipulation of the spin degree of freedom in an ABS hosted by a nanowire-based Josephson junction. 
The experiment used an InAs nanowire proximitized by aluminum leads, embedded in a superconducting loop. 
The loop was integrated as part of the inductor of a lumped-element readout resonator, in turn capacitively coupled to a microwave feedline for dispersive readout. 
Although the device was not actively initialized into an odd-parity state, occasional quasiparticle poisoning events allowed the system to randomly occupy the odd manifold, enabling access to the spin-$\frac{1}{2}$ states within the ABS, and thus serving as initialization.
To coherently manipulate the spin states, the authors employed a Raman driving scheme involving two microwave tones detuned from the direct ABS transition but resonant with a virtual transition pathway via the even-parity manifold.

\begin{figure*}[t]
\includegraphics[width=\textwidth]{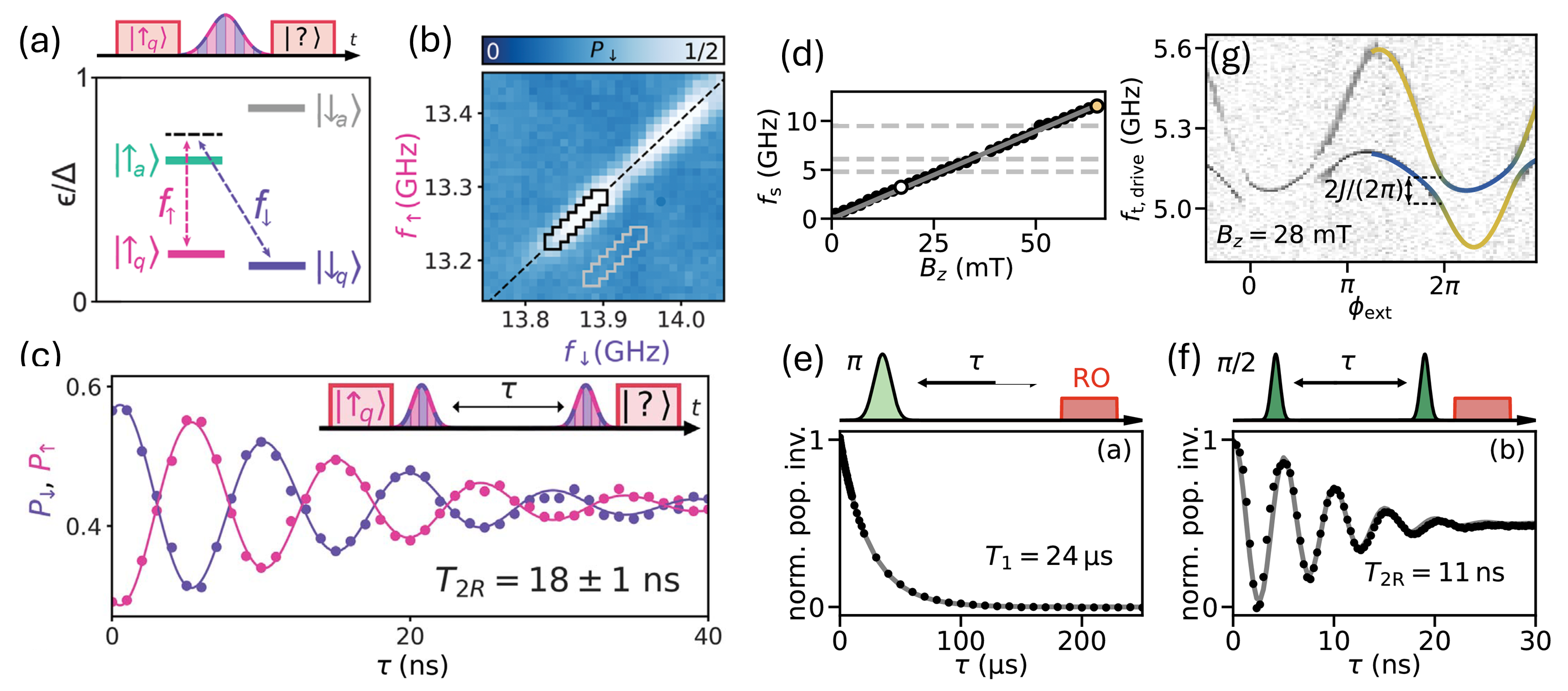}
\caption{ Coherent manipulation of Andreev spin qubits.
(a) Energies of the two spin states of the ASQ (pink and purple) and a higher-energy auxiliary doublet state (green), tuned to a $\Gamma$ energy configuration. 
Two microwave drives with frequencies $f_\downarrow$ and $f_\uparrow$ are equally detuned from the transitions between each of the spin states and the auxiliary doublet.
The two drives induce a Raman transition between the qubit states.
(b) Measurement of the Raman transition between the two spin doublet states of an SNS junction.  
Simultaneous Gaussian pulses with variable frequencies $f_\downarrow$, $f_\uparrow$ are applied to the ASQ initialized in the $\ket{\downarrow}$ state.
The black dashed line indicates $f_\downarrow - f_\uparrow$~=~610~MHz. 
(c) Ramsey measurement of the decoherence time of the ASQ. Oscillations are introduced by adding a phase proportional to $\tau$ to the final pulse. 
(d) Extracted spin-flip frequency of an ASQ defined on a SQDS Josephson junction, $f_{\rm s}$, versus the magnetic field applied along the nanowire $B_z$ (markers).
The data is fitted with a linear dependence (solid line), resulting in an effective Landé g-factor of $g^* = 12.7 \pm 0.2$.
(e, f) Qubit lifetime and Ramsey measurements of the qubit in (d), at $B_z$~=~65~mT. 
Solid lines indicate fits to the data
(g) Two-tone spectroscopy of the avoided crossing between the spin-flip transition of the ASQ (yellow) and a transmon qubit (blue), demonstrating coherent ASQ-transmon coupling. $B_z$~=~28~mT
Adapted and reprinted from Ref.~\citep{Hays2021,PitaVidal2023}.
} 
\label{fig:ASQ-expt}
\end{figure*}

A schematic of the Raman drive configuration is shown in Fig.~\ref{fig:ASQ-expt}(a). 
The lower and upper microwave tones are detuned from the even-parity transition but chosen such that their frequency difference matches the energy splitting between the spin states. 
Under this condition, a second-order transition is induced between the spin-up and spin-down states via the virtual intermediate level~\citep{Cerrillo2021}. 
The effectiveness of this approach is illustrated in Fig.~\ref{fig:ASQ-expt}(b), which shows the measured probability of finding the qubit in the spin-down (excited) state as a function of the two drive frequencies. 
A resonance condition is observed when the frequency difference matches the spin splitting, indicating successful driving of the spin.

Further confirmation of coherent control was obtained through time-domain measurements, such as Ramsey interference, shown in Fig.~\ref{fig:ASQ-expt}(c).
From these measurements, dephasing times ($T_2^*$) on the order of 50–100 ns were extracted, while the spin lifetime ($T_1$) was measured to be \SI{17}{\micro \second} in a separate measurement. 
These results demonstrated that coherent control of ABS spin states is feasible in realistic device settings and established Raman driving as a powerful technique for spin manipulation in superconducting-semiconducting hybrid systems.
The experiment also highlighted the sensitivity of the qubit to magnetic and electric noise, motivating further work to improve coherence through material and circuit engineering.

A complementary approach to ASQ control is shown in panels (d--g) of Fig.~\ref{fig:ASQ-expt}, based on the experiment reported in Ref.~\citep{PitaVidal2023}. 
This work addressed two key limitations of the earlier demonstration. 
First, instead of relying on stochastic quasiparticle poisoning for initialization, the junction was operated in a regime where a spinful ABS doublet became the ground state, due to charging effects in a superconductor--quantum dot--superconductor (SQDS) configuration. 
This increased the parity lifetime of the odd, with respect to the even, manifold, thus facilitating initialization into the computational space.

Second, coherent control of the spin degree of freedom was achieved by directly driving the spin-flip transition, rather than employing a Raman scheme, thus simplifying qubit control.
Different driving mechanisms have been comprehensively investigated theoretically, such as the spin-orbit mediated electric dipole spin resonance (EDSR) and supercurrent-mediated driving \citep{Park:PRB17,Fauvel2024,Paveski2024}. 
The driving mechanism observed experimentally is likely a combination of both.
The spin-flip frequency $f_{\rm s}$ was measured as a function of magnetic field aligned with the nanowire, revealing a linear dependence (Fig.~\ref{fig:ASQ-expt}(d)). 
The observed values of $T_2^*$  in this work (Fig.~\ref{fig:ASQ-expt}(e,f)) are comparable to those reported in Ref.~\citep{Hays2021}, suggesting that coherence was limited by environmental decoherence sources common to both implementations such as coupling to the nuclear spins~\citep{Hoffman2024}.

Additionally, Ref.~\citep{PitaVidal2023} demonstrates coherent coupling between the ASQ and a conventional transmon qubit. 
Two-tone spectroscopy reveals an avoided crossing when the spin-flip transition is brought into resonance with the transmon, as shown in Fig.~\ref{fig:ASQ-expt}(g). 
This result illustrates the potential of ASQs as a coupling element between spin-based and superconducting quantum processors.

An alternative approach to initialize the ASQ in the computational space has been investigated for scenarios where the spinful doublet states are not the lowest energy states of the system, as is often the case in SNS junctions with reduced charging energy or in certain gate configurations of SQDS junctions. 
As discussed, in such regimes, the system typically relaxes into the even-parity ground state, making deterministic initialization into the odd-parity manifold stochastic and uncommon. 
A promising solution to this limitation is the exploitation of parity polarization, a phenomenon whereby an effective imbalance between the occupation probabilities of the two parity sectors can be induced using microwave pulses. 
This effect was experimentally demonstrated in Ref.~\citep{Wesdorp2023}, where parity initialization fidelities up to 94 \% were achieved.
This technique enables initialization of the Andreev state within a fixed parity sector and offers a method to reset leakage out of the qubit subspace caused to quasiparticle poisoning.

Theoretical investigations have further analyzed the origin and robustness of parity polarization in nanowire junctions, establishing that the resonant microwave drive breaks a Cooper pair, and subsequently ionizes one of the levels, resulting in a change of the occupation parity \citep{Ackermann2023,kurilovich2024,Zatsarynna2024}.

A path for next-generation Andreev spin qubits lies in mitigating the dominant decoherence mechanisms identified in current platforms. 
Theoretical analysis suggests that dephasing in ASQs based on InAs nanowires is largely limited by magnetic noise arising from the nuclear spins of indium and arsenic \citep{Hoffman2024}. 
To address this limitation, an attractive strategy would be to implement ASQs in materials with low or zero nuclear spin content. 
One such candidate is isotopically purified germanium, which offers a nuclear-spin-free environment and has recently emerged as a versatile platform for hybrid superconducting-semiconducting devices.

Although coherent manipulation of spin states in Ge-based Josephson junctions has not yet been demonstrated, essential building blocks for ASQ implementation have already been realized. 
SNS junctions with high-transparency superconducting contacts \citep{Su2016,Delaforce2021,Ridderbos2019,Wu2024b,Hendrickx2018,Hendrickx2019,Vigneau2019,Aggarwal2021,tosato2023,Valentini2024} and quantum dot Josephson junctions have been demonstrated in planar Ge heterostructures~\citep{Lakic} (see Sec.~\ref{ss:germanium}). 
Furthermore, the integration of Ge quantum devices into cQED architectures has been achieved, with strong qubit-resonator coupling and coherent qubit operation reported in several recent works \citep{Valentini2024,Sagi2024,Hinderling2024}. The first theoretical articles discussing the proximity effect in Ge 2DHGs \cite{Pino2025-holes,Babkin2025-holes}are also very promising in terms of the new possibilities these systems offer for obtaining new functionalities in qubit platforms. Several related proposals share conceptual similarities with ASQs, including qubits based on Yu-Shiba-Rusinov states in single or double quantum dots coupled to superconducting leads~\citep{pavesic2022,Geier_PRR2024,steffensen2025}.

\subsubsection{Circuit-protected Andreev qubits}\label{ss:circuit-protected-AQ}
{A promising route towards more robust Andreev qubits is to combine the fermionic degree of freedom of the Andreev levels with the bosonic modes of the superconducting circuit in which the junction is embedded, so that protection is engineered at the circuit level, in the spirit of the Hamiltonian-protected qubits of Sec.~\ref{s:protected-qubits}. The key ingredient is the coupling between the Andreev degree of freedom and the supercurrent: through the spin-dependent Josephson potential of the junction, Eq.~\eqref{spin-Josephson}, the minimum of the circuit potential, and hence the plasmon wavefunctions, are displaced in phase space depending on the spin of the trapped quasiparticle. In a transmon circuit with a strong spin-supercurrent coupling, a flip of the quasiparticle spin can then only occur if it is accompanied by plasmons excitations, as dictated by the Franck-Condon principle. This Franck-Condon blockade exponentially suppresses spin relaxation at temperatures small compared to the plasmon energy~\citep{kurilovich2025}.

\begin{figure}[!htbp]
\centering
\includegraphics[width=0.8\linewidth]{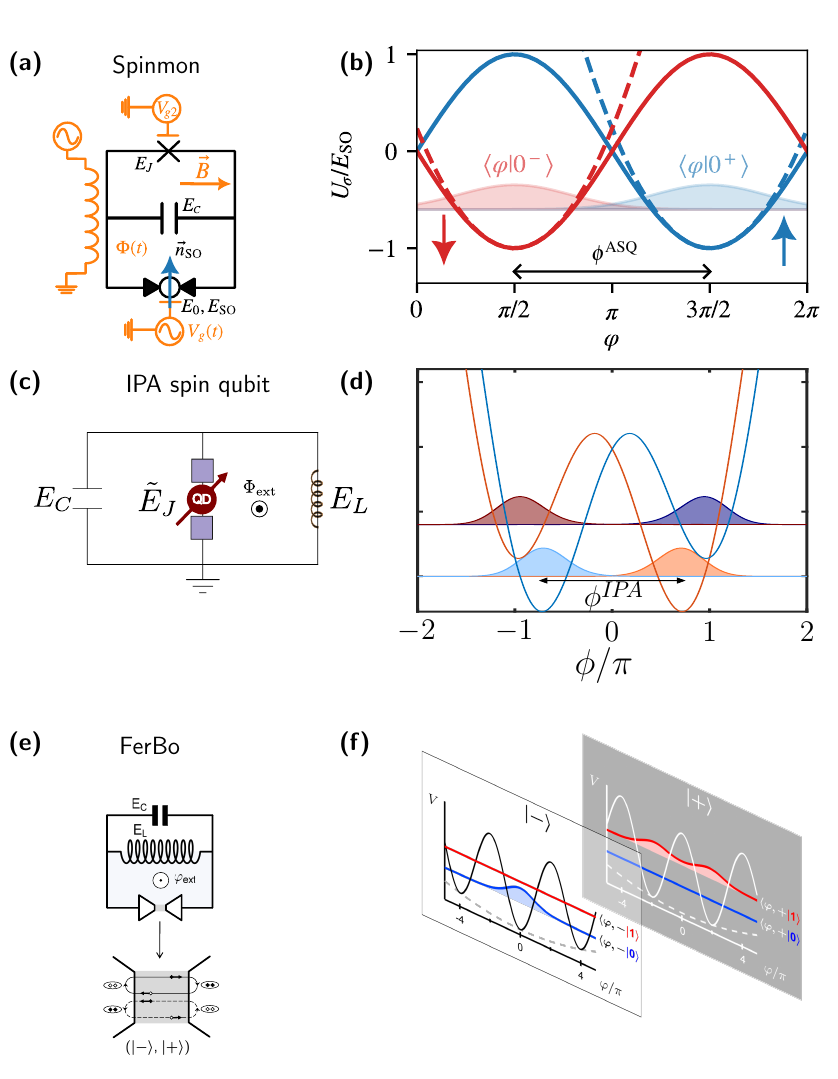}
\caption{Circuit-protected Andreev qubits. (a) Spinmon, i.e., a capacitively shunted ASQ junction (with energies $E_0$ and $E_{\rm SO}$), connected in parallel with a gate-tunable Josephson junction ($E_J$) for flux and gate-voltage control. (b) Spin-dependent potentials of the spinmon for spin $\uparrow$ (blue) and $\downarrow$ (red), whose minima, Eq.~\eqref{eq:ASQ_minima} with $E_0$ replaced by $E_0-E_J$, are separated by $\phi^{\rm ASQ}$ in phase space (see the definition after Eq.~\eqref{eq:ASQ_minima}), together with the wavefunctions of the two computational states. Adapted from Ref.~\citep{Tjernshaugen_arXiv2026}. (c) Inductively-protected Andreev (IPA) spin qubit, in which the ASQ is shunted by a linear inductance $E_L$. (d) Spin-dependent Josephson potentials (lines) and wavefunctions (filled curves) of the lowest-energy states of the IPA qubit, showing the localization of opposite spins in distinct wells separated by $\phi^{IPA}\sim 2\pi$, larger than the maximum separation $\phi^{\rm ASQ}\leq\pi$ in a conventional ASQ. Adapted from Ref.~\citep{delOlmo_arXiv2026}. (e) FerBo circuit: a fluxonium-like circuit in which the Josephson junction is replaced by a highly transmitting weak link hosting Andreev bound states $|-\rangle$ and $|+\rangle$. (f) The FerBo potential depends on the Andreev-level occupation: the ground state resides in the $|-\rangle$ manifold and the first excited state in the $|+\rangle$ one, both delocalized in phase. Adapted from Ref.~\citep{Caceres_arXiv2026}.}
\label{fig:circuit_protected_AQ}
\end{figure}

Such capacitively shunted ASQs have been termed spinmons~\citep{Tjernshaugen_arXiv2026}, see Fig.~\ref{fig:circuit_protected_AQ}(a,b). To achieve full control of the spinmon, Ref.~\citep{Tjernshaugen_arXiv2026} proposed connecting the ASQ junction hosting the trapped quasiparticle in parallel with an additional gate-tunable Josephson junction. The Josephson energy $E_J$ of the latter partially compensates the spin-independent term $E_0\cos\phi$ of the ASQ potential, Eq.~\eqref{eq:ESOpotential}, so that the separation between the spin-dependent potential wells, and hence the qubit splitting, can be tuned electrostatically via $\delta E=E_0-E_J$, while the loop formed by the two junctions allows for flux driving. The computational states are entangled states of the transmon and of the spin of the trapped quasiparticle, localized in spin-dependent wells of the Josephson potential, and their Kramers degeneracy is lifted by a Zeeman field. Full qubit control can be achieved either electrically, via electric-dipole spin resonance with the gate of the ASQ junction, or with an AC flux drive through the loop, and the qubit is predicted to be robust against charge noise, with flux noise as the main remaining relaxation channel.

In the inductively-protected Andreev (IPA) spin qubit~\citep{delOlmo_arXiv2026}, the ASQ is instead shunted by a large linear inductance, as in a heavy fluxonium (Sec.~\ref{ss:fluxonium}), see Fig.~\ref{fig:circuit_protected_AQ}(c,d). The inductive energy breaks the $2\pi$ periodicity of the spin-dependent Josephson potential and localizes the two spin states in distinct wells separated by $\phi^{IPA}\sim 2\pi$ in phase, up to twice the maximum separation $\phi^{\rm ASQ}\leq\pi$ in a conventional ASQ (see the discussion after Eq.~\eqref{eq:ASQ_minima}), so that the device behaves as two fluxoniums in the heavy regime, one for each spin. The overlap between the qubit wavefunctions, and therefore the relaxation rate, is suppressed by several orders of magnitude with respect to conventional ASQs, while the large anharmonicity allows for fast gates. Proximitized germanium with grAl contacts has been identified as a natural platform for this qubit~\citep{delOlmo_arXiv2026}: germanium offers negligible hyperfine coupling (Sec.~\ref{ss:germanium}), while grAl, whose high kinetic inductance has already been used to implement the superinductance of fluxonium qubits~\citep{Grunhaupt2019} and compact meandered superinductors~\citep{Kamenov2020} (see also Sec.~\ref{ss:fluxonium}), could provide both the superconducting contacts and the large shunt inductance required by the IPA qubit, together with a good resilience to magnetic fields.

A related idea, which does not require a trapped quasiparticle, is the FerBo (fermion-boson) qubit~\citep{Caceres_arXiv2026}, a fluxonium-like circuit consisting of a large inductance, a small capacitance and a highly transmitting Josephson weak link, described by the resonant-level model of Eq.~\eqref{eq:Josephson_resonant_level}, see Fig.~\ref{fig:circuit_protected_AQ}(e,f). The two qubit states belong to different Andreev manifolds of the weak link, which suppresses the charge matrix elements responsible for dielectric relaxation by several orders of magnitude, while, as in a fluxonium, their delocalization in phase space protects them against flux-noise dephasing. The FerBo thus avoids the usual tradeoff of the fluxonium, in which phase delocalization suppresses dephasing at the cost of an enhanced sensitivity to dielectric losses. Protection requires a high circuit impedance and a small asymmetry between the couplings of the level to the two leads, $\delta\Gamma/\Gamma\sim 10^{-2}$, conditions that are within reach of gate-tunable semiconductor weak links. Together with the Josephson-resonant-level and bifluxon circuits discussed in Sec.~\ref{s:protected-qubits}, these proposals illustrate how the hybridization of Andreev and circuit degrees of freedom opens a route towards protected hybrid qubits, whose experimental demonstration is an important near-term goal.}

\subsection{Coupling of Andreev qubits and scaling}

Coupling between Andreev qubits and their potential scalability are central considerations in the roadmap toward multi-qubit quantum information processing with Andreev state-based hybrid devices.
Both Andreev spin qubits (ASQs) and Andreev level qubits (ALQs) offer unique opportunities for qubit-qubit interactions owing to their microscopic nature and intrinsic coupling between qubit state and supercurrent. 
This section highlights different strategies for engineering such couplings and explores proposals for scalable architectures.

\begin{figure*}[h!]
\includegraphics[width=0.85\textwidth]{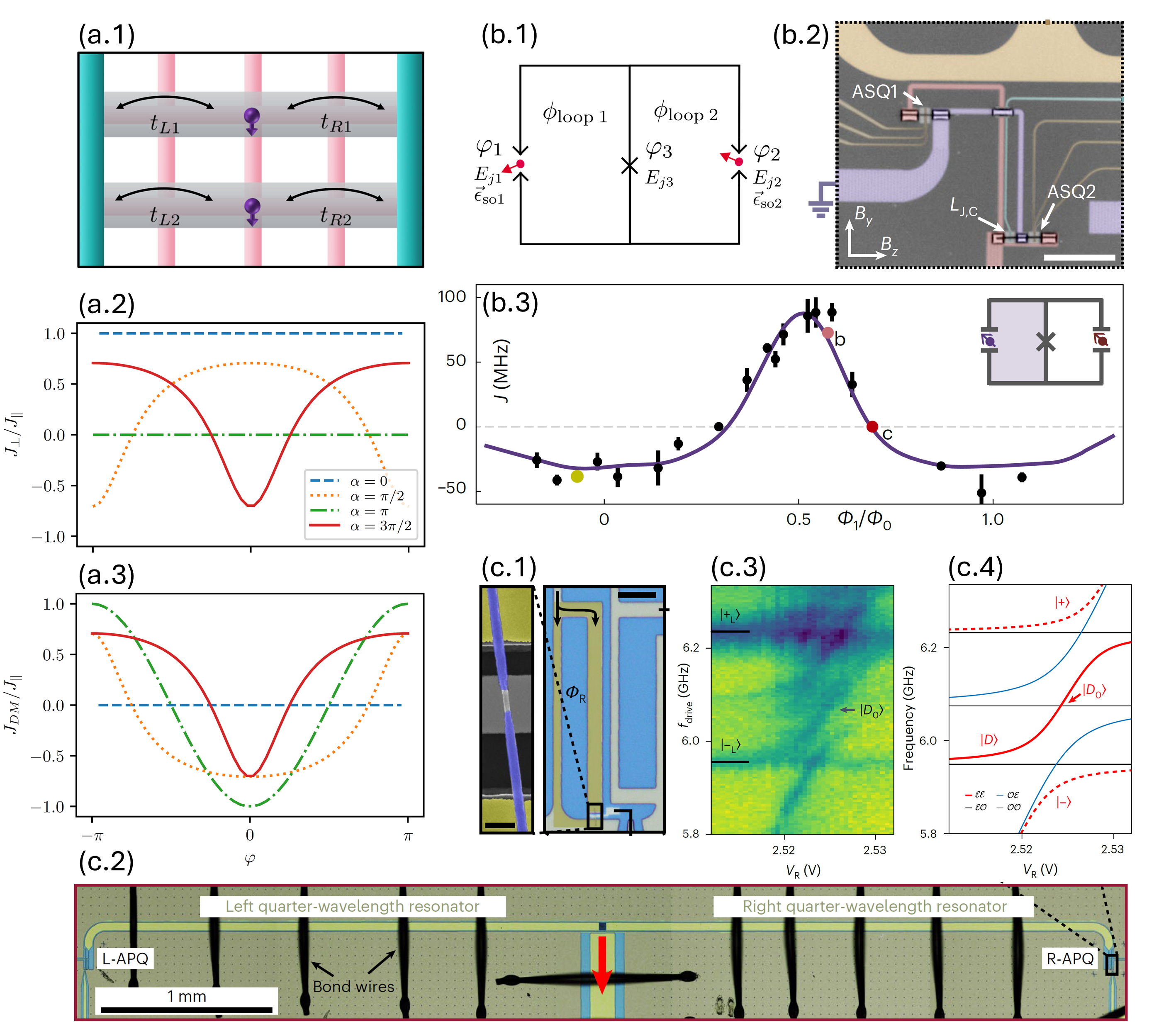}
\caption{ Approaches to coherent coupling between Andreev qubits.
(a) ASQ-ASQ coupling via crossed-Andreev processes enabled by wavefunction overlap.
(a.1) Two spinful Andreev levels (purple) weakly tunnel coupled ($t {jn}$) to two superconducting leads (teal).
The superconductors mediate the interaction between the two Andreev spin qubits.
(a.2) Spin exchange $J_\perp/J_{\parallel}$ and (a.3) Dzyaloshinskii-Moriya interaction strength $J_{\rm DM}/J$ for the system in (a.1) versus superconducting phase difference $\varphi$ and the SOI strength $\alpha$. 
When $\varphi= \pi \pm \alpha/2$ and $\alpha \neq 0$, the interaction is of pure Ising type. 
(b) Supercurrent-mediated coupling between distant Andreev spin qubits.
(b.1) Two superconducting loops, each interrupted by an ASQ junction. 
The two ASQs are connected in parallel to a conventional Josephson junction with Josephson energy $E_{j3}$.
(b.2) Experimental implementation of the circuit in (b.1).
The two ASQs are implemented with electrostatically-defined SQDS junctions on InAs nanowires and a transmon circuit is used for readout.
(b.3) Measured longitudinal spin-spin interaction for the device in (b.2) measured versus the flux through one of the two loops (highlighted in purple). 
(c) Microwave photon-mediated coupling between two distant Andreev level qubits. 
(c.1) (left) Scanning electron micrograph of one of the Andreev level qubits implemented in an InAs nanowire (grey) proximitized by aluminium (purple). 
(right) Optical image of the superconducting loop that contains the ALQ (yellow) inductively coupled to the resonator (light grey).
(c.2) Chip containing two capacitively coupled quarter-wavelength coplanar waveguide resonators (light green).
The end segment of each resonator is inductively coupled to one ALQ.
(c.3) Measured signal versus the electrostatic gate voltage controlling the frequency of the right ALQ, $V_{\rm R}$.
The frequencies of both ALQs are tuned in resonance with the antisymmetric mode of the resonator.
(c.4) Numerically calculated transition frequency spectrum as a function of $V_{\rm R}$. 
Different line colors correspond to different parity configurations, with $\mathcal{E}$ and $\mathcal{O}$ denoting the even and odd parities, respectively, of each ALQ.  
Resonance $\ket{D}$ (solid red line) corresponds to a two-qubit hybrid state with an antisymmetric superposition of the two ALQ excited states.
Adapted and reprinted from Ref.~\citep{Spethmann2022,Padurariu2010,PitaVidal2024,Cheung2024}.
} 
\label{fig:ASQ-coupling}
\end{figure*}

One avenue for coupling ASQs is through direct exchange interactions mediated by crossed-Andreev reflection. 
In this mechanism, two spinful Andreev states are weakly coupled to the same pair of superconducting leads, allowing for spin-spin interactions via virtual tunneling processes (Fig.~\ref{fig:ASQ-coupling}(a)).
Theoretical modeling has shown that such systems exhibit tunable spin exchange and Dzyaloshinskii-Moriya interactions, with symmetry and strength controlled by the spin-orbit coupling and the superconducting phase difference \citep{Spethmann2022}.
As shown in Fig.~\ref{fig:ASQ-coupling}(a.2, a.3), this approach allows for electric field and flux control of the spin-spin coupling type and strength, potentially enabling fast and addressable gate operations.

A different strategy exploits supercurrent-mediated coupling between spatially separated ALQs or ASQs. 
When two ASQs are embedded in separate superconducting loops that share a common Josephson element (see Fig.~\ref{fig:ASQ-coupling}(b.1)), the supercurrent flowing through the shared junction depends on the spin configuration of each Andreev qubit. 
This results in a longitudinal spin-spin interaction of the form $\sigma_{z,1} \sigma_{z,2}$ \citep{Zazunov2003,Chtchelkatchev2003,Padurariu2010}. 
An experimental implementation of this idea is shown in Fig.~\ref{fig:ASQ-coupling}(b.2), where two ASQs are implemented using gate-defined SQDS junctions and read out via a transmon circuit \citep{PitaVidal2024}. 
Measurements reveal a tunable longitudinal spin-spin coupling strength as a function of external magnetic flux, providing direct evidence of the interaction (Fig.~\ref{fig:ASQ-coupling}(b.3)).

Andreev level qubits (ALQs), which are based on even-occupancy states of the junction, have also been shown to be coherently coupled via shared microwave resonators. 
A resonator-mediated interaction between two ALQs was demonstrated with two superconducting loops inductively coupled to a common resonator pair, as illustrated in Fig.~\ref{fig:ASQ-coupling}(c). 
This coupling mechanism is similar in spirit to approaches used in transmon qubit arrays and quantum dot spin qubits \citep{Cheung2024}. 
Spectroscopy measurements showed hybridized two-qubit states, demonstrating the compatibility of these devices with photon-mediated long-range interactions \citep{Spethmann2022}.

\begin{figure*}[h!]
\includegraphics[width=\textwidth]{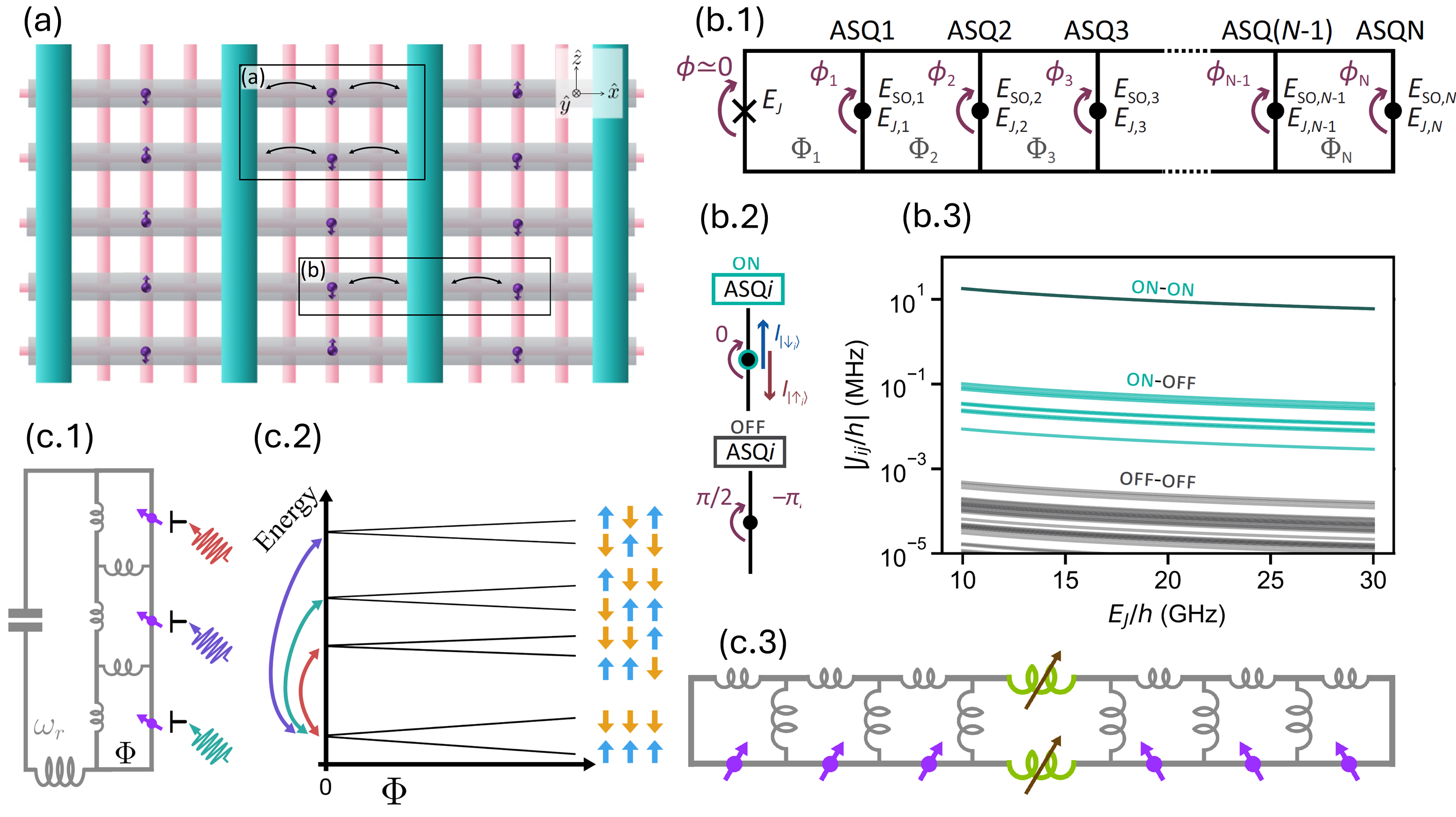}
\caption{ Scaled-up architectures with Andreev spin qubits.
(a) Scalable architecture of ASQs coupled via crossed-Andreev reflection.
(a.1) The setup consists of parallel nanowires (gray), orthogonal superconducting stripes (teal), and a crossbar design of electrostatic gates (pink) that form a rectangular array of quantum dots in the nanowires.
The outer electrostatic gates control the tunnel coupling to the superconductors. 
When certain tunnel barriers are lowered, the adjacent superconducting spin qubits interact pairwise with each other.
(b) All-to-all connected Andreev spin qubits.
(b.1) Circuit diagram of $N$ Andreev spin qubits connected in parallel to a coupling Josephson junction with Josephson energy $E_J$.
The circuit contains $N$ loops, with fluxes $\Phi_i$. 
The phase drop across ASQ$i$ is denoted as $\phi_i$ and is controlled by the combination of fluxes through all loops.
The phase drop across the coupling junction, $\phi$, goes to zero in the limit of $E_J$ being much larger than the ASQ Josephson energies. 
(b.2) Diagrams of two possible phase set points for an ASQ. 
When $\phi_i = \pm \pi/2$, the spin-dependent component of the supercurrent through the current vanishes and the qubit is labeled as OFF. 
When, instead, $\phi_i = 0, \pi$, the spin-dependent component of the supercurrent is maximal and the qubit is labeled as ON. 
(b.3) The absolute value of the longitudinal qubit-qubit coupling strength for $N=10$, $E_{{\rm SO},i}/h$~=~300~MHz, and $E_{J, i}/h$~=~0 for all ASQs.
Random offsets are added to the ideal flux-bias points to represent experimental inaccuracies. 
The dark green line indicates the coupling strengths between two qubits that are ON, $n$ and $m$, the light green lines indicate the (undesired) coupling strengths between either $n$ or $m$ and another qubit, and the gray lines indicate the (undesired) coupling strength between any other pair of qubits close to their OFF setpoint. 
(c) Architecture for partial quantum error correction with ASQs.
(c.1) Circuit for minimal bit-flip encoding with ASQs. 
The gates (black) include DC gate voltages (not shown) and AC drives (colored) on each ASQ.
(c.2) Energy levels as a function of flux, common to all loops.
The Kramers’ degeneracy point is at zero flux. 
Colored arrows correspond to single-spin transitions between the logical manifold (lowest) and the error manifolds.
(c.3) Circuit diagram for coupling two logical ASQ modules.  
The tunable inductors (green) can be used to turn on or off a spin-spin interaction between the
neighboring spins of two modules. 
The spin-spin interaction of the form $\sigma_{3,a,z} \sigma_{1,b,z}$ between the last spin of module $a$ and the first spin of module $b$ corresponds to a logical $ZZ$ interaction.
Adapted and reprinted from Ref.~\citep{Spethmann2022,Pitavidal2025,Lu2025b}.
} 
\label{fig:ASQ-scaling}
\end{figure*}

Beyond two-qubit coupling, proposals for scaling Andreev qubits to larger networks have also emerged. 
One approach involves using crossed-Andreev reflection in a two-dimensional crossbar array of nanowires and superconducting strips, with tunable tunnel couplings defined by electrostatic gates \citep{Spethmann2022}. 
As illustrated in Fig.~\ref{fig:ASQ-scaling}(a), such a layout enables pairwise interaction between ASQs on demand while keeping the architecture planar and potentially compatible with CMOS technology.

Another architecture, illustrated in Fig.~\ref{fig:ASQ-scaling}(b), considers an all-to-all connectivity scheme where multiple ASQs are connected in parallel to a common coupling junction.
In the limit where the coupling junction has a much larger Josephson energy than the individual ASQs, its phase remains fixed, enabling phase-mediated longitudinal spin-spin interactions between all qubits. 
Logical ON/OFF control of each qubit's interaction can be implemented by tuning the local phase difference, effectively programming the coupling graph \citep{Pitavidal2025}.

In parallel, new proposals are beginning to address quantum error correction using ASQs. 
An example is shown in Fig.~\ref{fig:ASQ-scaling}(c), where three ASQs encode a logical qubit protected against single bit-flip errors. 
Transitions between the logical and error manifolds are controlled by external flux and AC gate drives. 
This minimal encoding, together with tunable couplers between logical qubits, enables the construction of fault-tolerant units with reduced overhead \citep{Lu2025b}.

These diverse strategies—ranging from microscopic exchange to resonator-mediated and flux-controlled couplings—highlight the versatility of Andreev qubits for scalable quantum computing. 
Ongoing experimental advances in material platforms, coherence, and control circuitry will be critical for translating these concepts into large-scale quantum processors.

\section{Hole-based hybrid nanostructures }\label{ss:germanium} 

\subsection{Experimental progress}

In the past decade, p-type semiconductors have moved in the focus of attention as hole spins have emerged as a very interesting spin-qubit platform~\cite{PhysRevLett.95.076805,bulaev_electric_2007}, particularly in group IV materials, because they do not face the valley problem, they can be isotopically purified and fully electrically driven due to the strong spin-orbit interaction~\cite{Scappucci2021,Fang2023}.  In 2016, the first hole Loss-DiVincenzo spin-qubit~\cite{Loss1998} was realized in Silicon~\cite{Maurand2016}, and since then, significant progress has been made in improving coherence times by identifying sweet spots~\cite{Piot2022}. Furthermore, because of the large spin-orbit interaction, record spin-photon coupling strength has been reported~\cite{Yu2023}. In 2018, the first Ge spin-qubit was reported in one-dimensional nanostructures, called hut wires~\cite{watzinger_germanium_2018}. Ge has the advantage that the effective mass of the heavy-hole is light in the transport direction - significantly lighter than that of the holes in Si - thus reducing the nanofabrication constraints~\cite{Scappucci2021,Lodari,PhysRevB.109.075427,g29w-st3q}. This, combined with the very high material quality of planar Ge~\cite{Stehouwer2023,Stehouwer2025}, has led to significant progress in the past few years. Two-qubit gates and a 4-qubit device were demonstrated in 2020 and 2021, respectively~\cite{hendrickx_fast_2020,hendrickx_four-qubit_2021}. To date, two-dimensional sixteen-quantum-dot crossbars and ten spin-qubit arrays have been demonstrated, underlining the scalability potential of Ge/SiGe heterostructures~\cite{Borsoi2024,Wang2024,john2025}. In addition, single-qubit fidelities exceeding 99.9\% and two-qubit gate fidelities exceeding 99\% have been demonstrated~\cite{Lawrie2023,Wang2024}. Furthermore, a singlet-triplet qubit that can be operated at magnetic fields as low as 250$\mu$T has been demonstrated, making it compatible with superconducting technology~\cite{jirovec_singlet-triplet_2021}. 

For both Si and Ge holes, coherence times two to three orders of magnitude larger than those of III-V materials have been reported~\cite{Piot2022,Hendrickx2023}, making them an appealing platform for those hybrid devices for which the spin dephasing time is crucial, such as Andreev spin qubits. However, proximity-induced superconductivity in Si has been achieved only with very high B doping, an approach that is rather difficult to adapt to quantum devices~\cite{Bustarret2006,ProximitySilicon}. On the other hand, Ge, because of Fermi pinning, allows the creation of hybrid devices. 

\begin{figure*} 
\includegraphics[width=\textwidth]{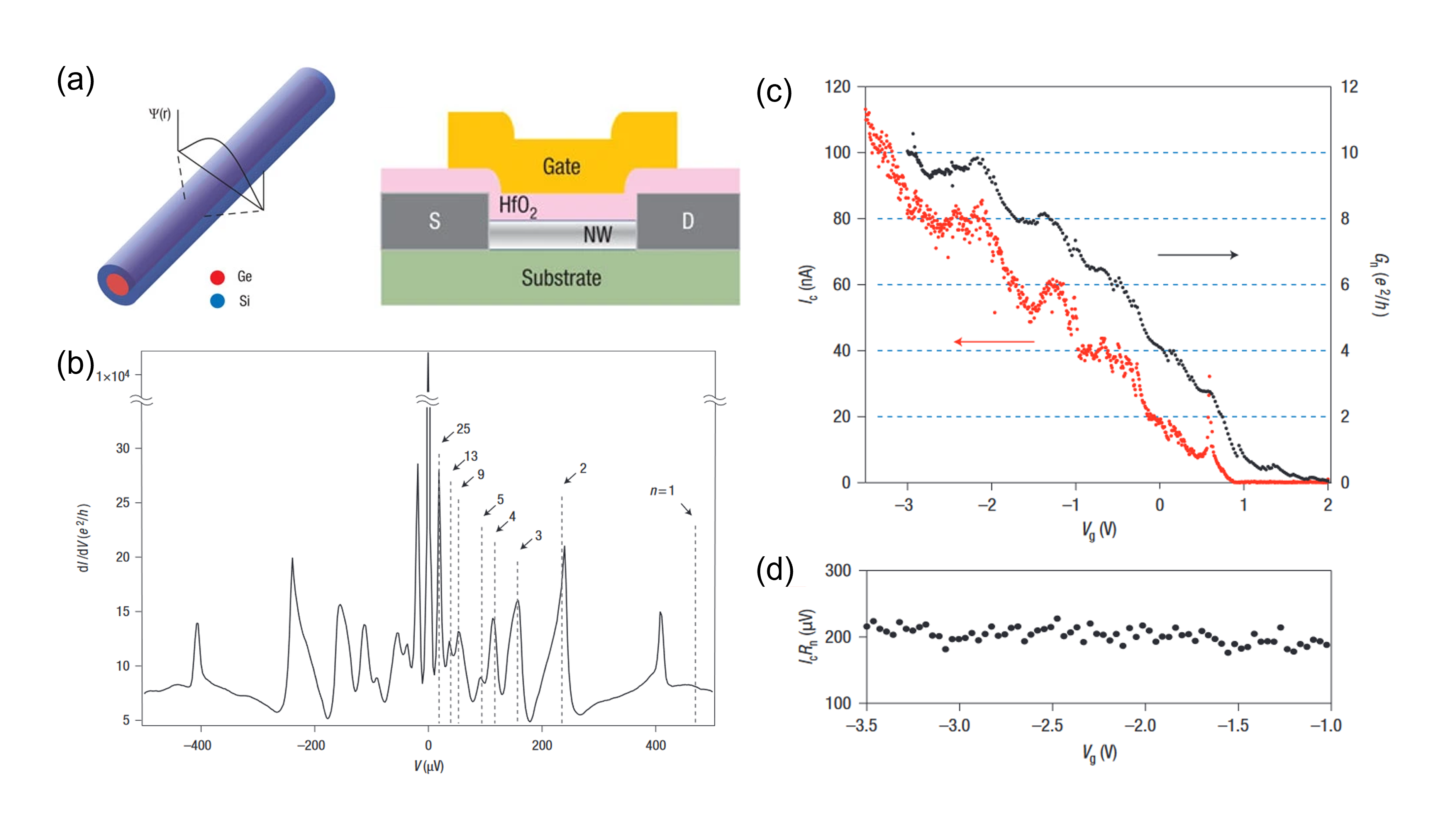}
\caption{(a) Schematic of a Ge/Si core-shell nanowire and a corresponding device with a top gate. (b) Differential conductance $dI/dV$ versus bias voltage revealing multiple Andreev reflection positions at $V_n=2\Delta/ne$. (c) Normal-state differential conductance $dI/dV$ (black) at $250m$~mT and critical current ($I_c$) (red) at zero field versus gate voltage ($V_g$). The stepwise increase in the measured quantities indicates conductance and superconducting critical current quantization. (d) $I_cR_N$ product versus $V_g$; the value of the $I_cR_N$ product differs from $\pi\Delta/\hbar$ by a factor of 3.6 suggesting that the measured current when the junction switches to the normal state is rather the smaller switching current ($I_s$) than the actual Josephson critical current.      Adapted and reprinted from Ref.~\cite{Xiang2006}.} 
\label{Fig:Xiang2006}
\end{figure*}

The first hole-based hybrid device was already reported in 2006 and was realized in the group of C. Lieber~\cite{Xiang2006}. For this, Ge/Si core-shell nanowires were used; such NWs confine a one-dimensional hole gas in the Ge core, which is confined and protected from oxidation by a thin Si shell (Fig.~\ref{Fig:Xiang2006}). The NWs were contacted by 30nm nm Al deposited by room temperature electron beam evaporation. The creation of ohmic contacts was attributed to a combination of the Fermi-level pining and unintentional annealing taking place during the device fabrication. 

\begin{figure*} 
\includegraphics[width=\textwidth]{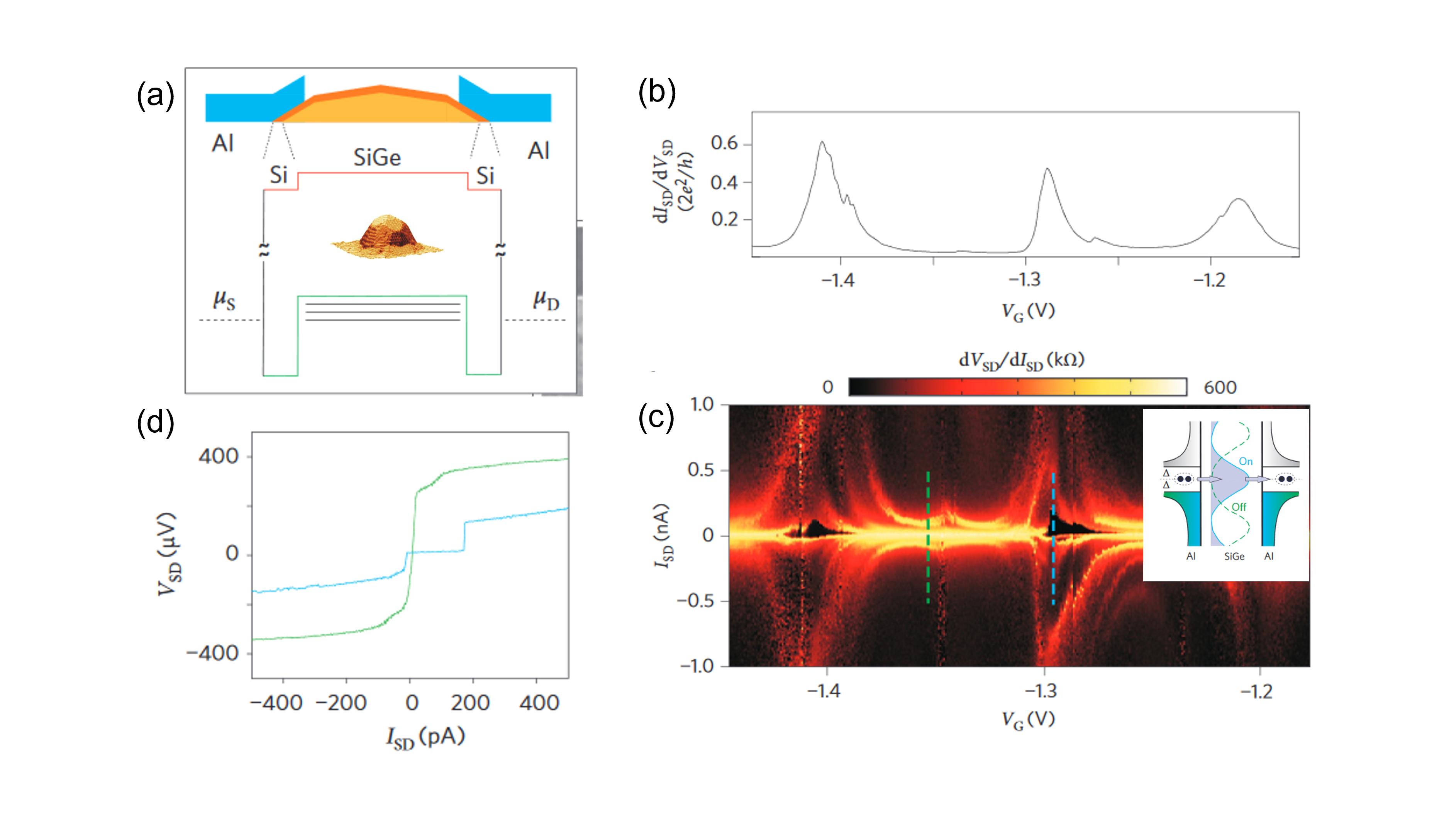}
\caption{(a) Top: Schematic of a self-assembled SiGe dome island hybrid device. A three-dimensional scanning tunneling microscopy picture of such a dome island is shown below. Bottom: Qualitative band diagram of the device with the thin Si cap grown on top of the SiGe island acting as a tunnel barrier and the electrochemical potential of the Al electrodes aligned within the valence band of the dome island. (b) Normal-state differential conductance $dI/dV$ at a $75$~mT out-of-plane field, showing Coulomb blockade oscillations. (c) Differential resistance versus current and backgate voltage for the same range as in (b) but for zero magnetic field. For the gate voltage values at which Coulomb peaks appear, zero state resistance is measured due to resonant Cooper pair tunneling, as shown in the inset. At gate voltage values corresponding to the Coulomb blockade regime, the device is resistive. (d) $V-I$ traces taken at the dashed green and blue lines in (c), illustrating the behaviour of the hybrid device on and off-resonance. The resonant switching current is below 200~pA. Adapted and reprinted from Ref.~\cite{Katsaros2010}.} 
\label{Fig:Katsaros2010}
\end{figure*}

Due to the high material and device quality, showing no Coulomb charging effects down to 5K, plateaus in the conductance measurements could be observed. The transfer of a hole-Cooper pair in a superconductor-semiconductor-superconductor device was demonstrated for the first time by cooling such devices to mK temperature. Switching currents exceeding 100~nA were measured. Furthermore,  multiple Andreev reflections were observed, from which a superconducting gap of $\Delta=235$~$\upmu$eV for the Al leads was extracted. In addition, measurements suggested the supercurrent quantization and $I_cR_n$ products of $200$~$\upmu$V were reported. This study demonstrated that Josephson field effect transistors are possible also for p-type semiconductors. 

A few years later, Ge self-assembled nanocrystals, dubbed dome islands, covered with a thin Si layer, were used for the realization of hybrid superconductor-quantum dot devices~\cite{Katsaros2010} (Fig.~\ref{Fig:Katsaros2010}). Also in this study Al was used as a superconducting contact. Inelastic cotunneling studies at $15$~mK revealed $\Delta=215$~$\upmu$eV for the Al leads. Resonant Cooper pair tunneling was demonstrated when the electrochemical potential of the quantum dot aligned with that of the source and drain leads, with switching currents of a few hundred pA. 

\begin{figure*} 
\includegraphics[width=\textwidth]{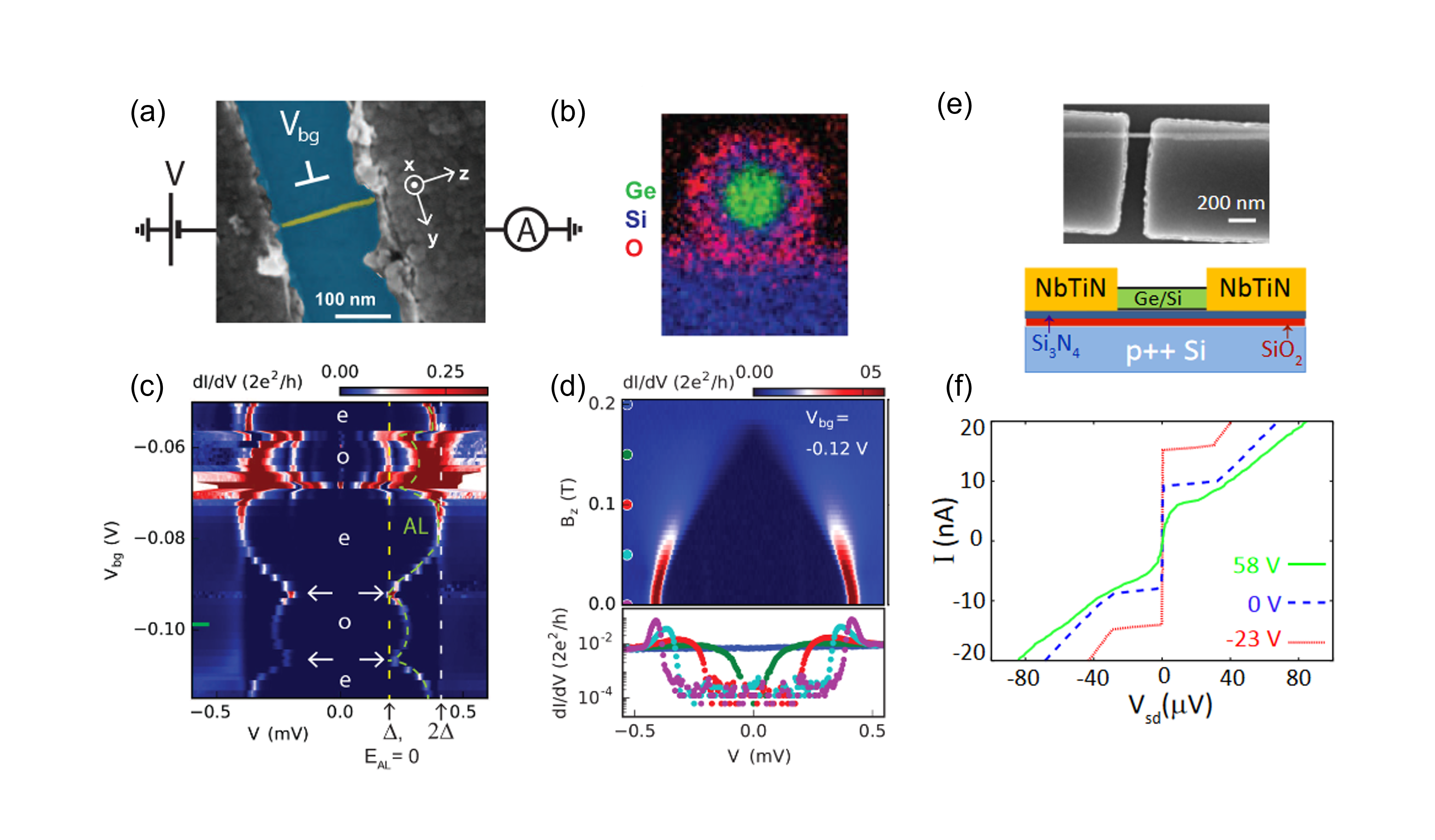}
\caption{(a) Scanning electron microscope image of a Ge/Si core/shell hybrid device with Al leads. (b) Energy dispersive X-ray spectroscopy of a cross-section transmission electron microscopy image of such a Ge/Si core/shell nanowire, highlighting the Ge core, Si shell, and the $SiO_x$ shell covering the nanowire. (c) Differential conductance $dI/dV$ versus bias voltage and backgate voltage, revealing ABS localized in the QD formed in between the leads as well as MAR. From the modulation of the ABS energy an even or odd QD modulation can be inferred. The Andreev level energy has an offset of $\pm\Delta$ due to the use of superconducting leads in the tunneling spectroscopy measurement. (d) Top: Induced superconducting gap measurement versus perpendicular magnetic field. Bottom: Line traces taken at the positions indicated in the upper plot. (e) Scanning electron microscopy image of a similar Ge/Si core/shell nanowire contacted with Ti/NbTiN leads. (f). Critical current measurements versus backgate voltage for such a device demonstrate the electric field modulation of the critical current. Adapted and reprinted from Ref~\cite{deVries2018} (panels (a)-(d)) and Ref.~\cite{Su2016} (panels (e)-(f)).} 
\label{Fig:VriesSu}
\end{figure*}

\begin{figure*} [h!]
\includegraphics[width=\textwidth]{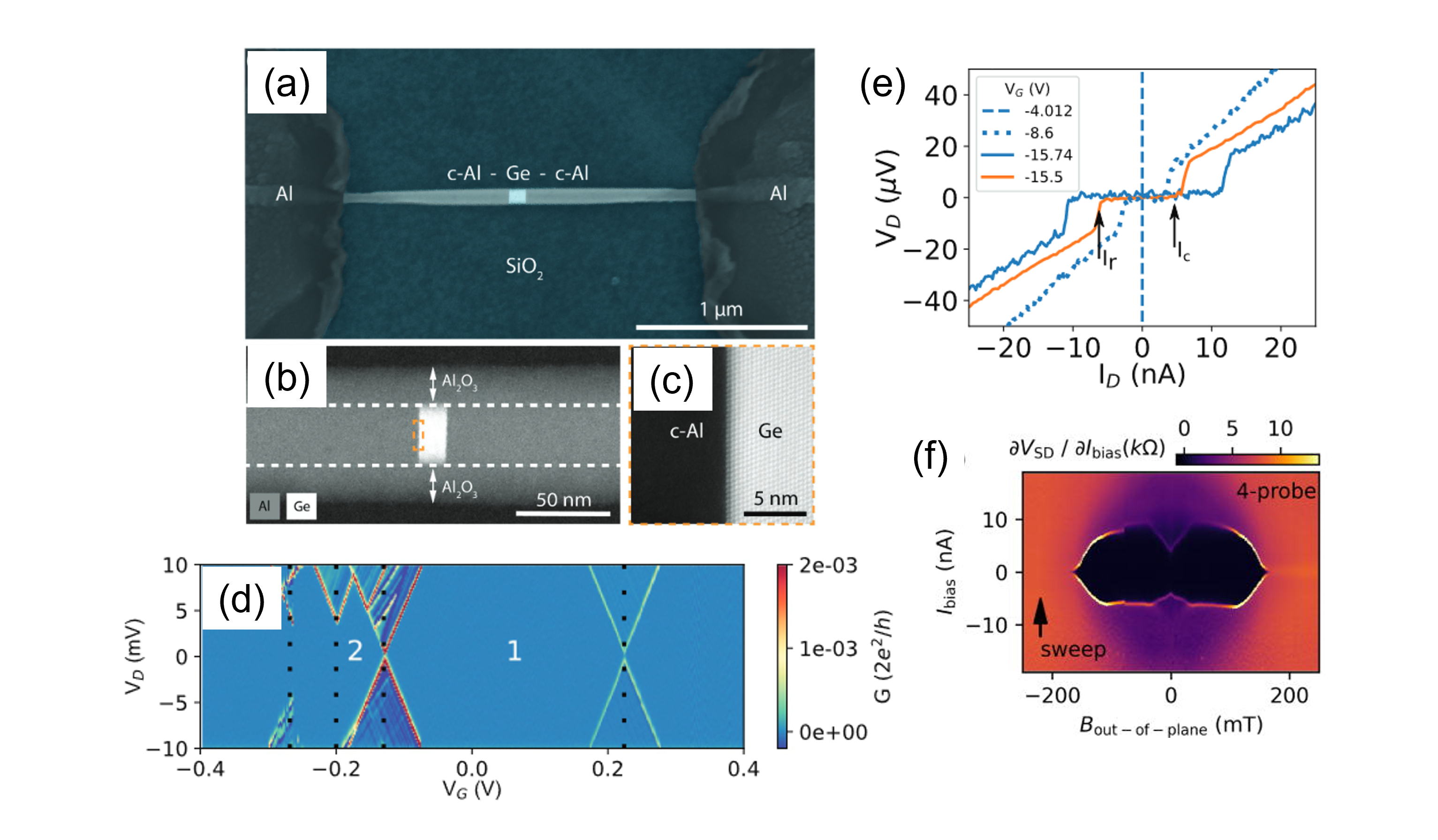}
\caption{Ge/Si core/shell hybrid devices. (a) Scanning electron microscopy image of an Al-Ge-Al nanowire device generated by rapid thermal annealing. (b,c) High resolution scanning transmission electron microscopy pictures demonstrating the abrupt Al-Ge interfaces. (d) Differential conductance $dI/dV$  versus bias and back gate voltage for such a Al-Ge-Al device in the weak coupling and in the few hole regime measured at $390$~mK. (e) Voltage versus current traces for different back gate voltages showing the onset of a supercurrent when the device is operated in the strong coupling regime. (f) Differential resistance $dV/dI$ measurement versus applied current and out-of-plane magnetic field for a Ge/Si core/shell NW Josephson junction device. The non-monotonic dependence of the switching current is attributed to better thermalization, which takes place during the generation of in-gap quasiparticles due to the applied magnetic field. Adapted and reprinted from Delaforce 2021 (panels (a)-(e))~\cite{Delaforce2021} and Wu 2024 (panel (f))~\cite{Wu2024b}.} 
\label{Fig:DelaforceWu}
\end{figure*}

In 2018, ultrathin Ge/Si core-shell NWs with Al leads were fabricated and intentionally annealed, causing interdiffusion~\cite{deVries2018}. Such a fabrication process led to the formation of a quantum dot within the Ge/Si core-shell wire and therefore the creation of Andreev bound states (Fig.~\ref{Fig:VriesSu}). A ground state transition from even to odd quantum dot occupation was demonstrated, and the superconducting gap was investigated, demonstrating a two-orders-of-magnitude lower subgap conductance than above-gap for superconductor - quantum dot - superconductor devices. In the same year, and again for Ge/Si core shell wires, Al leads Josephson field effect transistors were demonstrated, which could be tuned by the backgate voltage either into a high transparent regime with the supercurrent carried by multiple subbands or into a quantum dot regime with resonant supercurrents~\cite{Ridderbos2018}. Not only resonant supercurrents, but also supercurrents inside Coulomb diamonds were observed, which can take place via higher-order co-tunneling events~\cite{vanDam2006}. Gate-tunable supercurrent was also demonstrated in Ge/Si nanowires by using NbTiN as a contact and using Ti or Al for improving the contact transparency~\cite{Su2016} ((Fig.~\ref{Fig:VriesSu} (e)-(f)). These supercurrents could sustain magnetic fields exceeding 800 mT. However, the subgap conductance in this system was significant, as disordered superconductors, as contacts, and in contrast to Al, lead to a soft gap~\cite{Su2016}.

\begin{figure*} 
\includegraphics[width=\textwidth]{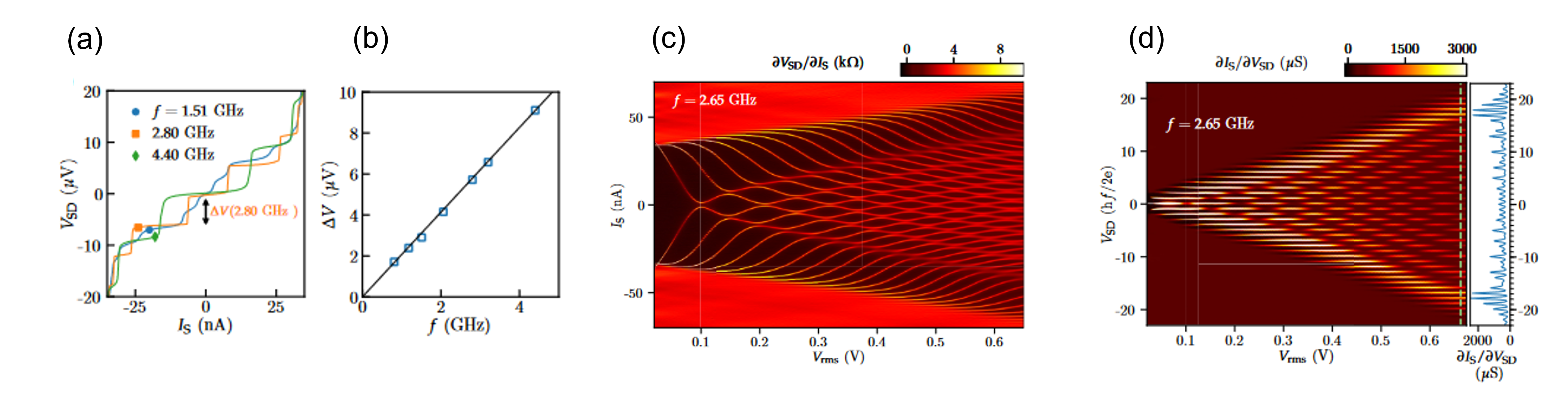}
\caption{Ge/Si core/shell hybrid devices. (a) Voltage versus current traces for different applied frequencies demonstrating the AC Josepshon effect in a hybrid Ge/Si core/shell device. (b) Extracted step height $\Delta V$ vs microwave frequency where the black line is a plot of $\Delta V = hf/2e$. (c) Differential resistance $dV/dI$ measurement versus applied current and microwave voltage demonstrating the appearance of Shapiro physics. (d) Same measurement as (c) but with current and voltage axes reversed before numerical derivation. The line trace on the right side demonstrates the presence of up to 23 Shapiro steps. Adapted and reprinted from Ref.~\cite{Ridderbos2019}.} 
\label{Fig:Ridderbos2019}
\end{figure*}

\begin{figure*} [h!]
\includegraphics[width=\textwidth]{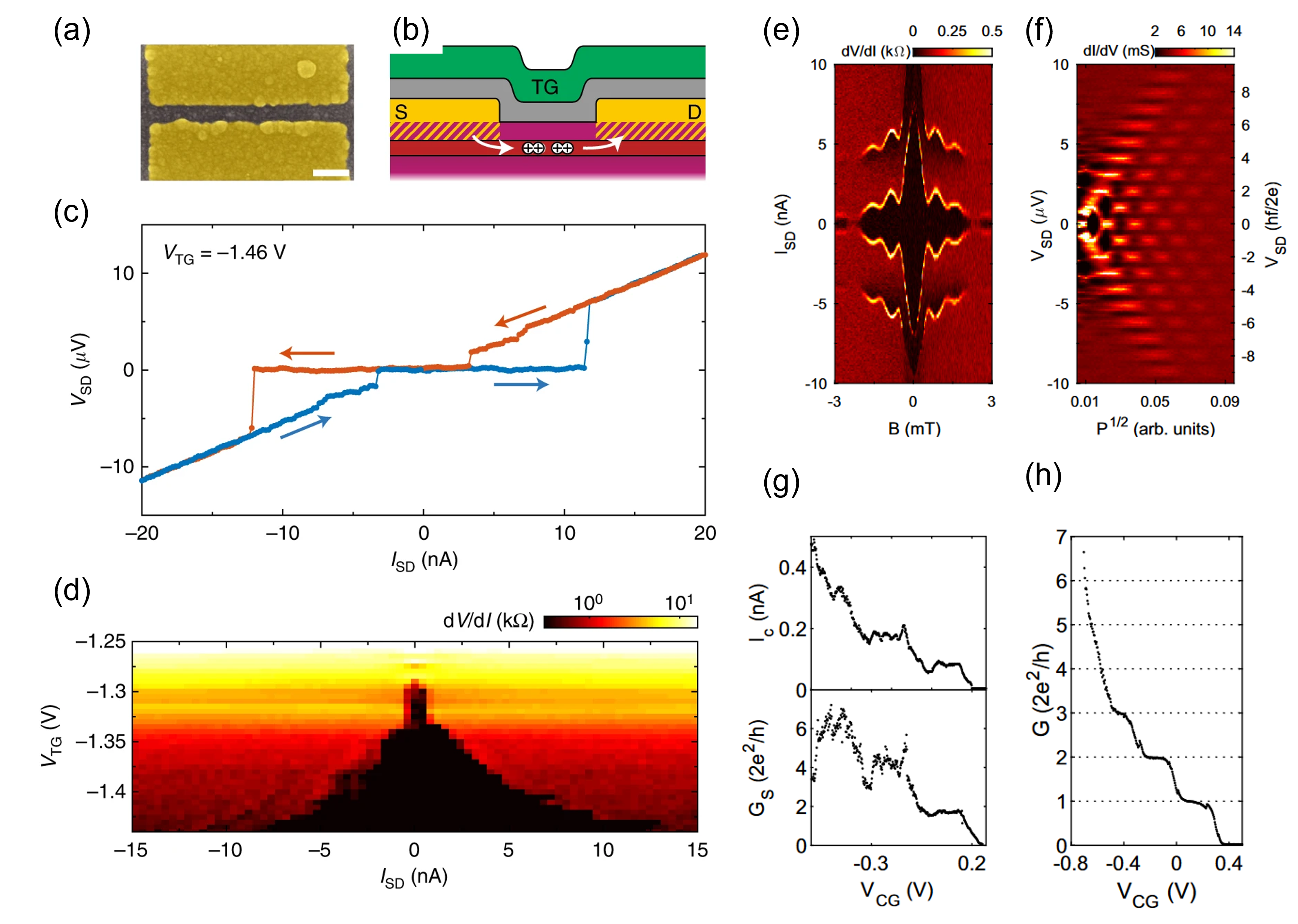}
\caption{Hybrid devices in planar Ge. (a) Scanning electron microscopy image of a Josephson field effect transistor device fabricated on a Ge/SiGe heterostructure. (b) Cross-section schematic of the device. The superconducting contacts are achieved by in-diffusion of Al presumably taking place during the atomic layer deposition of aluminum oxide at $300^oC$. The top-gate metal consist of 5/35nm of Ti/Pd. (c) Voltage versus current measurement demonstrating a switching current larger than 10nA. (d) Switching current modulation versus top gate voltage demonstrating the Josephson field effect behaviour of the hybrid device. (e) Differential resistance $dV/dI$ measurement versus applied current and out-of-plane magnetic field revealing a Fraunhofer-like modulation of the switching current. (f) Differential conductance $dI/dV$ measurement versus voltage and applied microwave excitation amplitude demonstrating Shapiro steps at $V = nhf/2e$. (g) Top (bottom) Switching current (subgap conductance) versus gate voltage demonstrating discretization of the switching current (steps exceeding conductance quantum values are cased by Andreev reflection) (h) Normal state conductance versus gate voltage at an out-of-magnetic-field value driving the Al leads to the normal state. Quantized conductance steps are visible. Adapted and reprinted from Ref.~\cite{Hendrickx2018} (panels (a)-(d)) and Ref.~\cite{Hendrickx2019} (panels (f)- (i)).} 
\label{Fig:Hendrickx20182019}
\end{figure*}

\begin{figure*}[h!]
\includegraphics[width=\textwidth]{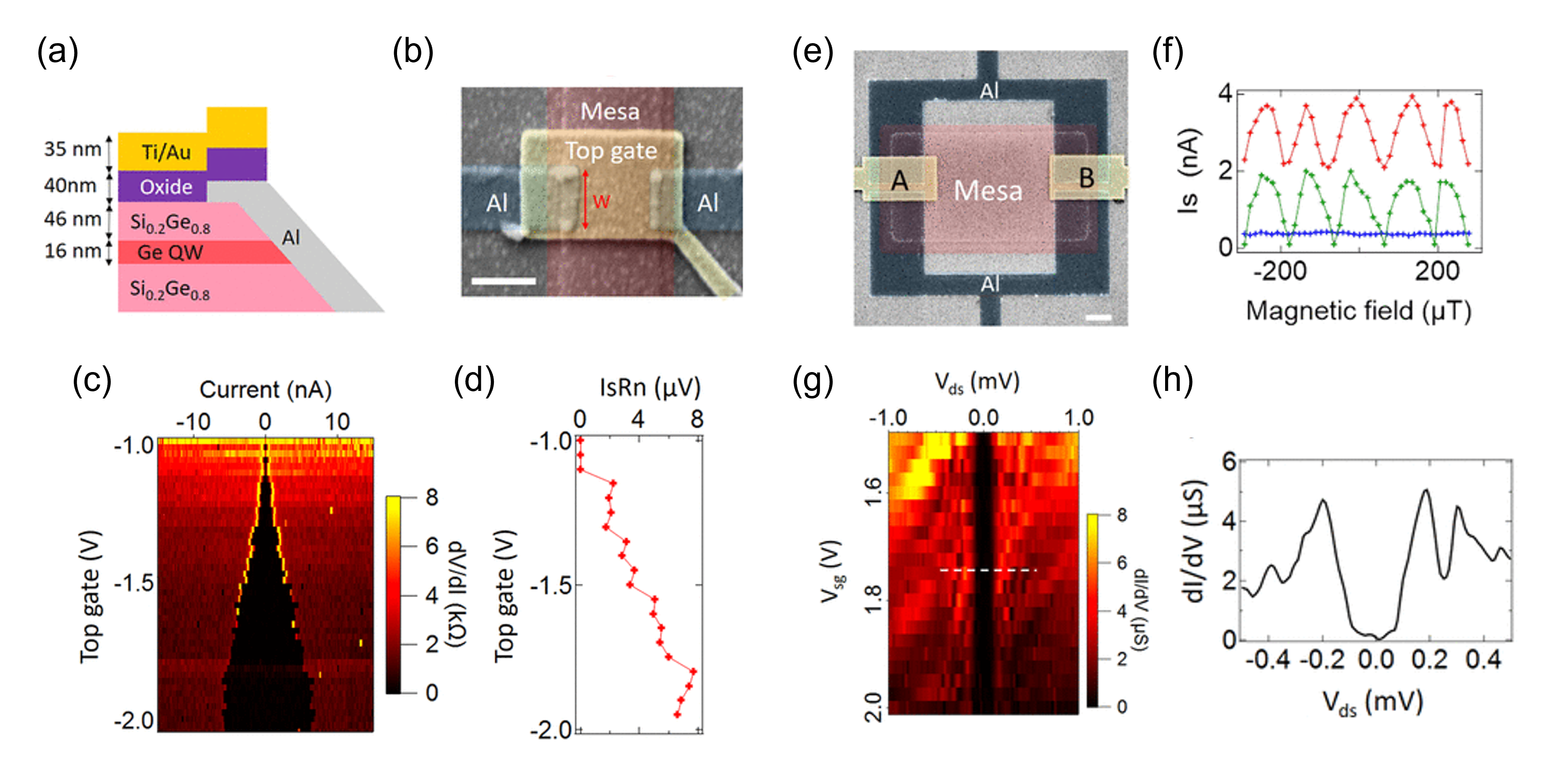}
\caption{(a) Schematic cross-section of a Ge quantum well contacted with Al from the side after a wet-etching process. (b) Scanning electron microscope image of a top-gated Josephson junction device. (c) Differential conductance $dI/dV$ measurement versus current and topgate voltage demonstrating electric field dependence of the switching current. (d)  $I_cR_N$ product versus $V_g$ significantly smaller to what was reported for Ge/Si core/shell hybrid devices~\cite{Xiang2006}. (e) Scanning electron microscopy picture of a SQUID device fabricated in planar Ge. (f) Switching current versus magnetic field for different gate voltage values. The blue trace corresponds to the case for which just one junction is on. (g) Tunneling spectroscopy with a superconducting contact versus side gate voltages which tune the tunnel barrier. (h) Line trace extracted from (g). An induced gap of about $\Delta^*=100$~$\upmu$eV was measured. Adapted and reprinted from Ref.~\cite{Vigneau2019}.} 
\label{Fig:Vigneau2019}
\end{figure*}

\begin{figure*}[h!]
\includegraphics[width=\textwidth]{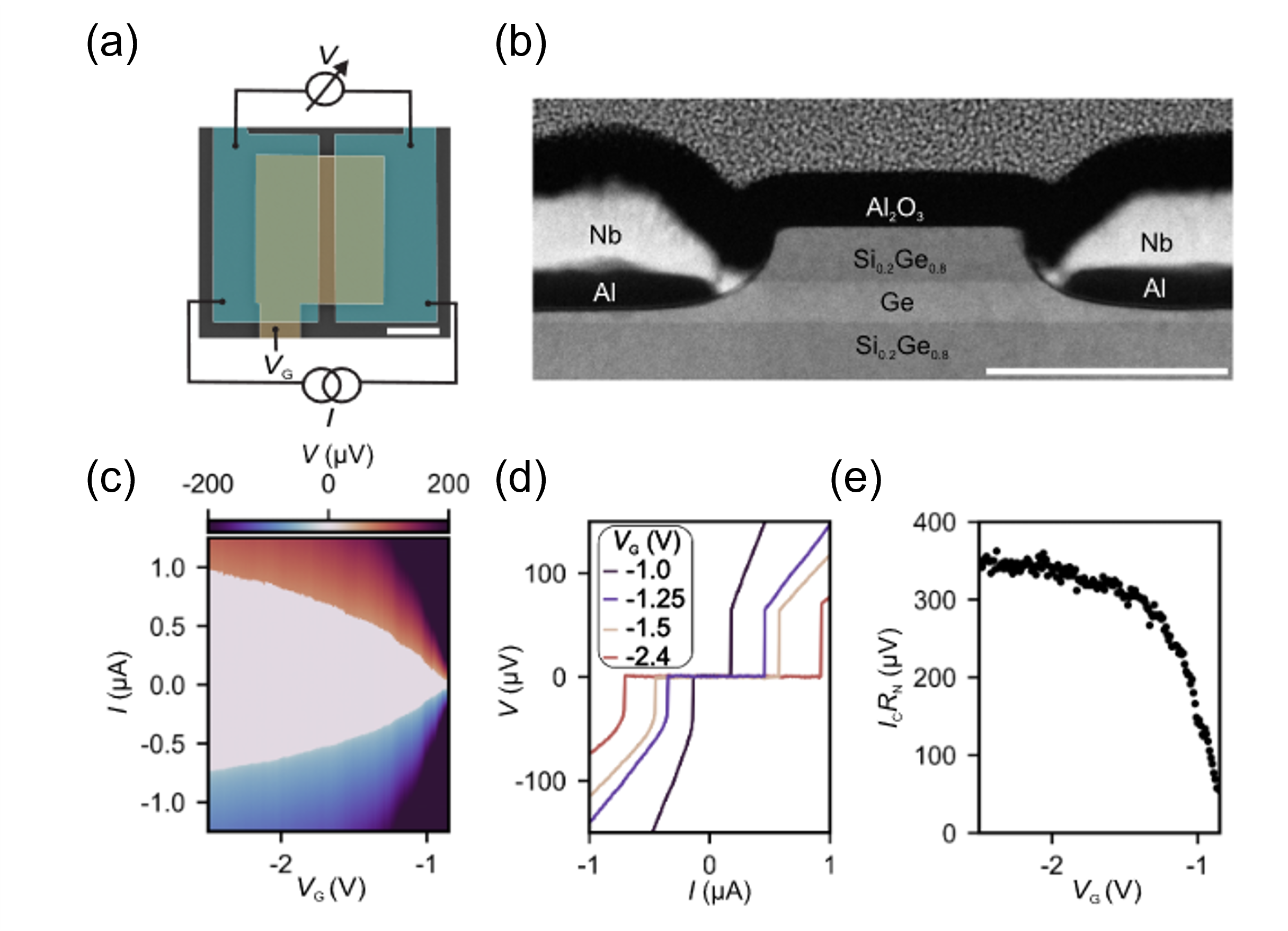}
\caption{(a) Scanning electron microscopy image of a Josephson field effect transistor device fabricated in planar Ge. (b) Cross-section scanning transmission electron microscopy picture showing the metal stack used for contacting the hole gas, with the Al layer directly in contact with the quantum well. (c) Voltage measured across the Josephson junction versus applied top gate voltage and current, demonstrating electrical tunability of the switching current. (d) Characteristic V versus I traces, extracted from c) for different gate voltages. (e) $I_cR_N$ product versus $V_g$ demonstrating values exceeding $300$~$\upmu$V. Adapted and reprinted from Ref.~\cite{Aggarwal2021}.} 
\label{Fig:Aggarwal2021}
\end{figure*}

\begin{figure*} [h!]
\includegraphics[width=\textwidth]{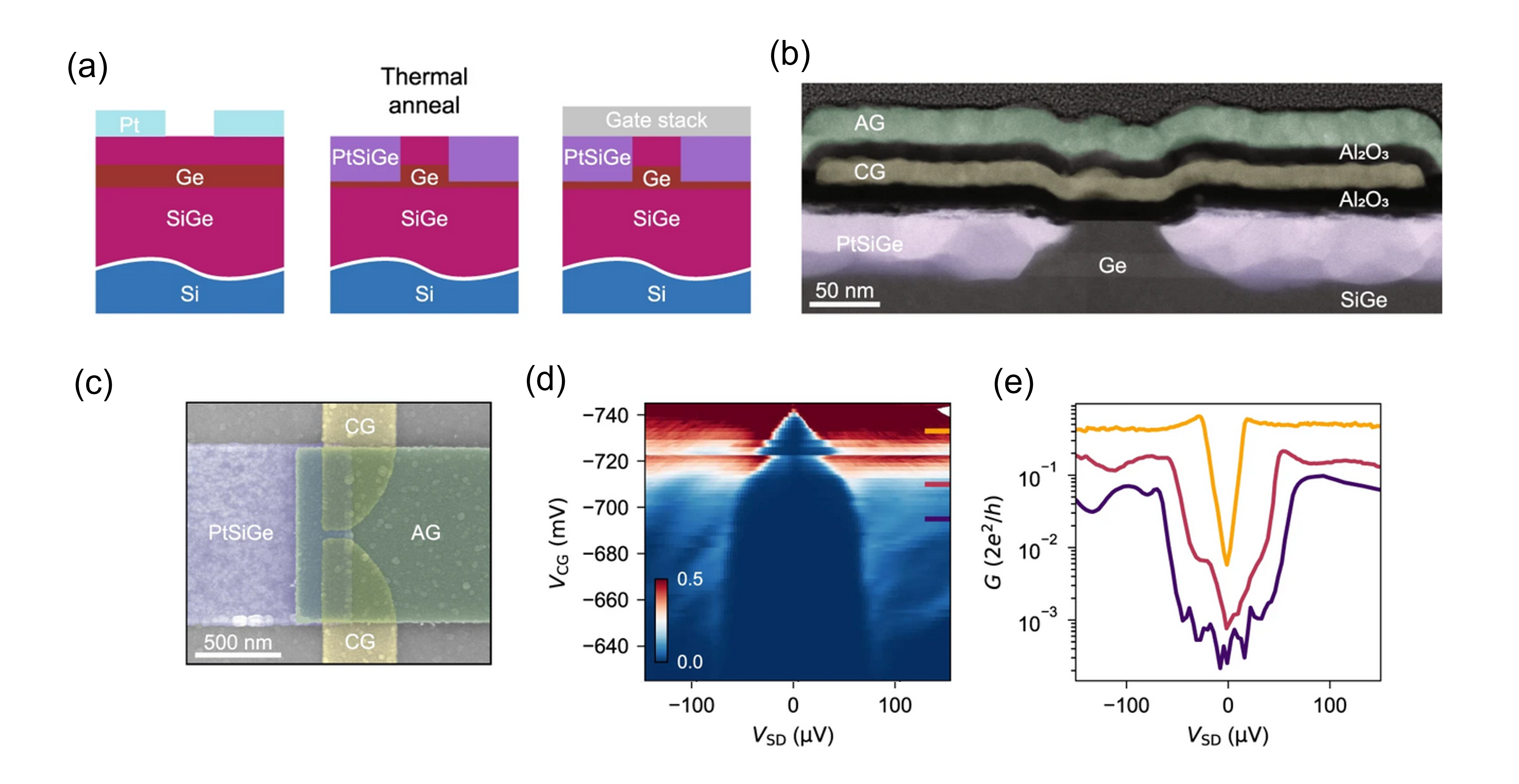}
\caption{(a) Schematics of the procedure used in order to create PtSiGe superconducting contacts to planar Ge. After the deposition of Pt a  15 min rapid thermal annealing at $400$~$^o$C in a halogen lamp under argon atmosphere leads to the formation of PtSiGe contacts. (b) Cross-section scanning transmission microscopy image showing the Ge QW contacted PtSiGe and two gate layers formed on top separated by aluminum oxide dielectric deposited by atomic layer deposition. (c) Scanning electron microscope image of the device used for performing tunneling spectroscopy measurements. The accumulation gate (AG) is used from accumulating holes while the constriction gates (CG) for forming the tunnel barrier. (d) Conductance measurement versus source drain bias and constriction gate voltages for measuring the induced gap in the two dimensional hole gas. e) Line cuts taken from d) showing a subgap conductance below $10^{-3}$ at a gate voltage of -733~mV and $\Delta^*=70$~$\upmu$eV. Adapted and reprinted from Ref.~\cite{tosato2023}.} 
\label{Fig:Tosato2023}
\end{figure*}

\begin{figure*}[h!]
\includegraphics[width=\textwidth]{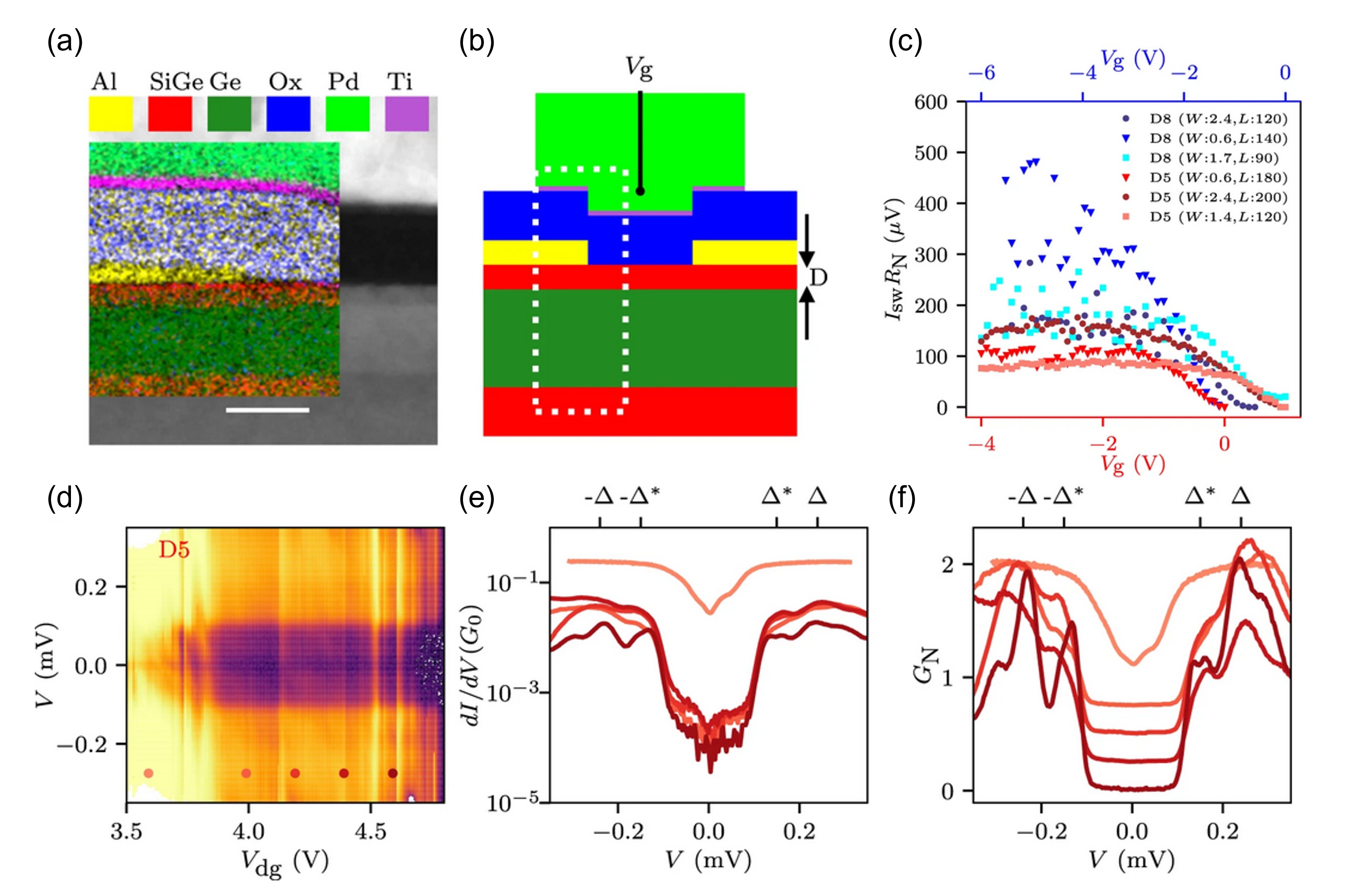}
\caption{(a) TEM image of an ultra shallow Ge QW with a 5nm $Si_{0.3}Ge_{0.7}$ cap contacted with a thin Aluminum layer deposited at $110$~K. The left inset displays energy dispersive x-ray data not revealing any Al in the $Si_{0.3}Ge_{0.7}$ and Ge QW. (b) Schematic of a JoFET device with the dashed rectangle corresponding to the inset in (a). (c) $I_{SW}R_N$ product versus $V_g$ for Ge QWs with 5nm (D5) and 8nm (D8) spacer thickness for different junction widths (W) and lengths (L). The largest measured value is close to $500$~$\upmu$V. (d) Differential conductance $dI/dV$ in logarithmic scale versus side gate voltage $V_{sg}$ and bias for extracting the induced superconducting gap for the 5nm spacer sample. (e) Line-cuts taken from (d) demonstrating a subgap conductance of about $10^{-4}$ for gate voltages above 4.5V. (f) Same traces as in d) but plotted in a normalized scale $G_N=(dI/dV)/G_{normal}$ to highlight the appearance of two peaks. The lowest one is attributed to the induced gap with $\Delta^*=150$~$\upmu$eV. The traces are shifted by $0.25 G_N$ with respect to each other. Adapted and reprinted from Ref.~\cite{Valentini2024}.} 
\label{Fig:Valentini2024}
\end{figure*}

\begin{figure*}
\includegraphics[width=\textwidth]{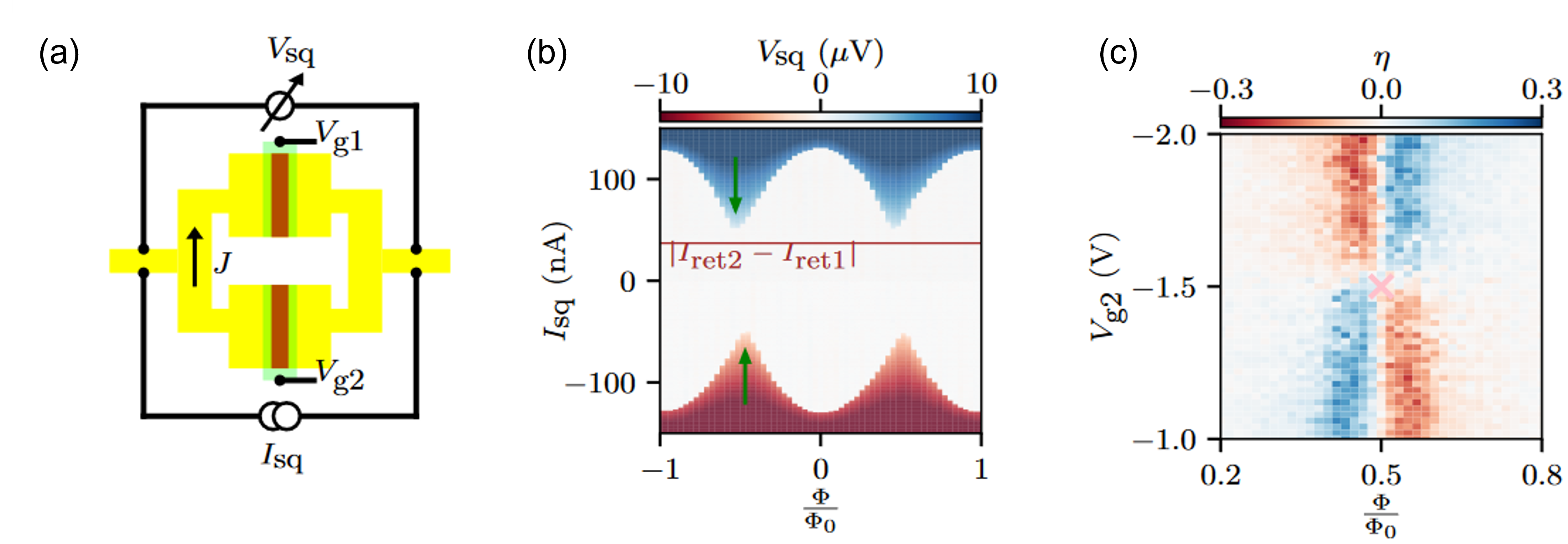}
\caption{(a) Schematic of a SQUID device realized in planar Ge. (b) Voltage $V_{sq}$ versus magnetic flux $\Phi$ and current $I_{sq}$. $I_{sq}$ was swept from positive values to zero and then from negative values to zero, representing therefore the retrapping currents. It is observed that the SQUID lobes are asymmetric with respect to the half flux, and the critical retrapping current at half flux is larger than the difference of the individual currents. Both effects are present due to the non-negligible inductance of the SQUID and the higher harmonics. The asymmetry of the SQUID lobes leads to the superconducting diode effect. (c) Gate and flux dependence of the superconducting diode efficiency. The efficiency is always zero at half flux quantum and can be inverted by changing the gate voltage value applied to the Josephson junctions. Adapted and reprinted from Ref.~\cite{Valentini2024}.} 
\label{Fig:ValentiniDiode}
\end{figure*}

\begin{figure*}[h!]
\includegraphics[width=\textwidth]{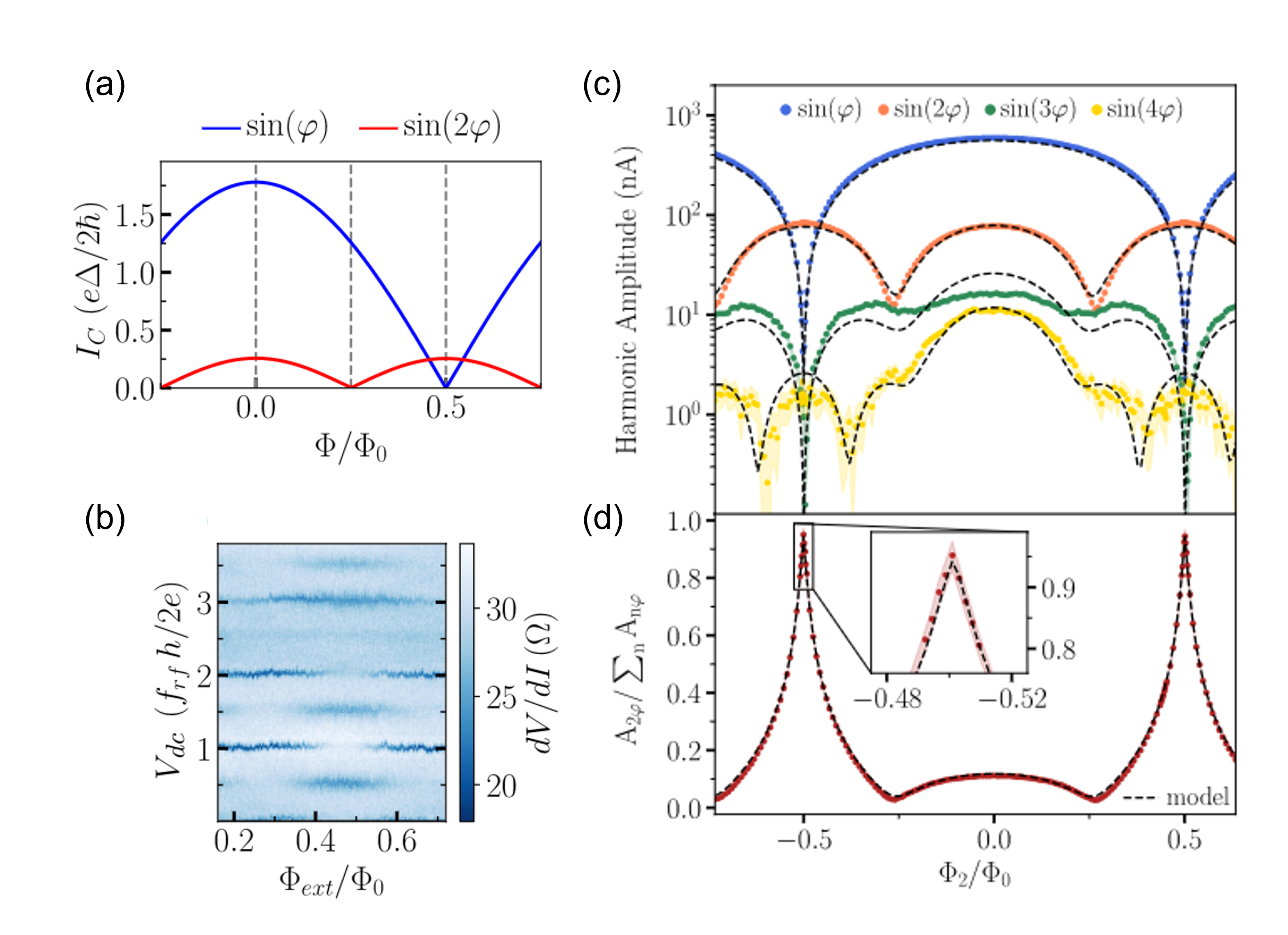}
\caption{ (a) Computed current phase relation for a symmetric SQUID device for a single-channel short junction model with a transparency of 0.7. The first two harmonics amplitudes are shown. For $\Phi/\Phi_o = 0.5$ the $\sin(\phi)$ term is suppressed while the $\sin(2\phi)$ has a finite value. (b) Differential resistance $dV/dI$ as a function of the normalized voltage and applied flux in the presence of microwave radiation and in the regime where the same critical current characterizes the two arms of the SQUID device. The appearance of fractional Shapiro measurements, which are a probe for a $\sin(2\phi)$ CPR, can be observed at half flux quantum. (c) First four harmonics amplitude for a balanced SQUID, integrated in a bigger SQUID, versus normalized flux showing that the first and third harmonic have a minimum at half flux quantum where the second and the fourth have their maximum value. (d) Ratio between the second harmonic amplitude and the sum of the four harmonic amplitudes versus flux demonstrating a value of about $95\%$ at half-flux quantum. Adapted and reprinted from Ref.~\cite{leblanc2023} (panels (a) - (b))  and Ref.~\cite{leblanc2023} (panels (c)-(d)).} 
\label{Fig:Shapiro}
\end{figure*}

\begin{figure*}[h!]
\includegraphics[width=\textwidth]{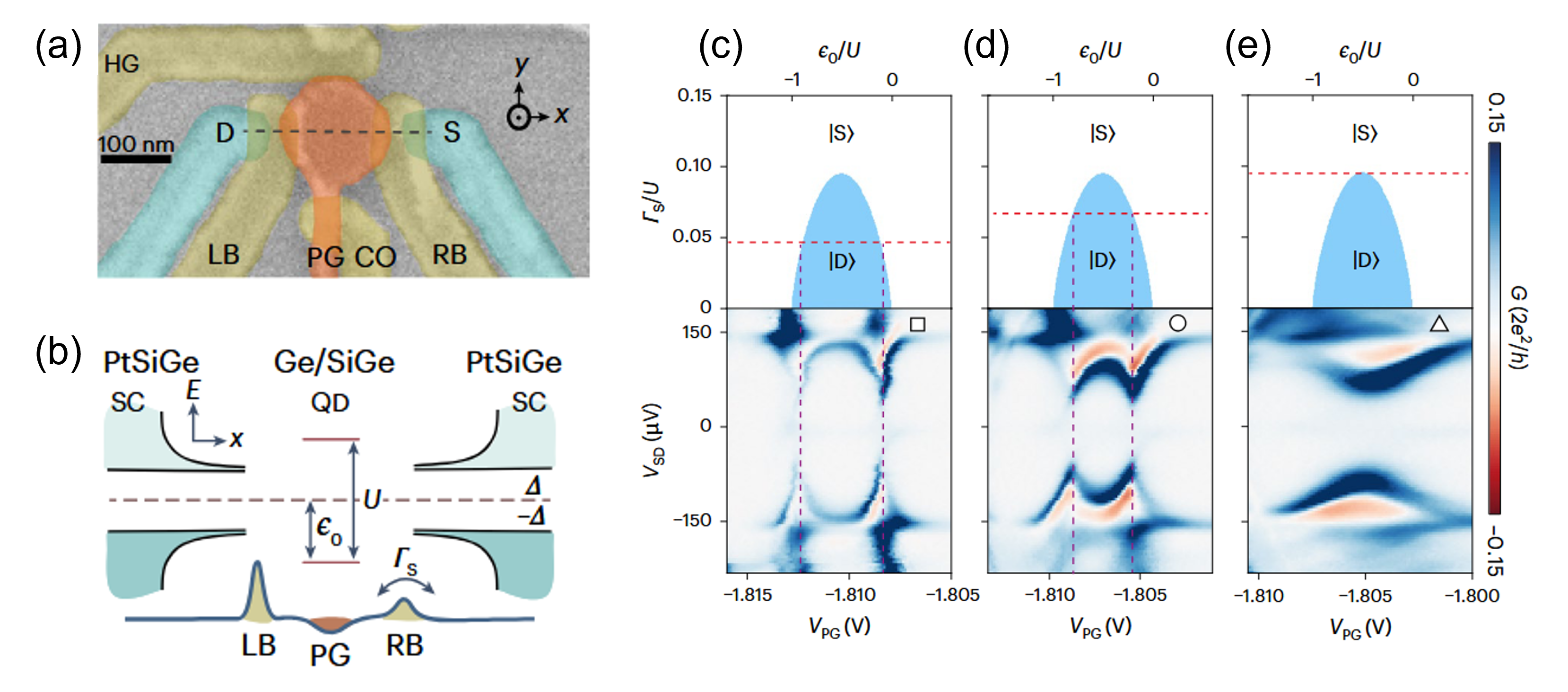}
\caption{ (a) Scanning electron microscope image of a proximitized quantum dot formed in a Ge/SiGe heterostructure. PtGeSi contacts are used as leads. (b) Schematic showing the energy diagram of the device. ((c)-(e))  (Bottom) Spectroscopy measurements for different voltage values of the plunger gate demonstrating the change of the ground state from doublet to singlet. (Top) Qualitative diagrams showing the singlet/doublet phase transition versus the strength of the tunnel coupling to the superconductor and the voltage value of the plunger gate. Adapted and reprinted from ~Ref.~\cite{Lakic}.} 
\label{Fig:ProximitizeddotsPlanarGe}
\end{figure*}

While the above-mentioned Ge/Si core/shell NW devices lead to the formation of QDs in cases of non-transparent interfaces, systematic annealing experiments were performed in the following years. These allowed to control the Al diffusion into the Ge core and therefore the size of the dot (Fig.~\ref{Fig:DelaforceWu}). This technique was also used to improve the contact transparency ~\cite{Kral2015,Sistani2019,Sistani2020,Ridderbos2019Nano,Delaforce2021}. Finally, such Ge/Si core shell nanowires have not only been used for demonstrating the DC Josephson effect but also for demonstrating the appearance of Shapiro steps upon RF illumination~\cite{Ridderbos2019}~ (Fig.~\ref{Fig:Ridderbos2019}). Such Shapiro steps appear when the frequency of the phase oscillations $\frac{2eV}{\hbar}$ matches the microwave irradiation frequency. Very recently, it was demonstrated that the critical current of such Ge/Si core-shell nanowire devices can be enhanced in the presence of a small magnetic field as this can lead to in-gap quasiparticles which facilitate the thermalization of the devices~\cite{Wu2024b} (Fig.~\ref{Fig:DelaforceWu} (f)). 

In parallel with the development of hybrid devices from self-assembled nanowires and nanocrystals, hybrid devices realized in Ge/SiGe heterostructures have also attracted considerable attention. This material platform, as explained above, has proven to be very interesting for hosting hole spin qubits~\cite{Scappucci2021} and offers advantages when moving to more complex geometries. It was shown~\cite{Hendrickx2018} that for a 22 nm buried quantum well, Al can be used for creating a Josephson field effect transistor (Fig.~\ref{Fig:Hendrickx20182019}). Al was diffused into the Ge quantum well during the fabrication of the device. While, the measured $I_cR_n$ products was one order of magnitude lower than what was reported for Ge/Si core/shell nanowires, the Fraunhofer effect, Shapiro steps and quantized supercurrent values were reported~\cite{Hendrickx2019}. In addition, to the creation of superconducting contacts to the Ge QW via annealing also, different approaches were investigated to improve the quality and properties of the devices. In one approach~\cite{Vigneau2019} a slanted mesa profile was created via wet chemistry and the Ge QW was directly contacted by Al (Fig.~\ref{Fig:Vigneau2019}). This allowed the creation of field-effect transistors and SQUID devices for holes. Furthermore, the first tunneling spectroscopy measurements were performed, indicating a soft gap with BCS-like peaks appearing at $\Delta^*=100$~$\upmu$eV. In a similar spirit, and a few years later, the Ge quantum well was directly contacted by Al after removing the SiGe spacer via a reactive ion etching process. In this study a bilayer was used namely Al/Nb~\cite{Aggarwal2021}. The goal of this study was to increase the superconducting gap and the magnetic field resilience (Fig.~\ref{Fig:Aggarwal2021}). By using this approach JoFET devices with $I_cR_n$ products exceeding $300$~$\upmu$V were reported. Also the critical field for in-plane magnetic fields reached 1.8T; such values are important as for planar Ge the confined states are of heavy-hole character with a small in-plane g-factor ~\cite{Jirovec2022}. In addition, highly tunable SQUID devices could be demonstrated, and from the current phase relation it was possible to extract an average transparency approaching 90~\% . 

Already, the early studies on planar Ge showed the potential of this platform for hybrid devices. The next important step was to demonstrate that it is possible to induce a hard gap, a key feature demonstrated in III-V materials~\cite{ChangNN:15}. In 2023 Tosato et al.~\cite{tosato2023} used a novel method to induce superconductivity into Ge. This time, Pt was used to form PtSiGe germanosilicide superconducting contacts via a thermally-driven solid-state reaction with the semiconductor heterostructure (Fig.~\ref{Fig:Tosato2023}). Tunneling spectroscopy revealed a subgap conductance two orders of magnitude lower than the above gap conductance. Furthermore, BCS-like peaks appeared  at $\Delta^*=70$~$\upmu$eV. In-plane critical fields exceeding 300mT were reported. Importantly, similar superconducting properties induced by PtSiGe were reproduced independently in subsequent studies~\cite{PRXQuantum.5.030337,Lakic}. In 2024, another work investigated the induced gap in planar Ge (Fig.~\ref{Fig:Valentini2024}). Inspired by the III-V community a thin layer of Al was deposited on a Ge/SiGe heterostructure with a SiGe spacer with a thickness varying between 5 and 7 nm. Josephson junctions and tunneling spectroscopy measurements were performed. While a hard gap was reported for a 5nm SiGe spacer, the gap already became soft for a spacer of 7nm. Interestingly, by measuring the $I_cR_n$ product, values near $500$~$\upmu$V were realized for the 7 nm spacer device, while for the 5 nm heterostructure with a hard gap, the values were significantly lower, suggesting that the $I_cR_n$ products do not necessarily represent a figure of merit for the quality of the realized devices. 

After the realization of high-quality hybrid JJ devices, the next goal was to create SQUIDs. Such hybrid SQUID devices have attracted more and more interest because they can be used not only to create supercurrent diodes but also as generators of non-sinusoidal current phase relations, which have gained interest for realizing protected qubits in hybrid semiconductor superconductor devices. The superconducting diode effect can appear either if the SQUID has a large inductance (when thin superconductors or disordered superconductors are used as superconducting leads) or when the current phase relation of the single junctions contains higher-order contributions. The origin of this effect can be understood from symmetry. A superconducting diode, i.e., a nonreciprocal critical current, $I_c^+\neq|I_c^-|$, requires that both time-reversal and inversion symmetries are broken. In a SQUID, time-reversal symmetry is broken by the magnetic flux $\Phi$, and inversion symmetry by the asymmetry between the two junctions. This is, however, not sufficient: for two junctions with purely sinusoidal current-phase relations and a negligible loop inductance, the total supercurrent, $I(\varphi)=I_1\sin\varphi+I_2\sin(\varphi+2\pi\Phi/\Phi_0)$, is again a phase-shifted sinusoid, whose maximum and minimum have the same magnitude, so that no diode effect arises. Higher harmonics in the current-phase relation, or a finite loop inductance, which makes the phase drop across the junctions depend on the circulating current, distort $I(\varphi)$ such that its maximum and minimum differ in magnitude, giving rise to a finite diode efficiency $\eta=(I_c^+-|I_c^-|)/(I_c^++|I_c^-|)$~\cite{Souto_PRL2022,Bozkurt_SciPost,Valentini2024,leblanc2023,Yan_arXiv2026}. Since $\Phi$ and $-\Phi$ are related by time reversal, $\eta$ is odd in flux and vanishes at $\Phi=0$ and $\Phi=\Phi_0/2$, while its sign can be inverted by tuning the relative weight of the harmonics with gate voltages, see Fig.~\ref{Fig:ValentiniDiode}(c). 

In fact, typically, both the inductance and high harmonics in the current-phase relation contribute to the superconducting diode effect, as recently demonstrated in a SQUID realized in planar Ge by using thin Al (Fig.~\ref{Fig:ValentiniDiode}). Gate-tunable Josephson diodes based on the same interferometric principle have also been demonstrated in InAs 2DEGs~\cite{Ciaccia_PRR2023}, and their microwave response has recently been studied in proximitized bilayer graphene interferometers~\cite{Yan_arXiv2026}. By investigating the superconducting diode effect versus flux, one can tune the device into a regime where the first harmonic contributions vanish due to destructive interference at half flux quantum~\cite{Valentini2024}. At such a configuration, the supercurrent is carried by pairs of Cooper pairs. Shapiro measurements were used to further elucidate the CPR periodicity. For a standard sinusoidal CPR under microwave irradiation, the current-voltage characteristics develop voltage steps, the Shapiro steps, when $V = s \frac{h f_{\textrm{ac}}}{2 e}$, where $s = 0,1,2,\dots$ and $f_{\textrm{ac}}$ is the frequency of the signal sent to the device. If the CPR becomes $\propto \sin \left(2  \varphi \right)$, signalling tunneling of pairs of Cooper pairs, steps at half-integer values also appear, i.e. $s = 0,0.5,1,\dots$. Such measurements have been performed in independent studies on planar Germanium~\cite{Valentini2024,leblanc2023} and fractional Shapiro steps could be observed (Fig.~\ref{Fig:Shapiro}).  Furthermore, in a more advanced study, double SQUID devices have been realized in order to investigate the CPR of the SQUID as with a simple SQUID device, one obtains information just about the CPR of the Josephson junction with the smallest switching current, ideally one order of magnitude lower than that of the second junction. For a double SQUID device a regime could be reached where the $ \sin \left(2  \varphi \right)$ component accounts for more than 95 per cent of the total supercurrent~\cite{leblanc2024gate}.

In 2025, the first proximitized quantum dot in planar Germanium has been realized~\cite{Lakic}. By using Pt and creating PtGeSi superconducting contacts, a single quantum dot could be created and Andreev-bound states could be studied~ (Fig.~\ref{Fig:ProximitizeddotsPlanarGe}).  This opens up the path towards combining spins confined in quantum dots with superconductivity for realizing Cooper pair splitters and Andreev spin qubits, in a material platform with long spin coherence times.

\begin{figure*}[h!]
\includegraphics[width=\textwidth]{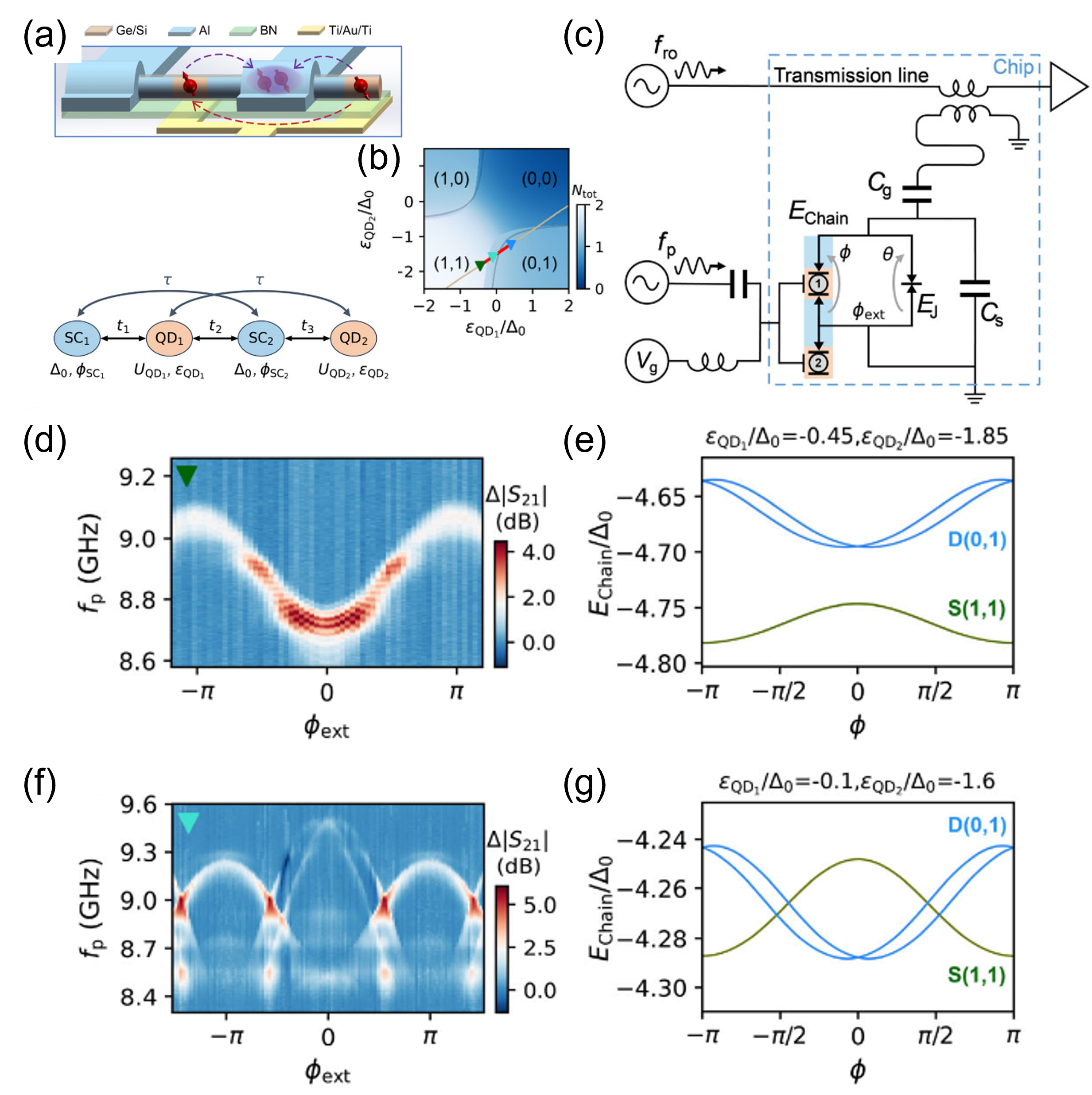}
\caption{ (a) Schematic of the four-site SC-QD chain (top) and the model (bottom) used for the analysis. (b) Calculated stability diagram in the crossed Andreev reflection-dominated regime.  (c) Schematic of the cQED setup used to measure the parity of the chain. (d) and (f) show the two-tone measurements versus external flux for extracting the transmon frequency. (e) and (g) show the calculated potential energy of the chain, which is consistent with the experimental observations. In (f), both states are detected simultaneously as the energy difference between the two states is smaller than the thermal fluctuation energy. Adapted and reprinted from Ref.~\cite{zhuo2025measurementparitydependentenergyphaserelation}.} 
\label{Fig:cQEDGeNanowires}
\end{figure*}

Finally, very recently, the first cQED experiments involving proximitized Ge QDs have been realized~\cite{zhuo2025measurementparitydependentenergyphaserelation}. A four-site superconductor - QD1 - Superconductor - QD2 chain was created in Ge/Si core/shell nanowires by using Al as superconducting material~(Fig.~\ref{Fig:cQEDGeNanowires}). This chain was integrated into a SQUID and operated in the transmon regime in order to read out the parity of the chain. This was feasible as the flux dependence of the transition energy of the transmon depends on the potential energy of the superconductor-QD chain. By performing two-tone measurements, the spectrum of the transmon for different fluxes, and therefore the parities of the chain, could be investigated, an important step in view of the recent progress in the realization of artificial Kitaev chains, described in Sec.~\ref{s:bottom-up}. 

Further progress in the field includes a recent demonstration \cite{fabris2026granularaluminuminducedsuperconductivity} showing that granular aluminum (grAl), composed of nanoscale aluminum grains embedded in an amorphous oxide matrix, can overcome the limitation of small in-plane $g$-factors. By depositing grAl on Ge/SiGe heterostructures, the authors achieved a hard superconducting gap with BCS peaks at 305 $\mu$eV and magnetic-field resilience in both the in-plane and out-of-plane directions. This enabled the resolution of Zeeman splitting of YSR states beyond 50 $\mu$eV (12 GHz). Leveraging this robustness, the approach also revealed unique signatures of hole physics and demonstrated tunability of the $g$-tensor.

\subsection{Theoretical progress}
Despite all these recent experimental advances demonstrating the feasibility of inducing superconductivity in Ge-based hybrid systems, the theoretical framework describing the proximity effect in these platforms have remained comparatively underdeveloped until recently. In analogy with conventional treatments of proximity-induced superconductivity in electron systems, the minimal theoretical approach involves augmenting the effective low-energy Hamiltonian of the topmost valence band with a constant pairing term, as proposed in Ref.~\cite{laubscher2024majorana}. This phenomenological strategy can be extended to include static pairing amplitudes in both heavy-hole and light-hole subbands, leading to emergent features not captured by single-band models~\cite{adelsberger2023microscopic}.

However, such simplified models inherently neglect the interband coupling and hybridization effects that are crucial in multiband valence-band systems, particularly in the presence of strong spin-orbit interaction and band mixing. As a result, they fail to capture the full complexity of the induced superconducting correlations, especially the momentum- and spin-dependent structure of the proximity-induced pairing.

A more rigorous microscopic treatment involves modeling the hybrid structure as a coupled system of semiconductor and superconductor degrees of freedom, linked via momentum-conserving tunneling terms. This approach has been shown to yield nontrivial renormalization of effective parameters, such as the $g$-factor in the heavy-hole sector~\cite{Luethi2023}, and provides a more accurate description of the induced pairing symmetry and its dependence on the underlying band structure.

From a symmetry perspective, it has been argued that direct induction of superconducting correlations is symmetry-allowed only in the conduction band~\cite{moghaddam2014exporting}, with the valence band acquiring superconducting character only through higher-order perturbative processes. Nevertheless, real devices exhibit structural inhomogeneities, interface roughness, and disorder that generically break spatial and point-group symmetries. These symmetry-breaking effects can relax the selection rules that constrain proximity-induced pairing, thereby enabling direct coupling between the superconductor and valence-band states beyond the idealized scenarios considered in symmetry-based arguments.

Consequently, a comprehensive theoretical description of the proximity effect in Ge-based hybrids must go beyond simplified constant-pairing models and incorporate the full multiband structure, interface physics, and disorder-induced symmetry breaking. Such an approach is essential for accurately predicting and engineering superconducting properties in next-generation quantum devices based on group-IV semiconductors.

Two recent theoretical studies \cite{Pino2025-holes,Babkin2025-holes} have adopted a comprehensive microscopic approach by employing the full $8 \times 8$ $\mathbf{k} \cdot \mathbf{p}$ Kane model (8KP) to describe the proximity effect in hole-based semiconductor–superconductor hybrid systems. These works systematically derive the effective low-energy theory of a proximitized semiconductor by incorporating all symmetry-allowed superconducting pairing terms that emerge from direct coupling between the superconductor and the various conduction and valence bands.

From this general framework, an effective theory can be extracted for hole states in strained germanium heterostructures with vertical confinement, which is particularly relevant for experimental platforms based on 2DHGs. The inclusion of all relevant bands reveals that the induced pairing structure is significantly more intricate than the conventional $s$-wave scenario. The resulting proximity-induced superconductivity exhibits nontrivial momentum dependence, including interband coupling terms and contributions arising from cubic Rashba spin–orbit interactions.

Building upon this foundation, a numerically supported effective theory has been developed, enabling the derivation of analytical expressions that describe various types of superconducting correlations. These include unconventional singlet and triplet pairing mechanisms with momentum-dependent structure, as well as distinct Zeeman and Rashba spin–orbit contributions. The explicit dependence of these effects on external electric and magnetic fields allows for concrete experimental predictions, such as the formation of Bogoliubov Fermi surfaces, and the onset of gapless regimes induced by strong in-plane magnetic fields. 

Another interesting feature is the emergence of exotic forms of superconductivity including pairings in the heavy hole sector of the form $\tilde\Delta\approx \Delta_{\mathrm{HH}} \frac{i(\delta_{h}\sigma_z-g_{xz}\mu_B B_z\sigma_x)}{\sqrt{\delta_{h}^2+(g_{xz}\mu_B B_z)^2}}$, with $g_{xz}$ characterizing the out-of-plane Zeeman term, perpendicular to the Rashba spin-splitting, due to $B_z$. Interestingly, this effective pairing resembles the familiar result of a proximitized Rashba semiconductor~\cite{Alicea_RPP2012,Leijnse_Review2012,beenakker2013search,Aguado_Nuovo2017,Lutchyn_NatRev2018} (see subsection \ref{Sec:top-down}), with the main difference of the cubic-in-momentum dependence of $\delta_{h}$ suggesting an effective f-type superconducting pairing.

These findings underscore the necessity of moving beyond simplified single-band or constant-pairing models to accurately capture the multiband and spin–orbit-coupled nature of hole systems in group-IV semiconductors. The resulting effective theory provides a more faithful description of the induced superconducting correlations and their dependence on strain, confinement, and interface properties.

These theoretical developments not only advance the understanding of proximity-induced superconductivity in multiband hole systems, but also establish a solid foundation for the design of next-generation spin-based quantum devices. In particular, the ability to characterize and control unconventional pairing symmetries, spin–orbit interactions, and their dependence on external fields provides a versatile framework for engineering Andreev spin qubits that could exploit the spin degree of freedom in novel and tunable ways. For example, Ref. \cite{hoffman2025} discusses various Josephson junction geometries to optimize qubit frequencies above experimental resolution. Furthermore, the theory introduced in Refs. \cite{Pino2025-holes,Babkin2025-holes}  enables the potential identification of parameter regimes where spin-selective Andreev bound states can be stabilized, including configurations with nontrivial topological character or momentum-dependent pairing textures. These insights pave the way for the realization of electrically controllable, highly coherent spin qubits with enhanced gate fidelities, leveraging the intrinsic advantages of group-IV semiconductors.
\section{Majorana states in semiconductor-superconductor nanostructures }\label{s:bottom-up}
Topological states constitute an ideal approach for the encoding and processing of quantum information. MZMs, appearing at the end of topological superconductors, are ideal for this task. They are non-local quasiparticles that feature protection against local sources of noise. In this section, we review these properties and discuss the recent progress in this direction.
\subsection{Majorana zero modes}
\label{s:MZMs}
The BdG description of superconductivity that described in Sec.~\ref{s:superconducting} provides a natural framework for engineering and understanding Majorana quasiparticles. Starting from a transformation similar to Eq. \eqref{BdG-transf}, we can write a Nambu spinor keeping the four possible components%
\begin{eqnarray}
\label{Nambu-spinor}
\hat\Psi=\left( \begin{array}{ccc}
 c_{\uparrow} \\
c_{\downarrow}\\  c^\dagger_{\downarrow} \\
-c^\dagger_{\uparrow}\end{array} \right).
\end{eqnarray}
This Nambu spinor can be interpreted as an operator version of the Majorana wave function. This connection arises from the fundamental particle-hole symmetry of superconductors, which imposes the so-called Majorana {\it pseudo-reality} condition \cite{PhysRevB.81.224515}, $\hat\Psi({\bf r})=P\,\hat\Psi({\bf r})=C\,\hat\Psi^*({\bf r})$, where $P$ and $C$ are the particle and hole conjugation operations.
It is important to stress, however, that this does not imply that every BdG quasiparticle is a Majorana fermion. Particle-hole symmetry maps a BdG eigenstate with energy $E$, created by the quasiparticle operator $\gamma^\dagger_E$, onto an eigenstate with energy $-E$, such that $\gamma^\dagger_{E}=\gamma_{-E}$. For $E\neq 0$, these two solutions describe one and the same physical excitation: a coherent superposition of electron and hole components which is \emph{not} its own antiparticle, $\gamma^\dagger_E\neq\gamma_E$. This is the case for the generic BdG quasiparticles and Andreev bound states discussed in the previous sections. In other words, while $\hat\Psi$ satisfies the Majorana condition, generic systems involve the coupling of several Nambu spinors that breaks the Majorana properties. This distinction between the individual properties of Nambu spinors and full systems poses a fundamental challenge for the direct observation of Majorana properties in superconducting quasiparticles.

A viable approach to circumvent this issue is to engineer systems with zero-energy solutions of the BdG equations, which remain self-conjugate under particle-hole symmetry. Indeed, a self-conjugate quasiparticle, $\gamma=\gamma^\dagger$, can only arise from a solution that is mapped onto itself by particle-hole symmetry, which requires $E=0$. Only such zero-energy, self-conjugate BdG solutions, when spatially isolated from each other, correspond to MZMs. In the literature, these states are also referred to as Majorana bound states (MBSs); for consistency, we use the term MZM throughout this Review. Owing to BdG symmetry, the energy spectrum must be symmetric, {\it i.e.} for every eigenvalue $E_n$ there is an eigenvalue $-E_n$, which implies that if pairs of MZMs are present in the system, any smooth deformation of the Hamiltonian cannot split them from zero energy, unless they hybridize with each other. 

\subsection{The Kitaev chain model}
\label{s:Kitaev_model}
Arguably, the Kitaev chain simplest model that supports well-separated MZMs~\cite{Kitaev2001,Kitaev2003}. This is a lattice model that describes a 1-dimensional p-wave superconductor. Specifically, the model considers $N$ spinless fermionic sites, with a Hamiltonian given by
\begin{equation}
    \mathcal{H}_{\rm Kitaev}=\sum_{n=1}^N \epsilon_n c_n^\dagger c_n+\sum_{n=1}^{N-1}\left(t_n c^\dagger_{n+1} c_n+\Delta_n c_{n+1} c_n +{\rm H.c.}\right)\,,
    \label{H_Kitaev}
\end{equation}
where $t$ and $\Delta$ are the hopping and pairing amplitudes between the nearest-neighbor sites. Despite its simplicity, the Kitaev model contains the relevant ingredients to host exponentially-localized Majorana states at the ends. In order to reveal the Majorana character of the ground-state wavefunctions, it is convenient to make the change of variable
\begin{equation}
    \gamma^A_n = c_n+c^\dagger_n\,,\qquad \gamma^B_n = i\left(c_n-c^\dagger_n\right)\,,
    \label{Majorana_op_def}
\end{equation}
which satisfy
\begin{equation}
    \left[\gamma^{(A,B)}_n\right]^\dagger =\gamma^{(A,B)}_n\,, \quad \left[\gamma^{(A,B)}_n\right]^2=1\,,\quad \left\{\gamma^A_n,\gamma^B_{n'}\right\}=2\delta_{n,n'}\,,
\end{equation}
\begin{figure*}
\begin{center}
\includegraphics[width=0.9\textwidth]{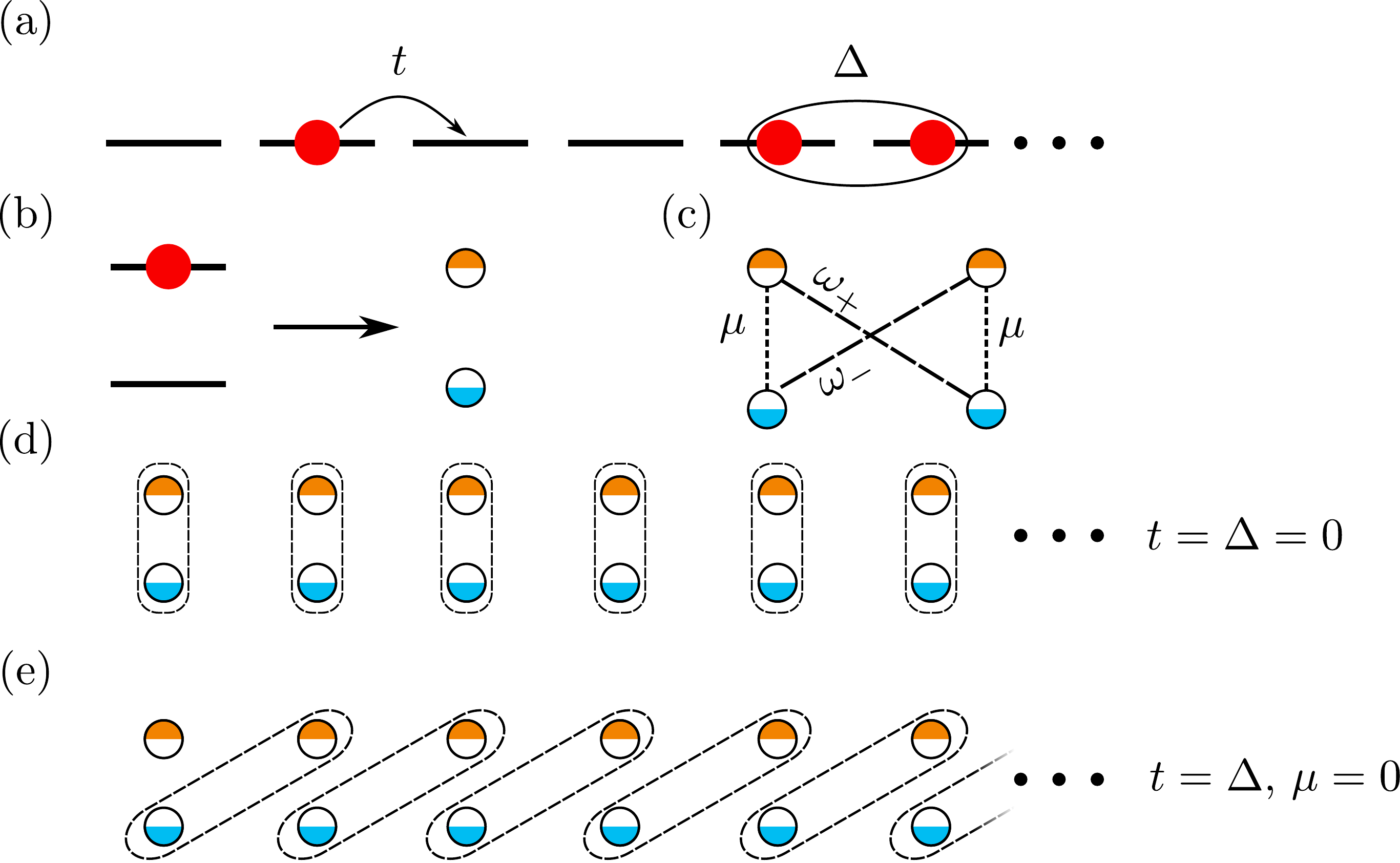}
\end{center}
\caption{(a) Sketch of the Kitaev chain. (b) Transformation  between of the fermion operators into the Majorana ones (half-filled circles). (c) Coupling between Majorana operators as described by the transformed Hamiltonian~\eqref{H_Kitaev_transformed}, with $\omega_{pm}=\Delta\pm t$. (d) Paradigmatic trivial situation, with local coupling between Majorana operators. (e) Paradigmatic topological situation, with non-local coupling between Majorana operators, leaving unpaired Majoranas at the end.} 
\label{Fig:X}
\end{figure*}
with all the other anti-commutators being zero, {\it i.e.} $\{\gamma^A,\gamma^A\}=\{\gamma^B,\gamma^B\}=0$. Using the operators defined in Eq.~\eqref{Majorana_op_def}, the Kitaev chain model \eqref{H_Kitaev} can be written as
\begin{equation}
    \mathcal{H}_{\rm Kitaev}=\frac{-i}{2}\sum_{n=1}^{N}\epsilon_n \gamma^A_n \gamma^B_n+\frac{i}{2}\sum_{n=1}^{N-1}\left(\omega_+\gamma^B_n\gamma^A_{n+1}+\omega_-\gamma^A_n\gamma^B_{n+1}\right)\,,
    \label{H_Kitaev_transformed}
\end{equation}
with $\omega_\pm=\Delta\pm t$. In this form, it becomes clear that $\epsilon$ coupled Majorana operators locally, while $\omega_{\pm}$ couples them between neighboring sites. The role of these two parameters become evident when considering two limiting scenarios. For $t_n=\Delta_n=0$, the local Majorana states at the same site couple, forming local fermions. In the charge basis, the Hamiltonian is given by
\begin{equation}
    \mathcal{H}_{\rm Kitaev}(t_n=\Delta_n=0)=\sum_{n=1}^N \epsilon_n c^\dagger_nc_n\,.
\end{equation}
This situation is identified as the ``trivial'' case, characterized by local fermions. The situation is a bit more interesting in the other limiting situation, $\epsilon_n=0$ and $t_n=\pm\Delta_n$. In this case, Majorana couple between neighboring sites, leading to
\begin{equation}
    \mathcal{H}_{\rm Kitaev}(t_n=\Delta_n)=\frac{i}{2}\sum_{n=1}^{N-1}t_n\gamma^B_n\gamma^A_{n+1}\,.
    \label{H_Kitaev_diagonal}
\end{equation}
Despite its apparent simplicity, the Hamiltonian contains important new physics compared to the trivial case. First, the fermions are non-local, delocalized between neighboring sites, with creation and annihilation operators given by $d_n=(\gamma^B_n+\gamma^A_{n+1})/2$ and $d^\dagger_n=i(\gamma^B_n-\gamma^A_{n+1})/2$. Importantly, there is one fermionic state that has ``disappeared'' from the Hamiltonian, illustrated by the sum going up to $N-1$ instead of $N$ as in the original Hamiltonian, Eq~\eqref{H_Kitaev}. In particular, Majorana the operators $\gamma^A_1$ and $\gamma^B_N$ remain uncoupled and perfectly localized at the ends for $t_n=\Delta_n$ and $\epsilon_n=0$. These uncoupled Majorana states can be understood as the formation of a new fermionic state delocalized across the Kitaev chain,
\begin{equation}
    f=\frac{1}{2}\left(\gamma^A_1+i\gamma^B_N\right)\,,\qquad f^\dagger=\frac{1}{2}\left(\gamma^A_1-i\gamma^B_N\right)\,.
\end{equation}
The shape of the transformed Hamiltonian, Eq.~\eqref{H_Kitaev_diagonal}, ensures the commutation rules $[H,\gamma^A_1]=[H,\gamma^B_N]=0$. In addition, filling or emptying the $f$ fermionic level does not change the Hamiltonian, as it has zero energy. Therefore, unlike conventional superconductors, the states with even and odd number of fermions have the same energy. It is worth noticing that Majorana states are equal superpositions of electrons and holes. Therefore, local detectors coupled to only one side of the system cannot determine the fermion parity in the system. This also makes Majorana state immune to local perturbations that do not close the superconducting gap.

For arbitrary values of $\epsilon_n$, $t_n$, and $\Delta_n$, the situation is a bit more general, as the forming fermions cannot be understood as being fully local or non-local. In general, there is a mixture of the two limiting scenarios previously presented. Nevertheless, it is possible to identify two different quantum phases, characterized by the absence/presence of MZMs when taking the $N\to\infty$ limit. These two phases feature a different value for the $\mathbb{Z}_2$ topological invariant, given by $M=(-1)^\nu$, with $\nu$ being the number of decoupled Majorana pairs. These two phases are usually referred to as ``topologically trivial'' and ``topologically non-trivial'' phases (or simply as ``trivial'' and ``topological'', abusing language). The exponential localization implies always a finite overlap between MZMs in realistic systems, that will split their energy and can potentially undermine their coherence. Therefore, it is a central task to separate them spatially, by designing long systems or engineering states with ideal Majorana localization. These 2 strategies led to the {\it top-down} and the {\it bottom-up} approaches that we discuss in subsections~\ref{Sec:top-down} and~\ref{Sec:bottom-up}.

\subsubsection{Majorana non-abelian properties}
\label{seq:non-Abelian_MBSs}
Before discussing these two approaches, let us first comment on one of the most fascinating aspects of MZMs, that make them attractive from a fundamental point of view, but has also potential for quantum technologies: their non-abelian properties. Particles in 3 dimensions necessarily obey either the Bose-Einstein or the Fermi-Dirac distributions, that differ from the sign acquired by the wavefunction after two particles are exchanged, {\it i.e.} $\ket{\ldots, \,\Psi_j,\ldots,\,\Psi_k}=\pm\ket{\ldots, \,\Psi_k,\ldots,\,\Psi_j}$ for bosons/fermions. In low dimensions, these commutation relations can acquire a more general form. The so-called Abelian anyons provide the simplest extension of the exchange properties of bosons and fermions: their exchange leads to a phase shift of the system's wavefunction, $\theta$, that is not $0$ or $\pi$, $\ket{\ldots, \,\Psi_j,\ldots,\,\Psi_k}=e^{i\theta}\ket{\ldots, \,\Psi_k,\ldots,\,\Psi_j}$. Even more interesting is the case of non-Abelian anyons. A set of isolated anyons define a degenerate ground state manifold. Exchanging (or braiding) anyons allow to transform the ground state within the manifold. The result only depends on which anyons are exchanged and their direction, but not on specific exchange path, if they are kept separated. 

MZMs are non-abelian anyons with very specific exchange statistics, described by the unitary operator~\cite{PhysRevLett.86.268}
\begin{equation}
    U_{j,k}=\frac{1}{\sqrt{2}}(1+\gamma_j\gamma_{k})=e^{\pi\gamma_j\gamma_{k}/4}\,.
    \label{Eq:braid_operation}
\end{equation}
Hereafter, we drop the $A,\,B$ superindices, as they are not relevant for the discussion. To understand the properties of this exchange operator, let us consider the situation with $2N$ MZMs, labeled with $i=1,\ldots,2N$ and initialized pairwise, $\Psi=\ket{\ldots,p_{j,j+1},\ldots,p_{k,k+1}}$, where $\Psi$ denotes the initial wavefunction and $p_{j,k}$ is the fermion parity of $j,\,k$ Majorana pairs, defined as $p_{j,k}=1-2n_{j,k}=-i\gamma_j\gamma_k$, where $n_{j,k}$ is the occupation number of the fermionic mode formed by the Majorana pair $(j,k)$. Exchanging MZMs belonging to the same pair leads to a global phase
\begin{eqnarray}
U_{j,j+1}\Psi=e^{i\pi p_{j,j+1}/4}\,\Psi,
\end{eqnarray}
where $U$ is defined in Eq.~\eqref{Eq:braid_operation}. The case where two MZMs belonging to different pairs are exchanged is more interesting
\begin{eqnarray}
U_{j+1,k}\ket{\ldots,p_{j,j+1},\ldots,p_{k,k+1}}=\frac{1}{\sqrt{2}}\left[\ket{\ldots,p_{j,j+1},\ldots,p_{k,k+1}}+i(-1)^s\ket{\ldots,-p_{j,j+1},\ldots,-p_{k,k+1}}\right]\,,
\label{Eq:braiding_wavefunction}
\end{eqnarray}
where $s=0,1$, depending on the total parity of the Majorana pairs between $j+1$ and $k$. Equation~\eqref{Eq:braiding_wavefunction} illustrates that braiding performs quantum operations on the ground state manifold. However, the set of gates that can be implemented by braiding is not sufficient for universal quantum computation, and braiding needs to be complemented by at least one operation that is not topologically protected~\cite{Karzig_PRX2016}.

\subsection{Top-down approach}\label{Sec:top-down}
\begin{figure*}[h!]
\begin{center}
\includegraphics[width=0.8\textwidth]{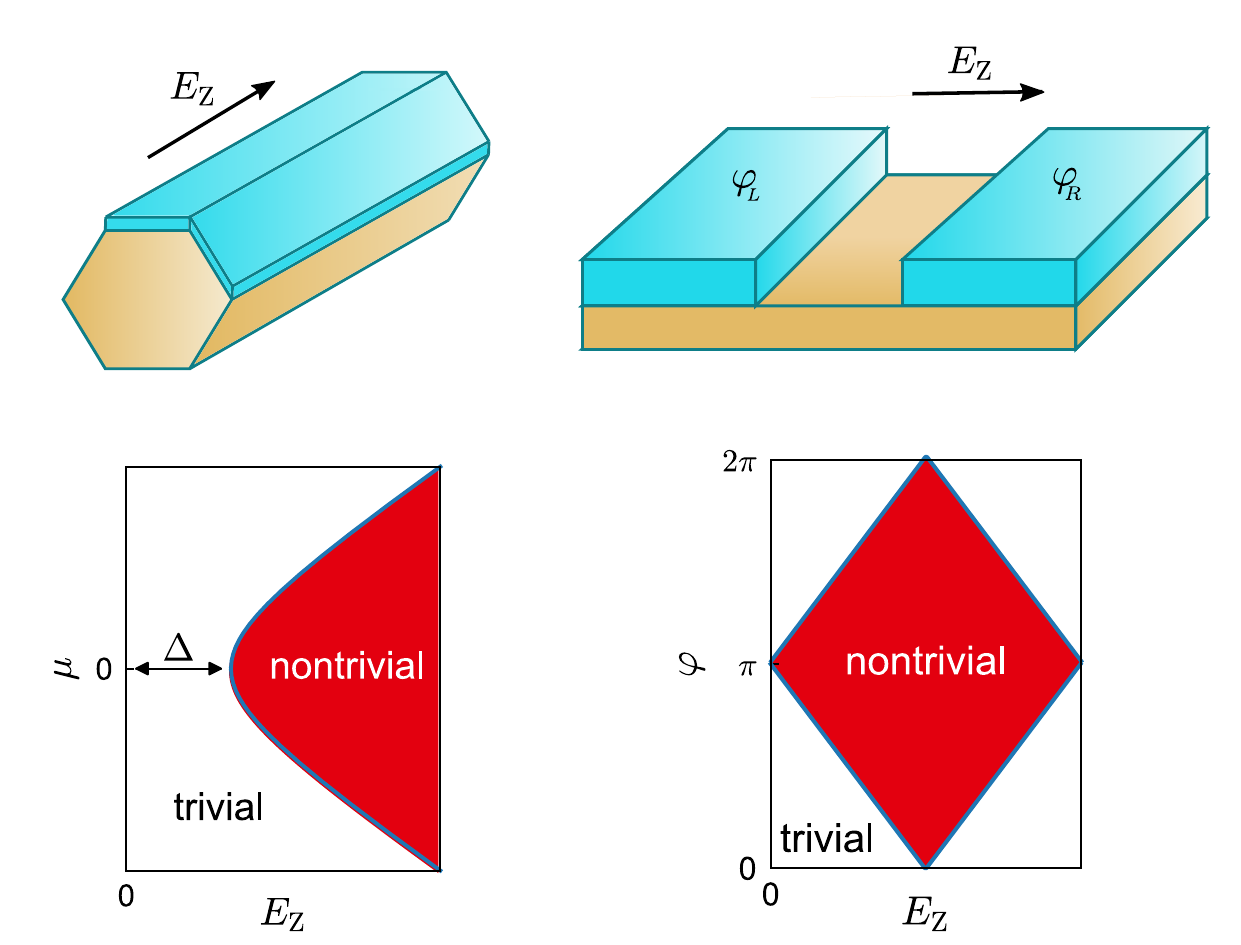}
\end{center}
\caption{Sketch of two of the most studied semiconductor-superconductor platforms for Majorana physics. (a) Semiconductor nanowire (orange) partially covered by a superconducting shell. (b) Planar Josephson junction, where the phase difference between the superconductors, $\varphi=\varphi_L-\varphi_R$, can tune the topological transition. Down: corresponding phase diagrams in terms of the applied Zeeman field versus chemical potential $\mu$ and phase difference $\phi$, respectively.} 
\label{Fig:Majorana-Platforms}
\end{figure*}

Before moving on to discuss bottom-up approach, the main focus of this section, we make a short detour to briefly discuss previous attempts to study MZM in topological superconductors (for comprehensive reviews on various aspects of this topic, see Refs.~\cite{Alicea_RPP2012,Leijnse_Review2012,beenakker2013search,Aguado_Nuovo2017,Lutchyn_NatRev2018,Prada_review,MarraReview_JAP,Tanaka2024-fk,Tanaka2012-hx,Stanescu2013-cr,Sato2016-bs,Mizushima2016-ls,Sato2017-xo}). Their experimental realization in platforms such as proximitized nanowires, vortices in topological superconductors, and magnetic chains of atoms, among others~\cite{Flensberg2021}, has fueled intense research efforts, also with the motivation of developing a platform for fault-tolerant quantum computing~\cite{NayakReview,MarraReview_JAP}.
This interest was arguably initiated by the possibility of engineering topological superconductors by combining different materials, starting from the breakthrough work of Fu and Kane, who proposed s-wave induced superconductivity in a topological insulator to obtain MZMs~\cite{Fu2008}. Soon after, this idea was extended to semiconductor-superconductor heterostructures, where the competition between s-wave induced superconductivity in the semiconductor and the applied magnetic field generates the conditions to obtain topological superconductivity~\cite{Oreg2010,Lutchyn2010} and MZMs in one-dimensional nanowires. This idea, and some similar approaches, have been explored in different experimental platforms derived from the concept of topological superconductors designed in a quasi-one-dimensional nanostructure based on hybrid semiconductor-superconductor devices. Fig. \ref{Fig:Majorana-Platforms} shows sketches of the two main  experimental platforms in 1D and 2D (top) and their corresponding phase diagrams (bottom). 
In the one-dimensional case, the topological phase transition occurs when
\begin{equation}
    E^2_{z}=\Delta^2+\mu^2\,,
    \label{Eq:topo_condition}
\end{equation}
with $E_z$ being the Zeeman spin-splitting in the semiconductor, $\Delta$ the induced s-wave pair amplitude, and $\mu$ the chemical potential in the semiconductor, see also Fig. \ref{Fig:Majorana-Platforms}b. The previous expression determines the superconducting gap closing, condition for the topological phase transition to occur. The spin-orbit coupling is essential to avoid the full spin-polarization, that would cause a suppression of superconductivity, leading to a gap re-opening in the topological phase. Even though the spin-orbit coupling strength does not enter in the topological condition in Eq.~\eqref{Eq:topo_condition}, it sets the value of the topological gap and, therefore, of the topological protection. For superconductivity to survive, it is essential that the parent gap in the superconductor is still preserved after the closing of the induced gap in the semiconductor.  

Over the past decade, numerous experiments have reported signatures consistent with the presence of MZMs in semiconductor-superconductor heterostructures. The most prominent among these is the observation of zero-bias conductance peaks~\cite{Mourik12012}, often accompanied by temperature and gate dependence in line with theoretical expectations \cite{Nichele_PRL2017}. Additional studies involving quantum dots have provided insights into the spatial localization of low-energy states \cite{Deng_Science2016,deng2018nonlocality}, while non-local transport measurements have probed correlations between the ends of nanowires and the BCS charge of subgap states \cite{Danon_PRL2020,Menard_PRL2020}. These efforts have culminated in protocols aimed at identifying the topological regime~\cite{Pikulin_arxiv2021}, including gap closing and reopening, with recent claims of successful implementation \cite{Aghaee_PRB2023,Aghaee_Nature2025,Aghaee_arxiv2025}.

Floating superconducting islands have added new dimensions to the study of MZMs, enabling gate-tunable ground state splitting and parity control. Experiments have demonstrated long-range coherent transport, exponential suppression of energy splitting with island length, and parity-dependent conductance features \cite{Whiticar_NatCom2020,Vaitiekenas_Science2020,Shen_NatComm2018}. However, even in these systems, unintentional quantum dot formation and Yu-Shiba-Rusinov physics can produce zero-bias peaks unrelated to topology \cite{Lee2014,Valentini_Science2021}.

Full-shell nanowires, {\it i.e.} semiconducting nanowires entirely surrounded by a superconductor, are another way to decrease the required external field to reach the topological regime. The narrow superconducting cylinder feature the so-called Little-Parks effect~\cite{Little_PRL1962}: a set of superconducting lobes that close when the magnetic field is half a flux quantum across the nanowire section, with a maximal gap when the flux is an integer number of the flux quantum~\cite{Vaitiekenas_Science2020,Vekris_SciRep2021,Ibabe_NatCom2023,Razmadze_PRB2024}. Signatures of low-energy states were found in the first lobe, where the phase of the superconducting parameter winds $2\pi$ and describes a fluxoid~\cite{Vaitiekenas_Science2020}, consistent with theory predictions~\cite{Penaranda_PRR2020,PhysRevB.109.115428}. Later experiments found that near-zero energy subgap states of non-topological origin could explain the measured low-energy features in the data~\cite{Valentini_Science2021,Valentini_Nature2022}. Apart from the low-energy state, additional subgap states can appear inside the Little-Parks lobes~\cite{PhysRevB.107.155423,PhysRevLett.134.206302}, analogous to the Caroli-de Gennes-Matricon states observed in type II superconductors. Different aspects of the Josephson effect in junctions based on full-shell nanowires have been discussed in Refs \cite{Giavaras2024,Paya2025a,Paya2025b}. Interesting aspects of these junctions include the flux modulation of the critical current  due to the Little-Parks effect \cite{Giavaras2024,Paya2025a}, including a characteristic skewness towards large fluxes, which is inherited from the skewness of Caroli-de Gennes-Matricon subgap states \cite{Paya2025a}. Furthermore, the flux modulation of supercurrents allows for the design of novel supercurrent valves \cite{Paya2025b} in junctions with different shell radii, as well as protected qubits \cite{Giavaras2025} based on  $\cos(2\varphi)$ Josephson couplings (see subsection \ref{s:protected-qubits}).

The superconducting phase difference between superconductors offers a way to engineer topological phases without large magnetic fields, for reviews see Refs.~\cite{Schiela2024,Lesser_JoPD2022}. The simplest proposal in terms of superconductors is composed of a planar semiconductor proximitized by two superconductors in a Josephson junction geometry~\cite{PhysRevX.7.021032,PhysRevLett.118.107701}. In this case, the superconducting phase difference $\varphi$ can be used as an additional experimental tuning parameter, e.g. by current biasing the junction or by flux biasing a superconducting loop in which the junction is embedded. Specifically, this proposal exploits the gap closure in transparent bulk Josephson junctions close to $\varphi=\pi$. At this point, a small Zeeman field can induce a topological phase transition (Fig. \ref{Fig:Majorana-Platforms}b, bottom).
Planar Josephson junctions have been also investigated experimentally in Refs.~\cite{ren2019topological,fornieri2019evidence,Dartiailh2021,banerjee2023control,banerjee2023local}, finding low-energy states.

Although significant experimental progress has been made, the interpretation of signatures as evidence for topological MZMs remains highly controversial. More than a decade after the first reports, the field continues to grapple with the fact that trivial low-energy states—arising from disorder, smooth confinement, or quantum dot formation—can mimic the local signatures of Majoranas, see for example \cite{Prada_PRB2012,Kells_PRB12,Vuik_SciPost19,Pan_PRR20}. These so-called quasi-Majoranas or Andreev zero modes challenge the reliability of local probes and have led to a re-evaluation of what constitutes definitive evidence. Recent theoretical developments using non-Hermitian topology have shown that many of these trivial zero modes are topologically equivalent to true MZMs in terms of their local observables, further blurring the distinction \cite{Avila_ComPhys2019}.

Nevertheless, this strategy—based on identifying local and non-local signatures of MZMs in semiconductor–superconductor nanowires—continues to be actively pursued by major groups in the field, including the Microsoft team.  Recent efforts include the development of the {\it topological gap protocol}~\cite{Pikulin_arxiv2021} to identify the topological regime via simultaneous detection of subgap states and gap reopening signatures \cite{Aghaee_PRB2023}, as well as a single-shot interferometric parity readout in InAs–Al hybrid devices, recently reported \cite{Aghaee_Nature2025}. In this work, the authors report the use of an interferometric setup to detect shifts in quantum capacitance that depend on the parity state of the system. Parity measurements achieve a signal-to-noise ratio of 1 within approximately 3.6 microseconds, with an error rate near 1$\%$, and the system maintains state stability for durations exceeding 1 millisecond under magnetic fields of around 2 Tesla.
A subsequent preprint \cite{Aghaee_arxiv2025} reports measurements on a so-called Majorana tetron device implemented in a superconductor–semiconductor heterostructure. This architecture consists of two parallel superconducting nanowires together with a trivial superconducting backbone. The authors describe single-shot interferometric parity measurements across two distinct loops, corresponding to Pauli-X and Pauli-Z operations on the tetron geometry, and report differences in lifetimes for these measurement types, discussing their implications for error correction and device stability.
While these results represent significant technical progress, their interpretation in terms of topological protection remains under scrutiny, and the distinction between trivial and topological zero modes is still debated. Specifically, some researchers have raised concerns about the presence of a topological superconducting phase \cite{legg2025commentinterferometricsingleshotparity}, citing issues related to the reliability of the topological gap protocol \cite{Aghaee_PRB2023}. These discussions reflect ongoing efforts within the community to assess the robustness of experimental evidence for topological quantum operations.

Alternative measurements involve measuring current-current correlation functions~\cite{Zocher_PRL2013,Haim_PRL2015,Haim_PRB2015,Lu_PRB2016,Ridley_arXiv2025}, interferometric signatures~\cite{Sau_PRB2015,Chiu_PRB2018,Hell_PRB2018,Wang2018-gr,Liu_PRL2019,Whiticar_NatCom2020}, and Josephson-based signatures~\cite{Kitaev2001,Kwon2003-qi,Jiang2011,Heck_PRB2011,Pikulin2012-fv,SanJose2012,Dominguez_PRB2012,Houzet_PRL2013,Dominguez_PRB2017,Pico-Cortes_PRB2017,Chiu_PRB2019,Laroche2019-nt}.

In parallel, a growing segment of the community is shifting toward bottom-up approaches, where minimal Kitaev chains are engineered using coupled quantum dots, as we discuss in the following sections.

\subsection{Majorana states in double quantum dots (minimal Kitaev chain)}
\label{Sec:bottom-up}

Recently, the bottom-up approach to engineer Majorana states has emerged as a promising route, free from many of the uncertainties that are present in other platforms. The idea stems from the observation that perfectly localized Majorana states appear in the Kitaev model for certain parameters, already pointed out in Kitaev's original paper~\cite{Kitaev2001}, see also details in Sec.~\ref{s:Kitaev_model}. This is true even for 2 sites -- a minimal Kitaev chain. At the sweet spot, Majorana states in minimal Kitaev chains share all the local properties of the MZMs appearing in long topological systems, including, in principle, their non-Abelian exchange properties~\cite{tsintzis2023roadmap}. However, they are not topologically protected: any perturbation or deviation from the Majorana {\it sweet spot} makes the two Majoranas overlap and lifts the ground-state degeneracy already to first or second order in the perturbation, in contrast with the expected exponentially small splitting for spatially separated MZMs in the topological phase of a long chain. For this reason, these fine-tuned, finite-size states are usually referred to as poor man's Majorana states (PMMs)~\cite{Leijnse_PRB2012}.

{Throughout this Review, we therefore distinguish between two different situations: (i) MZMs in the topological phase of an extended system, which are spatially separated, exponentially localized, and whose degeneracy is protected against any local perturbation weaker than the topological gap; and (ii) poor man's Majorana states (or Majorana sweet spots) in Kitaev chains, which are finite-size, fine-tuned and not topologically protected. Longer quantum-dot chains interpolate between these two limits, see Sec.~\ref{Chapter3:sec:PMM_experimentsI}. When we refer to topologically protected Majorana qubits, we always refer to the former; qubits based on minimal Kitaev chains should be regarded as testbeds for Majorana-based encoding, readout and non-Abelian operations, rather than as topologically protected qubits.}

At the minimal level, the 2-site Kitaev chain is described by
\begin{equation}
	\mathcal{H}=\sum_{\alpha=L,R}\varepsilon_\alpha d^{\dagger}_\alpha d_\alpha + \left(t d^{\dagger}_L d_R + \Delta d_L d_R +{\rm H.c.}\right),
    \label{eq:PMM_Ham_simpleModel}
\end{equation}

\begin{figure*}[th!]
\includegraphics[width=0.95\linewidth]{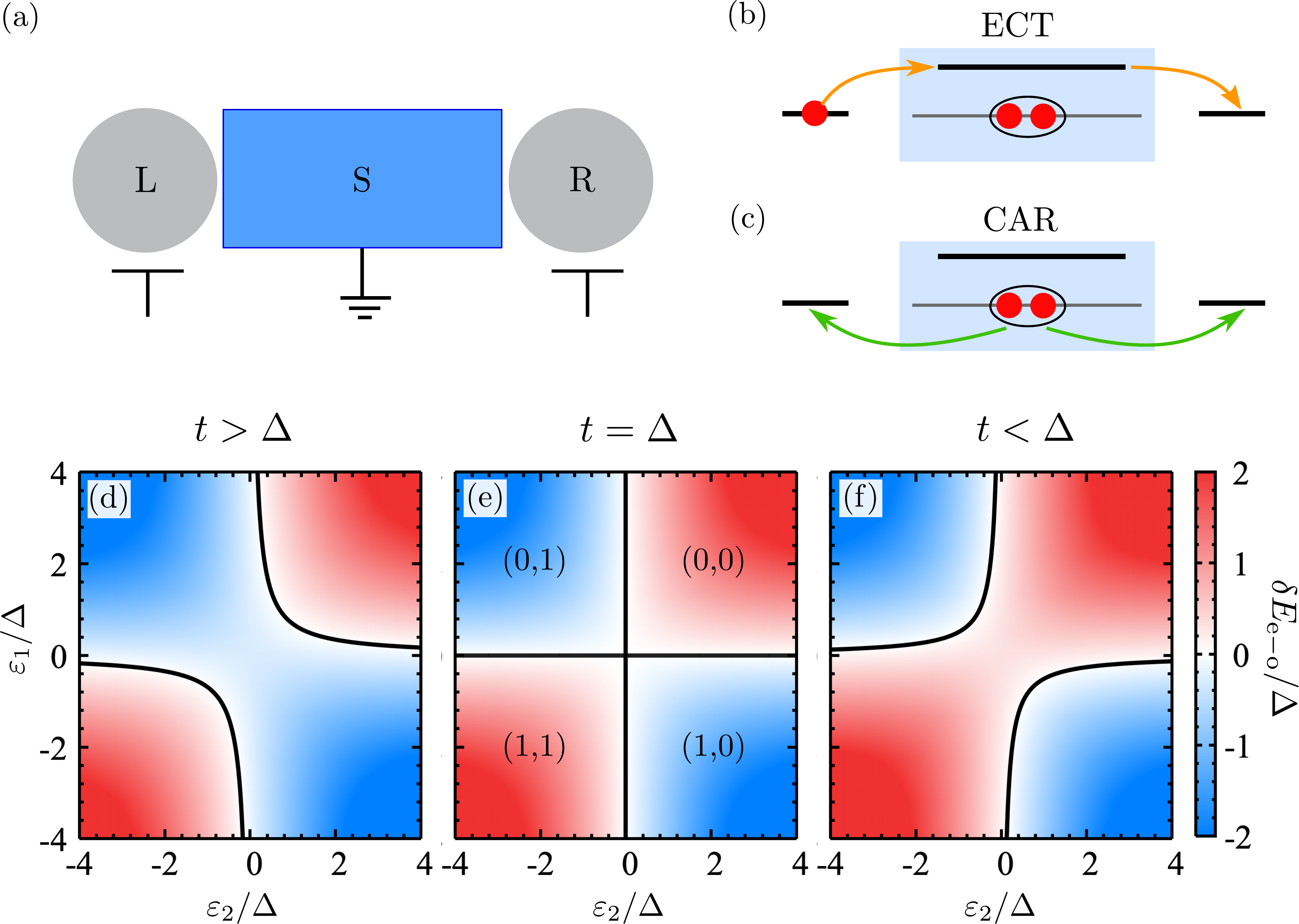}
\caption{(a) Sketch of a minimal Kitaev chain, composed by a narrow superconductor (blue) that mediates the coupling between two QDs (gray circles). The superconductor is grounded and can host subgap states (thick line) that allows for an electric tunability of CAR and ECT processes between the QDs~\cite{Liu_PRL2022}. (b) Schematic representations for the elastic cotunneling (ECT) process, where an electron (red circle) is transferred between the left and right dots occupying a virtual excited state in the superconductor (thick line in the middle segment). (c) Crossed Andreev reflection, where a Cooper pair splits (recombines) with one electron going (coming) from each QD. (d-f) Energy splitting between the even and odd parity states as a function of the energy of the left and right dots in a minimal Kitaev chain, described by Eq.~\eqref{eq:PMM_Ham_simpleModel} for $t>\Delta$, $t=\Delta$, and $t<\Delta$. Red/blue indicates an even/odd parity ground state and the black line indicates the even-odd degeneracy. Adapted and reprinted from Ref~\cite{Souto_chapter2024}.}
\label{fig:PMM_sketch}
\end{figure*}
this model can be written in BdG form as
\begin{equation}
\mathcal{H}=\Psi^\dagger\begin{pmatrix}
\varepsilon_L & t & 0 & \Delta\\
t & \varepsilon_R & -\Delta & 0\\
0 & -\Delta & -\varepsilon_L & -t\\
\Delta & 0 & -t & -\varepsilon_R
\end{pmatrix}\Psi\,,
\end{equation}
where we are using the Nambu spinor $\Psi=(d_L,d_R,d^{\dagger}_L,d^{\dagger}_R)$ and $t$ and $\Delta$ are chosen real. The model has a sweet spot for $t=\Delta$ and $\varepsilon_L=\varepsilon_R=0$, where the system has two eigenstates with energy $E=0$ and wavefunctions given by
\begin{equation}
    \psi_1=\frac{1}{\sqrt{2}}(1,0,1,0)^T\,,\qquad
    \psi_2=\frac{i}{\sqrt{2}}(0,1,0,-1)^T\,.
    \label{eq:PMM_wavefunctions}
\end{equation}
In second quantization, the operators associated with these two states are given by $\gamma_1=(d_L+d^{\dagger}_L)/\sqrt{2}$ and $\gamma_2=i(d_R-d^{\dagger}_R)/\sqrt{2}$, which are Hermitian operators [$(\gamma_{1,2})^\dagger=\gamma_{1,2}$] and describe Majorana states completely localized in the dots. 

The many-body basis provides a complementary point of view to the problem. This description accounts for the fermion occupation of each of the two QDs. Using the notation $\ket{n_Ln_R}$, with $n_{L(R)}$ denoting the occupation of the left/right QDs, the four possible states are $\ket{00}$, $\ket{11}$ and $\ket{01}$, $\ket{10}$ for the even and odd subspaces. In this language, the Hamiltonian in Eq.~\eqref{eq:PMM_Ham_simpleModel} becomes block-diagonal, as there are no terms mixing the two sectors with different fermion parity
\begin{equation}
\mathcal{H}=\begin{pmatrix}
0 & \Delta & 0 & 0\\
\Delta & \varepsilon_R+\varepsilon_L & 0 & 0\\
0 & 0 & \varepsilon_R & t\\
0 & 0 & t & \varepsilon_L
\end{pmatrix}\,.
\label{eq:H_PMM_1body}
\end{equation}
The even states $|00\rangle$ and $|11\rangle$ hybridize via $\Delta$ to form bonding and anti-bonding combinations $|e^{\pm}\rangle = (|00\rangle \pm |11\rangle)/\sqrt{2}$. Likewise, the odd  states $|10\rangle$ and $|01\rangle$ hybridize via $t$ to form $|o^{\pm}\rangle = (|10\rangle \pm |01\rangle)/\sqrt{2}$. These define two parity branches with distinct energy dispersions
\begin{eqnarray}
    E^{\pm}_e&=&\delta\pm\sqrt{\delta^2+\Delta^2}\,,\nonumber\\
    E^{\pm}_o&=&\delta\pm\sqrt{\mu^2+t^2}\,.
    \label{eq:PMM_energies}
\end{eqnarray}
$\mu = (\varepsilon_L-\varepsilon_R)/2$ defines the detuning axis and $\delta = (\varepsilon_L+\varepsilon_R)/2$ the common-mode axis, respectively.
When $t = \Delta$ and $\varepsilon_L=\varepsilon_R= 0$, the two many body ground states of different fermionic parity $|o^{-}\rangle$ and $|e^{-}\rangle$ become degenerate, namely $E^{-}_e=E^{-}_o$ see Fig. \ref{fig:PMM_sketch}e, giving rise to zero-energy excitations with Majorana character on each dot ~\cite{Leijnse_PRB2012}. It is straightforward to demonstrate that the local Majorana operators $\gamma_L=(d_L+d^{\dagger}_L)$ and $\gamma_R=i(d_R-d^{\dagger}_R)$ induce transitions between the two opposite parity ground states. 

The model described in Eq.~\eqref{eq:PMM_Ham_simpleModel} can be implemented in arrays of quantum dots coupled via short superconductors~\cite{Leijnse_PRB2012,Sau_NatComm2012,Fulga_NJP2013}. This geometry has been previously studied in the context of Cooper pair splitters, see for example Refs.~\cite{Recher_PRB2001,Beckmann_PRL2004,Russo_PRL2005,Hofstetter_Nature2009,Herrmann_PRL2010,Schindele_PRL2012,Das_NatCom2012,Fulop_PRL2015}. In the presence of strong Zeeman field in the QDs and spin-mixing ({\it e.g.} spin-orbit coupling), the middle superconductor mediates 2 kind of indirect coupling between the QDs: elastic cotunneling (ECT) where a single electron is transferred, occupying a virtual excited state in the superconductor; crossed Andreev reflection (CAR), where a Cooper pair splits (recombines) and the electrons end up (come from) in the QDs. 

A significant CAR amplitude requires the superconductor to be narrow (shorter or of the order of the superconducting coherence length). ECT and CAR amplitudes play the role of $t$ and $\Delta$ in the Kitaev model. One of these amplitudes is suppressed if the system has a well-defined spin quantization axis and the superconductor is spin singlet. Therefore, finding a Majorana sweet spot requires either spin-triplet superconductivity and/or spin-mixing contributions, like spin-orbit or different magnetization directions in the QDs.

On the other hand, achieving the Majorana sweet spot requires control on all parameters: levels of the dots ($\varepsilon_{L,R}$) and the relative amplitude between CAR and ECT ($t/\Delta$). Tuning the energy of individual dots can be done using external electrostatic gates that control the charge occupation of each dot, schematically shown in Fig.~\ref{fig:PMM_sketch}. The same applies to the tunnel amplitude between the superconductor and the QDs. However, tuning the relative amplitude of CAR and ECT is more challenging, as they scale in a similar way with external gates controlling the dots' properties. This issue was theoretically circumvented in Ref.~\cite{Leijnse_PRB2012} by considering non-collinear Zeeman fields in the dots, that may arise from an anisotropic Landé g-factor tensor in the dots. In this way, ECT and CAR amplitudes scale as $t=t_0\cos{\theta/2}$ and $\Delta=\Delta_0\sin{\theta/2}$, with $\theta$ being the angle between the spins and $t_0$ ($\Delta$) is the maximum ECT (CAR) amplitudes for parallel (anti-parallel) spin configurations in the dot.

\begin{figure*}[th!]
\includegraphics[width=0.9\linewidth]{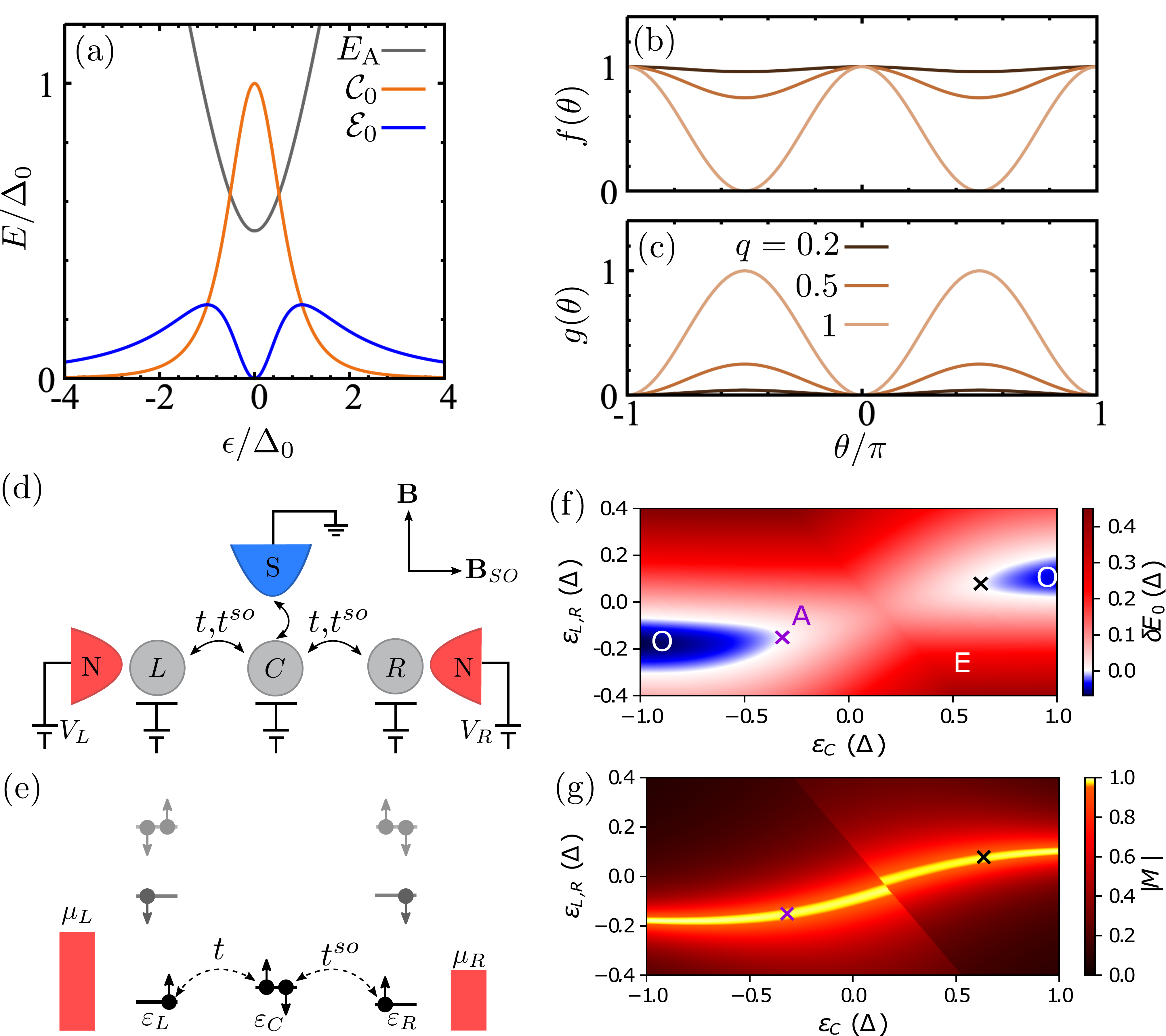}
\caption{CAR and ECT amplitudes, calculated using the model in Ref.~\cite{Liu_PRL2022}. (a) Gray line: energy of a subgap state, considering a parabolic model given in Eq.~\eqref{Chapter3:eq:parabolicABS}, with $\Delta_0=1$. The curve has been shifted down by $\Delta_0/2$. The orange and blue line show the amplitude for CAR and ECT amplitudes in Eq.~\eqref{Chapter3:eq:CAR_ECT_Wimmer}. (b) and (c) show the angle dependence of the $f$ and $g$ functions in Eq.~\eqref{Chapter3:eq:CAR_ECT_Wimmer} for different values of the spin precession, parameterized with $q$. (d) Sketch of the microscopic model of a Kitaev chain, formed by 3 tunnel-coupled QDs (gray dots), where the middle couples to a superconductor. (e) Energy levels of the dots. (f) Energy difference between the even and odd ground states, where red/blue indicates that the global ground state has an even/odd fermion parity. (g) Majorana polarization. Adapted and reprinted from Ref.~\cite{Liu_PRL2022} and \cite{Tsintzis2022}.}
\label{fig:microscopic_CAR_ECT} 
\end{figure*}

The spin-orbit coupling in semiconductor-superconductor and heavy compound superconductors provides a natural spin-mixing term for tunneling electrons. The spin-mixing term leads to finite ECT and CAR amplitudes for spin-polarized QDs when the spin-orbit and magnetic fields are non-collinear. This relaxes the need for non-collinear spin orientations of the dots. On the other hand, a subgap state in the central superconductor with energy $\epsilon<\Delta$, for example appearing due to confinement or accidental QDs in the superconductor, can dominate the coupling between the dots. The energy of this state, as well as the $u$ and $v$ BdG coefficients, are tunable using electrostatic gates. This tunability allows to control the relative amplitudes of ECT and CAR processes~\cite{Liu_PRL2022}. Up to the lowest order in the tunnel amplitudes, the CAR and ECT amplitudes are given by
\begin{eqnarray}
    t=f(\theta)\frac{t_Lt_R}{\Delta_0}\left(\frac{2uv}{E/\Delta_0}\right)^2\,,\quad\Delta=g(\theta)\frac{t_Lt_R}{\Delta_0}\left(\frac{u^2-v^2}{E/\Delta_0}\right)^2\,,
    \label{Chapter3:eq:CAR_ECT_Wimmer}
\end{eqnarray}
where $E$ is the energy of the ABS in the superconducting segment. In the infinite gap limit, 
\begin{equation}
    E=\sqrt{\epsilon^2+\Delta^{2}_0},
    \label{Chapter3:eq:parabolicABS}
\end{equation}
where $\Delta_0$ is the induced gap in the central region and $\epsilon$ determines the energy of the ABS, Fig.~\ref{fig:microscopic_CAR_ECT}(a), and is tunable through electrostatic gates. Here, the BdG coefficients are given by $u^2=1-v^2=1/2+\epsilon/2E$. For $\epsilon\gg0$ ($\epsilon\ll0$), the ABS is mostly an electron-like state with $u\approx1$ (a hole-like state with $v\approx1$), dispersing linearly with $\mu$. In this regime, the amplitude for ECT dominates and CAR vanishes as it depends on $u\,v$. On the other hand, the ABS has a minimum for $\epsilon=0$, where $u=v=1/\sqrt{2}$, where CAR peaks and ECT vanishes due to a destructive interference between second-order processes, see discussion in Ref.~\cite{Liu_PRL2022}. Therefore, the ECT and CAR amplitude intersects at two values as a function of the energy of the subgap state in the central superconductor, where Majorana sweet spots can appear. This enables gate and phase tunability of the different processes~\cite{Torres_SciPost2024}.

In this discussion, we have ignore the spin texture, described by the functions $f$ and $g$ in Eq.~\eqref{Chapter3:eq:CAR_ECT_Wimmer}, that account for the electron's spin-precession due to the spin-orbit coupling. These functions depend on $\theta$, the relative angle between the spin-orbit field and the local magnetization in the QDs. For parallel-polarized QDs, they are given by $f(\theta)=p^2+q^2\cos^2\theta$ and $g(\theta)=q^2\sin^2\theta$, with $p=\cos(k_{\rm so}\,L)$ and $q=\sin(k_{\rm so}L)$ describing the spin precession probability for an electron tunneling between the QDs. The definitions of $p$ and $q$ are reversed for QDs with anti-parallel polarization. Here, $p^2+q^2=1$, $k_{\rm so}$ being the spin-orbit wavevector and $L$ the length of the system.

When the spin-orbit field aligns with the QDs magnetization, $\theta=n\pi$, the functions $f=1$ and $g=0$, illustrated in~\ref{fig:microscopic_CAR_ECT}(b,c). In this scenario, CAR is suppressed, making it impossible to obtain a Majorana sweet spot with a finite gap to the excited states. This reflects the vanishing p-wave pairing amplitude when the Zeeman and spin-orbit fields align since $\Delta$ in Eq.~\eqref{Chapter3:eq:CAR_ECT_Wimmer} becomes zero. On the other hand, $\theta=\pi/2$ tends to boost CAR while keeping ECT amplitude finite. A finite exchange field does not affect to the presence of crossings where CAR and ECT are equal, see Ref.~\cite{Liu_PRL2022} for a discussion.

\label{Chapter3:sec:PMM_realistiModels}
Kitaev chains have been also studied using microscopic models that include many-body interactions in the QDs, a finite Zeeman splitting, therefore taking into account the 2 spin species, and the ABS in the central superconductor~\cite{Tsintzis2022}. A minimal model to describe this physics is the one sketched in Fig.~\ref{fig:microscopic_CAR_ECT}(d,e), which includes three spinful QDs, one of which is proximitized by a superconductor to describe the ABS. In particular, these models allow us to describe the strong QD-ABS coupling regime, where the gap to the excited states is enhanced and physics is described by the onset of YSR states~\cite{Liu_ComPhys2024,Alvarado_PRB2024}. In this limit, the notions of CAR and ECT are not well-defined, as high-order tunneling processes between the dots dominate the physics. For this reason, in the following, we will use a different notation to refer to the coupling strength between the even and odd fermion parity subspaces.

The microscopic models converge to the Kitaev chain with 2 sites in the limit of large Zeeman splitting in the QDs and weak QD-ABS coupling. In  this limit, the outer QDs can be occupied by 1 or 2 electrons for $|\varepsilon_{\alpha}|\ll E_{Z,\alpha}$ and the central one is in a superposition between 0 and 2 electrons. A second-order perturbation theory leads to two different couplings between $L$ and $R$ QDs: CAR and ECT~\cite{Liu_PRL2022}.

The characterization of Majorana sweet spots require introducing a figure of merit that accounts for the local Majorana character of the lowest energy states. The so-called Majorana polarization (MP) has been used in the past to characterize Majorana sweet spots in non-interacting~\cite{Sedlmayr_PRB2015,Glodzik_PRB2020,Awoga_PRB2024} and interacting systems~\cite{Aksenov2020,Tsintzis2022,Samuelson_PRB2024}. In the many-body formulation, MP is defined as
\begin{equation}
    M_\alpha=\frac{ W_{\alpha}^2-Z_{\alpha}^2}{ W_{\alpha}^2+Z_{\alpha}^2},
    \label{Chapter3:eq:MP_definition}
\end{equation}
where,
\begin{eqnarray}
    W_{\alpha}=\sum_\sigma\left\langle O\left|d_{\alpha\sigma}+d^{\dagger}_{\alpha\sigma}\right|E\right\rangle\,,\quad
    Z_{\alpha}=\sum_\sigma\left\langle O\left|d_{\alpha\sigma}-d^{\dagger}_{\alpha\sigma}\right|E\right\rangle\,,
\end{eqnarray}
with $\ket{O(E)}$ being the ground state wavevector with total odd (even) fermion parity. Equation~\eqref{Chapter3:eq:MP_definition} is valid for real Hamiltonians. Extensions for non-real Hamiltonians are provided in Ref.~\cite{Samuelson_PRB2024}. In an ideal sweet spot with perfectly localized Majoranas, $ W_{\alpha}=1$, $Z_{\alpha}=0$ and $W_{\bar{\alpha}}=0$, $Z_{\bar{\alpha}}=1$ for the two sides ($\alpha=L,R$ and $\bar{\alpha}\neq\alpha$). It means that $M_L=-M_R=\pm1$. The presence of additional Majorana components on the dot, due either to a non-perfect localization or a finite Zeeman field results on $|M_\alpha|<1$.

The local MP does not contain information about the non-local character of the Majorana wavefunctions. Therefore, having $|M_\alpha|\approx1$ does not imply a perfect Majorana localization on the outer $L/R$ QDs, as the Majorana wavefunctions can extend toward the center of the system without splitting the ground state degeneracy. A sweet spot with perfectly localized Majorana states would require having $W=1$ or $Z=1$ at each side. However, a state with a high MP still preserves the relevant Majorana properties, that implies that non-local experiments, including fusion and braiding~\cite{tsintzis2023roadmap}. Therefore, three conditions define high-quality Majorana sweet spots: (i) degenerate ground states with even and odd fermion parity; (ii) localized MZMs, characterized by a high MP value in both QDs; and (iii) substantial gap to the excited states.

Numerical simulations of the model have shown that the three conditions can only be met when the Zeeman field in the outermost QDs is sufficiently large (of the same order of magnitude or larger than the induced gap in the central superconductor)~\cite{Tsintzis2022}, see crosses in Figs.~\ref{fig:microscopic_CAR_ECT}(f,g). Coulomb interactions in the outer QDs boosts the effect of the magnetic field, increasing MP in sweet spots. Low MP solutions with degenerate ground states are possible when detuning the QDs from the sweet spots (effectively changing the ratio between $t$ and $\Delta$ of the Kitaev chain) or decreasing the Zeeman field, bringing the second spin species down in energy. The latter scenario does not necessarily imply a MP reduction, indicating the presence of Kramers-degenerate Majorana states with different spins~\cite{Bozkurt_arXiv2024}.

The addition of a second superconductor to the device provides an additional control parameter to tune the effective ratio between CAR and ECT~\cite{Torres_SciPost2024}, enabling Majorana sweet spots in serial double dots~\cite{Samuelson_PRB2024}. 

\subsubsection{Transport properties}
\label{Chapter3:sec:PMM_transport}
Electron transport is one of the most broadly used methods to characterized the electronic properties nanoscopic systems. In this section, we discuss different transport measurement proposals that can be used to characterize minimal Kitaev chains and to identify Majorana sweet spots. These measurements involve coupling the Kitaev chain to electrodes, that can be used to measure the local spectrum of the system.

\begin{figure}[t]
\includegraphics[width=1\linewidth]{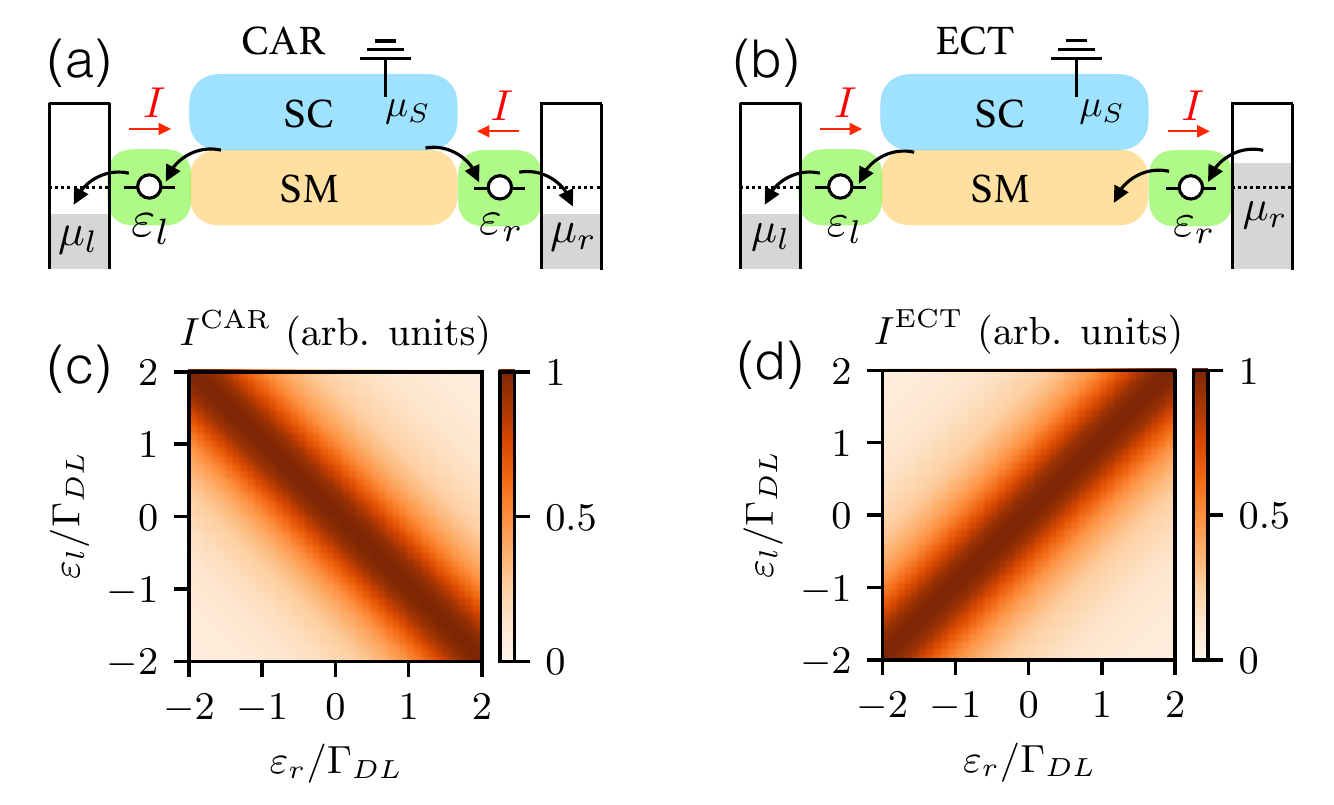}
\caption{ Top panels: description of the (a) CAR and (b) ECT contributions to the current through the leads. The bottom panels: numerical calculations using rate equations, showing the contributions of the CAR and ECT processes to the current. Adapted and reprinted from Refs.~\cite{Liu_PRL2022,Wang2022}.}
\label{fig:transportTheory_Kitaev}
\end{figure}

The first challenge consists of determining the points in parameter space where the coupling between the total even and odd parity states (CAR and ECT in the weak QD-ABS coupling) are equal. The experimental setup to measure these two coupling amplitude is shown schematically in the top panels of Fig.~\ref{fig:transportTheory_Kitaev}, taken from Ref.~\cite{Liu_PRL2022}. In the setup, two metallic leads couple to the outer QD dots, allowing for current to flow between the three different terminals (the two metallic electrodes and the grounded superconductor, blue rectangle in the figure). 

For $\varepsilon_L=-\varepsilon_R$, the Cooper pair splitting process (as well as the recombination) is resonant in energy, Fig.~\ref{fig:transportTheory_Kitaev}(a), since, due to energy conservation, CAR is resonant when $\varepsilon_L+\varepsilon_R=0$. The current direction in this case depends on the biasing conditions. The split electrons will flow out of the leads for $\mu_L<\varepsilon_L$ and $\mu_R<\varepsilon_R$. Under these conditions, a net current will cross the system, flowing from the superconductor and dividing equally to the left and right metallic leads. There is a corresponding process where electrons recombine into Cooper pairs. This process is also resonant for $\varepsilon_L=-\varepsilon_R$ and dominates for a large enough bias voltage in the leads, $\mu_L>\varepsilon_L$ and $\mu_R>\varepsilon_R$, resulting in a net current flowing from the metallic electrodes into the central PMM superconductor.

On the other hand, ECT requires the energy of the 2 QDs to be equal, $\varepsilon_L=\varepsilon_R$, Fig.~\ref{fig:transportTheory_Kitaev}(b). In this case, a finite current can flow between the leads if the bias is applied asymmetrically, so that electrons can tunnel on the right dot ($\mu_R>\varepsilon_R$) and leave to the left electrode ($\mu_L<\varepsilon_L$). This generates a current between the metallic leads, equal in magnitude but with opposite sign. The current direction changes if the biasing condition is reversed. In this picture, the central superconductor only mediates the coupling between the dots and does not contribute to the current through the system. This is only true if the energy of the two dots is small enough (smaller than the energy of the lowest ABS in the superconductor) to avoid the tunneling of quasiparticles between the superconductor and the QDs.

In the weak ABS-QD limit and large Zeeman field in the QDs, the current contributions are given by
\begin{eqnarray}
    I^{CAR}&=&\frac{e}{\hbar}\frac{\Gamma}{(\varepsilon_L+\varepsilon_R)^2+\Gamma^2}\left|\Delta\right|^2\,,\qquad {\rm for}\;\mu_{L,R}<0\,,\\
    I^{ECT}&=&\frac{e}{\hbar}\frac{\Gamma}{(\varepsilon_L-\varepsilon_R)^2+\Gamma^2}\left|t\right|^2\,,\qquad {\rm for}\;\mu_L<0<\mu_{R}\,,.
    \label{Chapter3:eq:PMM_Icar_Iect}
\end{eqnarray}
where the chemical potential of the superconductor is taken as $\mu_S=0$ and the tunnel coupling to the left and right leads are taken to be equal , $\Gamma_L=\Gamma_R\equiv\Gamma$. These contributions are shown in Figs.~\ref{fig:transportTheory_Kitaev}(c,d). The Majorana sweet spot locates at $\varepsilon_L=\varepsilon_R=0$, where both contributions to the current are maxima, leading to a peak in the local current.

\begin{figure}[t]
\includegraphics[width=1\linewidth]{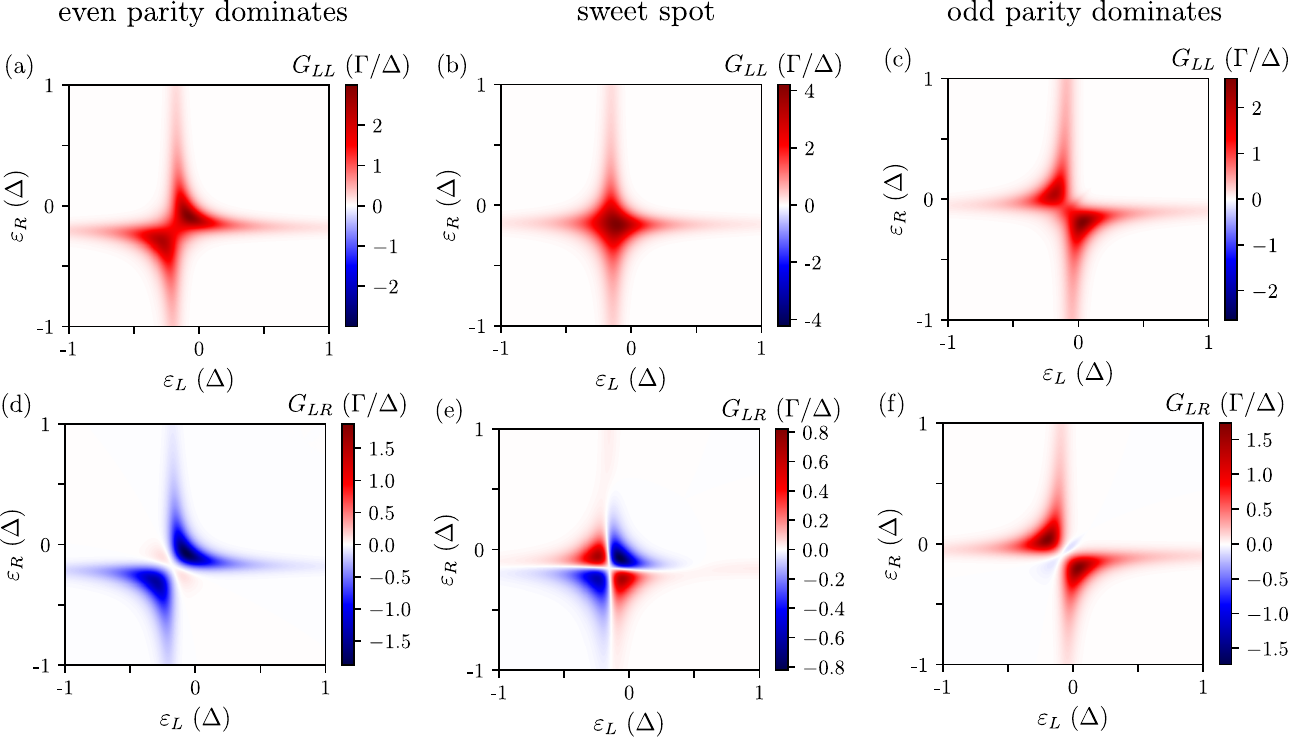}
\caption{Zero-bias conductance through a minimal Kitaev chain, as a function of the energy of the outer dots, which are attached to metallic electrodes, see Fig.~\ref{fig:microscopic_CAR_ECT}(d). Top panels show local conductance at the left lead, where the bias is applied symmetrically in the two leads. Lower panels show the non-local conductance, where bias is applied to right lead and current measured in the left one. From left to right, we show results for even parity-dominated regime (the coupling between the even states is stronger), the sweet spot (equal coupling within the two fermion parity sectors), and the odd parity-dominated regime. Reprinted from Ref.~\cite{Tsintzis2022} under CC-BY-4.0 license, \textcopyright  2022, The Author(s). Similar theoretical and experimental results were reported in Ref.~\cite{Dvir2023}, see also Sec.~\ref{Chapter3:sec:PMM_experimentsI}}
\label{fig:estabilityDiagrams_Kitaev2}
\end{figure}

On the other hand, the conductance allows to extract the spectrum of non-interacting systems. First, the local conductance allows to extract the energy difference between the lowest-energy states, as conductance peaks are expected for a voltage bias $eV=E_1-E_0$, with $E_0$ and $E_1$ being the energies of the 2 lowest states. Therefore, systems with a degenerate ground states lead to zero-bias conductance peaks. In Kitaev chains, this happens whenever the left and right QDs are on resonance, see upper panels Fig.~\ref{fig:estabilityDiagrams_Kitaev2}. Depending on the whether the coupling between even or odd states dominates (CAR and ECT in the weak QDs-ABS coupling regime), an avoided crossing appears between the zero-bias conductance lines that go through the anti-diagonal (diagonal). In contrast, the zero-bias conductance lines cross at the sweet spot, located at the maximum of the local conductance value.

The non-local conductance provides further information about the electron and hole components ($u$ and $v$ factors in the BdG language) of the lowest-energy state~\cite{Danon_PRL2020}, defined as 
\begin{equation}
    G_{\mu\nu}=\frac{\partial I_\mu}{\partial V_\nu}\,,
\end{equation}
with $\mu,\nu=L,R$. The lower panels of Fig.~\ref{fig:estabilityDiagrams_Kitaev2} show results for the non-local conductance, $G_{LR}$, through a minimal Kitaev chain at zero applied bias voltage. Depending on whether the coupling between the even/odd parity states dominate, the non-local conductance has a negative/positive sign. When both couplings are equal, the non-local conductance features an alternating negative-positive pattern, with the Majorana sweet spot located at the center of the pattern. Therefore, the Majorana sweet spot is characterized by a maximum in local conductance and a zero non-local conductance. The value of the conductance at the sweet spot is quantized~\cite{Alvarado_PRB2024}. Recent proposals suggested that transport through a normal and a superconducting lead allows to infer the local MP in Kitaev chains~\cite{Alvarado_arXiv25,Dourado_arXiv25}.

However, the differences between high and low-MP in the non-local conductance are sometimes faint, which makes it hard to distinguish between the two. For this reason, it is important to develop ways to identify high-MP sweet spots before proceeding to non-local experiments like braiding, Sec.~\ref{Sec:MajoranaBraid}. The local Majorana character of the wavefunction can be proven by coupling the system to an additional dot, as theoretically proposed and analyzed in Refs.~\cite{Prada_PRB2017,Clarke_PRB2017,Schuray_PRB2017} for Majorana wires. These proposals inspired experiments that showed similar patterns to the ones predicted by theory~\cite{Deng_Science2016,deng2018nonlocality}. In these references, they showed that a QD cannot split the ground state degeneracy when the dot couples to only one of the Majorana component. In contrast, the coupling between the QD can lift the ground state degeneracy if it couples to 2 Majorana states, leading to a  ``diamond-like'' shape, with a maximum splitting whenever the QD's energy aligns with the chemical potential of the superconductor. The situation is different if the two Majoranas overlap but only one couple to the additional dot. In this case, the ground state degeneracy is split, except when one of the additional QD's level align with the superconductor's chemical potential, leading to a ``bow tie-like'' pattern. In contrast, localized Majorana states lead to a robust zero-energy state, independent from the QD's energy. This idea has been recently extended to Kitaev chains, showing that an additional QD can be used to identify high-MP sweet spots~\cite{Souto_PRR2023,Benestad_PRB24}, see Ref.~\cite{Bordin_arXiv25} for an experimental demonstration.

Increasing the chain length to more sites increases protection against local fluctuations~\cite{Sau2012,Stenger_PRB2018,Svensson_PRB24,Dourado_PRB25,Dourado_arXiv25_2}, as experimentally demonstrated for 3-site Kitaev chains, see Sec.~\ref{Chapter3:sec:PMM_experimentsI}. However, Majorana sweet spots may evolve into trivial states when scaling up the system~\cite{Luethi_PRB25,Luethi_arXiv25}, highlighting the importance of tuning.

Recent works have proposed that Majorana states can also appear in QDs coupled via floating superconducting islands~\cite{Souto_PRB25,Nitsch_arXiv24}, allowing to explore the role of charge conservation in minimal Kitaev chains.

\subsubsection{Experiments in minimal Kitaev chains I: conductance}
\label{Chapter3:sec:PMM_experimentsI}
Early experiments studied CAR processes in a superconducting heterostructures, with the aim of generating a source for entangled electrons, so-called Cooper pair splitters~\cite{Recher_PRB2001,Beckmann_PRL2004,Russo_PRL2005,Hofstetter_Nature2009,Herrmann_PRL2010,Schindele_PRL2012,Das_NatCom2012,Fulop_PRL2015}. The experimental realization of these splitters require electron channels separated by a distance smaller than the superconducting coherence length, that is usually of the order of tens of nanometers to a micron for conventional BCS superconductors. In this context, QDs offer a fundamental advantage as their charging energy suppresses local Andreev reflections for a sufficiently small coupling to the central superconductor~\cite{Recher_PRB2001,Schindele_PRL2012}: processes where a Cooper pair tunnels to the same terminal.
\begin{figure}[h]
\includegraphics[width=1\textwidth]{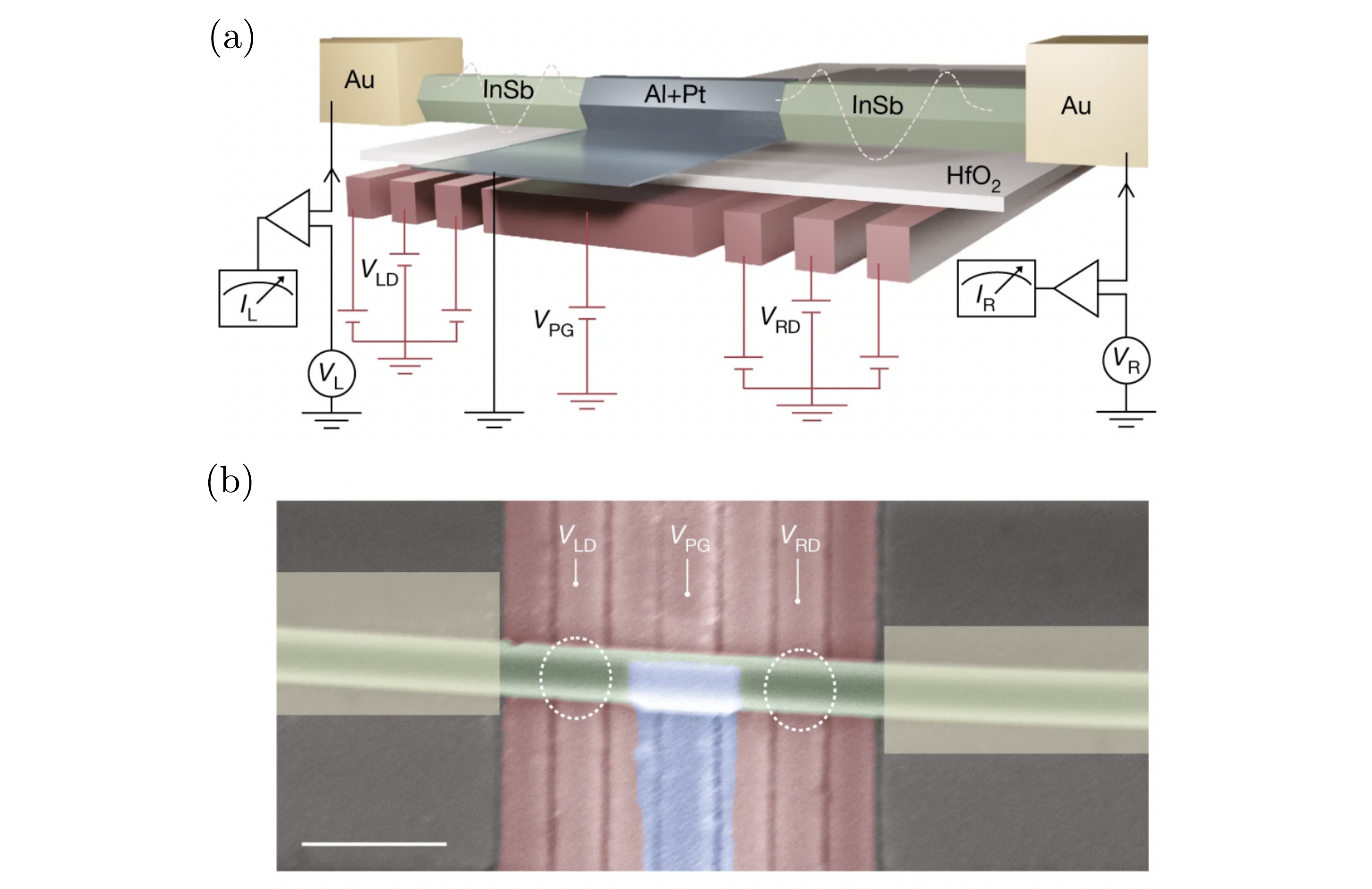}
\caption{Illustration of the PMM system used in Ref.~\cite{Dvir2023}. Two QDs are defined in an InSb nanowire and coupled via a central region that has proximity-induced superconductivity coming from a grounded superconductor. The outer Au leads are used for spectroscopy. The gates at the bottom (red) allow to tune the device into the desired configuration (b) False-colored scanning electron microscopy image of the device, before the fabrication of the N leads (schematically shown in yellow). The scale bar is 300 nm. Reprinted from Ref.~\cite{Dvir2023} with permission, \textcopyright  2023, The Author(s), under exclusive licence to Springer Nature Limited. All rights reserved.}
\label{fig:experiments_Kitaev2}
\end{figure}
\begin{figure}[t]
\includegraphics[width=1\linewidth]{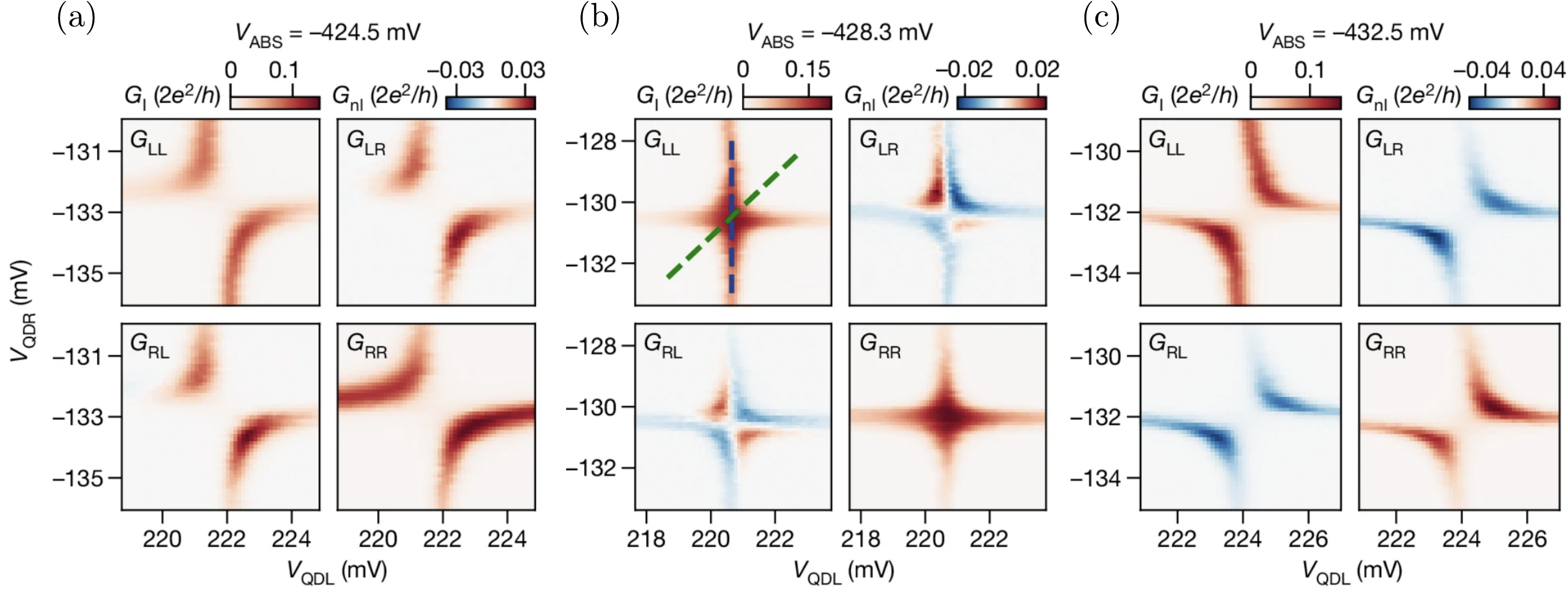}
\caption{Conductance matrices measured for different values of the gate controlling the semiconductor segment underneath the superconductor  ($V_{\rm PG}=$198, 210 and 218 mV, for panels (a-c). The three panels are representative measurements of the three possible regimes (ECT-dominated regime, close to the sweet spot with ECT and CAR amplitudes similar, and CAR-dominated regime) and in qualitative agreement with theory, see Fig.~\ref{fig:estabilityDiagrams_Kitaev2}. Adapted and reprinted from Ref.~\cite{tenHaaf_Nature2024}; qualitatively similar results were reported in Ref.~\cite{Dvir2023}.}
\label{fig:conductance_Kitaev2}
\end{figure}
As mentioned above (see also Ref.~\cite{Liu_PRL2022}), an Andreev bound state in the superconducting region can mediate the coupling between the electrons. These bound states, with energies below the superconducting gap, offer the dominant contribution for Cooper pairs tunneling to the dots due to their reduced energies. Therefore, they offer advantages, including the increase the Cooper pair splitting efficiency and the possibility of tuning the relative amplitudes between CAR and ECT by modulating the energy of the bound state, that can be achieved through electrostatic gates acting on the central superconductor. For this task, superconductor-semiconductor hybrid structures are advantageous, as they can exhibit subgap states that are gate tunable. Experiments have demonstrated the possibility of hybridizing a subgap state with a QD~\cite{Deng_Science2016,deng2018nonlocality,Poschl_PRB2022} and tuning the particle/hole component, {\it i.e.} the BCS charge of a subgap state, using electrostatic gates~\cite{Poschl_PRB2022_2}. These are required ingredients to realize PMM systems and tune them to the sweet spots where localized Majoranas appear.

Independent measurements on semiconductor-superconductor hybrid devices demonstrated CAR processes of Cooper pairs with the same spin in nanowires~\cite{Wang2022,Bordoloi_Nat2022,Bordin_PRX2023} and two-dimensional electron gases~\cite{Wang_Natcom2023}. This observation points towards the presence of equal-spin-triplet Cooper pairs due to the interplay between spin-orbit coupling and an external magnetic field \footnote{For interesting connections between odd-frequency superconductivity and triplet pairing \cite{RevModPhys.77.1321} and their role in the proximity effect, topics beyond the scope of this review, we refer the interested reader to, e.g., \cite{PhysRevB.70.012507,PhysRevB.87.104513,fksg-x8pr,Ahmed2025}.}. The existence of these Cooper pairs is an essential requirement to realize topological superconductivity. In particular, Refs.~\cite{Bordoloi_Nat2022,Wang_Natcom2023} determined the amplitudes for CAR and ECT using transport measurements across the device, as described in Sec.~\ref{Chapter3:sec:PMM_transport}. They demonstrated a fine-tuned regime of parameters where the amplitude of both processes is the same and where PMMs can appear. Additionally, a Cooper pair splitting efficiency around 90\% was demonstrated, that is over the threshold required for a Bell test experiment~\cite{Schindele_PRL2012}.

Additional local and non-local spectroscopic measurements in a PMM system were reported in Refs.~\cite{Dvir2023,tenHaaf_Nature2024,Zatelli_NatComm2024,Bordin_NatNano2025,tenHaaf_Nature2025}. The device in Ref.~\cite{Dvir2023}, sketched in Fig.~\ref{fig:experiments_Kitaev2}(a), is composed by a InSb nanowire fabricated using the shadow-wall lithography technique~\cite{Heedt2021,Borsoi2021}. The bottom gates (red) control the energy and the coupling of the QDs to the central superconductor (blue) and the metallic leads used for spectroscopy (yellow rectangle). The QDs are spin-polarized using an external magnetic field, boosted by the high Land\'e g-factor measured in InSb. The grounded superconductor at the center mediates CAR and ECT processes between the spin-polarized QDs. The central region hosts an Andreev bound state that mediates both processes and allows to tune their relative amplitude.

Measurements for the local and non-local conductance are shown in Fig.~\ref{fig:conductance_Kitaev2} \cite{Dvir2023}. The figure shows representative results for the three important regimes: $t>\Delta$ (left panels), $t<\Delta$ (right panels), and $t\approx\Delta$ (middle panel). Local spectroscopy (subpanels panels i and iv in every panel) reveals an avoided crossing between two levels. The direction of the avoided crossing along the diagonal/anti-diagonal reveals the dominance of CAR/ECT processes. In contrast, for $t\approx\Delta$, the levels seem to cross, showing a high conductance peak at the sweet spot. We also note a change on the sign of the non-local conductance between the left and right panels (subpanels ii and iii), while at the sweet spot (middle panels) there seems to be an alternation between positive and negative features, although with a dominance of the positive ones. 

In the simplest model proposed by Kitaev, the protection against deviations from the ideal sweet spot improves when additional dots are added to the chain. In particular, the truly topological phase is recovered in the limit of the chain being much larger than the localization length of the MZMs. In this thermodynamic limit, local perturbations cannot affect the system's global topological properties, as the non-local Majorana pairs are spatially isolated. This fundamental insight has motivated a concerted effort to build and probe minimal, multi-site Kitaev chains in semiconductor-superconductor platforms, serving as crucial stepping stones toward the scalable, topologically protected qubits required for quantum computation. Recent landmark experiments have realized such chains with three quantum dots (QDs) coupled via two superconducting segments~\cite{Bordin_PRL2024,Bordin_NatNano2025,tenHaaf_Nature2025,Bordin_arXiv25}. These three-site experiments have confirmed several key properties of emergent MZMs. Critically, it was demonstrated that increasing the system size from two to three sites results in a stronger protection of the Majorana zero-energy states against gate-induced detuning~\cite{Bordin_NatNano2025}, directly validating the core theoretical prediction of enhanced robustness with length. Furthermore, by employing an additional tunnel-coupled probe dot, the spatial localization of the Majorana wavefunctions could be directly measured, confirming their exponential decay into the chain~\cite{Bordin_arXiv25}. A comprehensive picture emerged from experiments probing the local density of states on all three sites simultaneously~\cite{tenHaaf_Nature2025}, which showed that the appearance of zero-bias peaks localized at the outer QDs is accompanied by a hard gap in the middle QD. The size of this bulk gap—and thus the degree of protection for the edge MZMs—was found to be tunable and sensitive to the superconducting phase difference across the chain, highlighting the role of phase biasing as a topological switch. Beyond static properties, these experiments also allowed researchers to directly visualize how the weight of the Majorana wave function can be systematically transferred between dots by tuning local potentials. This dynamic control provides a tangible route toward the braiding of Majoranas by physically moving them in a future, two-dimensional network of such chains. Collectively, these results represent a significant leap from minimal "poor man's Majorana" studies to the controlled engineering and diagnosis of Majorana states with enhanced robustness in a fully tunable multi-site system, establishing a foundational platform for probing non-Abelian statistics. We stress, however, that three-site chains remain far from the long-chain limit in which topological protection is expected.

Given the rapid experimental progress in realizing minimal Kitaev chains, there is now a concerted effort to understand how to scale these systems to achieve the robust topological protection promised by theory. Achieving longer, functional chains—comprising five, ten, or more quantum dots—presents a formidable tuning challenge. It requires the simultaneous optimization of dozens of interdependent gate voltages that control the charge states of individual QDs, the induced pairing potentials in superconducting segments, and the coherent tunneling couplings between all adjacent sites. This high-dimensional parameter space is fraught with capacitive cross-talk, and complex interdependencies, making manual tuning challenging.  With this motivation, approaches based on machine learning have been recently developed~\cite{Benestad_PRB24,Koch_PRA2023,Driel_arXiv2024}.

\subsubsection{Experiments in minimal Kitaev chains II: reflectometry and parity readout}
\label{Chapter3:sec:PMM_experimentsII}
Another interesting direction of research relates to using small Kitaev chains to create parity qubits. In order to achieve this, it is essential to estimate quasiparticle poisoning lifetimes in these systems and develop efficient readout methods of the fermionic parity $\hat P=i\gamma_1\gamma_2$ . 
\begin{figure*}[h!]
\centering
\includegraphics[width=\textwidth]{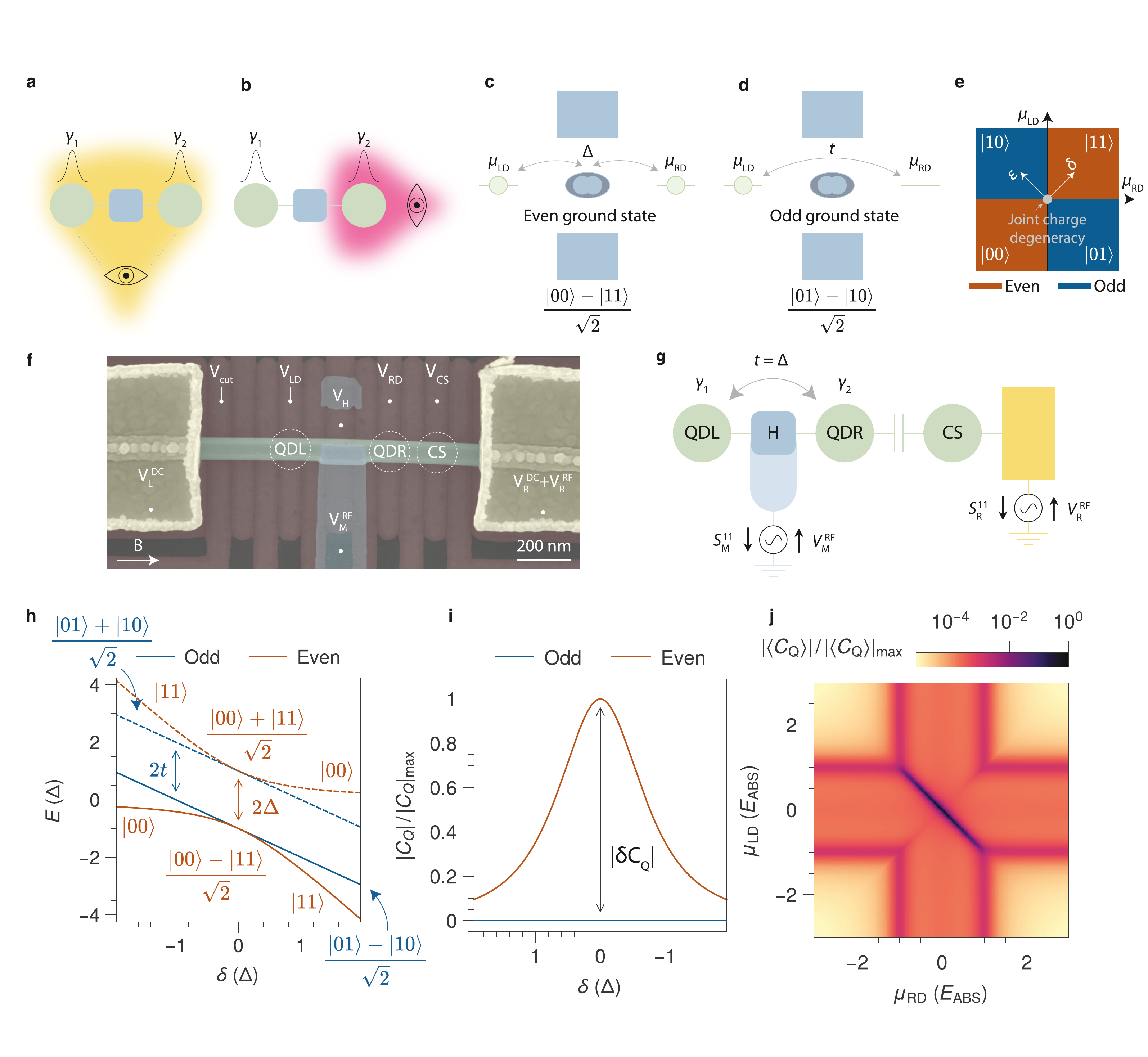}
\caption{(a) Illustration of two Majorana modes $\gamma_1$ and $\gamma_2$ on separate dots. A probe acting on both couples them and reveals their parity, while a probe acting on only one Majorana (b) does not. (c,d) Energy diagrams of the even and odd parity sectors. (e) Charge-stability diagram of two quantum dots, showing the detuning axis $\mu$ and common-mode axis $\delta$. Orange and blue sectors denote the even and odd parity ground states. The center marks the joint charge degeneracy, which is the PMM sweet spot when $t=\Delta$. (f) False-colored scanning electron micrograph of the device. The InSb nanowire (green) is proximitized by an Al strip (blue) and contacted by Cr/Au leads (yellow), with different dc voltages applied on Ti/Pd bottom gates (red) for defining the left/right quantum dots QDL, QDR and the charge sensor CS (dashed circles) and tuning the Kitaev chain. DC and RF voltages are applied to the three leads for biasing and reflectometry. (g) Schematic of the readout setup. The two QDs are coupled through a superconducting segment connected to an RF resonator, for readout via $S^{11}_M$. Likewise, the CS is read out via a second resonator $S^{11}_R$. (h) Energy dispersion of the even and odd  parity
states at zero detuning ($\mu = 0$), plotted as a function of the common-mode potential $\delta$. (i) Calculated
quantum capacitance for each parity branch, given by the second derivative of energy with respect to chemical potential, see Eq.\eqref{Cq}. (j) Simulation of the quantum capacitance around one charge degeneracy, the dominant anti-diagonal line corresponds to $\delta=0$. Reprinted from Ref.~\cite{vanloo2025singleshotparityreadoutminimal} with permission, \textcopyright  2026, The Author(s), under exclusive licence to Springer Nature Limited. All rights reserved.}
\label{fig:experiments_parity1}
\end{figure*}
The first experiments in this direction have recently demonstrated single-shot parity readout of a minimal Kitaev chain \cite{vanloo2025singleshotparityreadoutminimal}. These experiments introduce a measurement technique that distinguishes the fermionic parity of the
minimal Kitaev chain through its global quantum capacitance, which measures the response of the even- and odd-parity many-body ground states to gate voltage variations \cite{10.1063/5.0088229}. Specifically, a minimal Kitaev chain is probed by connecting
the middle superconductor to a multiplexed LC resonator, allowing RF reflectometry. The applied RF signal coupled to the SC induces small oscillations in its electrochemical potential relative to the QDs, effectively modulating the system along the common mode axis $\delta$. This results in a key difference between the parity branches: even states disperse nonlinearly due to hybridization via CAR, while the odd states change
linearly (see Fig \ref{fig:experiments_parity1}). This is reflected in the curvature of their energy levels and thus in their quantum
capacitance, which is proportional to the second derivative of energy with respect to chemical potential. 
\begin{equation}
C_Q(E^{-}_e) \approx \frac{e^2}{4}\frac{{\partial}^2E^{-}_e}{{\partial}\delta^2}\approx-\frac{ e^2}{4}\frac{\Delta^2}{\left(\delta^2+\Delta^2\right)^\frac{3}{2}}, \label{Cq}
\end{equation}
corresponding to a minimum around $\delta\approx0$ of width given by $\Delta$ \cite{vanloo2025singleshotparityreadoutminimal,PhysRevB.108.085437}, while the quantum capacitance of the odd states is
zero $C_Q(E^{-}_o)=0$ \footnote{We note that a small asymmetry in the level arms would also yield a signal for the odd state,
$C_Q(E^{-}_o)\approx -\frac{e^2}{4}\frac{t^2}{\left(\mu^2+t^2\right)^\frac{3}{2}}$
along the line $\mu\approx 0$. Such odd signal is however not seen in the experiments, which supports an interpretation based on symmetric level arms \cite{vanloo2025singleshotparityreadoutminimal}.}. Intuitively, the RF oscillation periodically shifts the balance between charge occupations in the QDs $|00\rangle$ and $|11\rangle$, inducing
charge motion between the quantum dots and the superconductor and producing a measurable quantum capacitance. In contrast, the odd states $|01\rangle$ and $|10\rangle$ are connected by tunneling that does not involve net
charge transfer to or from the superconductor, yielding zero quantum capacitance.
\begin{figure*}[h!]
\centering
\includegraphics[width=\textwidth]{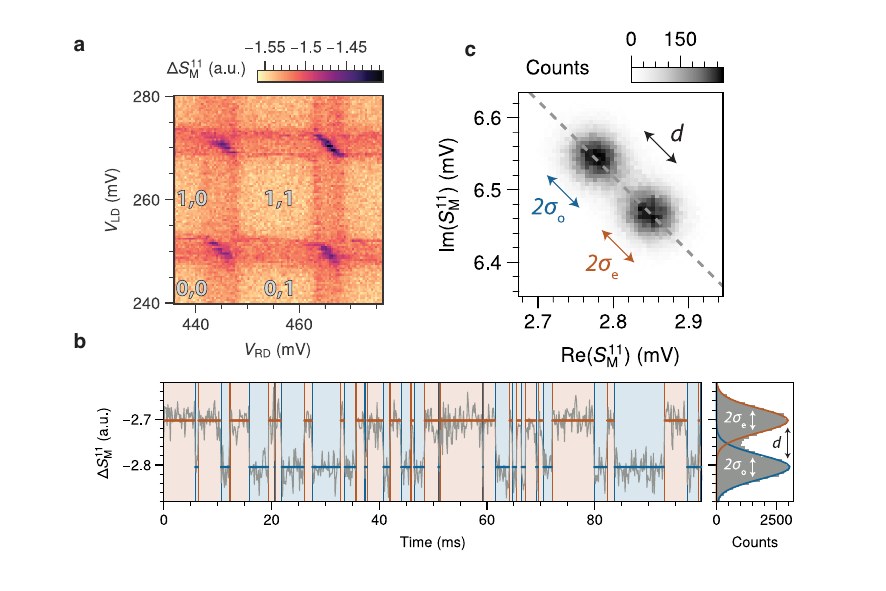}
\caption{(a) Charge-stability diagram with labeled charge states measured via the global quantum capacitance signal though changes in the resonator coupled to the superconductor, $S^{11}_M$. The dominant anti-diagonal lines correspond $\delta=0$, where the strength of CAR is maximal. (b) Random telegraph noise near one anti-diagonal resonance reflects real-time parity switching. For this trace, an integration time of $150 \mu s$ is used. Analysis of the data yields parity lifetimes $\tau_e = 1.82 \pm  0.05 ms$, $\tau_o = 1.88 \pm  0.04 ms$. The side panel displays count histograms along the principal
component axis. c) Histogram of complex $S^{11}_M$ from the time trace in panel (b). Reprinted from Ref.~\cite{vanloo2025singleshotparityreadoutminimal} with permission, \textcopyright  2026, The Author(s), under exclusive licence to Springer Nature Limited. All rights reserved.}
\label{fig:experiments_parity2}
\end{figure*}
The quantum capacitance measurements in Ref. \cite{vanloo2025singleshotparityreadoutminimal} therefore provide a direct, single-shot measurement of the non-local parity operator, therefore overcoming a key challenge in realizing Majorana-based qubits by enabling fast, high-fidelity parity readout. The power of this global approach was definitively proven by simultaneous local charge sensing (by performing RF reflectometry in another LC resonator coupled to the right lead): at the Majorana sweet spot where the even and odd parity states share identical local charge distributions, the charge sensor was completely blind to stochastic parity switches. In stark contrast, the global quantum capacitance signal showed clear telegraph noise, enabling real-time tracking of the parity state with a fidelity of $94\%$ 
and revealing parity lifetimes exceeding one millisecond, see Fig.\ref{fig:experiments_parity2}. 

This experiment not only satisfied the fundamental requirement of non-local measurement but also established a practical, high-fidelity readout that integrates seamlessly with existing gate-based architectures, unlike more complex interferometric methods that require significant device modifications. Specifically, similar functionality has recently been demonstrated using interferometric techniques in extended "top-down" hybrid nanowire devices \cite{Aghaee_Nature2025}, but those approaches require significant alterations to the device layout and introduce considerable control overhead. 

\subsubsection{Experiments in minimal Kitaev chains III: Majorana-based qubits}
\label{Sec:topoQubits}
In this subsection, we first introduce the basic concepts behind Majorana-based qubits and some of the qubit designs proposed in the literature, and then discuss the first experimental realization of a Majorana parity qubit in coupled minimal Kitaev chains~\cite{Zatelli_arXiv2026}.

MZMs offer a route toward fault-tolerant quantum information encoding and processing. The key characteristic is their non-local nature: quantum information in a Majorana qubit is stored in the collective state of a pair of spatially separated Majorana states. This non-local encoding offers inherent protection against local perturbations that do not close the superconducting gap, a significant advantage for quantum computing. Moreover, Majorana non-Abelian properties allow to perform some quantum operations in a protected way, as mentioned below in Sec.~\ref{Sec:MajoranaBraid}. We emphasize that this protection applies to spatially separated MZMs in the topological regime. The same qubit designs implemented with minimal Kitaev chains (see below) share the encoding and readout schemes, but not the topological protection.

In Majorana qubits, information is encoded in the fermion occupation of Majorana pairs, that can be in either the even or the odd fermion parity state. In a superconducting system, the total fermion parity is a good quantum number, invariant unless undesired quasiparticle poisoning events. Therefore, storing quantum information, in particular entangled states, require more than a single Majorana pair. Therefore, at least two pairs of Majorana states are needed to define a qubit. The Majorana qubit states can be defined as $\ket{0}=\ket{ee}$, $\ket{1}=\ket{oo}$ for the total even parity subspace, where $e/o$ denotes the even/odd fermion occupation of a given Majorana pair. These two states even parity have the same electron number and energy for decoupled MZMs, and differ from one Cooper pair that can split occupying the two non-local fermion modes defined by the Majoranas. There is another identical computational subspace with total odd fermion parity ($\ket{eo}$, $\ket{oe}$) that can be also used as a basis for computation.

Several devices have been proposed to encode quantum information in the parity degree of freedom, leveraging the non-local protection from MZMs. In the following, we discuss some of these proposals.

\begin{figure}[!htbp]
\centering
\includegraphics[width=0.92\linewidth]{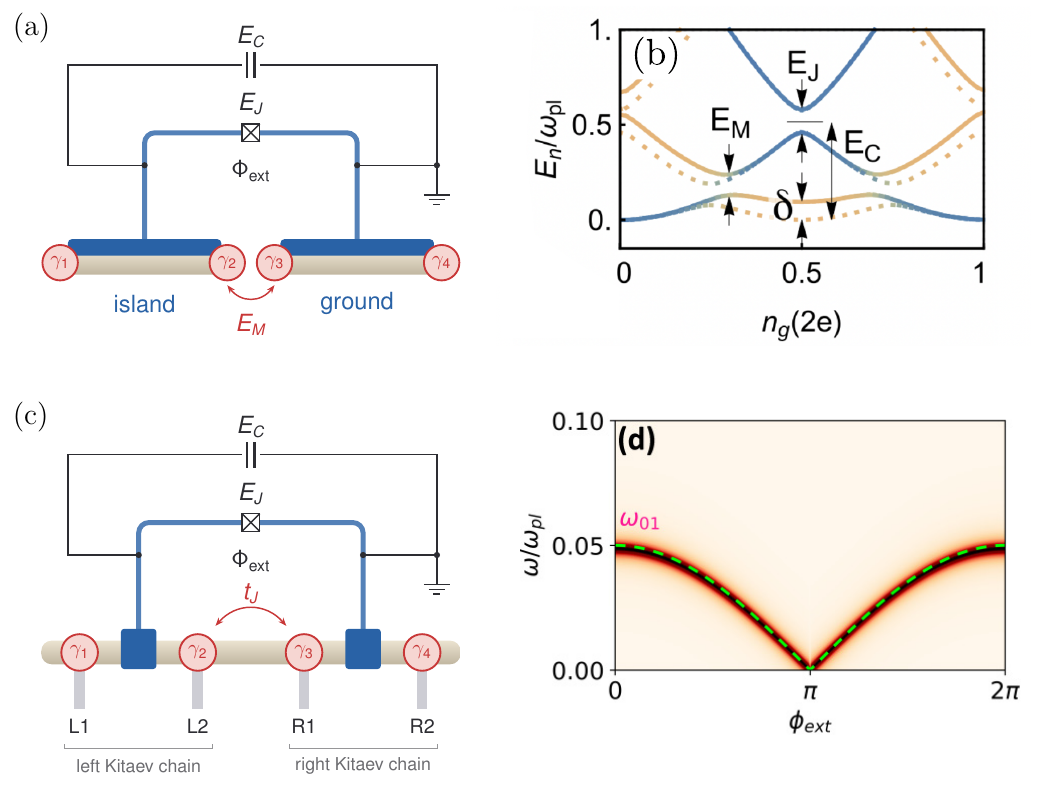}
\caption{(a) Sketch of a Majorana transmon qubit: a superconducting island is connected to ground through a topological Josephson junction formed by two semiconductor nanowires with a superconducting shell (blue), which host MZMs $\gamma_1$, $\gamma_2$, $\gamma_3$ and $\gamma_4$ at their ends. The inner MZMs, $\gamma_2$ and $\gamma_3$, are coupled with strength $E_M$ across the junction. The junction forms a SQUID loop, threaded by an external flux $\Phi_{\rm ext}$, with a reference Josephson junction ($E_J$), and the island is shunted to ground by a capacitance with charging energy $E_C$. (b) Energy spectrum of the system in the charging-energy dominated regime, known as Cooper pair box regime ($E_J/E_C=0.5$, $E_M/E_C\sim0.12$), and close to the topological phase transition for the wires. Blue/orange colors denote fermionic even/odd parities of the inner MZMs. (c) Minimal Kitaev-transmon (Kitmon) qubit, based on two minimal Kitaev chains, each formed by two QDs (L1, L2 and R1, R2) coupled through a superconducting segment and hosting the Majorana states $\gamma_1$, $\gamma_2$, $\gamma_3$ and $\gamma_4$. The inner QDs are tunnel coupled ($t_J$), and the junction is embedded in the same circuit as in (a). (d) Microwave absorption spectrum of the Kitmon qubit as a function of $\phi_{\rm ext}=2\pi\Phi_{\rm ext}/\Phi_0$ at the sweet spot, deep in the transmon regime ($E_J=50E_C$). The green dashed line corresponds to the analytical qubit frequency in Eq.~\eqref{eq:Kitmon_frequency}, which vanishes at $\phi_{\rm ext}=\pi$ as a consequence of the $4\pi$-periodic Josephson effect. Energies in units of the plasma frequency $\omega_{pl}=\sqrt{8E_JE_C}$. Panels (b) and (d) adapted and reprinted from Ref.~\cite{Avila2020a} and Ref.~\cite{Pino24}, respectively.}
\label{fig:Majorana_transmon}
\end{figure}

The Majorana transmon is an example of a qubit that exploits the extra degree of freedom introduced by four Majoranas, see Fig.~\ref{fig:Majorana_transmon}. This qubit is based on a topological Josephson junction formed by two tunnel-coupled MZMs~\cite{Avila2020a,Ginossar2014,Yavilberg15,Li2018,Hassler2011,Hyart2013,Keselman19,Avila2020,Smith2020,Lupo2022}. The minimal model for the system can be written as 
\begin{equation}
    \mathcal{H}=\mathcal{H}_C+\mathcal{H}_J+\mathcal{H}_M\,,
    \label{Eq:transmonMZM}
\end{equation}
where $\mathcal{H}_C=4E_C(n-n_g)^2$, with $n$ being the charge, $n_g$ a charge offset, and $E_C$ the charging energy of the island. The coupling between the two MZMs leads to a Josephson potential that is 4-$\pi$ periodic, $\propto \pm\cos(\phi/2)$~\cite{Kwon2003-qi}, with $\phi$ being the phase difference across the junction. The $\pm$ sign depends on the total parity branch, leading to a ground state with well-defined total fermion parity. The minimal model to understand the MZMs in the junctions is
\begin{equation}
    \mathcal{H}_M=iE_M \gamma_2\gamma_3\,,
\end{equation}
with $E_M$ being the coupling strength between the inner-most MZMs. The remaining states, including the ones above the gap, lead to a conventional Josephson term, $\mathcal{H}_J=E_J \cos(\phi)$. The Majorana transmon qubit can be manipulated and readout using microwave pulses, similar to what is currently done in conventional transmon qubits.

The Majorana transmon geometry has been recently studied in the context of minimal Kitaev chains, where analytic expressions for $E_M$ and $E_J$ have been derived~\cite{Pino24} and quantum operations have been studied~\cite{Pan_PRB2025}. {In this minimal Kitaev-transmon, or Kitmon, qubit~\cite{Pino24}, two minimal Kitaev chains, each formed by two QDs coupled through a superconducting segment, are connected through a weak link with tunnel coupling $t_J$, forming a minimal Majorana Josephson junction that is embedded in a transmon circuit, Fig.~\ref{fig:Majorana_transmon}(c). As in the nanowire-based Majorana transmon of Fig.~\ref{fig:Majorana_transmon}(a), the superconducting phase difference across the junction is controlled by an external flux $\Phi_{\rm ext}$ through a SQUID loop formed with a reference junction of Josephson energy $E_J$, and the circuit is described by the Hamiltonian $\mathcal{H}=4E_C(\hat n-n_g)^2-E_J\cos\hat\phi+V_{\rm DQD}(\hat\phi-\phi_{\rm ext})$, with $\phi_{\rm ext}=2\pi\Phi_{\rm ext}/\Phi_0$. Projecting the QD Josephson junction onto the low-energy subspace of the four Majoranas, and considering mirror-symmetric chains, its Josephson potential takes the compact form~\cite{Pino24}
\begin{equation}
V_{\rm DQD}(\phi)=\begin{pmatrix}
2E_o^- & E_M\cos(\phi/2)\\
E_M\cos(\phi/2) & 2E_e^-
\end{pmatrix},
\label{eq:Kitmon_potential}
\end{equation}
where $E_e^-$ and $E_o^-$ are the lowest even and odd many-body energies of each minimal Kitaev chain, see Eq.~\eqref{eq:PMM_energies}, and the off-diagonal term describes the coherent tunneling of single electrons across the junction mediated by the inner Majoranas. The latter gives rise to the $4\pi$-periodic Josephson effect discussed above, which, in a Kitmon, can be directly accessed by cQED spectroscopy. For two identical chains, and in the notation of Eq.~\eqref{eq:PMM_energies}, the Majorana coupling reads $E_M=t_Jt\Delta/\left[2\sqrt{(t^2+\mu^2)(\Delta^2+\delta^2)}\right]$, which is directly related to the Majorana polarization of the chains~\cite{Pino24}. Deep in the transmon regime, $E_J/E_C\gg1$, the phase is pinned close to $\phi_{\rm ext}$, and the lowest microwave transition, $\omega_{01}$, which defines the Kitmon qubit frequency, connects the two parity branches of Eq.~\eqref{eq:Kitmon_potential},
\begin{equation}
\hbar\omega_{01}\approx2\sqrt{\left(E_e^--E_o^-\right)^2+E_M^2\cos^2\frac{\phi_{\rm ext}}{2}}\,.
\label{eq:Kitmon_frequency}
\end{equation}
At the sweet spot, $E_e^-=E_o^-$, and the qubit frequency, $\hbar\omega_{01}\approx2E_M|\cos(\phi_{\rm ext}/2)|$, vanishes at $\phi_{\rm ext}=\pi$, a direct fingerprint of the $4\pi$-periodic Majorana-mediated Josephson coupling, see Fig.~\ref{fig:Majorana_transmon}(d). More generally, for $t=\Delta$ and $\mu=\pm\delta$, the qubit frequency is directly proportional to $E_M$, $\hbar\omega_{01}\approx t_J\cos(\phi_{\rm ext}/2)/[1+(\delta/\Delta)^2]$. Microwave measurements of $\omega_{01}$ as a function of flux and QD parameters would therefore allow extracting $E_M$, and hence the Majorana polarization of the junction, and distinguishing Majorana-mediated parity mixing from quasiparticle poisoning or from $4\pi$ phase slips due to QD resonances~\cite{Vakhtel2023}. In this sense, the Kitmon can be regarded as a minimal, Majorana-based, nonlocal parity pseudospin coupled to a transmon, a natural extension of the gatemons and Andreev spin qubits discussed in Sec.~\ref{s:qubits}. Note also that Eq.~\eqref{eq:Kitmon_potential} has the same structure as the effective Hamiltonian of the Majorana parity qubit discussed below, Eq.~\eqref{eq:H_parity_qubit}, where the inter-chain coupling also acquires a $\cos(\phi/2)$ dependence on the superconducting phase difference. In the parity qubit, however, $\phi$ is a control parameter set by the external flux threading the loop, whereas in the Kitmon it is promoted to a quantum variable, $\hat\phi$, conjugate to the island charge and governed by the circuit Hamiltonian. The Majorana parity is then coupled to the plasma mode of the circuit, which enables its readout and manipulation with standard cQED techniques. This parallels the coherent coupling between an ASQ and a transmon (Sec.~\ref{s:ASQ}, Fig.~\ref{fig:ASQ-expt}(g))~\cite{PitaVidal2023}: in the Kitmon, the fermion parity replaces the spin, and the term $E_M\cos(\hat\phi/2)$ couples parity and plasmon states, so that strong coupling and Jaynes--Cummings-like physics are also expected. The same idea could be extended to other circuits in which the phase is a quantum variable, such as fluxonium-like circuits~\cite{Manucharyan2009}, where the $4\pi$-periodic Majorana coupling would combine with the inductive energy of the loop, in close analogy with the Josephson resonant level and bifluxon circuits of Sec.~\ref{s:protected-qubits}~\cite{Vakhtel2024,Kalashnikov2020}.}

\begin{figure}
\includegraphics[width=1\linewidth]{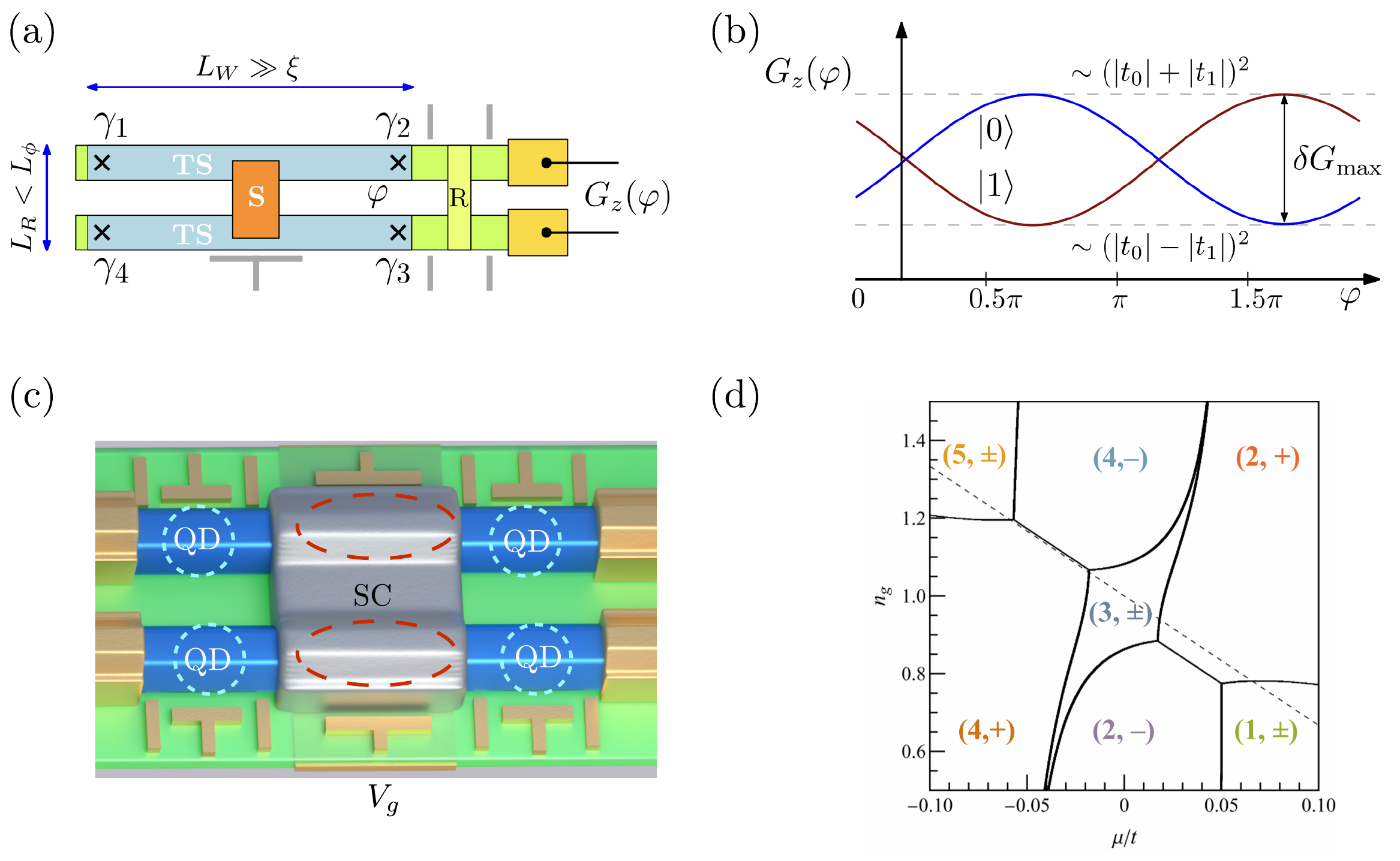}
\caption{(a) Sketch of the Majorana box qubit, where two topological superconductors (light blue) couple via a trivial superconductor (dark orange). The qubit readout can be done using conductance measurement via the leads attached to one end (yellow). (b) Conductance as a function of the flux. (c) Majorana box qubit based on minimal Kitaev chains, where 2 QDs couple to a superconducting island. (d) Phase diagram including the electron charge of the ground state.  Adapted and reprinted  from Ref.~\cite{Plugge_NJP2017} 
and Ref.~\cite{Nitsch_arXiv24}. }
\label{fig:Majorana_box_qubit}
\end{figure}

The Majorana box qubit is another geometry that can be used to encode and process quantum information. In this device, two topological superconductors couple via a trivial superconductor, forming an island~\cite{Plugge_NJP2017}, see Fig.~\ref{fig:Majorana_box_qubit}(a) for a sketch. The charging energy of the island splits the two fermion parity sectors. However, the presence of 4 MZMs allow to define a degenerate computational ground state, used as the qubit base. If we focus on the total even parity state,  the two relevant quantum states are $\ket{0}=\ket{ee}\otimes\ket{N_C}$ and $\ket{1}=\ket{oo}\otimes\ket{N_C-1}$, where $N_C$ denotes the number of Cooper pairs in the ground state. Therefore, the two states differ by a single Cooper pair that splits, where the two electrons occupy the low-energy state defined by the MZMs. The ground state degeneracy in the blockaded superconducting island plays the role of an effective spin. This spin can be used to demonstrate the so-called topological Kondo effect, that arises from the non-locality of the Majorana degrees of freedom~\cite{Beri_PRL2012}. 

Charge sensing on a pair of MZMs using, for example, the coupling to a QD, enables local parity readout that is sufficient to project the qubit state~\cite{Munk_PRR2020,Steiner_PRR2020,Schulenborg_PRB2021,Smith2020b}, or by measuring transport in an interferometry setup. The qubit initialization can be done using the current blockade through the system that fixes the parity of the system to a well-defined value~\cite{Nitsch_PRB2022}. The idea of fixing the parity of coupled Majorana wires inspired more complicated geometries that include several coupled wires~\cite{Karzig2017} and networks of wires for Majorana color codes~\cite{Litinski_PRB2018}. A recent work reported distinct parity lifetimes for different pairs in the MZM box qubit~\cite{Aghaee_arxiv2025}.

The Majorana box geometry has also been extended to minimal Kitaev chains, see Fig.~\ref{fig:Majorana_box_qubit}(c) for a sketch~\cite{Nitsch_arXiv24}. In this device, 4 QDs couple via a superconducting island. Differently from its nanowire counterpart, the MZMs appearing at the outermost QDs do not contribute to the total island charge. This leads to a complicated energy diagram, as shown by the ground state as a function of the island charge offset ($n_g$) and energy of the QDs ($\mu$) in Fig.~\ref{fig:Majorana_box_qubit}(d). Nevertheless, there is a regime where the device based on minimal Kitaev chains recover the behavior of the more conventional Majorana box qubit based on nanowires~\cite{Nitsch_arXiv24}.

\begin{figure*}[!htbp]
\centering
\includegraphics[width=\textwidth]{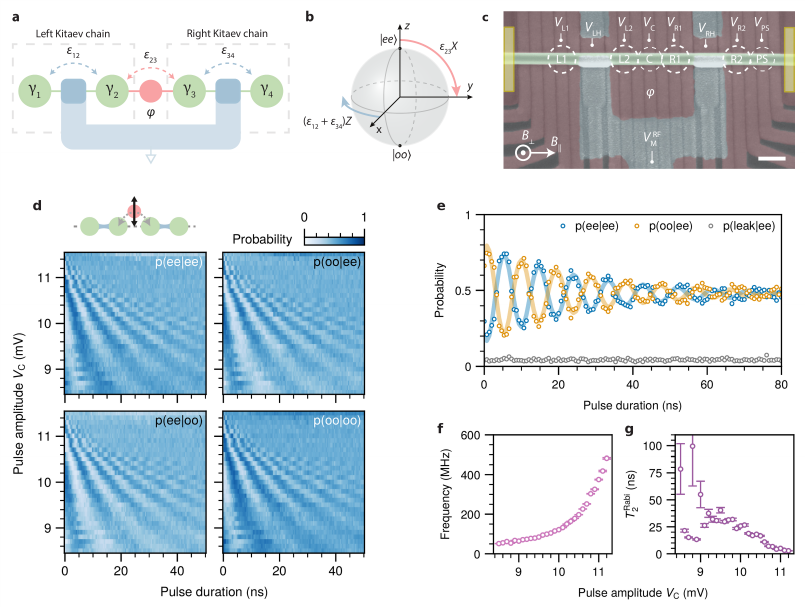}
\caption{Majorana parity qubit in coupled minimal Kitaev chains. (a) Schematic of a Majorana qubit formed by two minimal Kitaev chains. Four Majorana modes ($\gamma_1$, $\gamma_2$, $\gamma_3$, $\gamma_4$) encode the qubit, while a central quantum dot (red) mediates the inter-chain coupling. A superconducting segment (blue) fixes the superconducting phase $\phi$. (b) Bloch-sphere representation of the qubit in the global even-parity manifold. The poles correspond to $\ket{ee}$ and $\ket{oo}$. Rotations are generated by tuning the pairwise Majorana couplings $\varepsilon_{ab}$. Here, $X$ and $Z$ denote the Pauli matrices $\sigma_x$ and $\sigma_z$ of Eq.~\eqref{eq:H_parity_qubit}. (c) False-colored scanning electron micrograph of the device, with the InSb nanowire (green), the Al superconductor (blue) and the Ti/Pd bottom gates (red). Gate voltages define QDs in the nanowire (white dashed circles). Yellow boxes indicate Cr/Au normal contacts fabricated after imaging. Magnetic field orientations and the superconducting phase difference $\phi$ are indicated. Scale bar, 200~nm. (d) Conditional probabilities $p(ee|ee)$, $p(oo|ee)$, $p(ee|oo)$ and $p(oo|oo)$ as a function of the duration and amplitude $V_{\rm C}$ of the coupling pulse, showing coherent oscillations between the qubit states. (e) $p(ee|ee)$, $p(oo|ee)$, and leakage to the odd-parity manifold as a function of the pulse duration at $V_{\rm C}=9.8$~mV. (f) Oscillation frequency and (g) Rabi decay time $T_2^{\rm Rabi}$ extracted from panel (d) as a function of $V_{\rm C}$. Panels (a), (b), (c) correspond to Figs.~1(a), 1(b), 1(d), and panels (d)--(g) to Figs.~2(a)--2(d) of Ref.~\cite{Zatelli_arXiv2026}. Adapted and reprinted from Ref.~\cite{Zatelli_arXiv2026}.}
\label{fig:Majorana_parity_qubit}
\end{figure*}

{\paragraph{First experimental realization: a Majorana parity qubit in coupled minimal Kitaev chains} Very recently, Zatelli {\it et al.}~\cite{Zatelli_arXiv2026} reported the first coherent control of a qubit encoded in the fermion parity of Majorana states, using two minimal Kitaev chains coupled through a quantum dot (Fig.~\ref{fig:Majorana_parity_qubit}), similar to the device sketched in Ref.~\citep{tsintzis2023roadmap}. The device is defined in an InSb nanowire proximitized by a thin Al film, see Fig.~\ref{fig:Majorana_parity_qubit}(c). It contains two two-site Kitaev chains, each formed by two QDs coupled through a hybrid segment, which host four poor man's Majorana states: $\gamma_1$ and $\gamma_2$ in the left chain, and $\gamma_3$ and $\gamma_4$ in the right chain, Fig.~\ref{fig:Majorana_parity_qubit}(a). A central coupler QD mediates a gate-tunable coupling $\varepsilon_{23}$ between the two inner Majoranas, while the superconducting phase difference $\phi$ between the chains is controlled by the magnetic flux threading a superconducting loop. An additional parity-switching QD gives access to the odd-parity sector.

Since the total fermion parity is conserved, the four-Majorana system separates into a global even manifold, spanned by $\ket{ee}$ and $\ket{oo}$, and a global odd manifold, spanned by $\ket{eo}$ and $\ket{oe}$, where $e$ ($o$) denotes the even (odd) parity of each chain. As discussed above, each manifold defines a qubit, see the Bloch-sphere representation in Fig.~\ref{fig:Majorana_parity_qubit}(b). Within each manifold, the dynamics is described by the effective Hamiltonians~\cite{Zatelli_arXiv2026}
\begin{equation}
\mathcal{H}^{\rm even}=(\varepsilon_{12}+\varepsilon_{34})\sigma_z+\varepsilon_{23}\cos(\phi/2)\sigma_x\,,\qquad
\mathcal{H}^{\rm odd}=(\varepsilon_{12}-\varepsilon_{34})\sigma_z+\varepsilon_{23}\cos(\phi/2)\sigma_x\,,
\label{eq:H_parity_qubit}
\end{equation}
where $\sigma_x$, $\sigma_y$ and $\sigma_z$ are Pauli matrices acting within each parity manifold. We stress that these are not the Pauli matrices of a local degree of freedom, such as a spin, but are constructed out of non-local quantities, namely the fermion parities of the two chains. In the global even manifold, $\sigma_z=\ket{ee}\bra{ee}-\ket{oo}\bra{oo}$, $\sigma_x=\ket{ee}\bra{oo}+\ket{oo}\bra{ee}$ and $\sigma_y=-i\ket{ee}\bra{oo}+i\ket{oo}\bra{ee}$, while in the global odd manifold, $\sigma_z=\ket{eo}\bra{eo}-\ket{oe}\bra{oe}$, $\sigma_x=\ket{eo}\bra{oe}+\ket{oe}\bra{eo}$ and $\sigma_y=-i\ket{eo}\bra{oe}+i\ket{oe}\bra{eo}$. These Hamiltonians follow from the low-energy Hamiltonian of the four Majoranas, $\mathcal{H}_{\rm eff}=i\varepsilon_{12}\gamma_1\gamma_2+i\varepsilon_{34}\gamma_3\gamma_4+i\tilde\varepsilon_{23}\gamma_2\gamma_3$, where $i\gamma_a\gamma_b$ is the parity of the pair $(\gamma_a,\gamma_b)$, $\varepsilon_{12}$ and $\varepsilon_{34}$ are residual couplings between the Majoranas of the same chain, which vanish at the sweet spots, and $\tilde\varepsilon_{23}=\varepsilon_{23}\cos(\phi/2)$ is the inter-chain coupling mediated by the coupler QD. The intra-chain couplings therefore generate rotations around the $z$ axis of the Bloch sphere, while the inter-chain coupling, which transfers a fermion between the two chains, $\ket{ee}\leftrightarrow\ket{oo}$, generates rotations around the $x$ axis. The $\cos(\phi/2)$ dependence of the transverse term reflects the $4\pi$-periodic coupling of Majoranas across a junction, already discussed for the Kitmon qubit.

In the experiment, the odd manifold is accessed by flipping the parity of the right chain with the parity-switching QD. At the sweet spots of both chains, the two manifolds are expected to show identical dynamics, including the same Rabi frequency~\cite{tsintzis2023roadmap}, a property characteristic of isolated Majorana modes. Operating the qubit in both manifolds thus provides a useful diagnostic: unwanted couplings between non-neighboring Majoranas, such as $\varepsilon_{14}$, enter the transverse term with opposite signs in the two manifolds ($\varepsilon_{23}\pm\varepsilon_{14}$), so that different oscillation frequencies in the even and odd manifolds signal deviations from ideal, well-separated Majoranas~\cite{Zatelli_arXiv2026}.

The qubit is read out using the single-shot parity readout based on quantum capacitance discussed in Sec.~\ref{Chapter3:sec:PMM_experimentsII}. A radio-frequency resonator is coupled to the superconductor, and the signal is integrated for $500~\mu$s, much shorter than the parity lifetimes, which exceed $8$~ms. Each experiment consists of a readout--control--readout pulse sequence. During readout, the chains are detuned from their sweet spots and the coupler QD is far from resonance. During control, the chain QDs are pulsed to their sweet spots and the coupler QD is brought closer to resonance, which switches on $\varepsilon_{23}$ and induces rotations around the $x$ axis. Odd-parity states are mapped onto the even manifold before readout by pulsing the parity-switching QD through resonance.

Figure~\ref{fig:Majorana_parity_qubit}(d) shows the resulting conditional probabilities as a function of the duration and amplitude $V_{\rm C}$ of the coupling pulse. The chain parities oscillate coherently between $\ket{ee}$ and $\ket{oo}$, with a frequency that increases with $V_{\rm C}$. For $V_{\rm C}=9.8$~mV, the oscillations have a frequency $f=106.7\pm0.4$~MHz and decay with a characteristic time $T_2^{\rm Rabi}=26\pm2$~ns, while leakage to the odd-parity manifold remains around $5\%$, Fig.~\ref{fig:Majorana_parity_qubit}(e). The oscillation frequency is inversely proportional to the detuning of the coupler QD from resonance, Fig.~\ref{fig:Majorana_parity_qubit}(f), as expected for a coupling mediated by elastic cotunneling through the coupler QD. The decay time $T_2^{\rm Rabi}$ depends non-monotonically on $V_{\rm C}$, Fig.~\ref{fig:Majorana_parity_qubit}(g), which was interpreted as a crossover between two dephasing regimes~\cite{Zatelli_arXiv2026}: for strong coupling (fast oscillations), dephasing is dominated by charge noise on the coupler QD, whose charge dispersion increases near resonance, while for weak coupling (slow oscillations) it is dominated by noise acting on the chains, owing to the reduced protection provided by the weaker inter-chain coupling. Similar oscillations are observed in the odd manifold, and, at the Majorana sweet spot, the oscillation frequencies of the two manifolds nearly coincide ($f_{\rm even}=113.4\pm0.5$~MHz and $f_{\rm odd}=117.7\pm0.4$~MHz), as expected from Eq.~\eqref{eq:H_parity_qubit} for decoupled Majoranas within each chain. Furthermore, the oscillation frequency is periodic in the perpendicular magnetic field that controls $\phi$, and it is strongly suppressed near $\phi=\pi$, where the inter-chain coupling vanishes, in agreement with the $\cos(\phi/2)$ dependence in Eq.~\eqref{eq:H_parity_qubit}. Finally, detuning the chains away from their sweet spots ($t\neq\Delta$) provides a direct test of Eq.~\eqref{eq:H_parity_qubit}. Detuning a single chain suppresses the oscillation contrast and coherence in both manifolds, whereas a simultaneous antisymmetric (symmetric) detuning of the two chains, for which $\varepsilon_{12}+\varepsilon_{34}$ ($\varepsilon_{12}-\varepsilon_{34}$) remains small, preserves the oscillations only in the even (odd) manifold~\cite{Zatelli_arXiv2026}. These observations agree with a model of short, partially protected chains.

These experiments establish the first coherent control of a Majorana-based qubit, combining the single-shot parity readout of Sec.~\ref{Chapter3:sec:PMM_experimentsII} with gate- and phase-controlled couplings between Majorana states. In line with the discussion of Sec.~\ref{Sec:bottom-up}, the qubit is based on poor man's Majorana states and is therefore not topologically protected, and the current coherence times are limited to tens of nanoseconds during the driven evolution. Moreover, operations that couple Majoranas and lift the qubit degeneracy are not topologically protected, independently of the chain length~\cite{Zatelli_arXiv2026}. Extending these experiments to longer chains will make it possible to test whether the sensitivity of the idle qubit to perturbations decreases exponentially with the number of sites, as predicted, and to implement the fusion and braiding protocols discussed in the next subsection.
}

\subsubsection{Testing Majorana non-Abelian properties}
\label{Sec:MajoranaBraid}
The non-Abelian properties of MZMs constitute, together with the non-local encoding of quantum information and the resulting topological protection, one of their most characteristic features, see Sec.~\ref{seq:non-Abelian_MBSs} for an introduction. These properties emerge as a phase after the exchange of pairs of MZMs. In this subsection, we review different proposals to demonstrate MZM non-Abelian properties, many of those contained in a previous review~\cite{BeenakkerReview_20} for 1-dimensional topological superconductors.

Majorana fusion rules are a key concept in understanding the non-Abelian nature of Majorana states. These rules describe how a pair of MZMs interact or ``fuse''. The outcome of fusion results in a trivial (or vacuum) state or a non-trivial (fermionic) state. Mathematically, this can be represented as
\begin{equation}
    \sigma\times\sigma=I+\phi\,,
    \label{Eq:fusion}
\end{equation}
where $\sigma$ represents a Majorana fermion, $I$ the vacuum state, and $\phi$ the fermionic state. The probability of each possible fusion outcome depends only on the collective topological state of the system and is not determined by local properties. This ambiguity in the fusion outcome is a manifestation of the Majorana fermions' non-abelian properties.

A less formal but equivalent way to understand Eq.~\eqref{Eq:fusion} is that the fusion, {\it i.e.} coupling and measurement of a MZMs pair, has two possible outcomes corresponding to the occupation of the fermion spanned by the measured of the pair: empty ($I$) and full ($\phi$). The probability for each outcome depends on the joint state of the two MZMs. Therefore, the experimental demonstration of fusion rules requires the initialization of the system in a given basis and the measurement in a different one. The result of the fusion is topologically protected and insensitive to the details of the measurement. Different protocols have been proposed to demonstrate the MZM fusion rules~\cite{Bishara_PRB2009,Ruhman_PRL2015,Aasen2016,Rowell_PRA2016,Hell_PRB2016,Clarke_PRB2017,Barkeshli_PRB2019,Zhou_PRL2020,Zhou_NatCom2022,Souto_SciPost2022,Nitsch_arXiv2022}.

The ground state of a system composed by $N$ MZMs pairs is a space composed by $2^{N}$ linearly independent states. The ground state of the system corresponds to every pair being in the even parity state, the odd parity state, or a linear superposition. MZMs are non-Abelian anyons: the exchange of a set of Majorana transforms the state of the system between the different ground states that are locally indistinguishable. This operation is topologically protected and does not depend on the details on how it is performed, if MZMs remain well-separated and the exchange is adiabatic. Different set of exchanges of MZMs correspond to different unitary operations, allowing to perform topologically-protected operations by simply exchanging MZMs, although they need to be complemented by other operations for universal quantum computation~\cite{NayakReview}. The demonstration of MZM non-Abelian properties would constitute a direct proof of their topological origin. Below, we describe different Majorana braiding protocols introduced in the literature, see also Ref.~\cite{BeenakkerReview_20} for a review.

From the theory side, the most obvious way to demonstrate non-Abelian properties is to exchange MZMs in real space. In semiconductor-superconductor platforms, these states can be moved using external gates by either driving segments of nanowires between the trivial and the topological regimes, or by controlling the energy local energy of QDs in artificial Kitaev chains. Real-space MZM braiding requires extending the geometry out of plane, forming a T or a Y shape~\cite{Alicea2011,Harper_PRR2019}, adding an ancillary system to the linear geometry. In this way, one of the Majoranas can be moved out-of plane, avoiding MZM overlap. This protocol requires the ability of locally control the order parameter of the system, which may be challenging in the nanowire case, as it may require tens of gate electrodes. For this reason, simplified braiding schemes have been develop, where Majoranas are braid in parameter space instead.

\begin{figure}
\begin{center}
\includegraphics[width=0.9\linewidth]{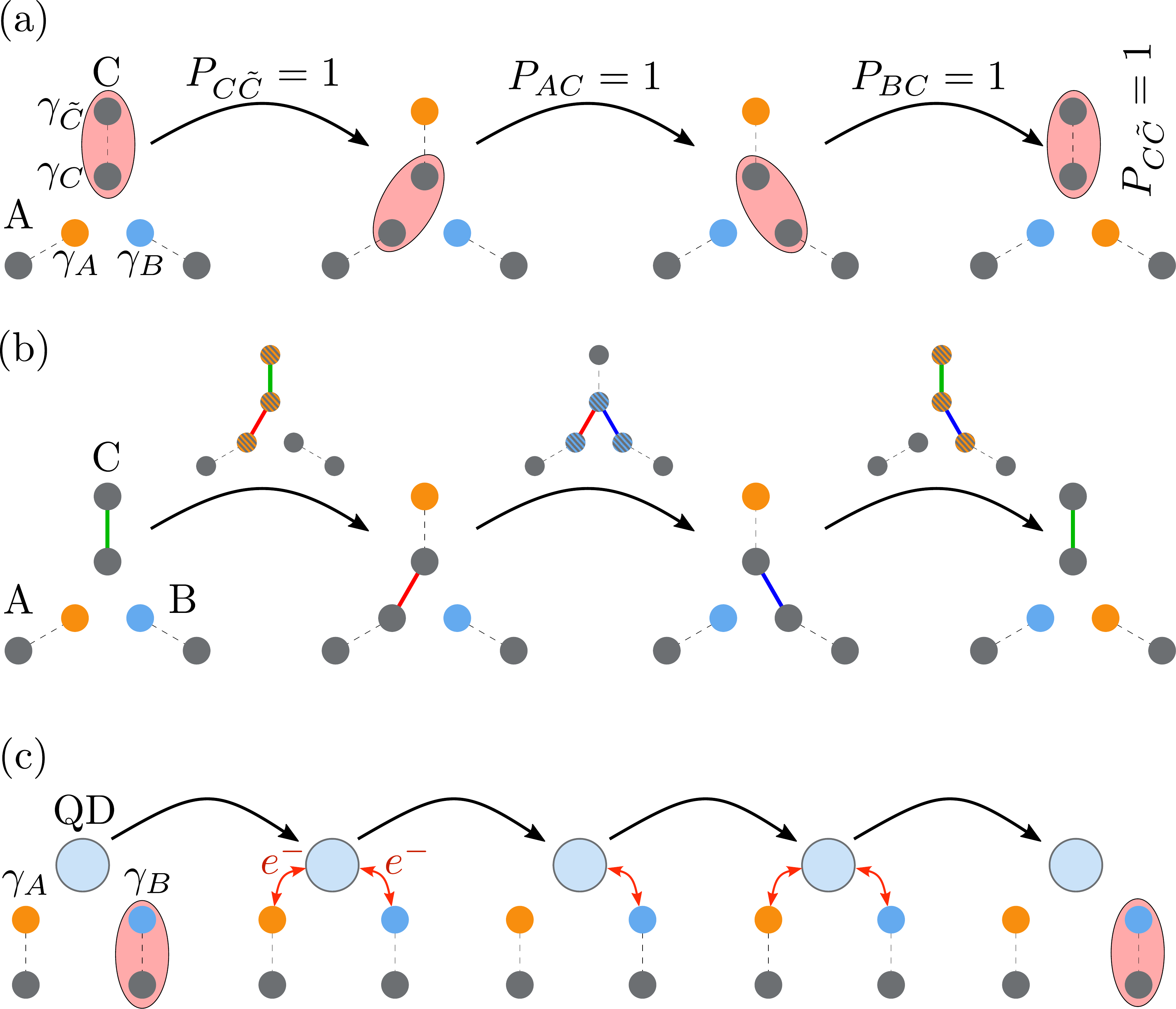}
\caption{(a) Sketch of the measurement-based braiding. The fermion parity of two nearby MZMs is projected and read at every step, $P_{jk}=i\gamma_J\gamma_k$, schematically represented by the red oval. After 4 successful measurements of the fermion parity, two MZMs are effectively braid. (b) Sketch of the hybridization-induced braiding. The system hosts on 3 MZMs initialized pairwise. By switching on and off the coupling between MZMs (thick colored lines) two of them are effectively braid, highlighted with orange and blue colors. (c) Sketch of the charge-transfer-based braiding protocol, where an electron is shuttled between a QD and either one or two MZMs.  }
\label{fig:MBS_braid_proposals} 
\end{center}
\end{figure}

Parity measurements provide a way to effectively braid a pair of MZMs~\cite{Bonderson_PRL2008,Karzig2017,Plugge_NJP2017}. The protocol is shown in Fig.~\ref{fig:MBS_braid_proposals}(a). The setup consists of 4 Majoranas, plus another two that are only used for initialization and readout (lower two MZMs in systems A and B). The protocol consists of measuring alternatively the parity of two MZMs at every step, enclosed by an ellipse in Fig.~\ref{fig:MBS_braid_proposals}(a). Mathematically, the measurement operation can be written as $M_{jk}=(1+P_{jk})/2$, with $P_{jk}=i\gamma_j\gamma_k$. It can be shown mathematically that the braiding operation for $\gamma_A$ (orange) and $\gamma_B$ (blue) is equivalent to the set of measurements $B_{AB}\propto M_{C\tilde{C}}M_{AC}M_{BC}M_{C\tilde{C}}$.

Conceptually, the hybridization-induced braiding (also known as ``three-point turn braid'') is similar to the spatial braiding of MZMs, see Fig.~\ref{fig:MBS_braid_proposals}(b) for a sketch. In this proposal, the coupling between one MZM and the three nearest-neighbors can be dynamically controlled (see thick colored lines). The protocol requires that only up to three MZMs are connected at the same time, see intermediate steps above the arrows in Fig.~\ref{fig:MBS_braid_proposals}(b). The three connected MZMs lead to a regular fermion and one unpaired MZM, whose position depends on the coupling strength. By adiabatically connecting and disconnecting the coupling between MZMs, the unpaired MZM localizes, effectively moving it. Three of these movements constitute a braid operation. Various versions of this protocol have been proposed and discussed in the literature for nanowires~\cite{Aasen2016,Hell_PRB2016,Clarke_PRB2017,Clarke_PRB2011,vanHeck2012,Burrello_PRA2013,Karzig_PRB2015,PhysRevLett.118.107701,Liu_SCP2021} and QD-based Kitaev chains~\cite{tsintzis2023roadmap,Boross_PRB2024,Nitsch_arXiv2025}. A simplified version of this protocol replaces the central 2 superconductors by a QD, being compatible with 1-dimensional structures~\cite{Liu_SCP2021,Luting_PRB2023,Miles_arXiv2025}.

Another way to test non-Abelian properties consists of shuttling single charges to a system composed by several MZMs~\cite{Flensberg_PRL2011,Krojer_PRB2022}, see Fig.~\ref{fig:MBS_braid_proposals}(c) for a sketch of the proposal. This charge can be provided by a QD whose energy is controllable using local gates. When the energy of the QD is adiabatically swept from below to above the system's Fermi level, one electron will be transferred between the QD and the system. In a successful protocol, the electron will be transferred to the zero-energy MZMs~\cite{Souto_PRB2020}. If the system couples to only one MZM (denoted by A), the operation can be understood as acting with the operator $C_A=\gamma_A$ to the state defined by the Majoranas. If the QD instead couples to 2 MZMs, the operator is written as $C_{A,B}=(\gamma_A+\gamma_B)/\sqrt{2}$. Therefore, the difference between applying these two operations in different orders, $C_AC_{A,B}=(1+\gamma_A\gamma_B)/\sqrt{2}$ and $C_{A,B}C_A=(1-\gamma_A\gamma_B)/\sqrt{2}$, is due to the non-Abelian properties of MZMs: the final wavefunction is the same (up to a global sign) after the two sequences if Majorana operators are replaced by fermion operators. In this picture, we have ignored the dynamical phase that arise from the unprotected nature of these operations. A protocol has been recently developed to echo away this phase~\cite{Krojer_PRB2022}.

In the proposals previously mentioned, the operations to demonstrate Majorana braiding have to be adiabatic, ensuring that the system remains in the ground state during the protocol. On the other side, the full protocol has to be faster than the characteristic time for single quasiparticles to tunnel into the system, so-called quasiparticle poisoning time. In semiconductor-superconductor systems, poisoning times in the range of $\mu$s to ms~\cite{Albrecht_PRL2017,Menard_PRB2019,Nguyen_PRB2023,PRXQuantum.5.030337} have been reported, also under relatively strong magnetic field~\cite{Aghaee_Nature2025}. 

\section{Outlook}\label{s:outlook}
The development of hybrid semiconductor-superconductor qubits represents a rapidly maturing frontier in quantum information science, offering a unique convergence of condensed matter physics, materials engineering, and quantum device design. As this review has outlined, the integration of gate-tunable semiconducting elements into superconducting circuits has enabled a new class of qubits with enhanced control, novel degrees of freedom, and, in some cases, the potential for intrinsic protection against decoherence.

{In the following, we first provide a comparative overview of the different hybrid qubit platforms discussed in this Review (Sec.~\ref{ss:comparison}), then summarize the main challenges that remain open across platforms (Sec.~\ref{ss:challenges}), and discuss how hybrid qubits could be integrated into scalable quantum processor architectures (Sec.~\ref{ss:architectures}). We close with a perspective on future research directions (Sec.~\ref{ss:future}).}

\subsection{Comparative overview of hybrid qubit platforms}\label{ss:comparison}
{Gate-tunable transmons are, arguably, the most mature platform: single- and two-qubit operations have been demonstrated~\cite{Casparis2016}, as well as operation in magnetic fields of the order of 1~T (Sec.~\ref{s:qubits}), but their coherence still lags behind that of state-of-the-art Al/AlO$_x$ transmons (Sec.~\ref{ss:challenges}). Recent developments have broadened the range of junction materials beyond InAs nanowires, including InAs 2DEGs, where losses are now being characterized systematically~\cite{Strickland2024}, germanium 2DHGs~\cite{Sagi2024,Kiyooka2024}, Ge/Si core/shell nanowires~\cite{Zhuo2023,Zheng2024}, and van der Waals materials such as graphene~\cite{Wang2019}. Very recently, an all-van-der-Waals transmon, in which a thin hBN layer between two NbSe$_2$ flakes provides both the Josephson coupling and the shunt capacitance, has reached $T_1\approx55~\mu$s and $T_{2E}\approx21~\mu$s, comparable to Al/AlO$_x$ qubits, with a much smaller footprint~\cite{Park_arXiv2026}. Although this device is not gate tunable, it demonstrates that van der Waals heterostructures can host high-coherence superconducting circuits, which makes van der Waals hybrid junctions a promising route towards more coherent gatemons. Hamiltonian-protected qubits have shown an order-of-magnitude enhancement of the lifetime close to the protected regime~\cite{Larsen2020}; reaching deep protection requires a precise control of the harmonic content of the junctions and large ratios $E_{\rm J,2}/E_C$.

Andreev qubits have reached coherent single-qubit control, supercurrent-mediated two-qubit coupling and coherent coupling to transmons~\cite{PitaVidal2023,PitaVidal2024,Pitavidal2025}, with lifetimes of $\sim 10~\mu$s and dephasing times typically below one microsecond (Table~\ref{tab:comparison}). The main near-term task is to identify and suppress the microscopic noise sources that limit their coherence, namely quasiparticle poisoning, charge noise and, for ASQs in III-V semiconductors, the nuclear-spin bath, for example by moving to materials with negligible hyperfine coupling, such as germanium (Sec.~\ref{ss:germanium}).

Majorana-based qubits are at a much earlier stage. Single-shot parity readout with parity lifetimes in the millisecond range has been demonstrated both in top-down nanowire devices~\cite{Aghaee_Nature2025,Aghaee_arxiv2025} and in minimal Kitaev chains~\cite{vanloo2025singleshotparityreadoutminimal}, and the first coherent control of a Majorana parity qubit has just been reported in coupled minimal Kitaev chains~\cite{Zatelli_arXiv2026}, with driven parity oscillations decaying on a timescale of tens of nanoseconds. The topological character of the top-down devices is still debated~\cite{legg2025commentinterferometricsingleshotparity}, whereas minimal Kitaev chains are, by construction, not topologically protected (Sec.~\ref{Sec:bottom-up}). Near-term goals include identifying the mechanisms that limit the coherence of these parity qubits, scaling Kitaev chains while keeping them tunable, and demonstrating measurement-based or braiding operations.

In short, gatemons and Andreev qubits are already functional qubits whose performance must be improved, whereas Majorana-based qubits promise a potentially much larger payoff but are at a much earlier stage of development. Much of the technology developed for the former, such as radio-frequency readout, field-compatible resonators and automated tuning, is directly transferable to Majorana devices. Table~\ref{tab:comparison} summarizes the current status of the different hybrid qubit platforms discussed in this Review, together with their main limitations and prospective roles.
}

\begin{sidewaystable}
\centering
\scriptsize
\renewcommand{\arraystretch}{1.25}
\begin{tabularx}{\textheight}{>{\raggedright\arraybackslash}p{1.9cm}>{\raggedright\arraybackslash}X>{\raggedright\arraybackslash}X>{\raggedright\arraybackslash}X>{\raggedright\arraybackslash}X>{\raggedright\arraybackslash}X}
\toprule
 & \textbf{Gatemon / gate-tunable fluxonium} (Sec.~\ref{s:qubits}) & \textbf{Hamiltonian-protected ($\cos 2\varphi$)} (Sec.~\ref{s:protected-qubits}) & \textbf{Andreev level qubit} (Sec.~\ref{s:ALQ}) & \textbf{Andreev spin qubit} (Sec.~\ref{s:ASQ}) & \textbf{Majorana-based qubits} (Sec.~\ref{s:bottom-up}) \\
\midrule
Qubit degree of freedom & Collective phase/charge mode of a circuit with a gate-tunable SNS junction & Parity of the number of Cooper pairs on an island with a $\pi$-periodic potential & Even-parity occupation of a single ABS & Spin of a single quasiparticle trapped in an ABS (odd parity) & Fermion parity of pairs of MZMs (at least four Majoranas per qubit) \\
Control & Microwave drive; frequency set by gate voltage, without flux lines & Gate and flux tuning to cancel the first harmonic; protected gates proposed~\cite{leroux2023} & Microwave drive; gate and flux tuning & Raman driving~\cite{Hays2021} or direct driving of the spin-flip transition~\cite{PitaVidal2023}; gate, flux and field tuning & Gate- and phase-controlled couplings between Majoranas; coherent parity rotations demonstrated with PMMs~\cite{Zatelli_arXiv2026}; braiding proposed; non-Clifford gates require unprotected operations \\
Readout & Dispersive (cQED) & Dispersive (cQED) & Dispersive, via state-dependent inductance & Dispersive, via spin-dependent supercurrent (resonator or transmon) & Parity readout via quantum capacitance or interferometry~\cite{Aghaee_Nature2025,vanloo2025singleshotparityreadoutminimal} \\
Two-qubit coupling & Capacitive or resonator-mediated; two-qubit operations demonstrated~\cite{Casparis2016} & Proposals~\cite{Schrade2022} & Resonator-mediated coupling observed spectroscopically~\cite{Cheung2024} & Supercurrent-mediated longitudinal coupling~\cite{PitaVidal2024}; ASQ-transmon coupling~\cite{PitaVidal2023}; CAR-mediated exchange proposed~\cite{Spethmann2022} & Not yet demonstrated; measurement-based schemes proposed~\cite{Plugge_NJP2017,Karzig2017} \\
Representative coherence (this Review) & $\mu$s-scale lifetimes; $T_2^*$ enhanced at high junction transparency~\cite{Bargerbos2020} & Tenfold increase of $T_1$ near the protected point~\cite{Larsen2020} & $T_1\approx 4$--$13~\mu$s, $T_{2E}\approx 0.4$--$0.6~\mu$s~\cite{Janvier2015,Hays2018} & $T_1\approx 17~\mu$s, $T_2^*\approx 50$--$100$~ns~\cite{Hays2021,PitaVidal2023} & PMM parity qubit: $T_2^{\rm Rabi}\approx 26$~ns~\cite{Zatelli_arXiv2026}; parity lifetimes $\sim 1$--$10$~ms~\cite{Zatelli_arXiv2026,Aghaee_Nature2025,vanloo2025singleshotparityreadoutminimal} \\
Magnetic field & Not required; compatible up to $\sim 1$~T & Not required & Not required & Optional (tens of mT used to split the spin states) & Required (Zeeman energy comparable to or larger than the induced gap) \\
Main limitations & Microwave losses and gate noise; coherence below that of SIS transmons & Fine tuning of junction harmonics; residual first harmonic & Quasiparticle poisoning; charge noise & Nuclear-spin noise (III-V); poisoning; parity initialization & Disorder; quasiparticle poisoning; magnetic fields; extensive tuning; demonstration of topological protection \\
Maturity & High: multiple materials, single- and two-qubit operations & Early: first signatures of protection & Intermediate: coherent control, two-qubit coupling & Intermediate: coherent control, two-qubit coupling & Early: single-shot parity readout; first coherent single-qubit control (PMMs) \\
Prospective role & Flux-line-free and field-compatible qubits and couplers; probes of hybrid junctions & Intrinsically protected qubits and memories & Compact qubits; probes of Andreev physics & Interfaces between spin and superconducting qubits; compact modules with all-to-all connectivity & Topologically protected memories and measurement-based topological quantum computing \\
\bottomrule
\end{tabularx}
\caption{Comparative overview of the hybrid qubit platforms discussed in this Review. Coherence figures correspond to representative experiments discussed in the main text and are not meant to be exhaustive.}
\label{tab:comparison}
\end{sidewaystable}

\subsection{Remaining challenges}\label{ss:challenges}
{Although the platforms discussed above differ in maturity, they share a set of common challenges: gate tunability comes with a sensitivity to charge noise; encoding information in individual subgap states makes Andreev and Majorana qubits vulnerable to quasiparticle poisoning; the semiconductor-superconductor interface introduces disorder and microwave losses; Majorana devices and some ASQs require magnetic fields; and device-to-device variability complicates calibration. We discuss these issues in turn.

\paragraph{Charge noise and electrostatic stability} Charge noise in gate dielectrics, surfaces and substrates affects all gate-tunable hybrid qubits, through the junction transmissions, the Andreev level energies, or the dot detunings and CAR/ECT ratio of Kitaev chains. Besides operating at gate sweet spots (Sec.~\ref{s:qubits}), exploiting high junction transparencies (Sec.~\ref{ss:reduced-charge dispersion}) and, in Kitaev chains, increasing the number of sites~\cite{Bordin_NatNano2025}, a systematic characterization of charge-noise spectra in hybrid devices, and of their relation to materials and gate stacks, is still largely missing.

\paragraph{Quasiparticle poisoning} Quasiparticle poisoning limits the lifetime of ALQs~\cite{Hays2018}, makes the initialization of ASQs stochastic unless the odd-parity state is stabilized~\cite{PitaVidal2023,Wesdorp2023}, and bounds the duration of any Majorana-based protocol (Sec.~\ref{Sec:MajoranaBraid}). Non-equilibrium quasiparticles are also a well-known problem in conventional superconducting qubits~\cite{Glazman-Catelani2021}, where they cause relaxation, dephasing and charge-parity switching~\cite{Martinis2009,Catelani2011,Wang2014,Vool2014,Serniak2018,Aumentado2004,Sun2012} and originate from stray infrared and ionizing radiation, including cosmic rays~\cite{Barends2011,Corcoles2011,Vepsaelaeinen2020,McEwen2021}. The mitigation toolbox developed there, which includes infrared shielding, gap engineering~\cite{Aumentado2004,Sun2012,Riwar2019}, normal-metal and vortex traps~\cite{Wang2014,Riwar2016}, phonon traps and downconversion~\cite{Henriques2019,Iaia2022,Bargerbos2023}, and optimized designs~\cite{Pan2022}, must however be adapted to hybrid qubits, where the problem is more severe: whereas in a transmon poisoning enters only through transition matrix elements and parity-dependent frequency shifts suppressed in $E_J/E_C$, a single quasiparticle trapped in a subgap level~\cite{PhysRevLett.106.257003,PhysRevB.89.104504,Zazunov2014} directly flips the parity that encodes an Andreev or Majorana qubit, a limitation that topological protection cannot remove~\cite{Rainis2012,Karzig2021}. Adapting these strategies to the constraints of hybrid devices, such as magnetic fields, and developing device-specific ones, such as field-induced quasiparticle trapping in nanowires~\cite{Uilhoorn2021}, is therefore a key task.

\paragraph{Materials, interfaces and disorder} Disorder in the semiconductor and at the semiconductor-superconductor interface (Sec.~\ref{sec:materials}), originating from charged impurities, interface roughness and an inhomogeneous coupling to the superconductor, remains the main obstacle for top-down Majorana platforms (Sec.~\ref{Sec:top-down}). It reduces, and can even close, the topological gap, breaks the topological phase into disconnected segments, and generates trivial low-energy Andreev states that mimic the transport signatures of MZMs~\cite{Prada_review}, which complicates the interpretation of experiments. Disorder is also a source of device-to-device variability for all hybrid qubits. Lossy semiconductor substrates also limit coherence in planar architectures; flip-chip approaches~\cite{Zellekens2022,Hinderling2023,Hinderling2024} and new material platforms such as germanium (Sec.~\ref{ss:germanium}) are promising routes to address this issue. Identifying and mitigating the microscopic sources of loss and noise in hybrid materials (two-level systems in oxides and gate dielectrics, surface states, and the nuclear-spin bath in III-V compounds~\cite{Hoffman2024}) remains a central task.

\paragraph{Magnetic fields} Majorana-based devices, and some ASQ implementations, require magnetic fields, which degrade standard superconducting circuits. Field-resilient circuit elements have, however, progressed considerably: NbTiN resonators maintain high quality factors in in-plane fields from several hundred mT up to several tesla~\cite{Samkharadze2016,Kroll2019,Muller2022}; lumped-element and fractal designs, together with vortex pinning and exclusion, further reduce field-induced losses~\cite{Graaf2012,Bothner2011,FeldsteinIBofill2022}; and granular aluminum (grAl), which has enabled high-impedance resonators, superinductors, fluxonium and transmon qubits~\cite{Grunhaupt2019,Maleeva2018,Grunhaupt2018,Kamenov2020,He2021,Winkel2019}, remains operational in fields up to $\sim 1$~T~\cite{Borisov2020}. Field-resilient superinductors are particularly relevant for the fluxonium-like protected circuits of Sec.~\ref{s:protected-qubits}~\cite{Vakhtel2024,Kalashnikov2020}, since junction-array superinductors (Sec.~\ref{ss:fluxonium}) are field sensitive; disordered superconductors~\cite{Samkharadze2016,Kamenov2020} and geometric superinductors~\cite{Peruzzo2020} are promising alternatives. Van der Waals superconductors offer a further route: together with the all-van-der-Waals transmon discussed in Sec.~\ref{ss:comparison}~\cite{Park_arXiv2026} and van der Waals hybrid junctions~\cite{Antony2021,Wang2019}, the expected field resilience and large kinetic inductance of atomically thin superconductors open prospects for compact, field-compatible hybrid circuits, including fluxonium-like protected qubits. Finally, approaches that reduce the required fields, such as phase-biased planar Josephson junctions~\cite{PhysRevX.7.021032,PhysRevLett.118.107701}, full-shell nanowires (Sec.~\ref{Sec:top-down}), or materials with large $g$-factors (Sec.~\ref{sec:g-factor}), are also important.

\paragraph{Long-term stability, variability and calibration} Slow drifts, charge jumps, gate hysteresis and device-to-device variability require each hybrid qubit to be characterized and periodically retuned. This is manageable for few-qubit experiments but becomes a major bottleneck for larger systems, in particular for Kitaev chains, where the number of parameters grows with the chain length. Automated and machine-learning-based tuning, as developed for semiconductor spin qubits, will therefore be essential. First steps in this direction have already been taken for minimal Kitaev chains, including machine-learned tuning from tunneling spectroscopy~\cite{Benestad_PRB24,Dourado_PRL2026}, Hamiltonian learning of the quantum-dot parameters~\cite{Koch_PRA2023}, and autonomous tuning to the sweet spot across different device platforms~\cite{Driel_arXiv2024}. Extending these tools to the simultaneous tuning and periodic recalibration of many qubits, and combining them with real-time feedback on parity or spectroscopic signals, is an important task for the coming years.

}

\subsection{From individual qubits to quantum processor architectures}\label{ss:architectures}
{Ultimately, all of the above challenges must be addressed simultaneously, and reproducibly, in multi-qubit devices, whereas most of the experiments reviewed here involve one or two qubits. Turning hybrid qubits into building blocks of quantum processors raises system-level questions that are already being addressed for conventional superconducting qubits~\cite{Krantz2019,Kjaergaard2020} and for semiconductor spin qubits~\cite{Burkard2023,Vandersypen2017}.

\paragraph{Integration with cQED} Since hybrid qubits are read out dispersively or via radio-frequency reflectometry, they can share the resonators, Purcell filters, parametric amplifiers, control electronics and frequency-multiplexed readout~\cite{Heinsoo2018} of superconducting processors, as already exploited for the parity readout of Kitaev chains~\cite{vanloo2025singleshotparityreadoutminimal}, and coexist with conventional qubits on the same chip~\cite{PitaVidal2023}. The main constraints are the microwave losses and magnetic-field requirements discussed in Sec.~\ref{ss:challenges}, for which flip-chip and three-dimensional integration, also used in large superconducting processors, offer a route to combine high-quality microwave circuitry with hybrid elements fabricated on dedicated substrates.

\paragraph{Control lines, wiring and cryogenic electronics} Replacing flux lines by electrostatic gates reduces on-chip dissipation and flux crosstalk, but each hybrid qubit requires several DC gate lines, and a Kitaev chain several gates per quantum dot and per superconducting segment. The resulting wiring overhead resembles that of semiconductor spin qubits, for which cryogenic CMOS electronics~\cite{Xue2021_cryoCMOS,Pauka2021_cryoCMOS} and shared-control (crossbar) architectures~\cite{Vandersypen2017} are being developed; since most gate voltages are static or slowly varying, they are well suited for such multiplexing, as illustrated by the crossbar architecture for ASQs of Fig.~\ref{fig:ASQ-scaling}(a)~\cite{Spethmann2022}. Packaging must also accommodate the heat load of the wiring~\cite{Krinner2019}, magnetic fields (when required) and radiation filtering.

\paragraph{Connectivity} State-of-the-art transmon processors rely mostly on nearest-neighbor interactions in planar lattices, mediated by fixed capacitive couplings or by flux-tunable couplers that enable fast, high-fidelity two-qubit gates with suppressed residual interactions~\cite{Yan2018,Sung2021,Arute2019,GoogleQAI2025_belowthreshold}. Longer-range couplers are now being developed to implement quantum low-density parity-check codes, which require a non-planar connectivity~\cite{Bravyi2024}. Hybrid qubits offer additional, physically different coupling mechanisms, which are tunable by gates or flux: resonator-mediated long-range coupling~\cite{Cheung2024}; supercurrent-mediated couplings which, in a parallel-loop geometry, can provide flux-programmable all-to-all connectivity~\cite{Pitavidal2025}; and direct CAR-mediated exchange between nearby Andreev spins~\cite{Spethmann2022}. For Majorana qubits, measurement-based architectures, in which qubits interact via joint parity measurements, replace conventional two-qubit gates~\cite{Plugge_NJP2017,Karzig2017}. Whether these schemes can be combined with low crosstalk and high fidelities in large arrays remains to be demonstrated.

\paragraph{Replacement or complementary building blocks?} In our view, hybrid qubits are unlikely to replace, in the near term, the Al/AlO$_x$ transmons that power current superconducting processors, whose coherence and fabrication reproducibility have been optimized over two decades and which have recently enabled quantum error correction below the surface-code threshold~\cite{GoogleQAI2025_belowthreshold}. Rather, we envision hybrid elements as complementary building blocks within future heterogeneous quantum processors: gate-tunable junctions as flux-free tunable couplers and frequency-tunable elements, and as field-compatible qubits for hybrid experiments; ASQs as compact interfaces between spin qubits and superconducting circuits; Hamiltonian-protected qubits as long-lived memories; and gate-tunable parametric amplifiers and resonators~\cite{Butseraen2022,Sarkar2022,Phan2023,Hao2024,Splitthoff2024,Strickland2023} as auxiliary components. The situation is different for Majorana-based qubits: if topological protection is demonstrated, they could form the basis of a qualitatively different architecture, with measurement-based operations and a potentially reduced error-correction overhead~\cite{Karzig2017}. Which of these scenarios materializes will depend on progress in addressing the challenges summarized in Sec.~\ref{ss:challenges}.
}

\subsection{Future directions}\label{ss:future}
{Following the first coherent control of a Majorana parity qubit in minimal Kitaev chains~\cite{Zatelli_arXiv2026}, one of the most immediate goals is the extension of these experiments to longer chains and, eventually, to parity qubits based on topologically protected MZMs. Beyond this, the unambiguous demonstration of topological protection and non-Abelian statistics remains the central open challenge of the field, and will rely on the progress in materials and interfaces discussed in Sec.~\ref{ss:challenges}.

Hamiltonian-protected qubits (Sec.~\ref{s:protected-qubits}) offer a complementary route towards passive error suppression, in which protection is engineered at the level of the circuit. Combining hybrid junctions with field-resilient superinductors, and embedding Majorana junctions in superconducting circuits, as in the Kitmon (Sec.~\ref{Sec:topoQubits}), could eventually merge circuit-level and topological protection.

Beyond quantum computing, hybrid qubits are also unique probes of mesoscopic superconductivity: they give time-resolved access to the dynamics of individual quasiparticles, spins and fermion parities in hybrid junctions, and engineered Kitaev chains provide a bottom-up route to explore topological superconductivity and its interplay with interactions and disorder. Several fundamental questions could be addressed with the techniques reviewed here. Among them are the microscopic origin and dynamics of non-equilibrium quasiparticles and their trapping in subgap states; the interplay between Coulomb interactions and superconductivity in quantum-dot junctions, from Yu-Shiba-Rusinov states to the Kondo regime and the associated singlet-doublet ($0$-$\pi$) quantum phase transitions~\cite{Martin-Rodero-Levy-Yeyati2011,Tanaka_2007,Zonda1,Estrada-2020,GarciaCorral2020,Yoshioka_JPSJ2000,Choi_PRB2004,Oguri_JPSJ2004,Lim_JPCM2008,Karrasch_PRB2008,Rodero_JPCM2012,Pillet_PRB2013,Luitz_PRL2012,Deacon2010,Lee2014,vanDam2006,Kadlecova2019,Jorgensen2007,Maurand2012}, which have now become accessible with microwave spectroscopy in cQED architectures~\cite{Bargerbos2023b,Wesdorp2024,Bargerbos2022,Fatemi2022,Kirsanskas2015,Hermansen2022}; the spin-orbit and Zeeman-split Andreev physics underlying the anomalous Josephson and superconducting diode effects~\cite{Bozkurt_SciPost,Valentini2024,Souto_PRL2022}; odd-frequency pairing associated with Majorana and Andreev states and its role in the anomalous proximity effect~\cite{Tanaka2012-hx,Tanaka2024-fk,PhysRevB.70.012507,RevModPhys.77.1321,PhysRevB.87.104513,fksg-x8pr,Ahmed2025}; and multiterminal Josephson junctions, whose Andreev spectra depend on several independent phases and can host synthetic topological phases~\cite{Riwar2016_NatCommun,Eriksson2017,Meyer2017}, and which are now being integrated into quantum circuits~\cite{PRXQuantum.5.020340}. In Majorana systems, the most fundamental open goals remain the demonstration of Majorana fusion rules~\cite{PhysRevB.108.085437} and non-Abelian braiding statistics, and the observation of genuinely non-local effects such as the topological Kondo effect~\cite{Beri_PRL2012}. Addressing these questions does not require full fault tolerance, and progress on them will, in turn, inform the design of better qubits.

In summary, hybrid semiconductor-superconductor qubits have evolved from proof-of-principle demonstrations into a diverse family of platforms, from functional gatemons and Andreev qubits to the first coherent Majorana parity qubits. Their future development will remain closely intertwined with progress in the physics and materials science of hybrid nanostructures.}

\section{Acknowledgments}
Work supported by the Horizon Europe Framework Program of the European Commission through the European Innovation Council Pathfinder Grant No. 101115315 (QuKiT); the Spanish Comunidad de Madrid (CM) ``Talento Program'' (Project No. 2022-T1/IND-24070); and the Spanish Ministry of Science, Innovation, and Universities through Grants CEX2024-001445-S (Severo Ochoa Centres of Excellence program), PID2025-169814NB-I00, PID2024-161156NB-I00, and PID2022-140552NA-I00.
Support from the CSIC Interdisciplinary Thematic Platform (PTI+) on Quantum Technologies (PTI-QTEP+), the NOMIS Foundation, the FWF Projects with DOI:10.55776/F86, DOI:10.55776/PAT7682124 and DOI:10.55776/P36507 is also acknowledged.

\bibliographystyle{elsarticle-num} 
\bibliography{bibliography.bib}

\end{document}